\documentclass[iop,revtex4]{emulateapj}
\usepackage{amsmath,amsthm,amssymb,latexsym,amsfonts}

\shorttitle{}
\shortauthors{Kim et al.}
\begin{document}
\title{The $\it{AKARI}$ 2.5--5.0 Micron Spectral Atlas of Type-1 Active Galactic Nuclei:
 Black Hole Mass Estimator, Line Ratio, and Hot Dust Temperature}

 \author{\textbf{Dohyeong Kim}\altaffilmark{1,2}, \textbf{Myungshin Im}\altaffilmark{1,2},
 \textbf{Ji Hoon Kim}\altaffilmark{1,2,3}, \textbf{Hyunsung David Jun}\altaffilmark{1,2},
 \textbf{Jong-Hak Woo}\altaffilmark{2},
 \textbf{Hyung Mok Lee}\altaffilmark{2}, \textbf{Myung Gyoon Lee}\altaffilmark{2},
 \textbf{Takao Nakagawa}\altaffilmark{4},
 \textbf{Hideo Matsuhara}\altaffilmark{4}, \textbf{Takehiko Wada}\altaffilmark{4},
 \textbf{Shinki Oyabu}\altaffilmark{5}, \textbf{Toshinobu Takagi}\altaffilmark{4}, 
 \textbf{Youichi Ohyama}\altaffilmark{6}, and \textbf{Seong-Kook Lee}\altaffilmark{1,2}}
 
 \altaffiltext{1}{Center for the Exploration of the Origin of the Universe (CEOU), 
 Astronomy Program, Department of Physics and Astronomy, Seoul National University, 
 Shillim-Dong, Kwanak-Gu, Seoul 151-742, South Korea}
 \altaffiltext{2}{Astronomy Program, Department of Physics and Astronomy, Seoul National University, Shillim-Dong, Kwanak-Gu, Seoul 151-742, South Korea}
 \altaffiltext{3}{Subaru Telescope, 650 North A'ohoku Place, Hilo, Hawaii 96720, U.S.A.}
 \altaffiltext{4}{Institute of Space and Astronautical Science, Japan Aerospace Exploration Agency, Sagamihara, Kanagawa 252-5210, Japan}
 \altaffiltext{5}{Graduate School of Science, Nagoya University, Furo-cho, Chikusa-ku, Nagoya, Aichi 464-8602, Japan}
 \altaffiltext{6}{Institute of Astronomy and Astrophysics, Academia Sinica, P.O. Box 23-141, Taipei 106, Taiwan}
 \email{dohyeong@astro.snu.ac.kr and mim@astro.snu.ac.kr}
  
\begin{abstract}
 We present 2.5--5.0\,$\mu$m spectra of 83 nearby ($0.002\,<\,z\,<\,0.48$) and bright ($K<14$mag)
 type-1 active galactic nuclei (AGNs) taken with the Infrared Camera (IRC) on board $\it{AKARI}$.
 The 2.5--5.0\,$\mu$m spectral region contains emission lines such as
 Br$\beta$ (2.63\,$\mu$m), Br$\alpha$ (4.05\,$\mu$m), and polycyclic aromatic
 hydrocarbons (PAH; 3.3\,$\mu$m), which can be used for studying the black hole (BH) masses and 
 star formation activities in the host galaxies of AGNs. The spectral region also
 suffers less dust extinction than in the ultra violet (UV) or optical wavelengths,
 which may provide an unobscured view of dusty AGNs.
 Our sample is selected from bright quasar surveys of Palomar-Green (PG)
 and SNUQSO, and AGNs with reverberation-mapped BH masses from \citet{peterson04}.
 Using 11 AGNs with reliable detection of Brackett lines, we derive the Brackett-line-based BH mass estimators.
 We also find that the observed Brackett line ratios can be explained with the commonly adopted physical conditions of the broad line region (BLR).
% For 11 AGNs with reliable detection of Brackett lines,
% we measure the Brackett line properties such as the line widths and
% luminosities, and use these quantities to derive Brackett line-based
% BH mass estimators and to investigate physical condition of
% broad line regions (BLRs).
 Moreover, we fit the hot and warm dust components of the dust torus by adding photometric data of SDSS, 2MASS, $\it{WISE}$, and $\it{ISO}$ to
 the $\it{AKARI}$ spectra, finding hot and warm dust temperatures
 of $\sim1100\,\rm{K}$ and $\sim220\,\rm{K}$, respectively, rather than the commonly cited hot dust temperature of 1500\,K.
\end{abstract}

\section{INTRODUCTION}

 An active galactic nucleus (AGN)
 is a bright central part of a galaxy that shines via the energy released by materials that falls toward the super massive black hole (SMBH). 
 During its active phase, the AGN luminosity can outshine that of the host galaxy, emitting an enormous amount of energy
 ($10^{43}$--$10^{48}~\mathrm{erg~s^{-1}}$; e.g., \citealt{woo02}) over a wide range of wavelengths from gamma-ray to radio.
 The spectrum of an AGN provides us with important information concerning the structure and the physics around the SMBH that powers the AGN activity.
 High-energy photons, such as X-ray and UV, probe the accretion disk around the SMBH that gives off hot thermal emission. 
 Optical photons trace the accretion disk or the broad and narrow line regions farther away from the SMBH.
 Meanwhile, infrared (IR) observations can trace the dust-heated radiation from the dusty torus or the star forming regions.
 Finally, jets can be probed in the radio.

 In recent years, the near-infrared (NIR) wavelength region has become an important window for understanding AGNs.
 In general, hot dust emission is known to be the main contributor to the NIR light from an AGN
 (e.g., \citealt{barvainis87,kobayashi93}), and many studies have been carried out
 using observations in the NIR to understand the dust torus structures, such as the covering factor \citep{mor09,mor11},
 dust torus size \citep{minezaki04,suganuma06},
 hot dust temperature ($T_{\rm HD}\sim 1500\,\rm{K}$; e.g., \citealt{barvainis87,kobayashi93,glikman06,landt11}),
 and the evolution of circumnuclear dust \citep{haas03,jiang10,hao12,jun13}.

 Additionally, the NIR wavelength region includes interesting emission/absorption line features 
 such as the hydrogen Paschen and Brackett lines, the polycyclic aromatic hydrocarbon (PAH) emission feature
 at 3.3\,$\mu$m, and the absorption lines from ${\rm H_{2}O}$ ice (3.1\,$\mu$m), ${\rm CO_{2}}$ (4.26\,$\mu$m), 
 and CO (4.67\,$\mu$m) gas. These line features can be used to understand the BLR size, 
 the BH mass ($M_{\rm BH}$), the star formation, and the materials surrounding the nuclear region as outlined below.
 
 Recently, NIR Paschen lines have been put forward as a useful $M_{\rm BH}$ estimator \citep{kim10,landt11,landt13},
 since they are much less affected by dust extinction than
 the common $M_{\rm BH}$ estimators using UV/optical lines \citep{vestergaard06,mclure04,greene05}. 
 A similar study can be extended to the Brackett lines which would be even less affected by the dust than the 
 Paschen lines. In the case of $E(B-V)$ = 2 mag (a value corresponding to the typical color excess of 
 dusty, red AGNs; e.g., \citealt{glikman07,urrutia09}), the Br$\alpha$, Br$\beta$, P$\alpha$, and P$\beta$ line 
 fluxes would be suppressed by factors of 1.31, 1.62, 2.16, and 3.97, respectively, assuming the galactic
 extinction law with $R_{V}=3.07$ \citep{mccall04}. In other words, the Br$\alpha$ line is three times 
 less affected by the dust extinction than the P$\beta$ line in this example. 

 The 3.3\,$\mu$m PAH feature has been suggested as a possible star formation indicator \citep{tokunaga91}, 
 and has been detected in many local AGNs. A number of studies suggest that
 there may exist a correlation between the starburst (traced by the 3.3\,$\mu$m PAH feature) and the AGN activity (traced, e.g., by
 5100\,$\rm{\AA}$ continuum luminosity, hereafter, $L_{\rm 5100}$), although the correlation may break down at very high IR luminosities
 \citep{oi10,woo12,imanishi11,kim12,yamada13}. Finally, molecular lines such as 
 ${\rm H_{2}O}$ ice, ${\rm CO_{2}}$, and CO arise from molecular clouds or the diffused interstellar medium in the line of sight toward the AGN \citep{spoon04},
 and they are useful tracers for the reservoir of molecular gas in the host galaxy and in the vicinity of an AGN.
 NIR hydrogen line ratios are also suggested as a useful means to understand the physical conditions of BLR \citep{ruff12}.
 
 Studies of Brackett lines and PAH features have a special importance for understanding the evolutionary stages
 of galaxies, especially the dusty AGN phase. Dusty, red AGNs are AGNs whose optical light is obscured by the foreground dust
 (see \citealt{kim10} for the demography of various types of red AGNs). They are suspected to be an intermediate population 
 in the evolution of galaxies, situated between the dust enshrouded star forming phase 
 (such as ultra luminous infrared galaxy, ULIRG) and the luminous AGN phase. 
 During the red AGN stage, SMBHs are expected to grow rapidly providing feedback to the star formation 
 of host galaxies \citep{hopkins05,li07}, 
 but heavy dust extinction in host galaxies prohibits observational studies of
 the activities occurring in red, dusty AGNs. Brackett lines and PAH features can possibly 
 serve as NIR spectral diagnostics that can explore the AGN and star formation activity of
 dusty, red AGNs due to the low extinction of the light in NIR. 

 While studies of NIR spectra at $\lambda <$ 2.5\,$\mu$m are common, studies based on the 2.5--5.0\,$\mu$m spectra of nearby AGNs have been scarce, 
 mainly due to the observational limitation caused either by atmospheric absorption or strong thermal background at $\lambda >$ 2.5\,$\mu$m on the ground.
 In order to better understand the 2.5--5.0\,$\mu$m spectra of AGNs, we conducted a spectroscopic study of 83 nearby AGNs using the $\it{AKARI}$ Infrared 
 Camera (IRC) \citep{murakami07,onaka07} as a part of the $\it{AKARI}$ mission program, Quasar Spectroscopic Observation with NIR Grism (QSONG; 
 \citealt{im10}). The $\it{AKARI}$ IRC covers the wavelength range 2.5--5.0\,$\mu$m with R $\sim$ 120 at 3.6\,$\mu$m and offers high sensitivity at 
 these wavelengths thanks to the cold and transparent environment in space. The scientific goals of our study are to (1) provide a spectral atlas of nearby, 
 bright AGNs that have been studied extensively in previous works; (2) establish the $M_{\rm BH}$ estimator using Brackett lines so that we can estimate 
 $M_{\rm BH}$ reliably by reducing the dust extinction effect (the Brackett-based $M_{\rm BH}$ estimator will be an improvement over Paschen line-based 
 estimators); (3) understand the correlation between the hot dust emission and other AGN properties; and (4) study the 3.3\,$\mu$m PAH feature in AGNs 
 to understand the connection between AGN and star-formation activities.
 In this paper, we will describe the sample characteristics, present the 2.5--5.0\,$\mu$m $\it{AKARI}$ spectra of 83 AGNs,
 derive the $M_{\rm BH}$ estimator using the hydrogen Brackett lines of 10 AGNs where the Brackett lines are well detected,
 and investigate the correlation between the hot dust emission and AGN properties. 
 The investigation of the PAH feature will appear in a future work.

\section{THE SAMPLE AND OBSERVATION}

\subsection{The Sample}

 When constructing the sample, we impose the following two criteria. 
 In order to place one of the redshifted Brackett lines (Br$\beta$ and Br$\alpha$)
 in the IRC spectroscopic wavelength range of 2.5--5.0\,$\mu$m, our sample is limited to AGNs at $z < 0.5$.
 In addition, we selected objects that are brighter than 14 mag in the $K$ band to gain S/N $>$ 10 for the continuum
 detection.

 Using the above criteria, we constructed a sample of 108 nearby, bright AGNs that are 
 culled from the three different AGN source catalogs below.
 The first source of the sample is a catalog of 35 bright and nearby type-1 AGNs with
 $M_{\rm BH}$ values measured from the reverberation mapping method \citep{peterson04,denney10,grier08}.
 The reverberation mapping method provides the reference $M_{\rm BH}$ values,
 which are used as the basis for deriving $M_{\rm BH}$ estimators using spectra taken at a single-epoch.
 Among the reverberation mapping sample, 31 AGNs were observed before $\it{AKARI}$ stopped 
 its operation in 2011.
 Note that two objects among these 31 AGNs, 3C\,273 and NGC\,7469, were observed through another 
 $\it{AKARI}$ mission program (AGNUL; PI: Takao Nakagawa).
 The AGNs that are selected from the reverberation mapping sample have a wide range of $M_{\rm BH}$ values,
 $10^{6.3}$--$10^{8.9} M_{\odot}$, and they are at $z = 0.002$ -- 0.29.
 The second source is the PG QSO catalog \citep{green86}, 
 since PG QSOs are optically very bright and have been extensively studied in previous works
 \citep{kaspi00,vestergaard02,vestergaard06}.
 We find 69 PG QSOs at $z = 0.03$ -- 0.48 to satisfy our selection criteria, excluding 
 those that are already included in the reverberation mapping sample.
 Among them, 48 were observed by $\it{AKARI}$. For their $M_{\rm BH}$ values, we use the values derived from the H$\beta$ line as
 listed in \citet{vestergaard06} where the $M_{\rm BH}$ values span the range of $10^{6.5}$--$10^{9.7} M_{\odot}$.
 Finally, we also supplement the sample by four nearby bright quasars from Seoul National University Quasar Survey in Optical (SNUQSO; \citealt{im07,lee08}).
 SNUQSO is a survey of bright quasars missed by previous quasar survey at high and low galactic latitude
 (b $\textless\mid20^{\circ}\mid$), and it provides additional bright QSOs to the sample.
 In total, 108 bright, low-redshift QSOs are selected, of which 83 were observed by $\it{AKARI}$

 Figures 1 and 2 show the basic characteristics of the sample.
 AGNs in our sample are located at a redshift range of $z = 0.002$ -- 0.48, and span over a wide range of
 $M_{\rm BH}$ ($10^{6.3}$--$10^{9.7} M_{\odot}$), bolometric luminosity ($L_{\rm bol}$; $10^{43.1}$--$10^{47.2}~\mathrm{erg~s^{-1}}$)
 and accretion rate (-2.35$\leq$ $\log$(Eddington ratio) $\leq$0.43).
 The $L_{\rm bol}$ values of the PG QSOs and reverberation-mapped AGNs are estimated from $L_{\rm 5100}$
 using a bolometric correction factor of 10.3 \citep{richards06}, 
 and their $L_{\rm 5100}$ values are taken from \citet{bentz09,vestergaard06}.
 For SNUQSOs, the H$\alpha$ luminosity from \citet{im07,lee08} is converted to $L_{\rm 5100}$ using Equation (1) of \citet{greene05},
 and the converted $L_{\rm 5100}$ is translated to $L_{\rm bol}$ with a correction factor of 10.3.
 All of the objects observed by $\it{AKARI}$ in our sample and their basic properties are presented in Table 1.
 We note that the $M_{\rm BH}$ values in Table 1 are calculated with a recently updated virial factor of 5.1 \citep{woo13}.

 \begin{figure}
 \centering
 \figurenum{1}
 \includegraphics[width=\columnwidth]{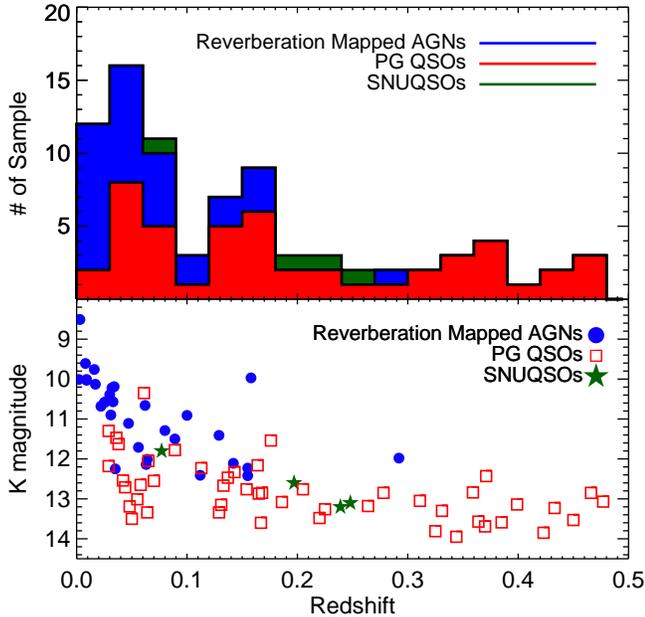}\\
 \caption{
 Top: redshift distribution of AGNs in our sample. The red, blue, and green histograms are for PG QSOs, reverberation
 mapped AGNs, and SNUQSOs, respectively. Bottom: redshift vs. $K$-band magnitude of our AGNs where the meaning of
 the colors is identical to the top panel. The $K$-band magnitudes include the host galaxy light. Note that PG QSOs are about
 two magnitude fainter than the reverberation mapped AGNs in apparent magnitude since they are in general at higher redshifts.}
 \end{figure}

\subsection{The Observation}

 The 2.5--5.0\,$\mu$m NIR spectra were obtained with the IRC infrared spectrograph
 \citep{onaka07} on the $\it{AKARI}$ satellite during phase 3 of
 the $\it{AKARI}$ operation \citep{murakami07}, i.e., our observation was carried out during 2008--2011
 after the helium cooling was over.
 We used the NG grism mode which provides a spectral resolution of ${\rm R=120}\,\lambda / \rm{3.6\, \mu m}$.
 The spectral resolution corresponds to the FWHM velocity resolution of
 2500 $\mathrm{km~s^{-1}}$ or $\sigma = 1062~{\rm km~s^{-1}}$ at 3.6\,$\mu$m which is adequate 
 for studying the broad emission lines
 of quasars and AGNs whose typical FWHM velocity widths are greater than 2000 $\mathrm{km~s^{-1}}$.
 Also, the wavelength coverage of 2.5--5.0\,$\mu$m enables us to sample
 Br$\beta$, Br$\alpha$, and PAH at low redshift.

 For most of the targets, three-pointing observation was performed
 where one-pointing observation is made of seven to nine 
 frames with 44 s of exposure each.
 For a few sources, 1, 2, or 5 pointings were assigned
 (total integration time of 396 to 1848 seconds) due to the observational constraints of the telescope.
 Each target was placed at the center of a square-shaped slit window
 (1\farcm0 $\times$ 1\farcm0 size)
 designed to avoid confusing the spectrum with neighboring objects.
 Table 1 summarizes the observation of each target.

\begin{figure*}
\begin{center}
\centering
\figurenum{2}
\includegraphics[scale=0.6]{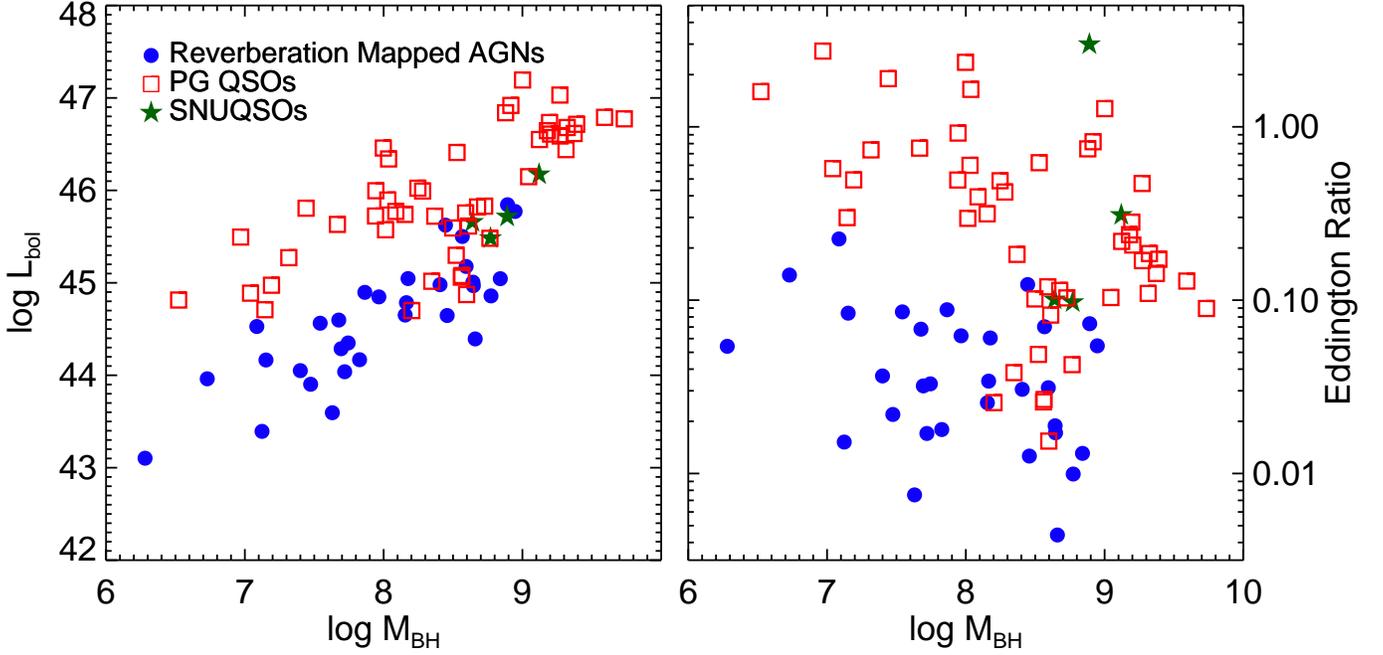}\\
 \caption{
 Left: $M_{\rm BH}$ values vs. $L_{\rm bol}$ values of AGNs in our sample. 
 The meaning of the symbols are identical to Figure 1.
 Our sample covers a wide range in $M_{\rm BH}$ ($10^{6.3}$--$10^{9.7} M_{\odot}$) and
 $L_{\rm bol}$ ($10^{43.1}$--$10^{47.2}~\mathrm{erg~s^{-1}}$).
 Right: $M_{\rm BH}$ values vs. Eddington ratios of our AGNs. Our AGN sample covers an extensive range of 
 BH accretion rates (-2.35 $\leq$ $\log$(Eddington ratio) $\leq$ 0.43).
 Note that PG QSOs and SNUQSO AGNs are intrinsically more luminous and have higher accretion rates
 than the reverberation mapped AGNs.}
 \end{center}
\end{figure*} 

\begin{figure*}
\figurenum{3}
\includegraphics[scale=0.6]{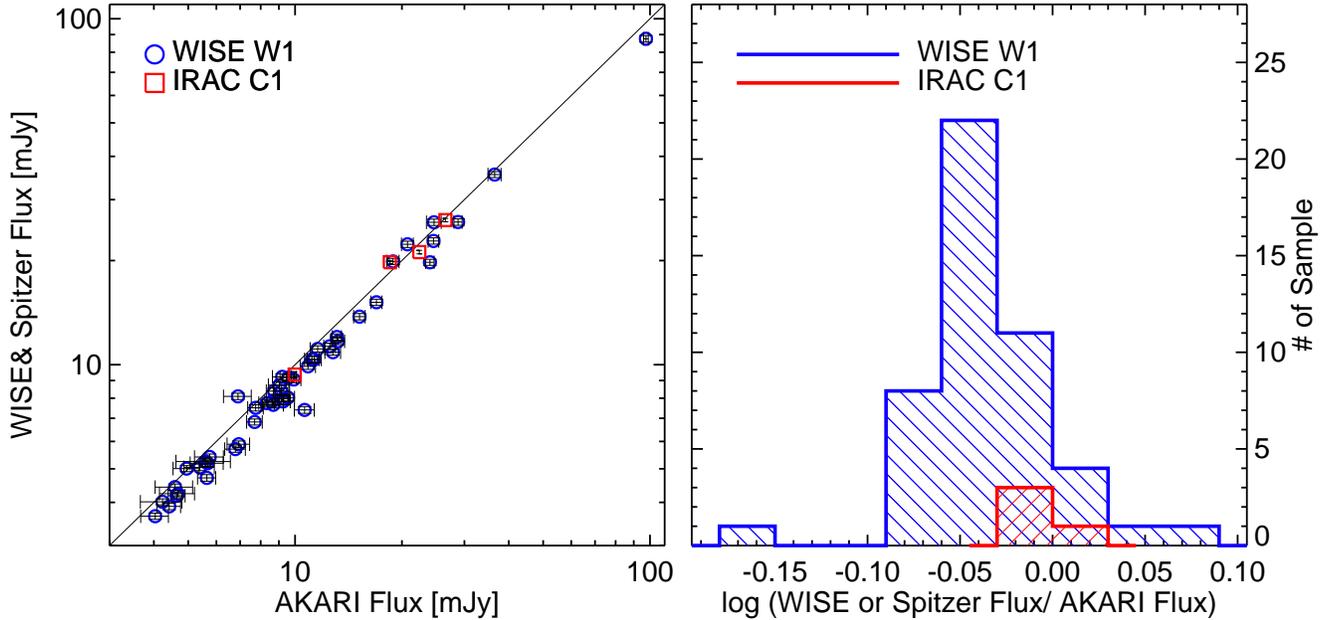}\\
\caption{
 Left: comparison of the $\it{WISE}$ $W$1 or $\it{Spitzer}$ IRAC $Ch$1 magnitudes vs. 
 the $\it{WISE}$ $W$1 and $\it{Spitzer}$ IRAC $Ch$1 magnitudes derived from the $\it{AKARI}$ spectra.
 The blue circles and the red squares indicate the $\it{WISE}$ $W$1 and the $\it{Spitzer}$ IRAC $Ch$1 fluxes, respectively.
 The black solid line is for the case where the $\it{AKARI}$, and $\it{WISE}$ or $\it{Spitzer}$ fluxes 
 are identical.
 Right: Distribution of the ratios of the $\it{WISE}$ or $\it{Spitzer}$ fluxes to 
 the $\it{AKARI}$ fluxes of PG QSOs.
 The blue and the red histograms 
 indicate the ratios for $\it{WISE}$ and the $\it{Spitzer}$ fluxes, respectively.}
\end{figure*}

\subsection{Data Reduction}

 The data reduction was performed using the standard data reduction software package
 ``IRC Spectroscopy Toolkit for Phase 3 data Version 20110301 (2011/03/01)'' \citep{ohyama07}, 
 but with an additional procedure for cosmic ray/hot pixel removal 
 to improve quality of the final spectrum. First, we used the standard IRC pipeline
 on each pointing data separately, up to the point before the extraction of the one-dimensional
 spectrum. Then, we removed hot or bad pixels in the stacked two-dimensional spectrum using L.A.Cosmic \citep{vandokkum01}
 in spectroscopic mode, since the $\it{AKARI}$ phase 3 images typically come with many hot pixels
 and we wanted to remove effectively the hot pixels.
 We set the hot or bad pixel detection limit to 4.5$\sigma$ for removal of the hot pixels.
 Then, a one-dimensional spectrum of each pointing for each object was extracted using an extraction aperture width of 5 pixels, 
 for which aperture correction is performed using the pipeline and assuming a point
 source PSF to correct for the missing flux due to the adoption of the finite aperture size.
 The extraction aperture width gives the optimal S/N since the FWHM of PSFs in the IRC data is about 5 pixels (7\farcs3).
 Note that the AGN spectrum can be contaminated by the host galaxy light,
 since the square-shaped slit window (1\farcm0) does not completely block the light from the extended regions of a host galaxy.
 Hence, we are effectively sampling the light from the AGN$+$host galaxy within
 an observational rectangular aperture of 7\farcs3 $\times$ 1\farcm0, which corresponds to physical scales of
 1.49 kpc $\times$ 12.3 kpc at $z=0.01$ and 32.5 kpc $\times$ 267 kpc at $z=0.3$
 where the long axis runs along the dispersion direction.
 In both cases, the aperture contains a significant portion of host galaxies,
 although we expect that the measured Brackett lines are dominated by emission from the nuclear region, as we describe in Section 4.1.
 Finally, the extracted one-dimensional spectra from all of the pointings of a source were combined. In this process,
 any remaining outlier pixels (hot or bad pixel features) in the one-dimensional spectrum were removed by 
 $\sigma$ clipping with 4.5$\sigma$ where the median value and the $\sigma$ of a data point were 
 determined including two adjacent data points.
 The wavelength and flux calibration was performed in the IRC pipeline and the expected accuracies of the
 wavelengths and fluxes are about 0.01\,$\mu$m and 10\% \citep{ohyama07}, respectively.

 To demonstrate the accuracy of the photometric calibration, we compare the fluxes of PG QSOs from the
 $\it{AKARI}$ spectra to $\it{WISE}$ $W$1 and $\it{Spitzer}$ IRAC $Ch$1 magnitudes.
 To simplify the comparison, we restrict the sample for the comparison to point-like
 sources (i.e., unresolved) in all of the $\it{AKARI}$, $\it{WISE}$, and $\it{Spitzer}$ images.
 Thus, we restrict the sample for the photometry comparison to those in the PG QSO sample.
 For this comparison, we used the $\it{WISE}$ $W$1 magnitudes measured with a profile-fitting method provided by the $\it{WISE}$ all-sky source catalog.
 However, for the $\it{Spitzer}$ IRAC $Ch$1 magnitudes, we measured the magnitudes using IRAC $Ch$1 
 PBCD imaging data that we obtained from the $\it{Spitzer}$ archive.
 In order to measure the magnitudes from the IRAC $Ch$1 PBCD imaging data,
 we used the IRAF daophot task with an aperture radius of 3$\times$FWHM.
 Among the PG QSOs, five objects (PG\,0007$+$106, PG\,0050$+$124, PG\,1100$+$772, PG\,1501$+$106, and PG\,1704$+$608)
 have IRAC $Ch$1 PBCD imaging data. 
 Excluding PG\,0050$+$124 which is too bright and saturated at the center of the IRAC image, 
 we used four objects from the $\it{Spitzer}$ archive for the comparison.
 Finally, using $\it{AKARI}$ spectra, we calculated the corresponding $\it{WISE}$ $W$1 
 and $\it{Spitzer}$ $Ch$1 magnitudes by
 applying appropriate filter transmission curves.
 Figure 3 shows the comparison of the fluxes from $\it{AKARI}$ with those from $\it{WISE}$ and $\it{Spitzer}$.
 The $\it{AKARI}$ fluxes are on average 10$\%$ larger than the $\it{WISE}$ fluxes,
 while the $\it{Spitzer}$ $\it{Ch}$1 fluxes are in good agreement with the $\it{AKARI}$ flux values.
 Although the reason for the systematic offset in the flux is not clear, 
 we find the 10$\%$ difference to be acceptable considering the
 general flux calibration uncertainty of $\it{AKARI}$ \citep{ohyama07}.

 The flux errors of the spectra from the pipeline appear somewhat larger than the fluctuation of the data points (e.g., Figure 7).
 To check the flux error values from the pipeline, we used independent flux measurements derived from several pointing observations
 (Section 2.2), and we checked the determined uncertainty of the flux at each wavelength by adopting the rms value of the flux measurements from
 each pointing data. We find that the error derived this way agrees with the $\it{AKARI}$ pipeline result, suggesting that
 the $\it{AKARI}$ pipeline error measurement reflects unknown fluctuations in the continuum level determination or sensitivity fluctuations of IRC.

\subsection{Construction of Composite Spectrum}
 In an attempt to increase the S/Ns of the AGNs without Brackett line detections, we created a composite spectrum of
 the 48 PG QSOs. The PG QSOs are less affected by host galaxy contamination
 than the reverberation-mapped sample due to bright luminosity ($L_{\rm bol}$: 
 $10^{44.6}$--$10^{47.2}~\rm{erg~s^{-1}}$; Figure 2).
 
 To construct the composite spectrum, we stacked, deredshifted, and flux-calibrated 
 spectra of PG QSOs in logarithmic scale.
 Considering PG QSOs have different luminosity ranges, a geometric mean was used to construct the composite spectrum.
 We built a wavelength grid with a pixel resolution twice the spectral resolution of IRC
 (${\rm R=120}\,\lambda / \rm{3.6\, \mu m}$).
 In each wavelength grid, the weighted mean flux of each object was stacked for the geometric mean flux.
 
 Figure 4 shows the composite spectrum of PG QSOs in $\lambda$ versus $F_{\nu}$.
 The location of key lines are also displayed in the figure. The composite spectrum is available in the supplemental material accompanying the article.
 Although we could not detect Brackett lines in individual PG QSO spectra,
 some of the NIR Hydrogen emission lines are now visible in the composite spectrum.

\begin{figure}
\begin{center}
\centering
\figurenum{4}
\includegraphics[width=\columnwidth]{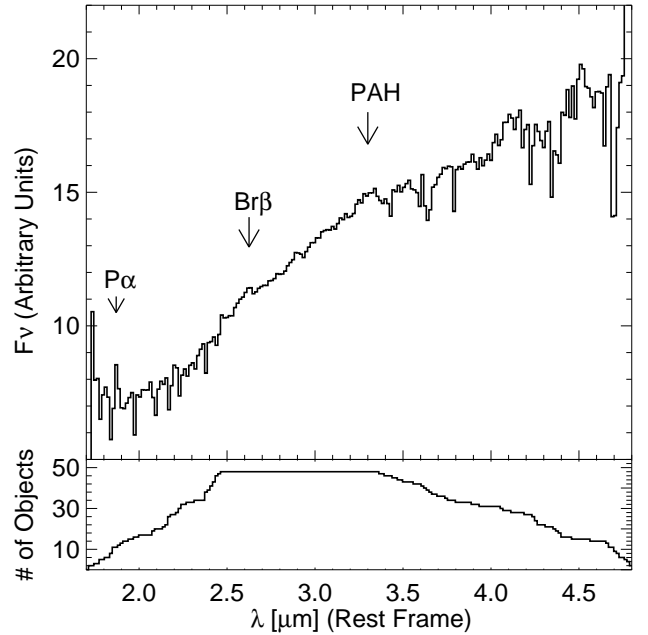}
\caption{
 Top: composite spectrum of the 48 PG QSOs in arbitrary $F_{\rm \nu}$ units.
 The composite spectrum indicates P$\alpha$, Br$\beta$, and PAH lines.
 Bottom: the number of spectra contributing for the composite spectrum.
}
\end{center}
\end{figure}

\section{$\it{AKARI}$ NIR SPECTRA}

 We show the $\it{AKARI}$ spectra from our program in Figure 5. The machine-readable form of the 
 spectra in Figure 5 are available.
 Several interesting lines are also indicated in Figure 5, such as the Hydrogen Brackett lines,
 3.3\,$\mu$m PAH, and other molecular/atomic lines.
 Br$\alpha$ and Br$\beta$ lines are identified in 9 and 7 objects at $S/N > 3$, respectively,
 and we mark the location of these lines in Figure 5 when they are identified.
 Potential detections are marked with a \textquotedblleft?\textquotedblright \,sign. On the other hand, the Paschen lines are identified in 5 objects.
 All of the Brackett line detections come from
 the reverberation mapping sample whose members are brighter than the QSOs in the other two samples. 
 The rather low detection rate of the Brackett lines reflect the reduced sensitivity 
 and the increased number of detector defects during the warm mission phase of $\it{AKARI}$.

\begin{figure}
\begin{center}
\figurenum{6}
\includegraphics[width=\columnwidth]{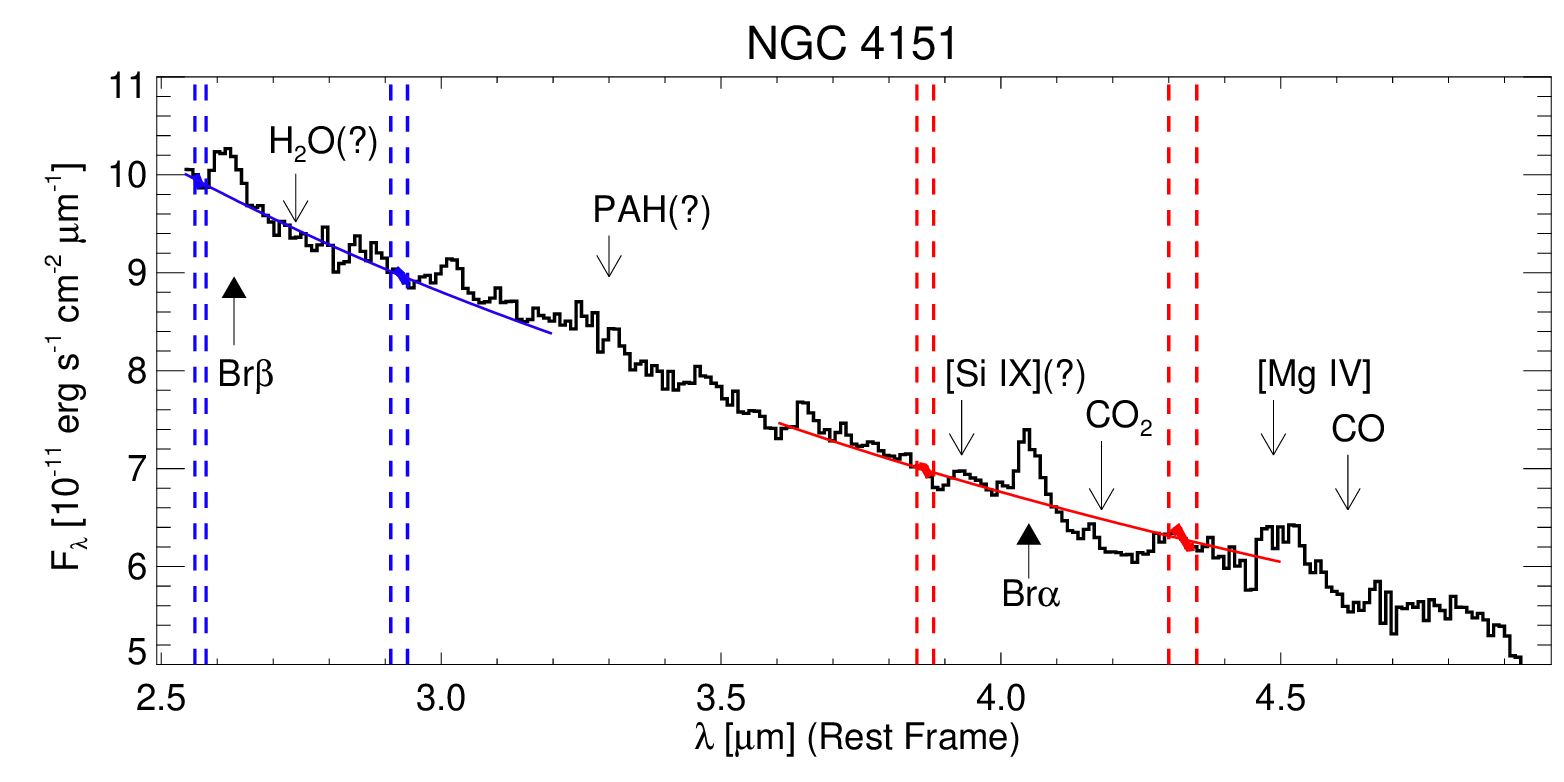}
\caption{
 NIR spectrum (2.5--5.0\,$\mu$m) of NGC\,4151.
 The 2.5--5.0\,$\mu$m wavelength window includes Br$\alpha$ and Br$\beta$
 lines and various molecular lines
 such as $\rm{H_{2}O}$, [\ion{Si}{9}], $\rm{CO_{2}}$, [\ion{Mg}{4}], and CO.
 The blue and the red solid lines indicate the continuum fit to the regions
 around the Br$\beta$ and Br$\alpha$ lines, respectively.
 The dashed lines indicate the wavelength range used for the continuum fitting.
}
\end{center}
\end{figure}

\section{BRACKETT LINES}

 Using the spectra we presented in the previous section, we measured the line fluxes and
 widths of Brackett lines and used them to estimate the $M_{\rm BH}$ values and the physical conditions
 of the gas in BLR. In this section, we will describe the measurements of the line fluxes and widths,
 present the Brackett-line based $M_{\rm BH}$ estimators, and discuss the BLR physical condition inferred 
 from the line luminosity ratios of the NIR Hydrogen lines.

\begin{figure*}
% \epsscale{0.8}
\figurenum{7}
\includegraphics[scale=0.42]{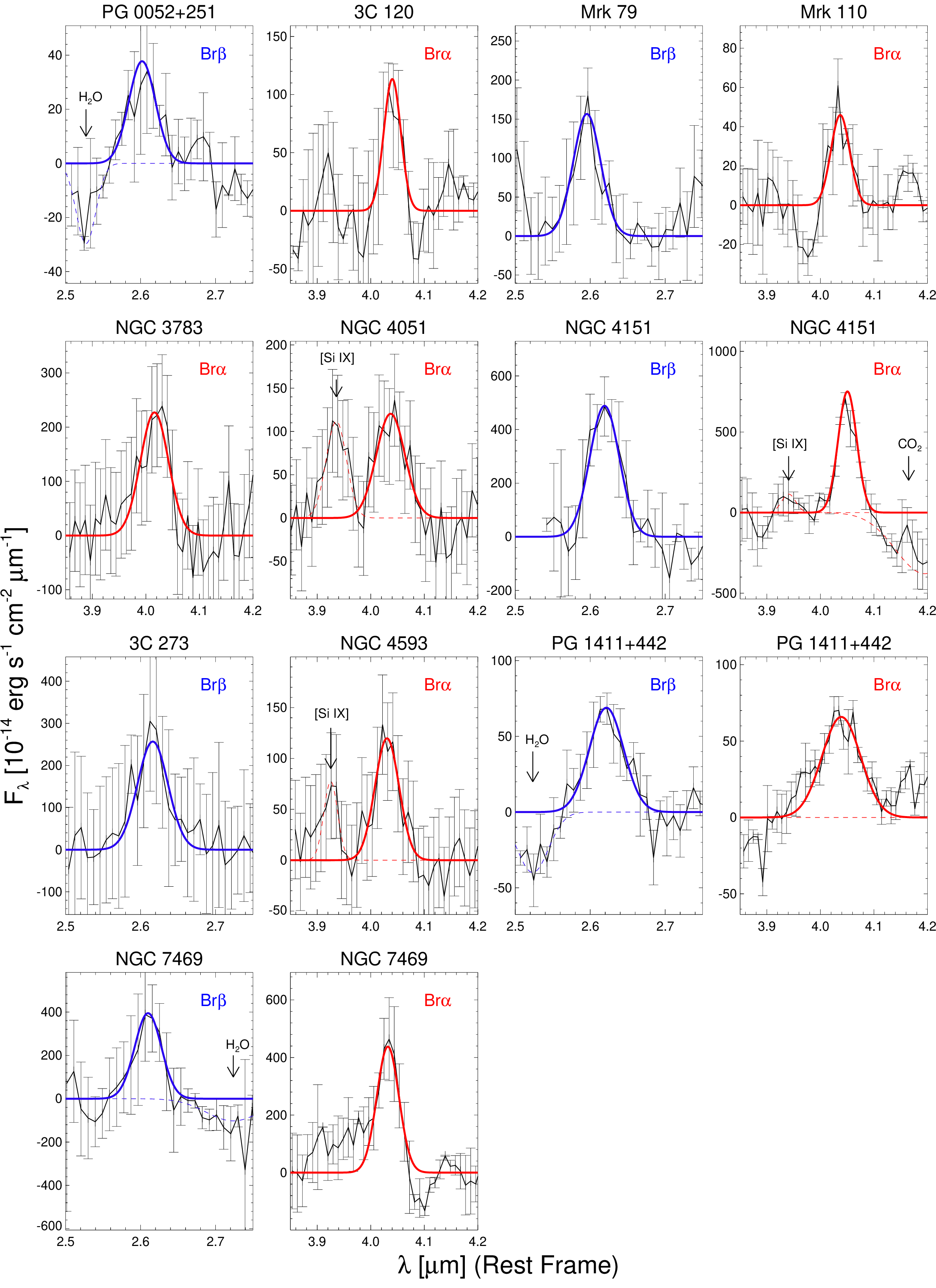}\\
 \caption{
 Gaussian fits of the Brackett lines of our sample.
 The black solid lines and bars indicate the observed spectra in rest frame and the associated errors.
 The blue and red lines indicate the fits for Br$\beta$ and Br$\alpha$, respectively.
 The dashed lines show the fits to the molecular/atomic lines around Brackett lines, such as
 $\rm{H_{2}O}$, [\ion{Si}{9}], and $\rm{CO_{2}}$ lines.
 The solid lines indicate the Brackett line component of the best-fit spectra.
}
\end{figure*}

\subsection{Brackett Line Luminosity and Width}
 
 In this subsection, we describe how the Brackett lines are fitted to derive line fluxes and widths.
 The fitting of the lines starts with the identification of the line in the deredshifted spectra. 
 As mentioned already in the previous section, we find Br$\alpha$ and Br$\beta$ 
 at $S/N > 3$ in 9 and 7 AGNs, respectively.
 After the line identification, the continuum around the Brackett lines were
 determined by fitting a linear function to both sides of each Brackett line.
 Although a NIR spectrum of a quasar is dominated by the black body radiation from the hot part of the dust torus,
 the linear fit of the continuum should provide a reasonable approximation of the local continuum
 around each emission line \citep{landt08,kim10}.
 However, determining the best wavelength region for the continuum is not a simple task, because
 many molecular absorption or atomic emission lines exist around the Brackett lines.
 Such lines are ${\rm H_{2}O}$ (2.53, 2.66, 2.73, 2.75--3.55\,$\mu$m), [\ion{Si}{9}]
 (3.93\,$\mu$m), ${\rm CO_{2}}$ (4.26\,$\mu$m) and CO (4.67\,$\mu$m).
 Hence, we chose the continuum regions that are least affected by such lines.
 The wavelength regions adopted for the continuum determinations are listed in Table 2.

 After continuum subtraction, we fit the Brackett lines and molecular/atomic lines simultaneously.
 Fitting of the molecular/atomic lines near the Brackett lines is necessary, since the fitting of 
 the Brackett lines can be influenced by the presence of molecular/atomic lines such as the
 ${\rm H_{2}O}$ absorption lines at 2.53 and 2.66\,$\mu$m
 (for Br$\beta$), and the [\ion{Si}{9}] and $\rm CO_{2}$ gas line at 3.93 and 4.1--4.4\,$\mu$m (for Br$\alpha$).
 Figure 6 shows a 2.5--5.0\,$\mu$m spectrum of NGC\,4151, which includes Br$\alpha$,
 Br$\beta$, ${\rm H_{2}O}$, [\ion{Si}{9}], and ${\rm CO_{2}}$ lines.
 The Blue and red solid lines are the continuum levels used for the Br$\beta$ and Br$\alpha$
 emission line measurements. The region bracketed by the dashed lines indicates the wavelength range used for
 the continuum fitting of each Brackett line.

 We used a single Gaussian function to fit each line considering
 the low spectral resolution of the $\it{AKARI}$ spectroscopy,
 setting the central wavelength, the line width, and the line flux to be free parameters.
 An interactive data language (IDL) procedure, \texttt{MPFITEXPR} \citep{markwardt09}, was used to fit the lines.
 The fit provides the line-of-sight velocity dispersion, $\sigma$, and fluxes of the Brackett lines.
 Note that the narrow emission line does not significantly affect the line flux and $\sigma$ measurements.
 As a test, we set a model spectrum with a broad ($\sigma$: 850 -- 2500 $\mathrm{km~s^{-1}}$) and a narrow line component ($\sigma$: 170 $\mathrm{km~s^{-1}}$)
 at the $\it{AKARI}$ spectral resolution. We adopt the flux ratios of the narrow to broad component flux to be
 0.115 and 0.227 for Br$\beta$ and Br$\alpha$, respectively.
 These ratios are derived by first taking a median H$\alpha$ narrow/broad component flux ratio
 in \citet{landt08} for five objects (Mrk\,110, Mrk\,79, NGC\,7469, NGC\,4593, and NGC\,4151),
 and then converting the H$\alpha$ flux to the Brackett flux
 based on the computation from the CLOUDY code (version 10.00; \citealt{ferland98}) for the BLR condition
 ($\alpha$ = -1.0, $n$ = $10^9$ $\rm{cm}^{-3}$, and $U$ = $10^{-1.5}$; \citealt{kim10}) and the narrow line condition
 ($n$ = $10^{4.5}$ $\rm{cm}^{-3}$ and $U$ = $10^{-2}$; \citealt{cox00}).
 We fitted the degraded model emission line with a single Gaussian function. The measured flux and $\sigma$ values are
 108\% (104 -- 115\%) and 95\% (90 -- 98\%) compared to the original broad component properties,
 which would affect $M_{\rm BH}$ by a small amount, $\sim$0.03 dex.
 It is also confirmed that the narrow emission lines in the NIR do not affect the line flux 
 and $\sigma$ measurements as much ($<10\%$) in a relatively low-resolution spectra (e.g., \citealt{kim10}).

 \begin{figure*}
 \figurenum{8}
 \begin{center}
\includegraphics[scale=0.6]{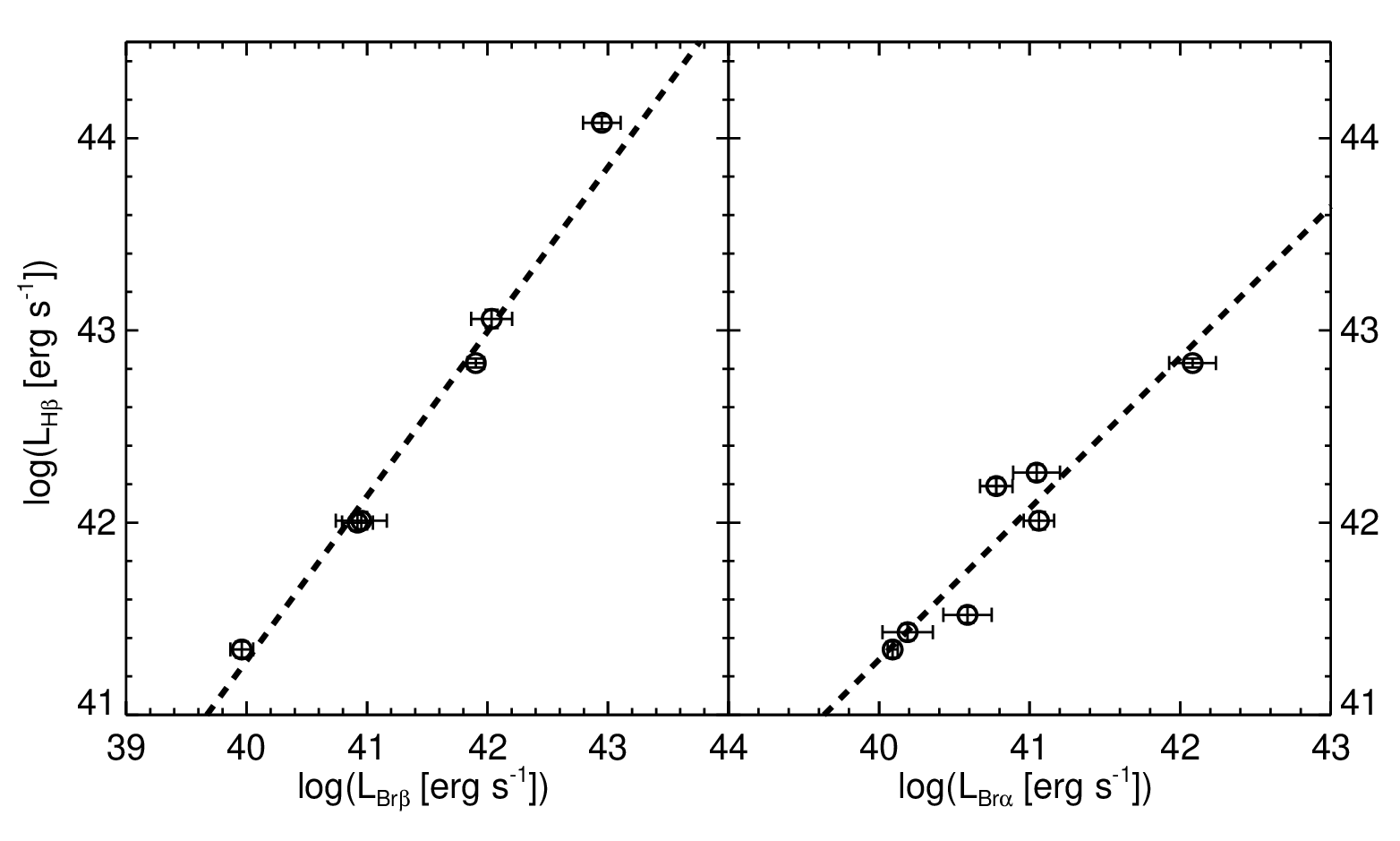}
 \caption{
 Comparison of the line luminosities of the Brackett lines
 vs. the H$\beta$ line. The dotted line indicates the best-fit correlation 
 between the luminosities of the H$\beta$ and the Brackett lines.
}
 \end{center}
 \end{figure*}

 \begin{figure*}
 \figurenum{9}
 \begin{center}
\includegraphics[scale=0.6]{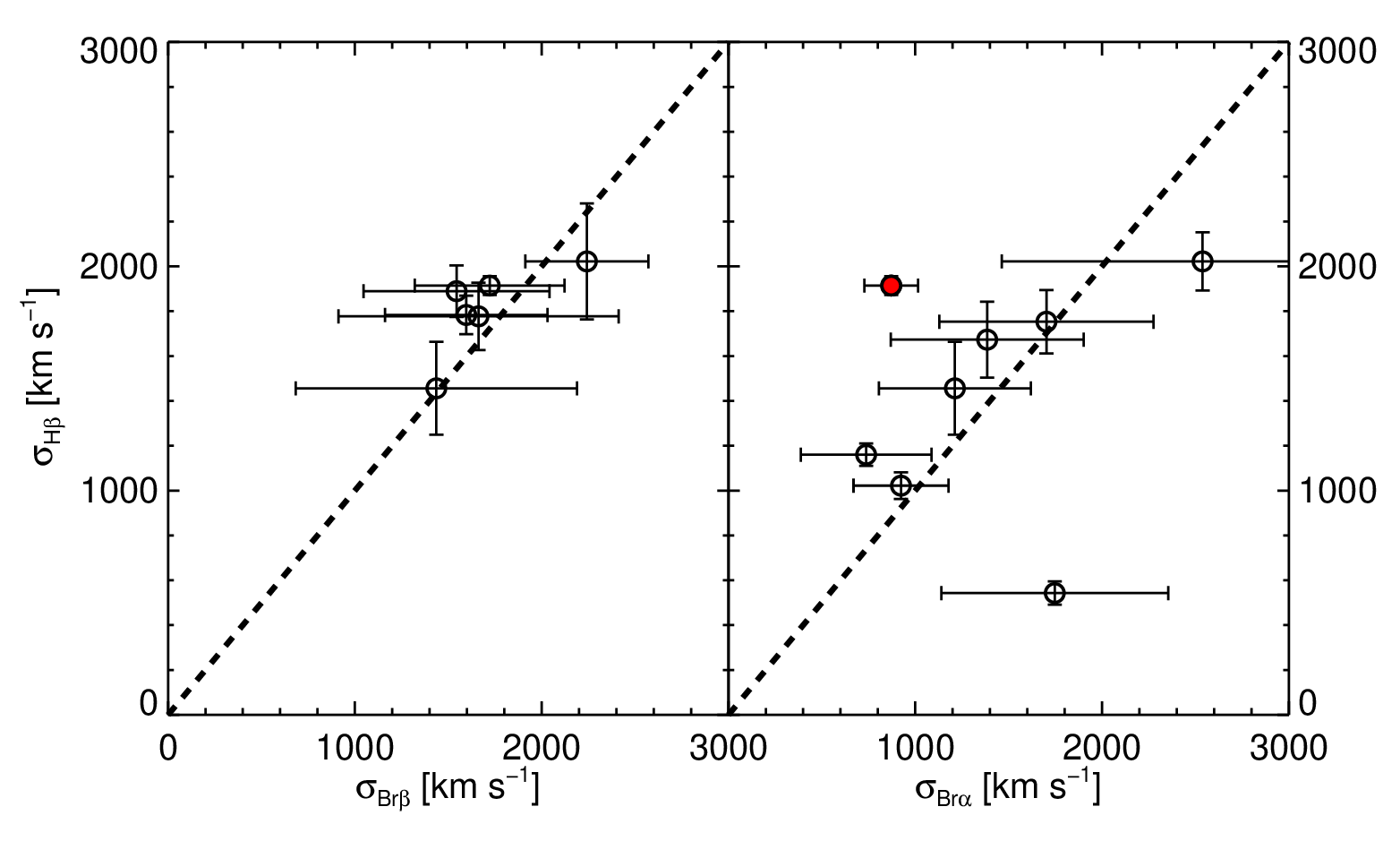}
 \caption{ 
 Comparison of the $\sigma$ widths of the Brackett lines vs. those of the 
 H$\beta$ line. The dashed line indicates the case where the $\sigma$ value of the H$\beta$ and
 the Brackett lines are identical. 
 The weak correlation for Br$\alpha$ is likely to be due to low S/N and
 the $\rm CO_{2}$ absorption feature right next to the Br$\alpha$ line.
 The $\rm CO_{2}$ absorption is the most prominently seen in the spectrum of NGC\,4151,
 which is represented by the red filled circle.}
 \end{center}
 \end{figure*}

 For further analysis in the following subsections, we exclude objects with
 substantial $\sigma$ errors ($>$60\% of the measured $\sigma$ value) or measured Brackett line FWHM (2.35\,$\sigma$) 
 less than the spectral resolution (${\rm R=120}\,\lambda / \rm{3.6\, \mu m}$).
 The excluded cases are
 Mrk\,110 (Br$\beta$) and Ark\,120 (Br$\alpha$).
 Finally, eight Br$\alpha$ and seven Br$\beta$ lines are selected and used for further investigation of the line properties.
 All of these are from the reverberation mapping sample, since the PG QSOs are generally too faint 
 for a reliable fitting of the lines (e.g., Figure 1).
  
 Note that the selected Brackett lines show a small systematic shift in the wavelength of $\delta\lambda=(\lambda_{\rm measured}-\lambda_{\rm lab})$
 $\sim -0.014$\,$\mu$m ($\langle \delta\lambda_{\rm Br\alpha} \rangle =-0.015$\,$\mu$m and $ \langle \delta\lambda_{\rm Br\beta} \rangle =-0.013$\,$\mu$m).
 The Br$\alpha$ and Br$\beta$ lines are shifted by a similar amount in wavelength
 ($\delta\lambda_{\rm Br\alpha}-\delta\lambda_{\rm Br\beta} = -0.001\pm0.005$), suggesting that the wavelength
 shift applies to the entire spectral range. The shift is a bit larger than the expected wavelength offset error
 in the $\it{AKARI}$ IRC of 0.01\,$\mu$m \citep{ohyama07}. 
 The origin of the systematic offset is unclear, although it seems to be instrumental.
 A potential source of the offset is molecular/atomic 
 absorption affecting the peak wavelength of the emission lines. 
 However, if this were true, then we would expect a stronger offset in Br$\alpha$ than in Br$\beta$ 
 since Br$\beta$ is generally much less affected by molecular absorption features than Br$\alpha$, 
 but this was not the case.
 
 The formal fitting errors are found to be about 27\% (between 7\% and 49\%) in $\sigma$ and 28\% (between 6\% and 46\%)
 in flux, due to low S/N and the necessity of simultaneously fitting the other lines in the vicinity.
 Another source of error comes from the determination of the continuum.
 We varied the wavelength regions for the continuum fit from the values noted in Table 2
 to see how strongly the changes in the continuum
 level affect the line fitting results. We find that the variation in the continuum level
 changes the derived $\sigma$ and flux values by 13\% and 8\% respectively. Overall, combining
 these two errors, we obtain the error in $\sigma$ and flux to be 33\% and 30\%, which corresponds to 0.30 dex in $M_{\rm BH}$.
 
 Energy sources of the Brackett lines could be BH activity or star formation activity.
 Considering the large slit aperture of the IRC and the extraction width we adopted,
 the measured Brackett line fluxes may include the star formation contribution from the host galaxy. 
 To determine the dominant energy source of the Brackett lines,
 we compared the expected $L_{\mathrm{Br\alpha}}$ from star formation to the measured $L_{\mathrm{Br\alpha}}$.
 The relationships $L_{\mathrm{3.3PAH}}/L_{\mathrm{IR}}\sim10^{-3}$ \citep{mouri90,imanishi02} and
 $L_{\mathrm{Br\alpha}}/L_{\mathrm{IR}}\sim10^{-4}$ \citep{imanishi10} are expected for starburst galaxies,
 suggesting that the luminosity of the Br$\alpha$ line from star formation is 10\% of the 3.3\,$\mu$m PAH 
 line luminosity.
 Among our seven Br$\alpha$-detected AGNs, the 3.3\,$\mu$m PAH emission line is detected in only two objects
 (NGC\,4051 and NGC\,7469) using the extraction width adopted in this study (Kim et al. in prep).
 The expected star formation contributions of the Br$\alpha$ line based on the 3.3\,$\mu$m PAH line are 8\% and 46\%
 for NGC\,4051 and NGC\,7469, respectively.
 This assumes the above 10\% flux ratio of $L_{\mathrm{Br\alpha}}$ to $L_{\mathrm{3.3PAH}}$,
 but the relation contains a large scatter of a factor of two to three. 
 Therefore, we conclude that the contribution of star formation to the Bracket emission lines is
 negligible (no or weak detection of PAHs) for most of the AGNs in our sample.
 NGC\,7469 is an exceptional case with a very strong 3.3\,$\mu$m PAH.
 Even in this case, the star formation contribution is less than half for Br$\alpha$, but we caution 
 possible over-estimation of the line flux (134\%) and under-estimation of the line width (80\%) for NGC\,7469.
 
 The measured values of the $\sigma$ and luminosities of the Brackett lines and
 the corresponding H$\beta$ properties are presented in Table 2.
 Note that the $\sigma$ values are corrected for the instrumental resolution by subtracting
 the instrumental resolution in quadrature ($\sigma^2=\sigma_{\rm obs}^2-\sigma_{\rm inst}^2$) from the
 observed values. The derivation of the line luminosities assumes a standard
 $\Lambda$CDM cosmology of $H_{0}$=70 km s$^{-1}$ Mpc$^{-1}$,
 $\Omega_{m}$=0.3 and $\Omega_{\Lambda}$=0.7 (e.g., \citealt{im97})

\subsection{Correlation between Balmer and Brackett Lines}

 In this section, we will investigate how the properties of the Brackett lines correlate with those of
 the Balmer lines. If the two lines arise from BLRs similar to each other,
 then we expect their line properties to correlate well. This would justify the use of 
 the Brackett lines as an $M_{\rm BH}$ estimator (e.g., \citealt{kim10}), just like the Balmer lines 
 that are commonly used as $M_{\rm BH}$ estimators \citep{greene05,greene07}.
 For this comparison, we take the average of the multi-epoch $\sigma$ values and the luminosities of the H$\beta$ line
 from previous studies \citep{kaspi00,marziani03,peterson04,vestergaard06,landt08}.
 When there is a large discrepancy in the $\sigma$ values ($> \times 1.5$) of H$\beta$ and H$\alpha$,
 we take the average of the two. Such a case occurred for only one object (PG\,1411$+$442).
 Note that we could not find the H$\beta$ line luminosity data for NGC\,4051, 
 so we only adopted a $\sigma$ value from \citet{peterson04} for NGC\,4051.

 Figures 8 and 9 show the correlations between the Brackett lines
 and the H$\beta$ line in luminosity and $\sigma$.
 In Figure 8, we compare the luminosities of the Brackett lines with those of the H$\beta$ line.
 The Pearson correlation coefficients are 0.991 and 0.947 for Br$\beta$ and Br$\alpha$, respectively.
 By performing a linear bisector fit to these points, we find that the two quantities correlate with
 each other as

 \begin{equation}
 \begin{split}
 \log \left( \frac{L_{\mathrm{H\beta}}}{10^{42}~\mathrm{erg~ s^{-1}}} \right) & = \left( \mathrm{0.99\pm 0.05} \right) \\
 & + \left( 0.85\pm 0.05 \right) \log \left( \frac{L_{\mathrm{Br\beta}}}{10^{42}~\mathrm{erg~ s^{-1}}} \right),
 \end{split}
 \end{equation}
 and
 \begin{equation}
 \begin{split}
 \log \left( \frac{L_{\mathrm{H\beta}}}{10^{42}~\mathrm{erg~ s^{-1}}} \right) & = \left( \mathrm{0.86\pm 0.11} \right) \\
 & + \left( 0.79\pm 0.07 \right) \log \left( \frac{L_{\mathrm{Br\alpha}}}{10^{42}~\mathrm{erg~ s^{-1}}} \right).
 \end{split}
 \end{equation}

 In Figure 9, the dashed line indicates a perfect correlation between the $\sigma$ values of the Brackett and H$\beta$ lines.
 The figures show a reasonable correlation between the Br$\beta$ and H$\beta$ line widths, but a weak correlation between
 the Br$\alpha$ and H$\beta$ line widths. The Pearson correlation coefficients are 0.735 and 0.273 for Br$\beta$ and Br$\alpha$, respectively.
 The weak correlation in $\sigma$ for Br$\alpha$ is likely due to low S/N and the difficulty in fitting some of the Br$\alpha$ lines.
 The zodiacal background increases rapidly at $\lambda >4$ $\mu$m, causing the background noise to increase for Br$\alpha$.
 Furthermore, the existence of the CO$_{2}$ absorption feature right next to Br$\alpha$ makes it challenging to provide a good fit.
 ${\rm CO_{2}}$ absorption is the most prominent in the spectrum of NGC\,4151, which also seems to be
 a significant outlier in Figure 9.The exclusion of the NGC\,4151 point improves the $\sigma$ correlation for Br$\alpha$,
 with the correlation coefficient of 0.468.

 We also compared the Brackett line luminosities to the BLR radii.
 For the radii of BLR, we adopted values from the time lags in the variability between the H$\beta$ 
 and the $L_{\rm 5100}$ from \citet{bentz13}.
 Figure 10 shows the correlation between the Brackett line luminosities and the BLR radii. 
 The best-fit relations are

 \begin{equation}
 \begin{split}
 \log \left( \frac{R_{\mathrm{BLR}}}{\rm lt-days} \right) & = \left( \mathrm{0.82\pm0.08} \right) \\
 & + \left( 0.62\pm0.05 \right) \log \left( \frac{L_{\mathrm{Br\beta}}}{10^{40}~\mathrm{erg~ s^{-1}}} \right),
 \end{split}
 \end{equation}
 and 
 \begin{equation}
 \begin{split}
 \log \left( \frac{R_{\mathrm{BLR}}}{\rm lt-days} \right) & =\left(\mathrm{0.75\pm0.03} \right) \\
 & + \left( 0.66\pm0.06 \right) \log \left( \frac{L_{\mathrm{Br\alpha}}}{10^{40}~\mathrm{erg~ s^{-1}}} \right).
 \end{split}
 \end{equation}

 \begin{figure}
 \figurenum{10}
 \begin{center}
\includegraphics[width=\columnwidth]{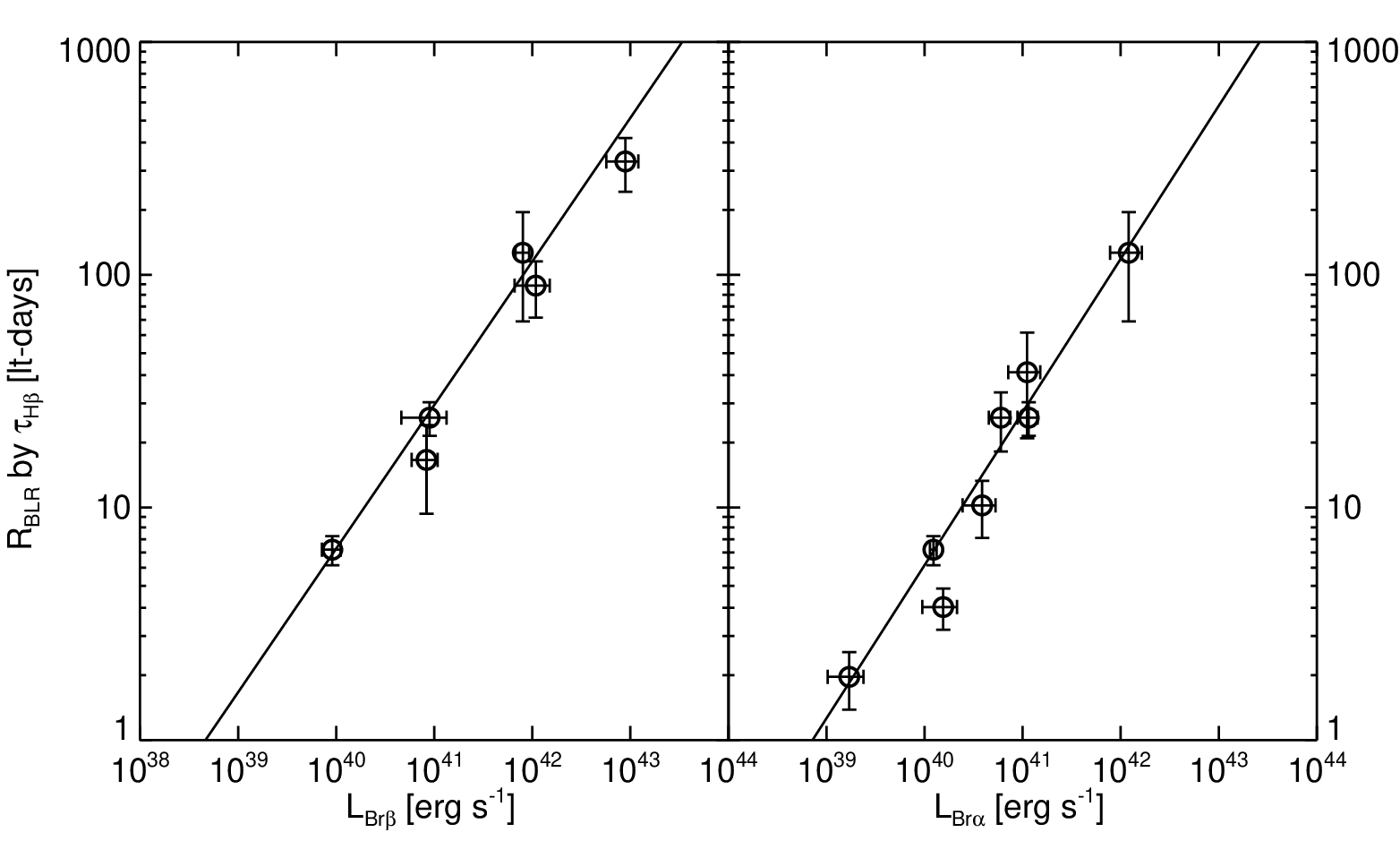}
 \caption{
 Comparison of the line luminosities of the Brackett lines
 vs. the BLR sizes. For the BLR size, we use the time lag in the variability 
 between the H$\beta$ and $L_{\rm 5100}$ from \citet{bentz13}.
 The solid line indicates the best-fit correlation between the luminosities of
 the Brackett lines and the BLR sizes.
 }
 \end{center}
 \end{figure}

\subsection{BH Mass Estimators with Brackett Lines}

 In this section, we present $M_{\rm BH}$ estimators that are based on the Br$\beta$ and
 Br$\alpha$ lines. The idea here is that the radius and the velocity of the BLR can be
 inferred from the luminosity and $\sigma$ of the Brackett lines, similar to 
 the $M_{\rm BH}$ estimators based on the Balmer of the Paschen lines. 
 Mathematically, we need to find three unknown parameters $a$, $b$, and $c$ 
 in the following equation:
 
\begin{equation}
 \log(M)= a + b\log(L)+ c \log(\mathrm{\sigma}).
\end{equation}

 For simplicity, we fix the exponent of the velocity term, $c$ to 2, as expected in the virial theorem.
 The exponent of the luminosity term, $b$,
 is expected to be 0.5 theoretically and empirically, i.e., $R_{\mathrm{BLR}} \sim L^{0.5}$
 \citep{dibai77,kaspi00,greene05,kim10}. However, we will treat $b$ as a free parameter.
 In order to derive the $M_{\rm BH}$ estimators, we used the \texttt{MPFITEXY} routine \citep{williams10}
 to fit a linear relationship in the logarithmic scale as in Equation (5).
 \texttt{MPFITEXY} returns the fitted parameters with the uncertainties of the parameters and an intrinsic scatter in the fitted relation.
 The routine has been tested against several other similar routines, and has shown to return reliable outputs \citep{park12}.
 As a result, we obtain the following relations:

\begin{equation}
\frac{M_{\rm BH}}{M_{\odot}}=10^{6.61\pm0.23}\,\left( \frac{L_{\mathrm{Br\beta}}}{10^{40}~\mathrm{erg~ s^{-1}}}\right) ^{0.68\pm0.13}\,\left(\frac{\mathrm{\sigma_{Br\beta}}}{10^{3}~\mathrm{km~s^{-1}}}\right) ^{2}, 
\end{equation}
and
\begin{equation}
\frac{M_{\rm BH}}{M_{\odot}}=10^{6.68\pm0.20}\,\left( \frac{L_{\mathrm{Br\alpha}}}{10^{40}~\mathrm{erg~ s^{-1}}} \right) ^{0.66\pm0.21}\,\left( \frac{\mathrm{\sigma_{Br\alpha}}}{10^{3}~\mathrm{km~s^{-1}}}\right) ^{2}.
\end{equation}

 In Figure 11, we plot the $M_{\rm BH}$ from the reverberation mapping method
 versus the $M_{\rm BH}$ from the Brackett-line-based estimators, i.e., the values from Equations (6) and (7).
 The rms scatters of the data points (including measurement errors) with respect to the best-fit correlations of
 Equations (6) and (7) are 0.229 dex (Br$\beta$) and 0.407 dex (Br$\alpha$).
 The intrinsic scatter of Equation (7) is 0.284 dex, but the intrinsic scatter of Equation (6) is not determined
 since the fit did not require intrinsic scatter. The value of the intrinsic scatter is expected to be small ($\lesssim$ 0.2 dex).

 \begin{figure*}
 \begin{center}
 \centering
 \figurenum{11}
 \includegraphics[width=\columnwidth]{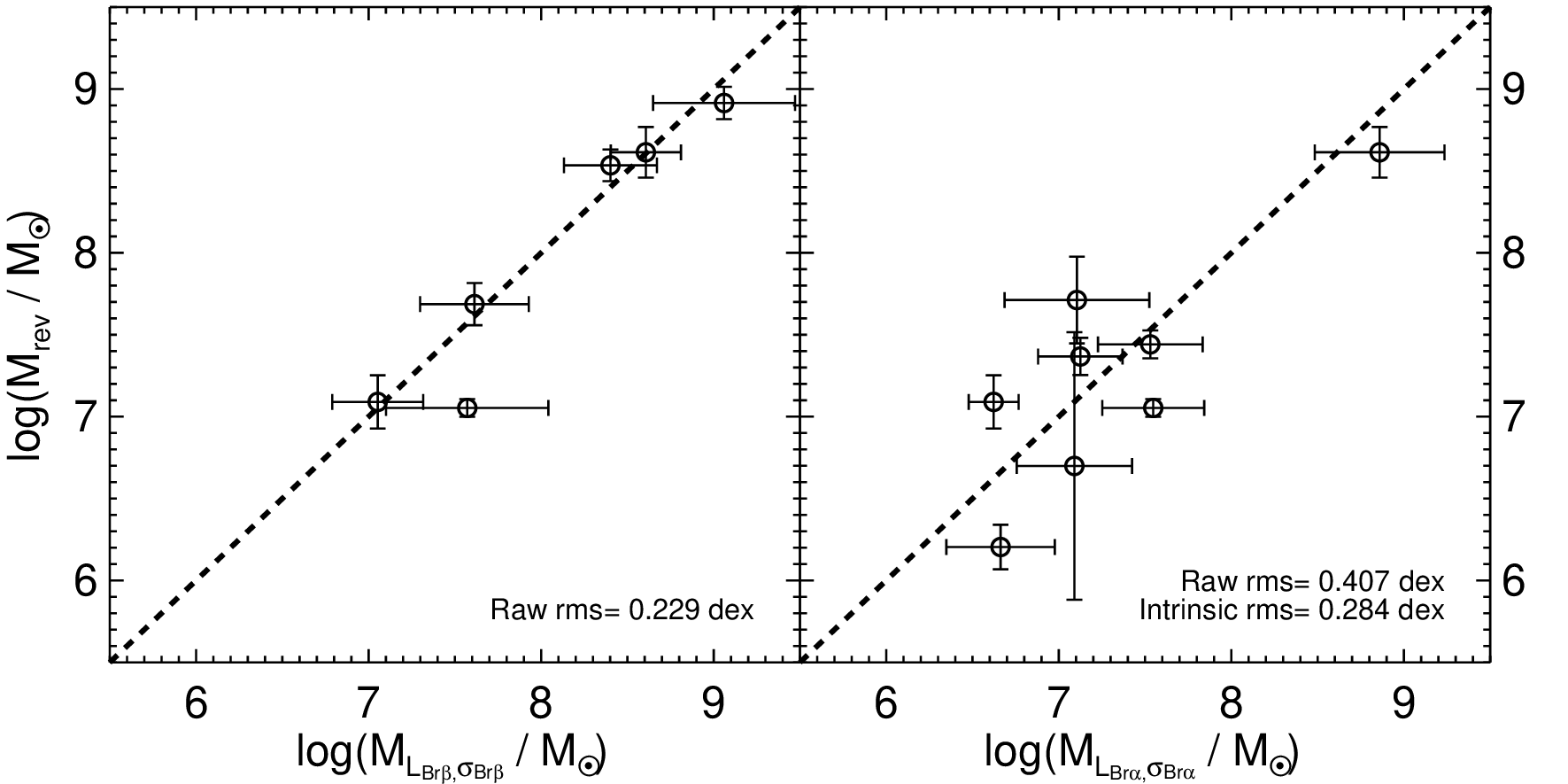}
 \caption{
 Comparison of the Brackett-line-based $M_{\rm BH}$ vs. $M_{\rm BH}$ derived from the reverberation mapping method.
 The Brackett-line-based $M_{\rm BH}$ is derived by using the $L$ and $\sigma$ of the Brackett lines
 as Equations (6) and (7). The dashed line indicates a line where $M_{\rm BH}$ from Brackett lines and the reverberation mapping methods are identical.
 In each panel, we indicate the intrinsic and the raw (no correction of the measurement errors) scatters of the points with respect to the dashed line.}
% However, the intrinsic scatter of Br$\beta$ is not determined because of that chi-square value is less than 1.}
 \end{center}
 \end{figure*}
 
\begin{figure*}
\begin{center}
\centering
\figurenum{12}
\includegraphics[width=\columnwidth]{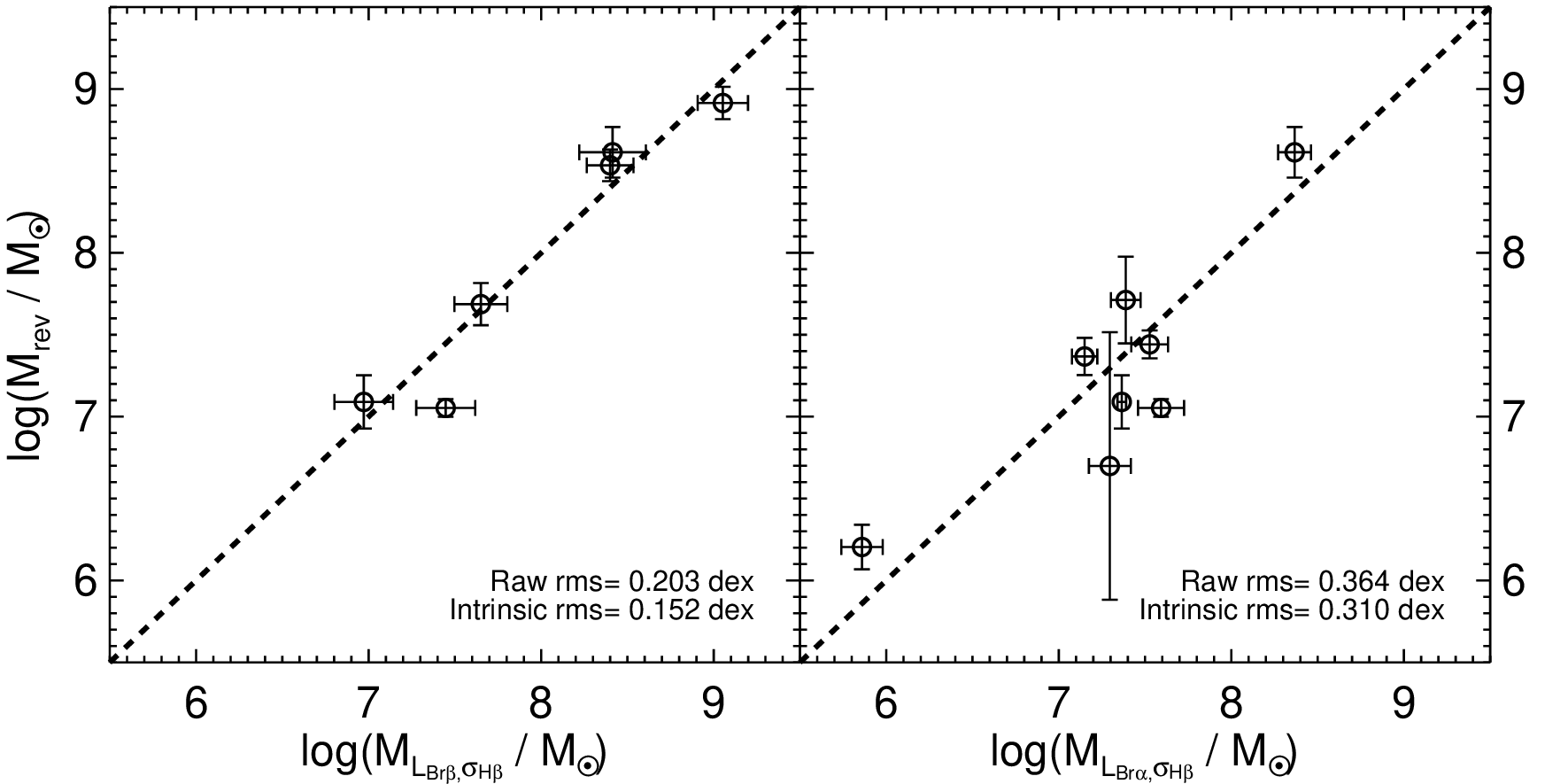}
\caption{
 Comparison of $M_{\rm BH}$ from Equations (8) and (9) vs. $M_{\rm BH}$ derived from the reverberation mapping method.
 The Equations (8) and (9) are established by using the $L_{\rm Br\alpha,\beta}$ but adopting
 ${\rm \sigma_{H\beta}}$. The meaning of the line is identical to Figure 11.}
\end{center}
\end{figure*}

\begin{figure*}
\begin{center}
\centering
\figurenum{13}
\includegraphics[width=\columnwidth]{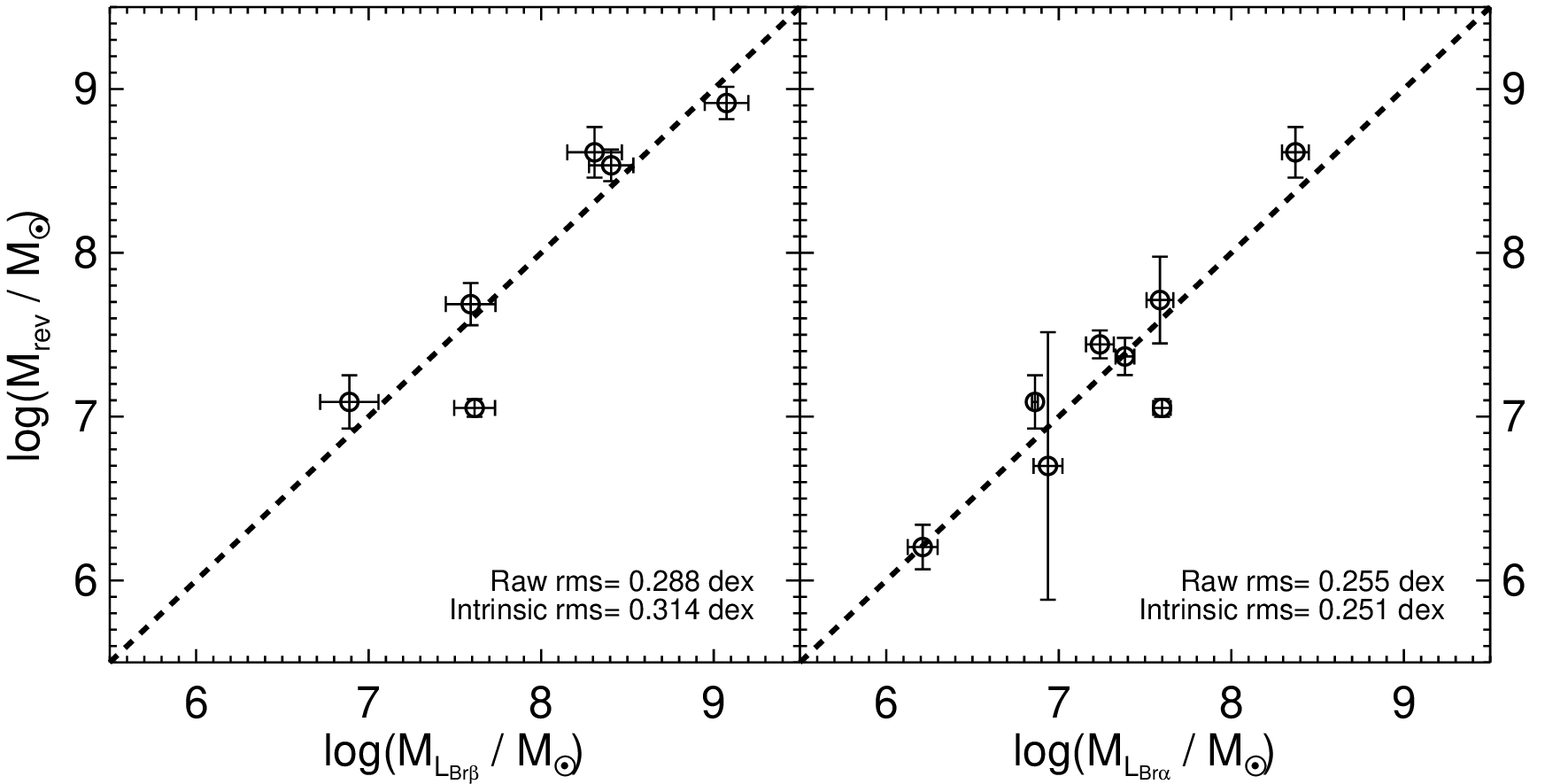}
\caption{
 Comparison of $M_{\rm BH}$ from Equations (10) and (11) vs. $M_{\rm BH}$ derived from the reverberation mapping method.
 The Equations (10) and (11) are established by using only $L_{\rm Br\alpha,\beta}$.
 The meanings of the line is identical to Figure 11.}
\end{center}
\end{figure*}

 Since the $\sigma$ values are rather poorly determined from the Brackett lines but we expect
 that they correlate closely with the $\sigma$ values of the H$\beta$, we also derived 
 the $M_{\rm BH}$ estimators using $L_{\mathrm{Br\alpha,\beta}}$ but adopting 
 $\mathrm{\sigma_{H\beta}}$ as a proxy for $\mathrm{\sigma_{Br\alpha,\beta}}$. The 
 $\mathrm{\sigma_{H\beta}}$ values are taken from \cite{peterson04}.
 Since dust extinction in a host galaxy can reduce its line luminosity significantly
 but not as much for the line width, this is a plausible path to estimate $M_{\rm BH}$ of red AGNs
 whose $\mathrm{\sigma_{H\beta}}$ can be determined to a good accuracy, but its H${\beta}$ luminosity
 is quite uncertain due to an unknown amount of dust extinction.
 In this process, we also fixed the exponent of the velocity term, $c$, to 2.
 The resultant $M_{\rm BH}$ estimators using $L_{\mathrm{Br\alpha,\beta}}$ and $\mathrm{\sigma_{H\beta}}$
 are given below and also in Figure 12 as

\begin{equation}
\frac{M_{\rm BH}}{M_{\odot}}=10^{6.44\pm0.17}\,\left( \frac{L_{\mathrm{Br\beta}}}{10^{40}~\mathrm{erg~ s^{-1}}} \right) ^{0.72\pm0.10}\,\left( \frac{\mathrm{\sigma_{H\beta}}}{10^{3}~\mathrm{km~s^{-1}}} \right) ^{2},
\end{equation}
and
\begin{equation}
\frac{M_{\rm BH}}{M_{\odot}}=10^{6.76\pm0.17}\, \left( \frac{L_{\mathrm{Br\alpha}}}{10^{40}~\mathrm{erg~ s^{-1}}} \right) ^{0.48\pm0.17}\,\left( \frac{\mathrm{\sigma_{H\beta}}}{10^{3}~\mathrm{km~s^{-1}}} \right) ^{2}.
\end{equation}

 The raw rms scatters (including measurement errors) with respect to a perfect correlation between
 the reverberation-mapping-based $M_{\rm BH}$ and $M_{\rm BH}$ for Equations (8) and (9) are
 0.203 dex (Br$\beta$) and 0.364 dex (Br$\alpha$), while the intrinsic scatters are 0.152 dex (Br$\beta$)
 and 0.310 dex (Br$\alpha$). However, the intrinsic scatters are not better than the values for
 the purely Brackett-based $M_{\rm BH}$ estimators in Equations (6) and (7).

 Since the $\sigma$ values of Brackett lines, especially for Br$\alpha$, show a weak correlation, we test whether or not
 the $\sigma$ values of Brackett lines are adding useful information to the $M_{\rm BH}$ estimators.
 For this test, we derive $M_{\rm BH}$ estimators using only Brackett line luminosities:

\begin{equation}
\frac{M_{\rm BH}}{M_{\odot}}=10^{6.92\pm0.26}\,\left( \frac{L_{\mathrm{Br\beta}}}{10^{40}~\mathrm{erg~ s^{-1}}} \right) ^{0.73\pm0.15}, 
\end{equation}
and 
\begin{equation}
\frac{M_{\rm BH}}{M_{\odot}}=10^{6.79\pm0.15}\,\left( \frac{L_{\mathrm{Br\alpha}}}{10^{40}~\mathrm{erg~ s^{-1}}} \right) ^{0.76\pm0.14}.
\end{equation}

 Figure 13 shows the $M_{\rm BH}$ values from the reverberation mapping method
 versus the $M_{\rm BH}$ values from the $M_{\rm BH}$ estimators using only $L_{\rm Br\beta}$ or $L_{\rm Br\alpha}$, i.e., the values from
 Equations (10) and (11).
 We find that the intrinsic scatters in these relations are 0.314 dex (Br$\beta$), and 0.251 dex (Br$\alpha$). 
 Therefore, for Br$\beta$, the intrinsic scatter in the $M_{\rm BH}$ estimators becomes smaller as we move from the luminosity-only estimator
 to the full Br$\beta$ luminosity$+\sigma$ estimators, lending support to the usefulness of Equation (6).
 On other hand, the same cannot be said for the Br$\alpha$-based estimators. This suggests that Br$\alpha$ $\sigma$ measurement is
 more challenging with the current data, and the evaluation of the Br$\alpha$-based $M_{\rm BH}$ estimators is needed with data sets with better sensitivity.
 Considering that the Eddington ratio of AGNs used for Equations (10) and (11) are between
 0.015 and 0.225, Equations (10) and (11) could be useful when (i) 
 the Brackett line luminosities are available but not for reliable $\sigma$ values and (ii)
 the Eddington ratio of the sample is within a range of 0.015 to 0.225.

\subsection{Line Luminosity Ratios}

 We investigate the physical conditions of BLR by comparing the observed line luminosity ratios of the Balmer
 through Brackett lines to the line luminosity ratios computed from the CLOUDY code (version 10.00; \citealt{ferland98}). 
 The line luminosity ratios are computed by varying three physical parameters -- the shape of the ionizing
 continuum ($\alpha$ = -1.0, -1.5), the ionization parameter
 ($U$ = $10^{-5.5}$ -- $10^{0.5}$),
 and the hydrogen density ($n$ = $10^{9}$ -- ${10}^{13}$ $\rm{cm}^{-3}$).
 The hydrogen number densities are commonly taken as $n \simeq 10^{9}$ -- $10^{10}$ cm$^{-3}$ for
 type-1 AGNs \citep{davidson79,rees89,ferland89},
 but there are studies suggesting a high hydrogen 
 number density ($n > 10^{12}$\rm{cm}$^{-3}$) with a low ionization parameter
 of $U = 10^{-5}$ (e.g., \citealt{ruff12}) for BLR.

 Figure 14 shows the observed line luminosity ratios (with respect to H$\beta$) as a
 function of wavelength for seven AGNs, for which we have at least one Brackett to Balmer line ratio 
 and an additional line ratio of either Paschen or Brackett lines. 
 The line luminosities are corrected for galactic extinction with the York Extinction Solver \citep{mccall04}, adopting $E(B-V)$ values 
 from \citep{schlafly11} and $R_{V} = 3.07$.
 No attempt is made to correct for internal dust extinction.
 Note that the Paschen line ratios are taken from \citet{kim10}.
 The mean line ratios of the seven AGNs are $L_{\rm Br\alpha}$/$L_{\rm Br\beta} = \rm 1.37 \pm 0.31$,
 $L_{\rm Br\beta}$/$L_{\rm H\beta} = 0.073 \pm 0.010$, and $L_{\rm Br\alpha}$/$L_{\rm H\beta} = 0.081 \pm 0.014$.

\begin{figure*}
\begin{center}
\centering
\figurenum{14}
\includegraphics[scale=0.4]{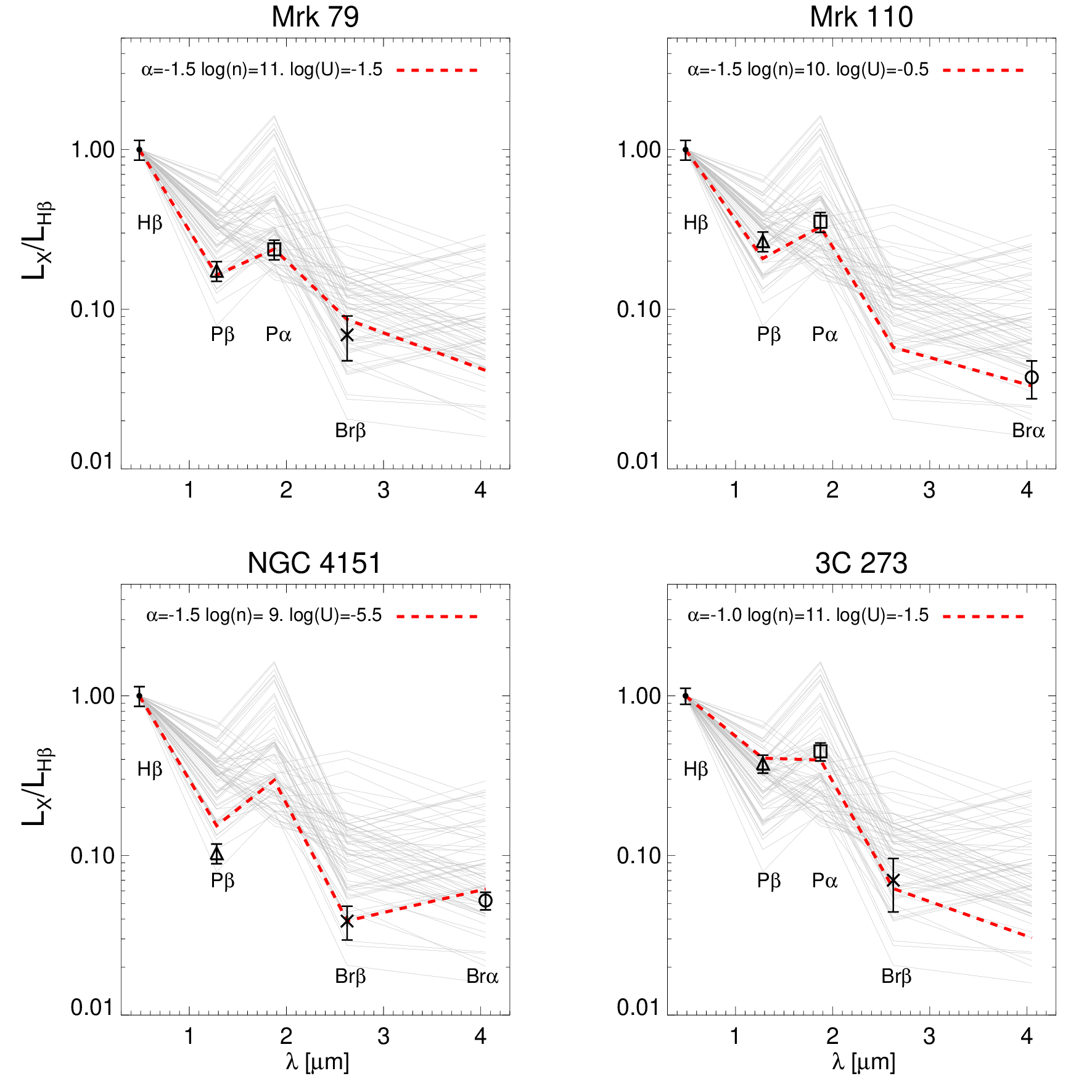}
\caption{
 Line luminosity ratios of Paschen and Brackett with respect to H$\beta$.
 The crosses and the circles represent AGNs in the reverberation mapped sample 
 with the Brackett line luminosity measurements.
 When available,
 we also plot the line luminosity ratios for P$\alpha$ (squares) and P$\beta$ (triangles)
 with respect to H$\beta$ from \citet{kim10}. 
 The gray lines represent the expected line ratios from the CLOUDY code, assuming various
 parameters with different combinations of $\alpha = -1.0 \sim -1.5$,
 $n = 10^{9\,\sim\,13}$ $\rm{cm}^{-3}$, and $U = 10^{-5.5\,\sim\,0.5}$. 
 In each panel, we indicate the most likely set of the parameters, with the CLOUDY model predictions
 for such a parameter set in the red dashed line.
}
\end{center}
\end{figure*}

\begin{figure*}
\begin{center}
\centering
\figurenum{14}
\includegraphics[scale=0.4]{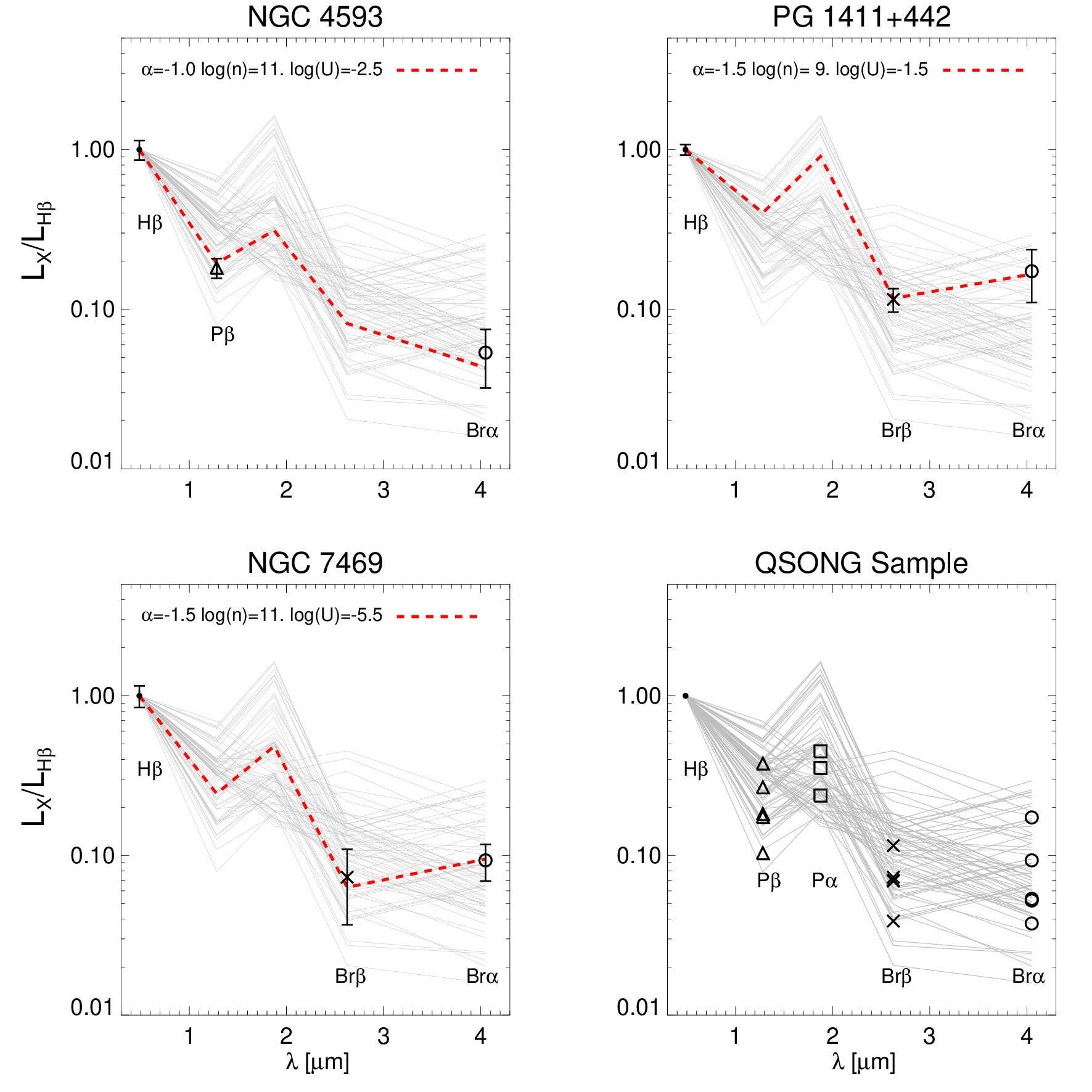}
\caption{
 Continued
}
\end{center}
\end{figure*}

 In order to find the most likely physical condition for each object, 
 we searched for a parameter set that minimizes $\chi^{2}$, which is a function
 of line ratios such as $L_{\rm P\beta}$/$L_{\rm H\beta}$, $L_{\rm P\alpha}$/$L_{\rm H\beta}$,
 $L_{\rm Br\beta}$/$L_{\rm H\beta}$, and $L_{\rm Br\alpha}$/$L_{\rm H\beta}$ following the approach by \citet{ruff12}:
\begin{equation}
\chi^{2} = \sum_{i=1}^N \frac{(R_{{\rm observed},i} -R_{{\rm model},i})^{2}}{\sigma_{i}^{2}},
\end{equation}
 where $N$ is the number of line luminosity ratios, and two types of $R_{i}$ are the line luminosity
 ratios from either observation (when with a subscript, \textquotedblleft observed\textquotedblright) or the CLOUDY model (with
 a subscript, \textquotedblleft model\textquotedblright), and $\sigma_{i}$ is the error in the measured line ratio.
 The best-fit parameter sets are indicated for each object in Figure 14, and are summarized in Table 3.
 The best-fit line luminosity ratios are also plotted as a
 red dashed line in Figure 14.

\begin{deluxetable}{cccc}
\tabletypesize{\scriptsize}
\tablecolumns{4}
\tablewidth{0pt}
\tablenum{3}
\tablecaption{BLR Conditions\label{tbl3}}
\tablehead{
\colhead{Object Name}&	\colhead{$\alpha$}&	\colhead{n}&				\colhead{U}\\
\colhead{}&				\colhead{}&			\colhead{[$\rm{cm}^{-3}$]}&		\colhead{}\\
}
\startdata
Mrk\,79&		-1.5&	$10^{11}$&	$10^{-1.5}$\\
Mrk\,110&	-1.5&	$10^{10}$&	$10^{-0.5}$\\
NGC\,4151&	-1.5&	$10^{9}$&	$10^{-5.5}$\\
3C\,273&		-1.0&	$10^{11}$&	$10^{-1.5}$\\
NGC\,4593&	-1.0&	$10^{11}$&	$10^{-2.5}$\\
PG\,1411$+$442&	-1.5&	$10^{9}$&	$10^{-1.5}$\\
NGC\,7469&	-1.5&	$10^{11}$&	$10^{-5.5}$\\
\enddata
\end{deluxetable}

 The CLOUDY models can reproduce the observed line luminosity ratios, under commonly 
 cited BLR physical conditions of $n = 10^{9}$ -- $10^{11}$ cm$^{-3}$ and $U$ = $10^{-2.5}$ -- $10^{0.5}$,
 suggesting that most of the objects with Brackett line detections do not require a more exotic condition
 ($n > 10^{12}$ $\rm{cm}^{-3}$ and $U < 10^{-5}$), such as that found in \cite{ruff12}. One exception is
 NGC\,7469, for which we find a best-fit parameter set of $n = 10^{11}$ $\rm{cm}^{-3}$ and $U = 10^{-5.5}$. 
 This is the case where the Brackett line luminosity may be affected by the star formation component
 (possibly by as much as 50\%), and thus the interpretation of the line luminosity ratio result cannot be taken
 too seriously for this object.

\section{DUST COMPONENT}

 In this section, we present hot ($>1000\,\rm{K}$; e.g., \citealt{barvainis87,glikman06})
 and warm ($\sim200\,\rm{K}$; e.g., \citealt{netzer07,deo09}) dust temperatures
 and luminosities of the PG QSOs and the reverberation-mapped AGNs, and we investigate
 the correlation between hot dust luminosity ($L_{\rm HD}$) and $L_{\rm bol}$.
 For this purpose, we restrict our analysis to luminous AGNs with $L_{\rm{bol}} > 10^{45.7} \, \mathrm{erg~s^{-1}}$ 
 for which the host galaxy light contribution is less than 10\% at 0.51\,$\mu$m \citep{shen11}, and even smaller at NIR.
 
 To study the AGN dust components, we assembled a multi-wavelength data set from the optical to MIR by obtaining photometry of the objects
 from SDSS, 2MASS, $\it{WISE}$ ($W$3 and $W$4), and $\it{ISO}$ (7.3\,$\mu$m: \citealt{haas03}). These multi-wavelength photometric data 
 are combined with the $\it{AKARI}$ spectra, and we fit them with a SED model that is composed of 
 a single power law and a double black body radiation component as
 \begin{equation}
 F_{\nu}=C_{0}\nu^{\alpha}+C_{1}B_{\nu}(T_{\rm{HD}})+C_{2}B_{\nu}(T_{\rm{WD}}),
\end{equation}
 ${\noindent}$where $\alpha$ is the slope of the underlying power-law continuum which dominates the optical light,
 $B_{\nu}$ is the Planck function as a function of $\nu$, 
 $T_{\rm{HD}}$ and $T_{\rm{WD}}$ are the hot and warm dust temperatures, and 
 $C_{0}$, $C_{1}$, and $C_{2}$ are the normalization constants of each component.
 The fitting result gives us the luminosities and the temperatures of the hot/warm dust components.

\begin{figure*}
\begin{center}
\centering
\figurenum{15}
\scalebox{1.1}{\plotone{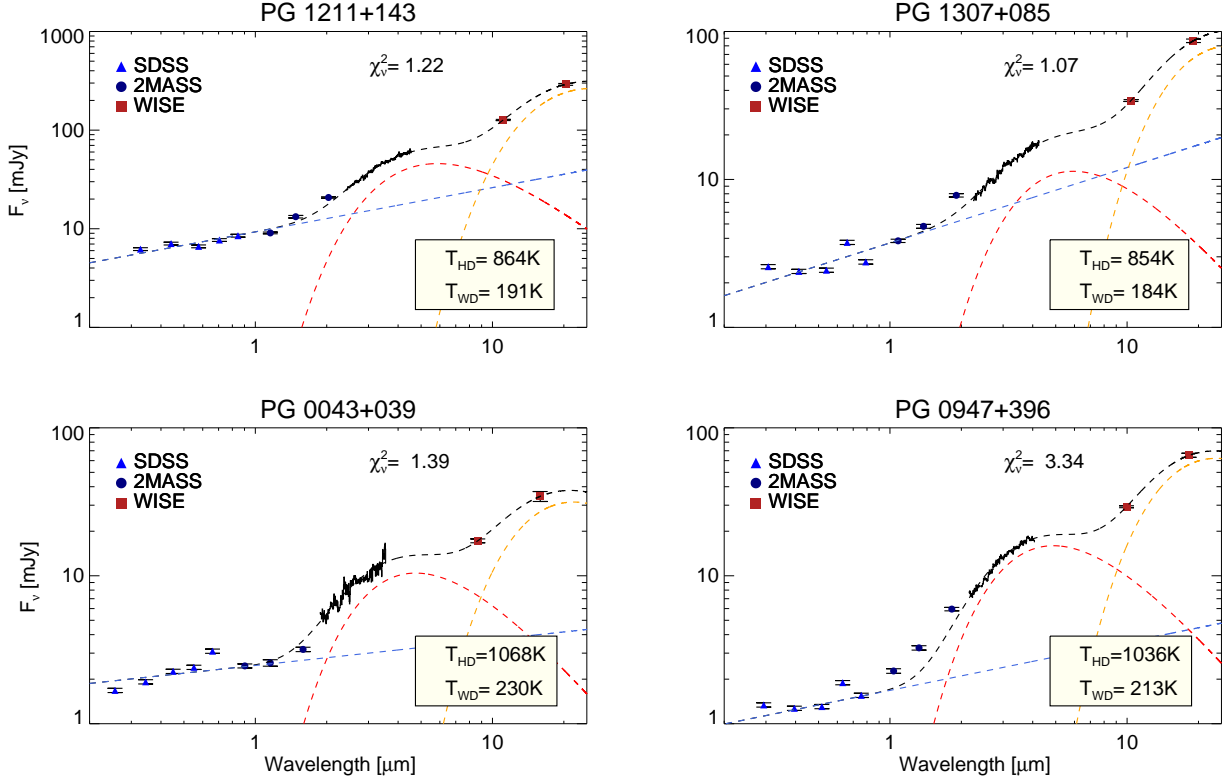}}
\caption{Photometric and spectroscopic SED. The SEDs are shown in rest-frame.
Triangle, circle, square, and star indicate photometric data point from SDSS, 2MASS, $\it{WISE}$ ($W$3 and $W$4), and $\it{ISO}$ (7.3\,$\mu$m), respectively.
The black solid line denotes $\it{AKARI}$ spectrum. The dashed lines are fitted results.
The blue, red, and yellow dashed lines mean power-law, hot, and warm dust components, respectively, and the black dashed line is the sum of all components.
%DK140805
The measured hot and warm dust temperatures are indicated in the lower right-hand corner.
The dust properties are taken only if $\chi_{\nu}^{2} < 5$, and the $\chi_{\nu}^{2}$ of each sample is indicated at top of each figure.
The PG\,1322$+$659 is excluded by the $\chi_{\nu}^{2}$ limit, and the SED of the PG\,1322$+$659 is indicated in last panel of this figure.
}
\end{center}
\end{figure*}

\begin{figure*}
\begin{center}
\centering
\figurenum{15}
\scalebox{1.1}{\plotone{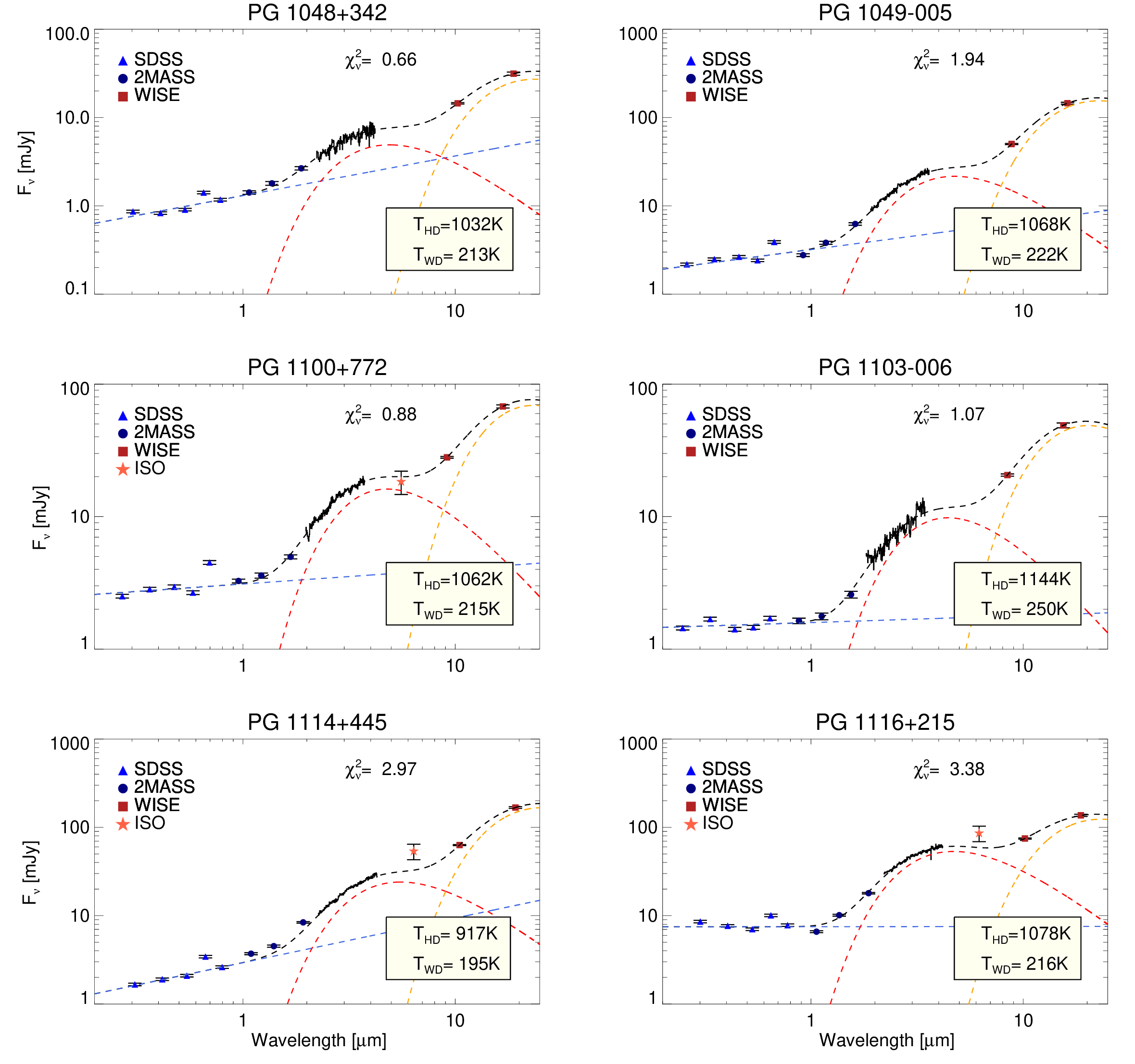}}
\caption{Continued}
\end{center}
\end{figure*}

\begin{figure*}
\begin{center}
\centering
\figurenum{15}
\scalebox{1.1}{\plotone{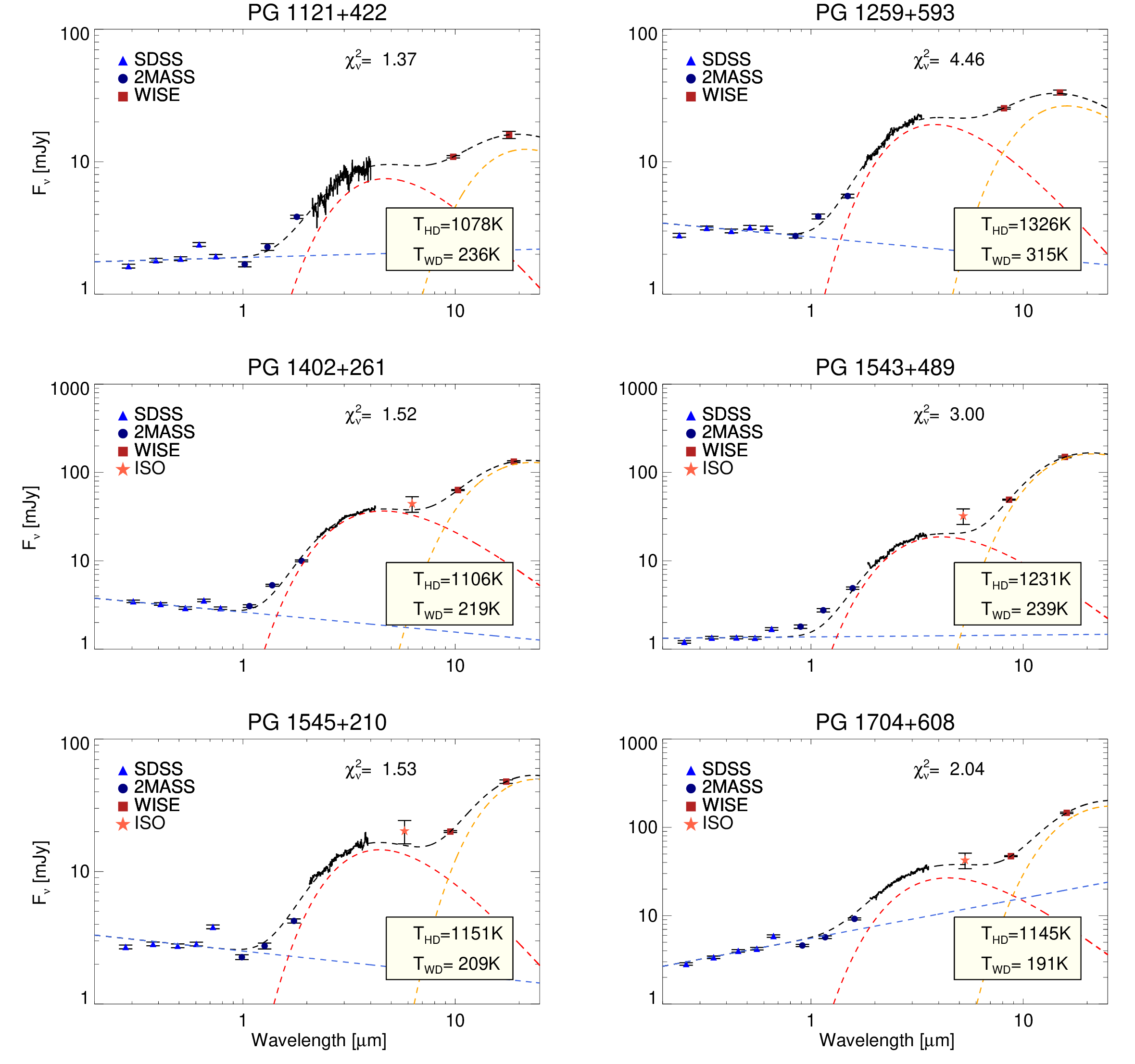}}
\caption{Continued}
\end{center}
\end{figure*}

\begin{figure*}
\begin{center}
\centering
\figurenum{15}
\scalebox{1.1}{\plotone{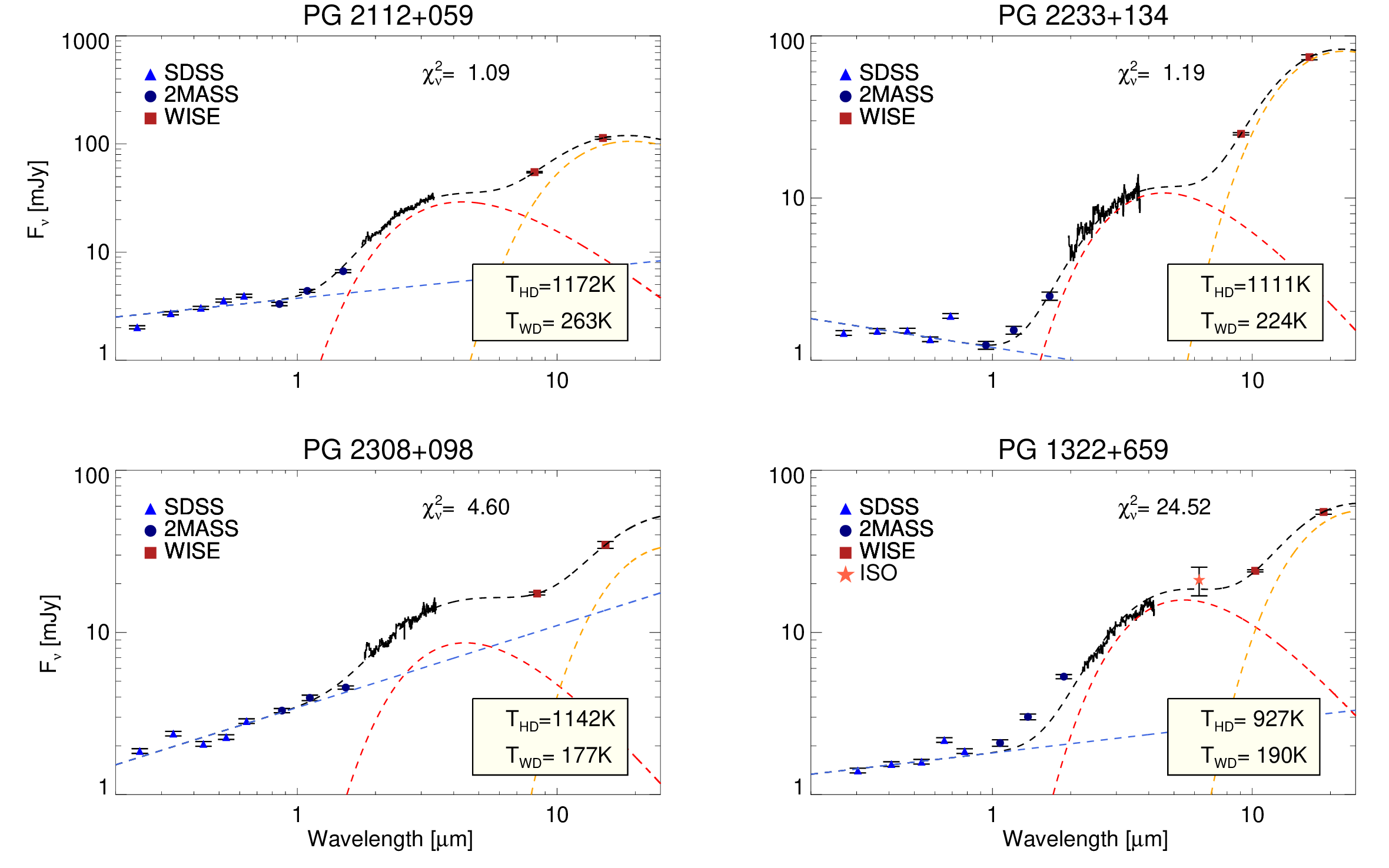}}
\caption{Continued}
\end{center}
\end{figure*}

\begin{deluxetable*}{ccccccc} %DK140805 include alpha
\tabletypesize{\scriptsize}
\tablewidth{0pt}
\tablenum{4}
\tablecaption{Dust Components Fitting Results\label{tbl4}}
\tablehead{
\colhead{Objects}& \colhead{$T_{\rm HD}$}& \colhead{$T_{\rm WD}$}&	\colhead{$\log(L_{\rm HD}$)}&
\colhead{$\log(L_{\rm WD}$)}&	\colhead{$\rm CF_{HD}$}&	\colhead{$\alpha$}\\
\colhead{}&	\colhead{[K]}&	\colhead{[K]}&	\colhead{[$\rm erg~s^{-1}$]}&	\colhead{[$\rm erg~s^{-1}$]}&
\colhead{}&	\colhead{}
}
\startdata
PG\,1211$+$143&	864$\pm$5&		191$\pm$2&	45.27$\pm$0.01&	45.38$\pm$0.02&		0.28&	-1.55$\pm$0.02\\
PG\,1307$+$085&	854$\pm$13&		184$\pm$3&	45.27$\pm$0.03&	45.45$\pm$0.03&		0.18&	-1.49$\pm$0.02\\
PG\,0043$+$039& 1068$\pm$14&	230$\pm$9&	46.22$\pm$0.02&	46.04$\pm$0.07&		0.23&	-1.83$\pm$0.03\\
PG\,0947$+$396&	1036$\pm$2&		213$\pm$3&	45.78$\pm$0.00&	45.68$\pm$0.03&		0.60&	-1.68$\pm$0.01\\
PG\,1048$+$342&	1032$\pm$20&	213$\pm$5&	45.06$\pm$0.03&	45.12$\pm$0.04&		0.25&	-1.55$\pm$0.04\\
PG\,1049$-$005&	1068$\pm$6&		222$\pm$2&	46.48$\pm$0.01&	46.65$\pm$0.02&		0.43&	-1.68$\pm$0.03\\
PG\,1100$+$772&	1062$\pm$9&		215$\pm$3&	46.20$\pm$0.02&	46.14$\pm$0.03&		0.29&	-1.89$\pm$0.04\\
PG\,1103$-$006&	1144$\pm$17&	250$\pm$5&	46.33$\pm$0.03&	46.37$\pm$0.03&		0.35&	-1.95$\pm$0.05\\
PG\,1114$+$445&	917$\pm$2&		195$\pm$2&	45.56$\pm$0.00&	45.73$\pm$0.02&		0.51&	-1.50$\pm$0.01\\
PG\,1116$+$215& 1078$\pm$1&		216$\pm$3&	46.17$\pm$0.00&	45.84$\pm$0.03&		0.37&	-2.00$\pm$0.02\\
PG\,1121$+$422& 1078$\pm$15&	236$\pm$9&	45.55$\pm$0.02&	45.11$\pm$0.06&		0.21&	-1.95$\pm$0.04\\
PG\,1259$+$593& 1326$\pm$1&		315$\pm$8&	46.81$\pm$0.00&	46.32$\pm$0.04&		0.41&	-2.15$\pm$0.01\\
PG\,1402$+$261& 1106$\pm$3&		219$\pm$2&	45.95$\pm$0.00&	45.79$\pm$0.02&		0.64&	-2.23$\pm$0.01\\
PG\,1543$+$489& 1231$\pm$2&		239$\pm$2&	46.58$\pm$0.00&	46.81$\pm$0.02&		0.83&	-1.98$\pm$0.01\\
PG\,1545$+$210&	1151$\pm$8&		209$\pm$3&	46.03$\pm$0.00&	45.82$\pm$0.03&		0.29&	-2.17$\pm$0.04\\
PG\,1704$+$608& 1145$\pm$3&		191$\pm$2&	46.63$\pm$0.01&	46.67$\pm$0.03&		0.36&	-1.55$\pm$0.01\\
PG\,2112$+$059& 1172$\pm$7&		263$\pm$4&	46.91$\pm$0.00&	46.82$\pm$0.03&		0.52&	-1.75$\pm$0.03\\
PG\,2233$+$134& 1111$\pm$7&		224$\pm$3&	46.08$\pm$0.01&	46.26$\pm$0.02&		0.40&	-2.25$\pm$0.05\\
PG\,2308$+$098& 1142$\pm$10&	177$\pm$17&	46.29$\pm$0.02&	46.08$\pm$0.15&		0.19&	-1.50$\pm$0.03\\
\enddata
\end{deluxetable*}

 We note three technical points that we took into account during the fit.
 First, we excluded SDSS photometry between 0.6 and 0.7\,$\mu$m in the rest frame to avoid the inclusion of the H$\alpha$ line which 
 generally has a large equivalent width and produces a bump in the SED that deviates from a power law
 which is assumed as the continuum shape in the optical.
 Second, we increased the errors of the SDSS data points to represent the deviation of the optical continuum shape from a power law.
 Here, the deviation is caused by the intrinsic fluctuations in the continuum spectra due to
 Fe emission complex, emission lines other than H$\alpha$, and the host galaxy light.
 Without increasing the error, a fit to the SDSS photometry points
 with a simple power-law function ends up with a rather
 large reduced $\chi^{2}$ values due to very small original SDSS photometry errors.
 From a composite spectrum of a quasar \citep{glikman06}
 placed at $0 \leq z \leq 0.4$, we find the intrinsic fluctuation of the optical spectrum with respect to a simple featureless power law to be
 $\sigma_{m} = 0.035$ mag, and this value is added in quadrature to the 
 original SDSS photometry error in each band.
 Third, we calculate the reduced chi-square, $\chi_{\nu}^{2}$ of the fit,
 and the hot and warm dust properties are taken seriously only if $\chi_{\nu}^{2} < 5$.
 One object (PG\,1322$+$659) is excluded by the $\chi_{\nu}^{2}$ limit for which we had difficulty fitting
 the shape of the $\it{AKARI}$ spectrum (the last object in Figure 15).

 Through the procedure above, we obtain the temperatures and luminosities of the hot and warm dust components of 19 AGNs (2 reverberation mapped AGNs and 17 PG QSOs).
 The fitting results are presented in Figure 15, showing that both the hot and warm dust components exist and contribute to the NIR and MIR spectra of AGNs.
 However, four objects (PG\,1114$+$445, PG\,1116$+$215, PG\,1543$+$489, and PG\,1545$+$210) show an excess of 7.3\,$\mu$m $\it{ISO}$ photometry.
 This excess can hint at the existence of an intermediate-temperature dust component or a continuous dust temperature.

 The SED fit provides formal fitting errors of 0.8\% in $T_{\rm HD}$ and 3.1\% in $L_{\rm HD}$, where
 we obtain a small error in $T_{\rm HD}$ due to the strong constraints provided by the $\it{AKARI}$ spectra.
 The errors in Table 4 indicate the formal fitting errors.
 Since the changes in the assumptions that went into the fitting model could influence the result
 more than the formal fitting errors, we explored how the derived quantities could be affected by
 changing the following conditions in the fit as an attempt to estimate more realistic errors:
 (1) fixing the power-law slope of the continuum, $\alpha$, in Equation (13) to a value that deviates by five times
 the formal error from the fit; (2) the reduction of the fluxes of the $\it{AKARI}$ data by 10\%,
 since $\it{AKARI}$ fluxes could be overestimated by that amount (Figure 3), and
 (3) the subtraction of the host galaxy contribution.
 The mean absolute differences in the fitted quantities from the change in the power law slope are 3.3\% ($T_{\rm HD}$; between 0.6\% and 8.7\%) and 11\%
 ($L_{\rm HD}$; between 2.6\% and 36\%), and from the flux of $\it{AKARI}$ data are 2.7\% (between $<$1\% and 6.5\%) in $T_{\rm HD}$
 and 15\% (between 11\% and 22\%) in $L_{\rm HD}$. The reduction in the $\it{AKARI}$ flux leads to the reduction in $T_{\rm HD}$ and $L_{\rm HD}$.
 We note that the change of the power-law slope and the $\it{AKARI}$ flux value keeps the $\chi_{\nu}^{2}$ value to less than five, except for one object.
 We estimated the contribution of the host galaxy light in $L_{\rm 5100}$ \citep{peterson04,vestergaard06}
 using Equation (1) of \citealt{shen11}, and subtracted the host galaxy light assuming an elliptical galaxy SED shape \citep{bruzual03}
 with the conditions of solar metallicity, $\tau=0.1$ Gyr, $\rm t = 10$ Gyr, and a Chabrier IMF. We examined how the fitting result changes
 with the subtraction of the host galaxy component, and find mean differences of 1.5\% (between $<$1\% and 7.5\%)
 in $T_{\rm HD}$ and 7.5\% (between $<$1\% and 22\%) in $L_{\rm HD}$, where $T_{\rm HD}$ decreases and $L_{\rm HD}$ increases in most cases.
 With these considerations, the error of $T_{\rm HD}$ is realistically $\sim$4\%, while the error in $L_{\rm HD}$
 could be as much as $\sim$14\%. For the warm dust component, we have only two photometric data points for most AGNs
 and a single point in some AGNs; we caution that the fitted result have larger errors and can be affected more easily
 by the fitting method than the hot dust component.

 The left panel of Figure 16 shows $L_{\rm bol}$ versus $T_{\rm HD}$. The $L_{\rm bol}$ values are estimated from
 $L_{\rm 5100}$ using a bolometric correction factor of 10.3 \citep{richards06}, and the $L_{\rm 5100}$ values are taken from
 \cite{bentz09} and \cite{vestergaard06}. The $L_{\rm bol}$ and the $T_{\rm HD}$ values have a positive correlation as
\begin{equation}
 \frac{T_{\rm HD}}{1800\,\mathrm{K}} =10^{-0.231\pm0.003} \left( \frac{L_{\rm bol}}{10^{46}~\mathrm{erg~ s^{-1}}} \right)^{0.061\pm0.005},
\end{equation}
 and the Pearson correlation coefficient is 0.690.
 We find that the mean $T_{\rm HD}$ and $T_{\rm WD}$ are $1083\,\rm{K}$ and $221\,\rm{K}$, respectively,
 which is lower than the commonly cited $T_{\rm HD}$ of 1500\,K (e.g., \citealt{barvainis87,elvis94}), and somewhat lower than the
 value quoted in a recent study (1260\,K, from \citealt{glikman06}).
 The difference does not seem to arise from technical parts of the fitting method. When we
 measure the $T_{\rm HD}$ using a composite spectrum of \cite{glikman06} with our method,
 we find the result to be identical to that of the previous study (i.e., $1260\,\rm{K}$).
 Also, we estimate $T_{\rm HD}$ values of our sample using the methods of \cite{glikman06},
 and the measured $T_{\rm HD}$ values used in this way are consistent with our results.

 The right panel of Figure 16 shows the hot dust luminosities versus the $L_{\rm bol}$ values for the 19 AGNs.
 For the bolometric luminosity, we use $L_{\rm 5100}$ from our
 SED fitting multiplied by a bolometric correction factor of 10.3 \citep{richards06}.
 In the figure, we also indicate lines with three different values for the 
 covering factor of the hot dust component ($\rm CF_{HD}$). 
 Here, the covering factor is given by the ratio of the luminosity from 
 the hot dust component to $L_{\rm bol}$ ($\rm CF_{HD}$=$L_{\rm HD}/L_{\rm bol}$; e.g., \citealt{maiolino07}).
 We find that $L_{\rm HD}$ correlates well with $L_{\rm bol}$, 
 and their mean covering factor is found to be 0.38. 
 The measured mean covering factor is similar to several previous results \citep{sanders89,elvis94,roseboom13}.
 However, we note that there is no significant downturn in $\rm CF_{HD}$ with $L_{\rm bol}$, in contrast to previous results which reported that
 $\rm CF_{HD}$ has an anti-correlation with $L_{\rm bol}$ at the high luminosity end of $10^{46.5}~\mathrm{erg~s^{-1}}$
 \citep{wang05,maiolino07,treister08,mor11,mor12}. This discrepancy is quite 
 possibly due to the low number statistics in our data, considering that
 the anti-correlation found by other groups has a large scatter.

\begin{figure*}
\begin{center}
\centering
\figurenum{16}
\scalebox{1.1}{\plotone{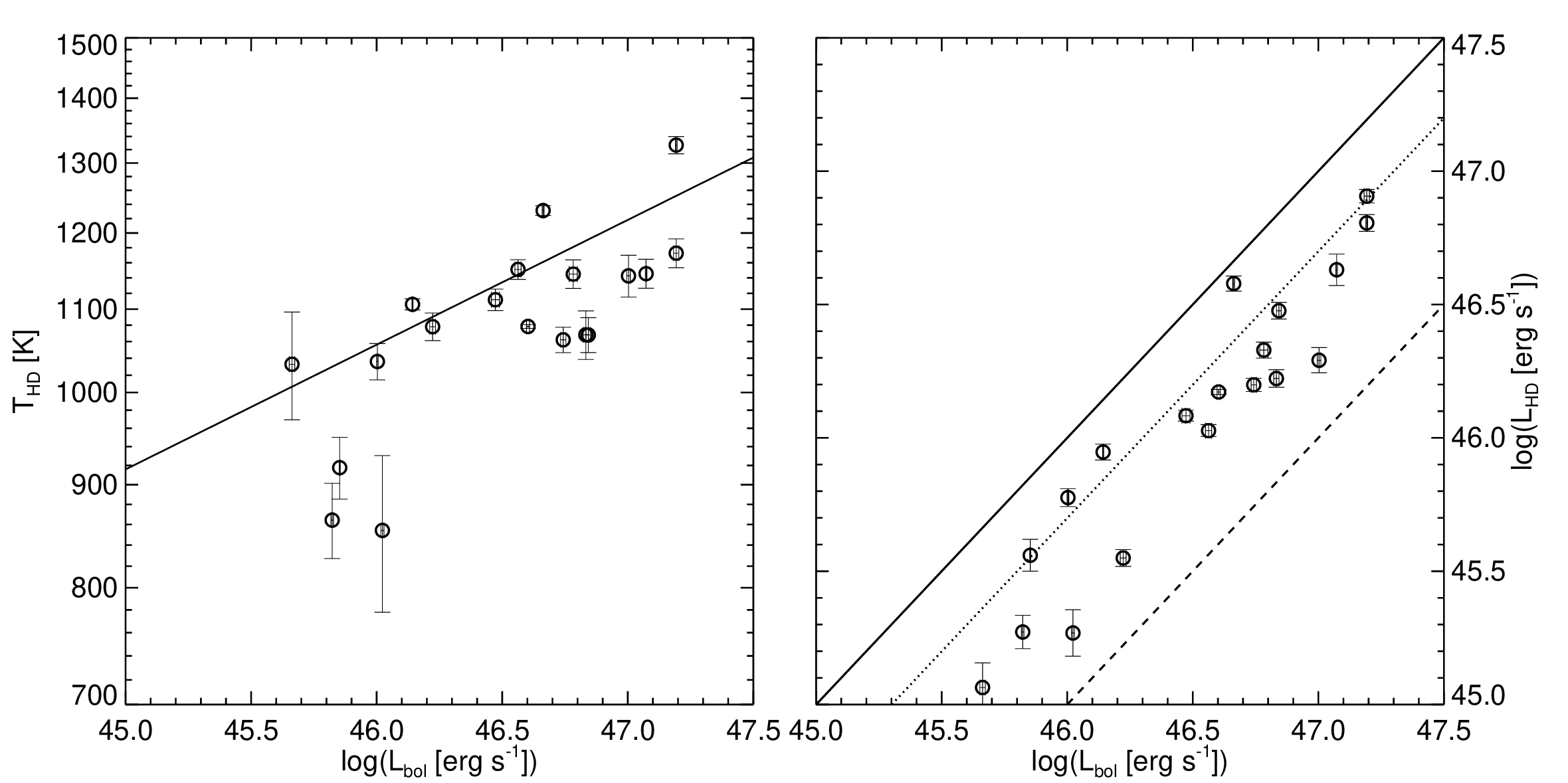}}
\caption{Left: $L_{\rm bol}$ vs. $T_{\rm HD}$ of the 19 AGNs. The solid line
indicates the best-fit correlation.
Right: comparison between $L_{\rm bol}$ and $L_{\rm HD}$.
The solid, dotted, and dashed lines denote $\rm CF_{HD}=$ 1.0, 0.5, and 0.1, respectively, and
the measured mean $\rm CF_{HD}$ of the 19 AGNs is 0.38.}
\end{center}
\end{figure*}

 \section{SUMMARY}
 
 Using the $\it{AKARI}$ IRC in grism mode, we obtained 2.5--5.0\,$\mu$m spectra
 of 83 nearby (0.002$<$ z $<$0.48), bright ($K$ $<$14 mag) type-1 AGNs. 
 The Brackett lines are identified in 11 AGNs in the sample, allowing us to derive $M_{\rm BH}$
 estimators that are based on Br$\alpha$ and Br$\beta$ line properties.
 We find the Br$\beta$-based $M_{\rm BH}$ estimator using the measured line width and luminosity so as to be useful
 with an intrinsic scatter of $\sim$0.1 dex or less with respect to the $M_{\rm BH}$ values from the
 reverberation mapping method. On the other hand, the Br$\alpha$-based estimator shows a larger scatter and
 no strong dependence on $\rm{\sigma_{Br\alpha}}$, demonstrating the difficulty of $\rm{\sigma_{Br\alpha}}$ measurements
 and the need for future data to refine the estimator.
 However, both $M_{\rm BH}$ estimators can potentially be applied to derive $M_{\rm BH}$ values of dusty AGNs
 due to the low extinction in the Brackett line wavelengths. 
 The continuum shape is well traced for all of the objects, and we derived
 the temperatures and luminosities of the hot and warm dust components
 of 19 luminous AGNs ($L_{\rm{bol}} > 10^{45.7}~\mathrm{erg~s^{-1}}$) for which host galaxy contamination is expected
 to be small ($<$ 10\%). The use of the $\it{AKARI}$ spectra provides the advantage 
 of tracing the NIR continuum densely, thus improving the dust temperature 
 measurements. Our results show the hot dust temperature to be about 
 $1100\,\rm{K}$, which is lower than the commonly cited value of 1500\,K but broadly consistent with most recent 
 hot dust temperature measurements. The covering factor of the hot dust
 component is also derived with a mean $\rm CF_{HD}$ of 0.38, consistent with previous
 measurements.
 The $\it{AKARI}$ spectral atlas and the tabulated spectral measurements 
 can be downloaded from this article, revealing
 the rarely studied spectral range of 2.5--5.0\,$\mu$m for low-redshift AGNs.

\begin{deluxetable}{ccc}
\tabletypesize{\scriptsize}
\tablewidth{0pt}
\tablenum{5}
\tablecaption{Spectrum of Mrk 335\label{tbl6}}
\tablehead{
\colhead{$\lambda$}&	\colhead{$f_{\lambda}$}&	\colhead{$f_{\lambda}$ Uncertainty}\\
\colhead{[$\mu$m]}&	\colhead{[$\rm{erg~s^{-1}~cm^{-2}~\mu m^{-1}}$]}&	\colhead{[$\rm{erg~s^{-1}~cm^{-2}~\mu m^{-1}}$]}\\
}
\startdata
2.543&  3.500E-11&  4.653E-13\\
2.552&  3.363E-11&  4.595E-13\\
2.562&  3.207E-11&  3.482E-13\\
2.572&  3.133E-11&  2.153E-13\\
2.581&  3.106E-11&  3.536E-13\\
2.591&  3.118E-11&  5.891E-13\\
2.601&  3.128E-11&  2.988E-13\\

\enddata
\tablecomments{This table display only a part of spectrum of Mrk 335. The entire spectra of 83 AGNs and the composite spectrum of 48 PG QSOs
are available as a tar file in the electronic version of $Astrophysical~Journal~Supplement~Series$.}
\end{deluxetable}

\acknowledgments

 This work was supported by the Creative Initiative Program of the National Research Foundation of Korea (NRF),
 No. 2008-0060544, funded by the Korea government (MSIP).
 D.K. acknowledges the fellowship support from the grant NRF-2014-Fostering Core Leaders of
 Future Program No. 2014-009728 funded by the Korean government.
 This work was also supported by grant No. 2012R1A4A1028713 (M.G.L. and H.M.L.), grant No. 2012R1A2A2A01006087 (J.H.W.),
 from the National Research Foundation of Korea (NRF) grant funded by the Korea Government (MSIP).
 This research is based on observations with $\it{AKARI}$, a JAXA project with the participation of ESA.
 We thank the anonymous referee for useful comments.

\section{Appendix}
 We provide reduced 2.5--5.0\,$\mu$m spectra of 83 AGNs and the composite spectrum of 48 PG QSOs. Table 5
 is an example spectrum of Mrk\,335. The full version of the reduced spectra is available in machine readable table form.

\pagebreak
%%%%%%%%%%%%%%%%%%%%%%%           REF            %%%%%%%%%%%%%%%%%%%%%%%%%%%%%%

% % %figure % % % % % % % % % % % % % % % % % % % % % % % % % % % % % % % % % % % % %

\begin{figure}
\figurenum{5}
\includegraphics[width=12cm,height=4cm]{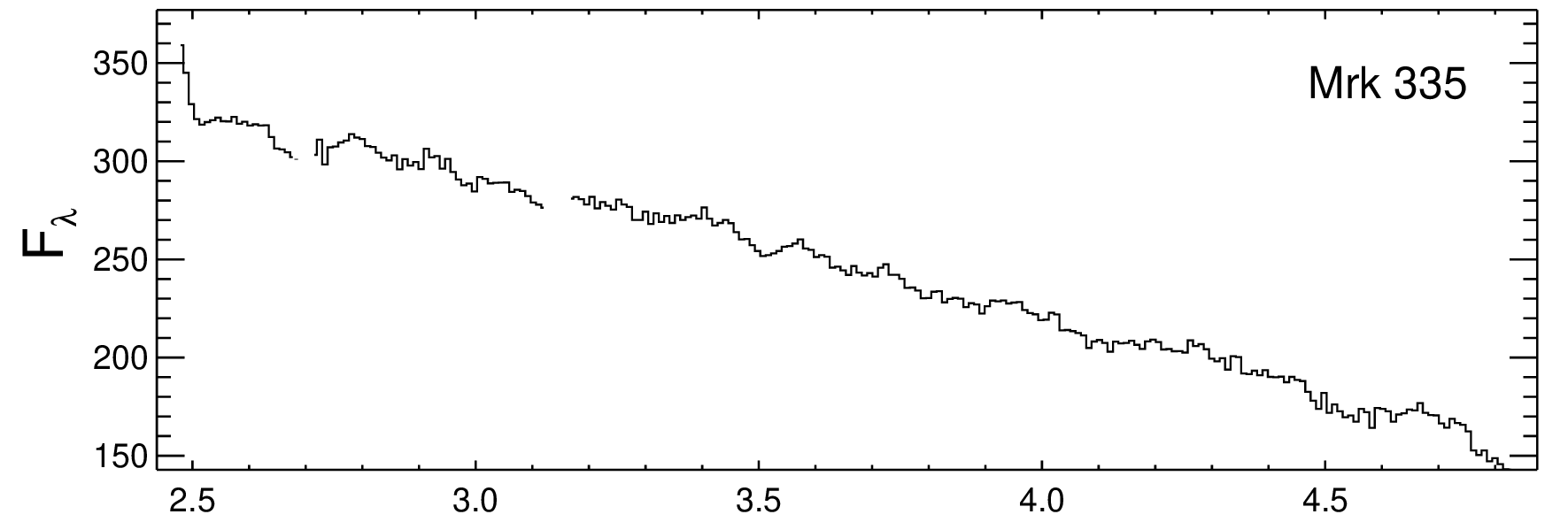}
\includegraphics[scale=0.20]{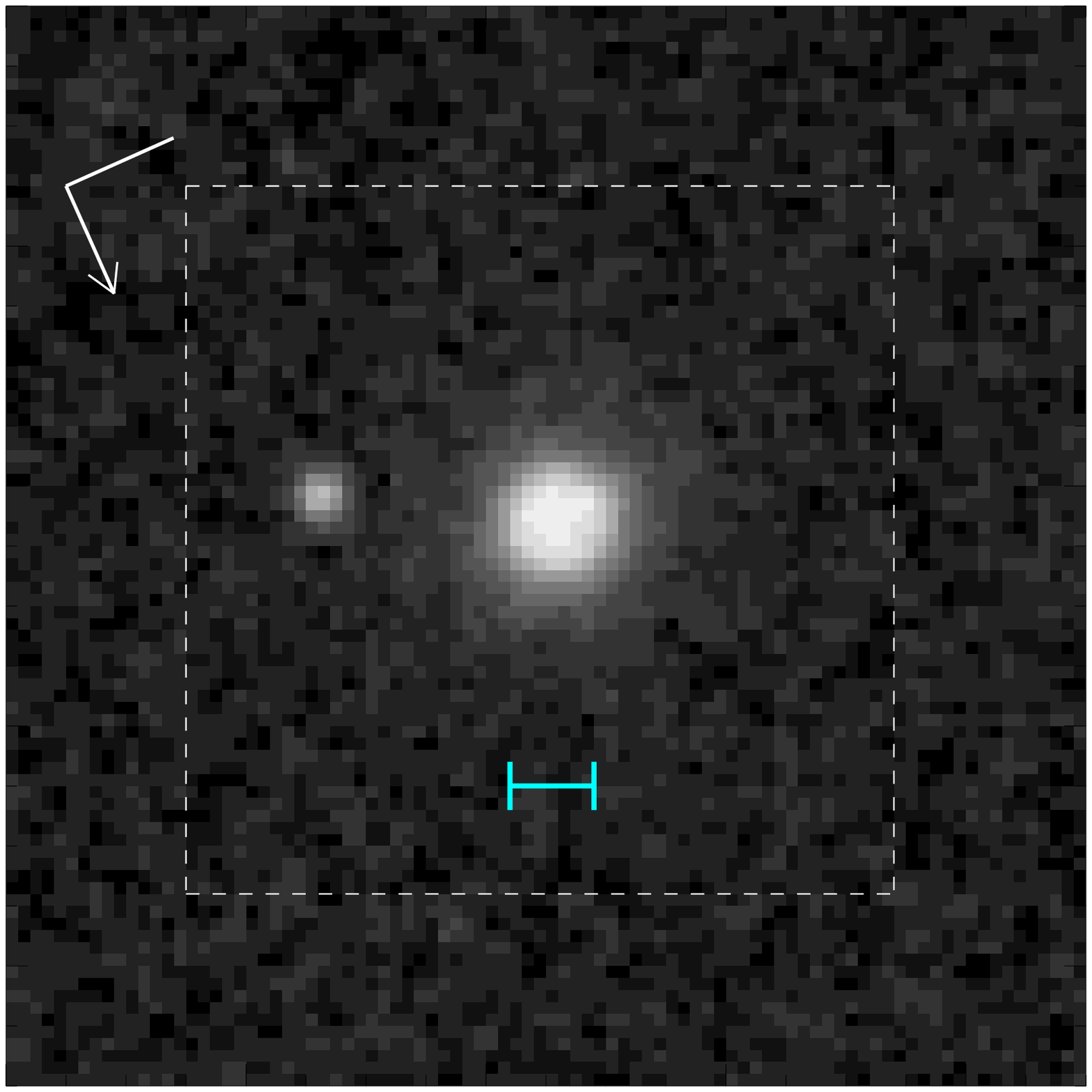}\\
\includegraphics[width=12cm,height=4cm]{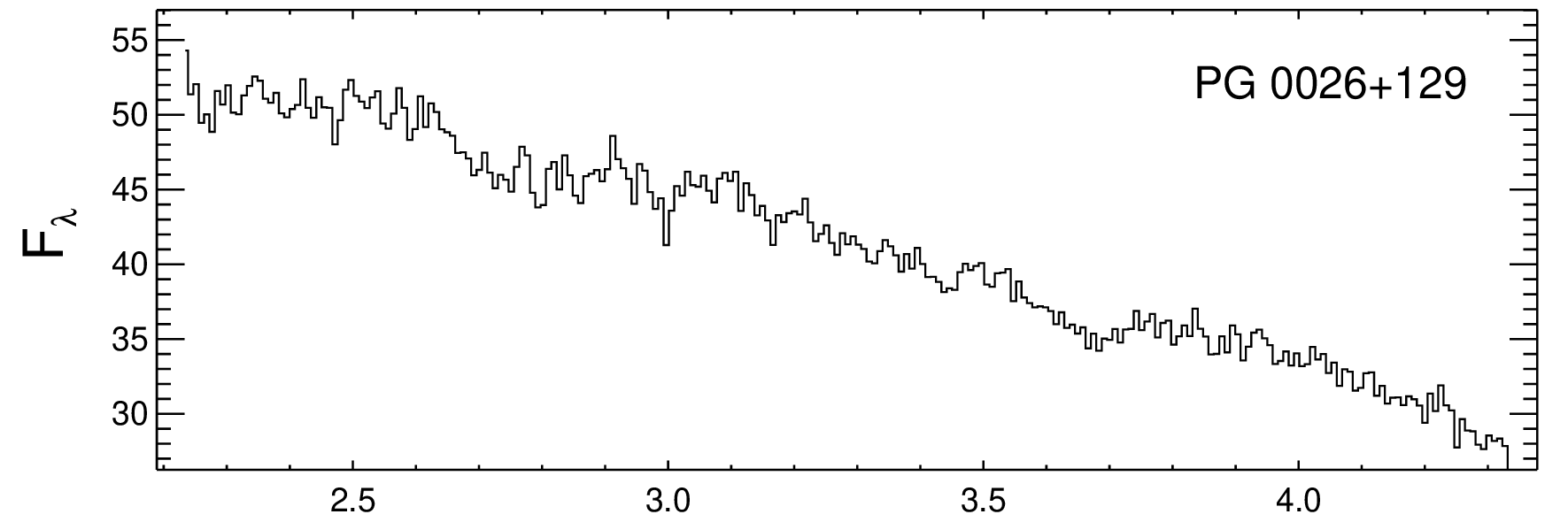}
\includegraphics[scale=0.20]{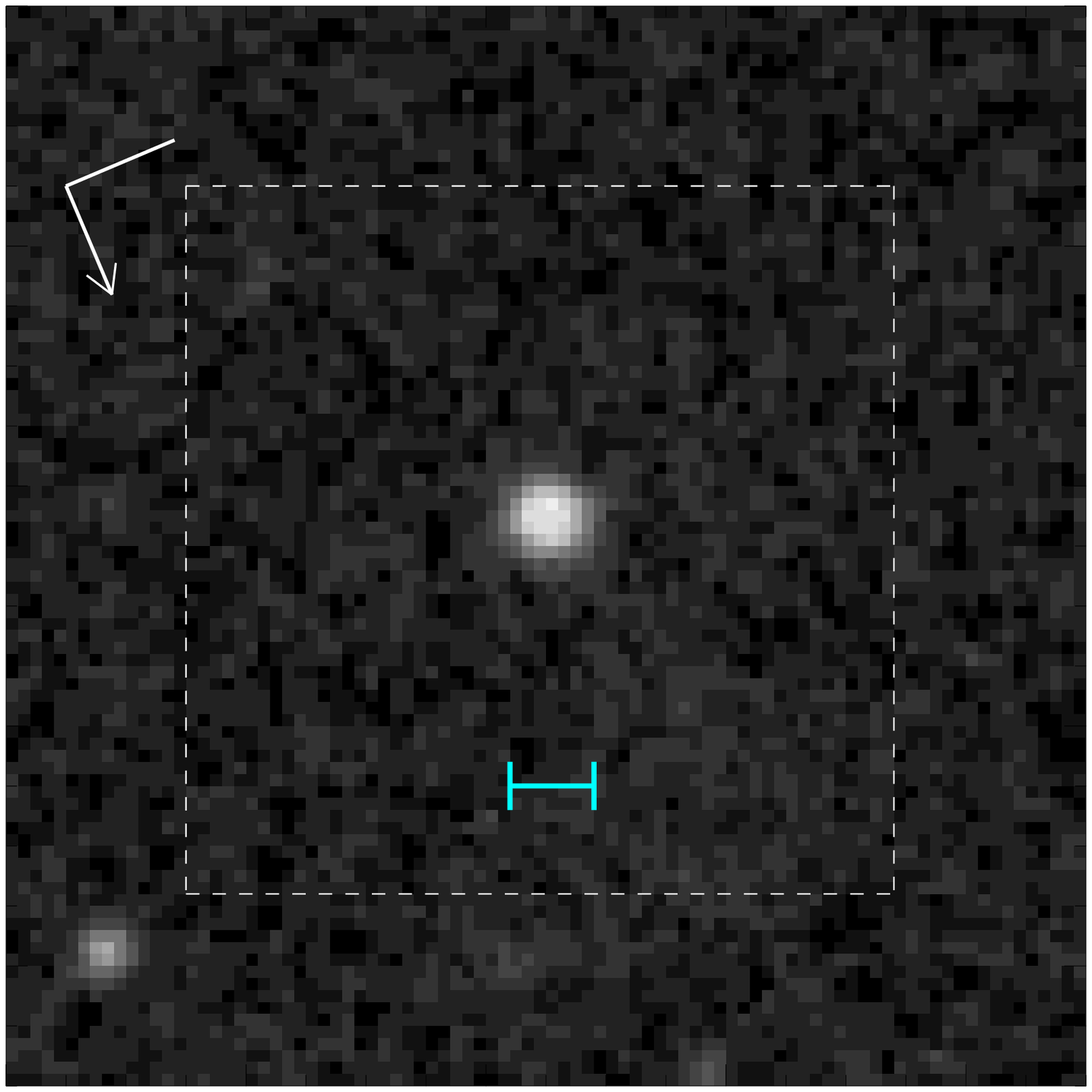}\\
\includegraphics[width=12cm,height=4cm]{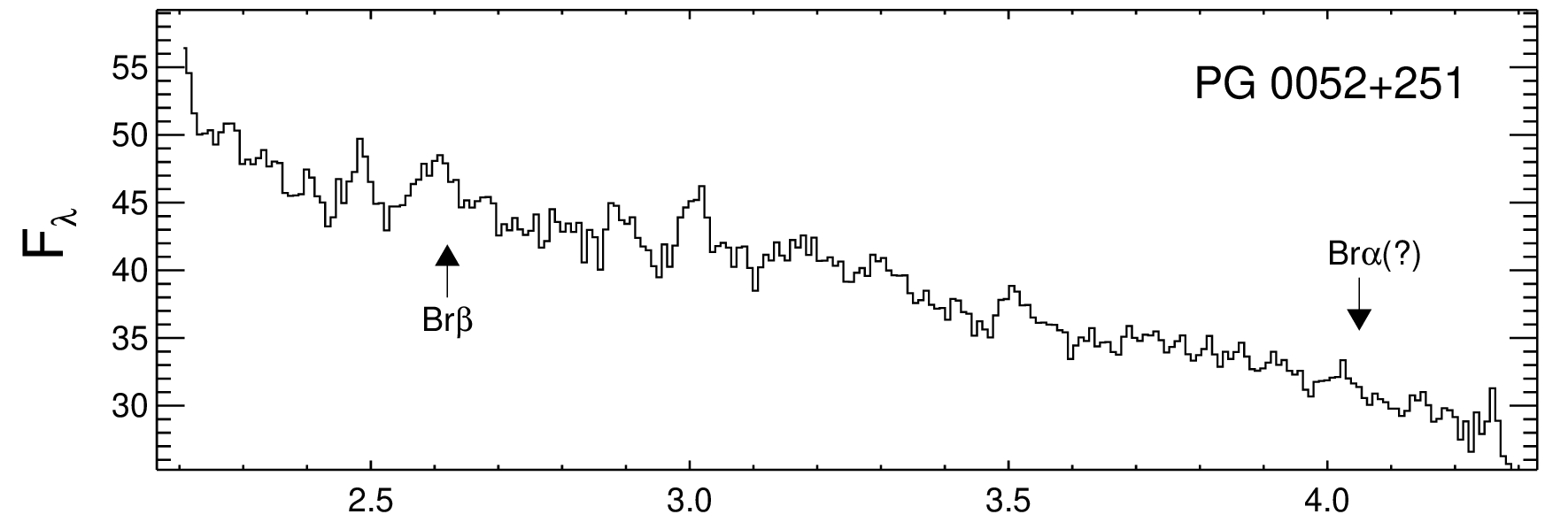}
\includegraphics[scale=0.20]{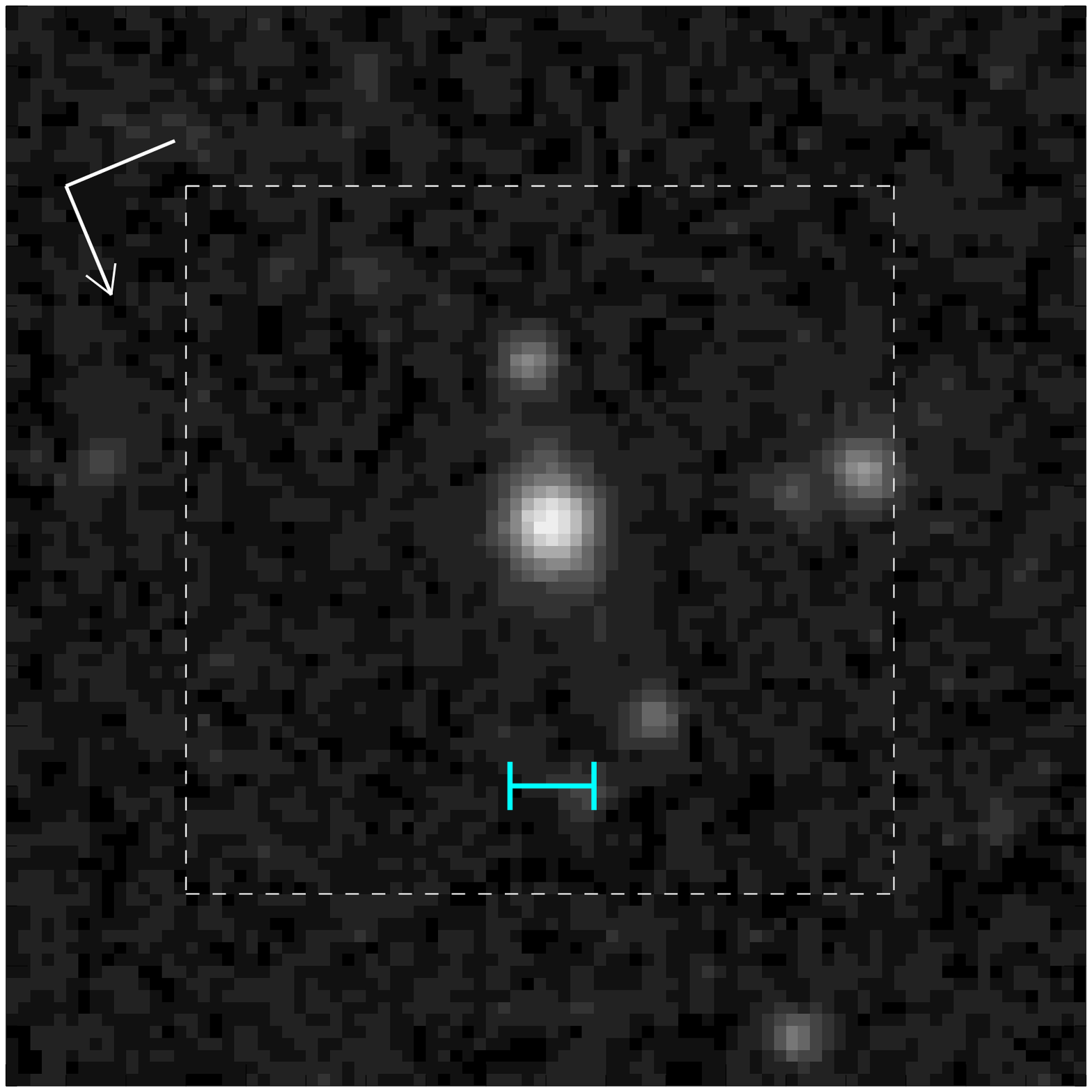}\\
\includegraphics[width=12cm,height=4cm]{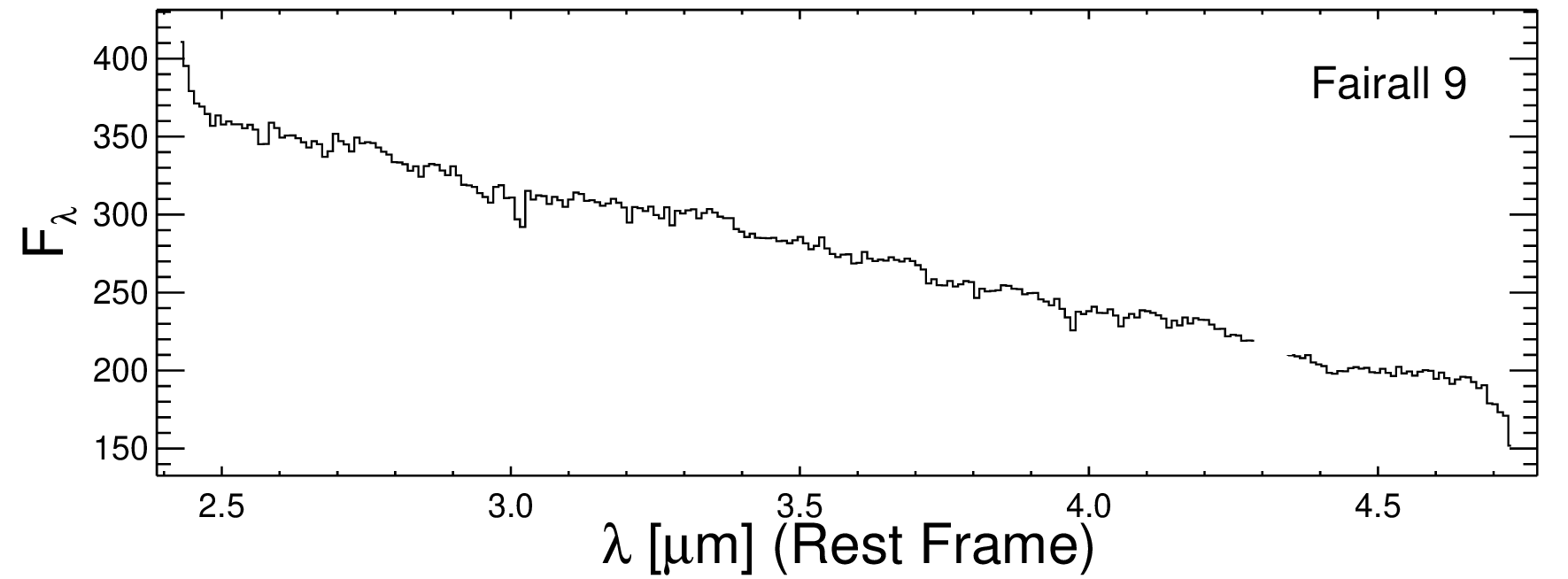}
\includegraphics[scale=0.20]{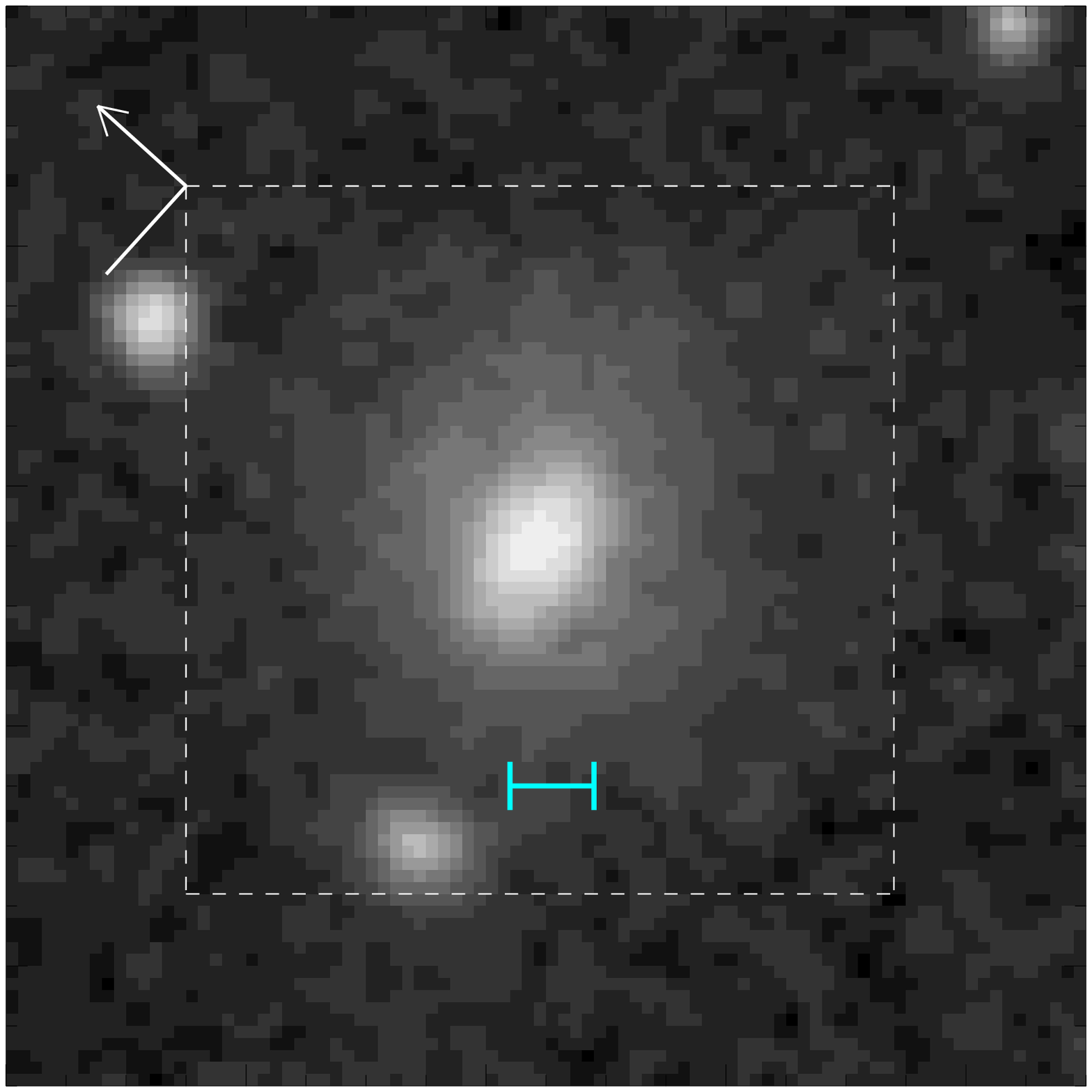}\\
\caption{Spectra and images of AGNs in the QSONG sample. The left panels show de-redshifted 2.5--5.0\,$\mu$m spectra and the spectral regions affected by remaining hot or bad pixels are masked out. The ordinate is in units of $\mathrm{10^{-13}~erg~s^{-1}~ cm^{-2}~ \mu m^{-1}}$. Several hydrogen lines (P$\alpha$, Br$\beta$, and Br$\alpha$) and molecular lines (PAH, ${\rm CO_{2}}$, and CO) are marked on the spectra. When the line detection is not obvious, the line is marked with a question mark. The right panels indicate object images from the digitized sky survey (DSS). The white dashed line box and the cyan solid bar indicate 1\farcm 0 $\times$ 1\farcm 0 slit window and 7\farcs3 width of extraction aperture, respectively. The arrow at the top left denotes north.}
\end{figure}
\clearpage

\begin{figure}
\figurenum{5}
\includegraphics[width=12cm,height=4cm]{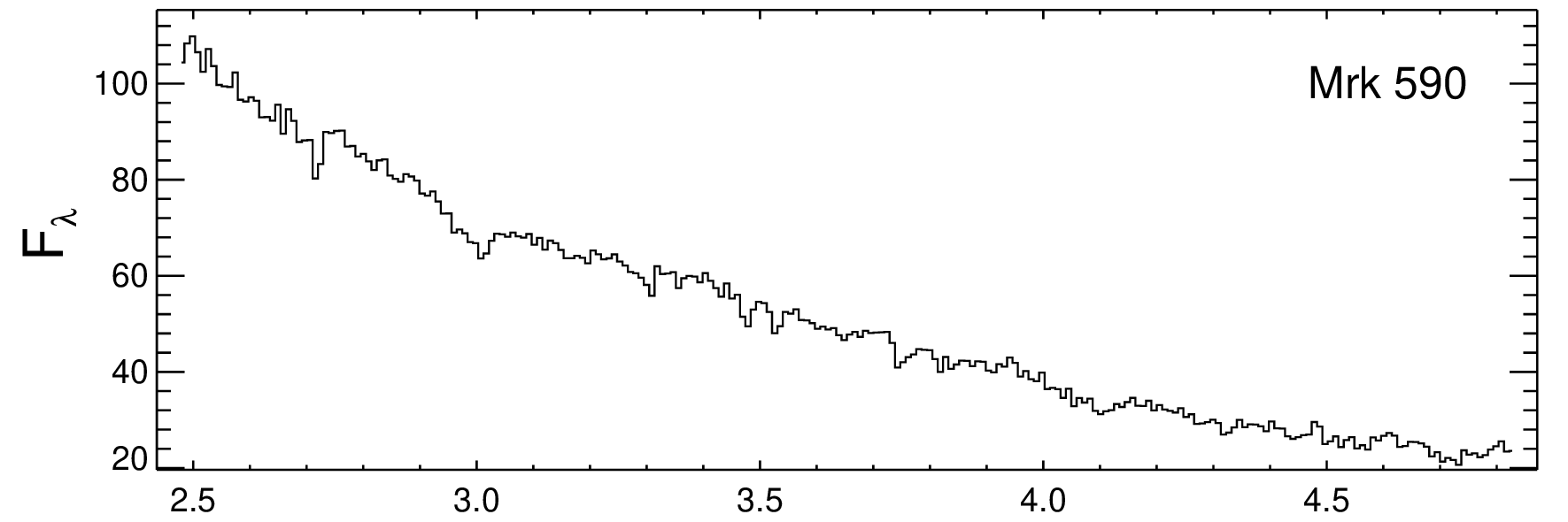}
\includegraphics[scale=0.20]{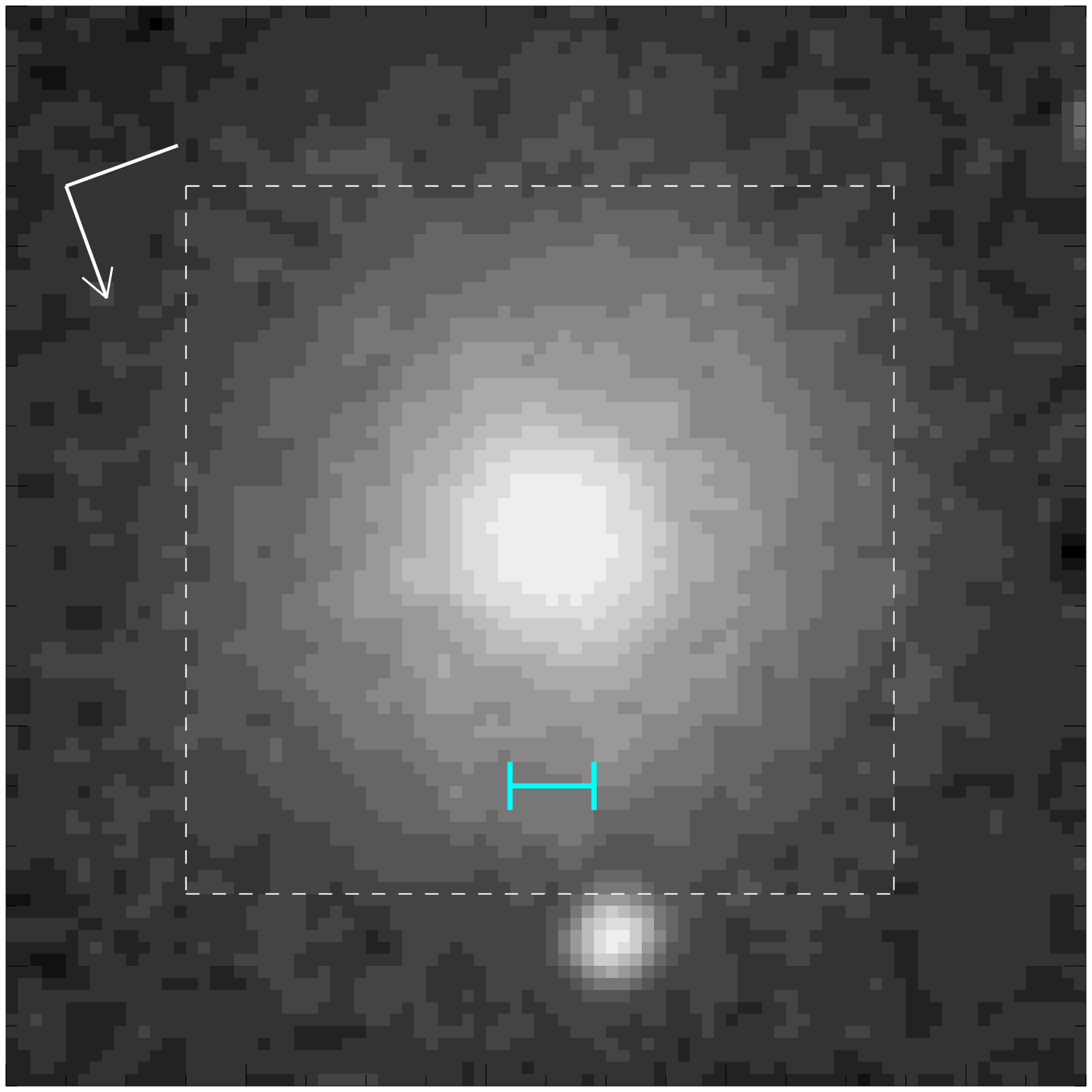}\\
\includegraphics[width=12cm,height=4cm]{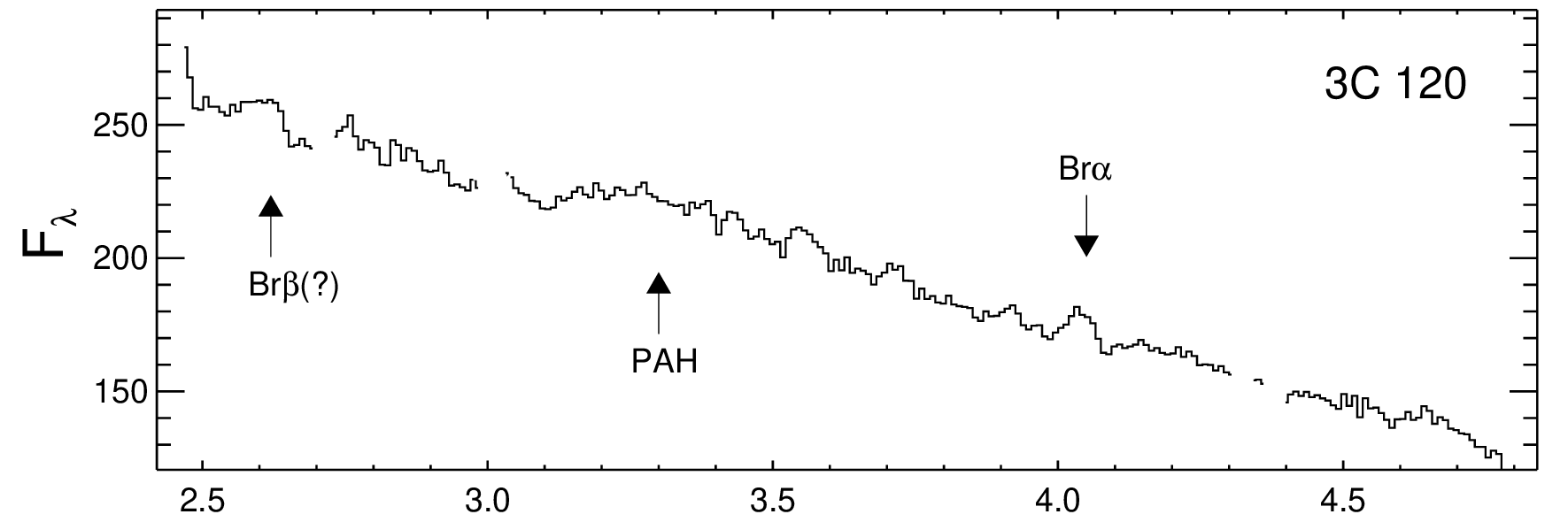}
\includegraphics[scale=0.20]{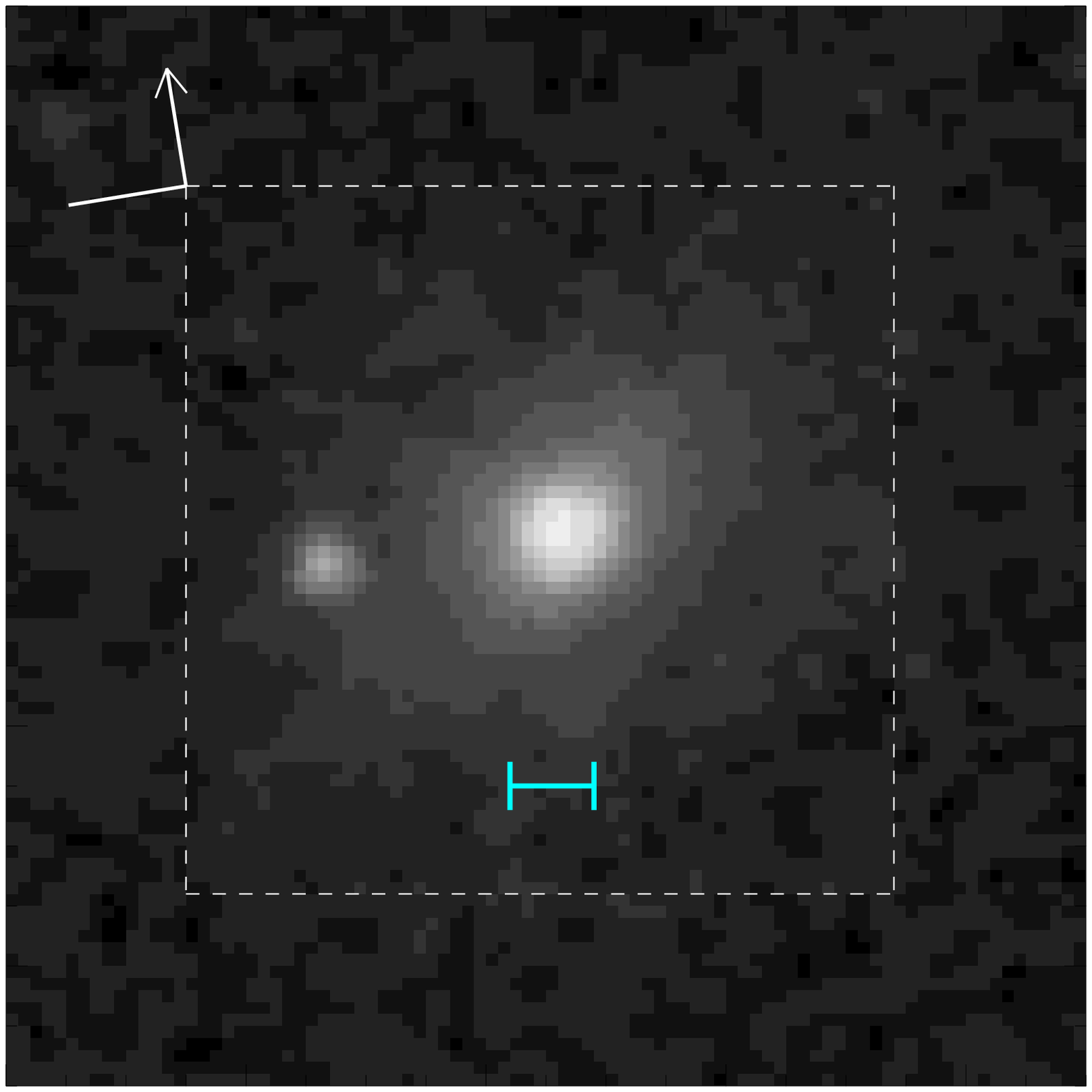}\\
\includegraphics[width=12cm,height=4cm]{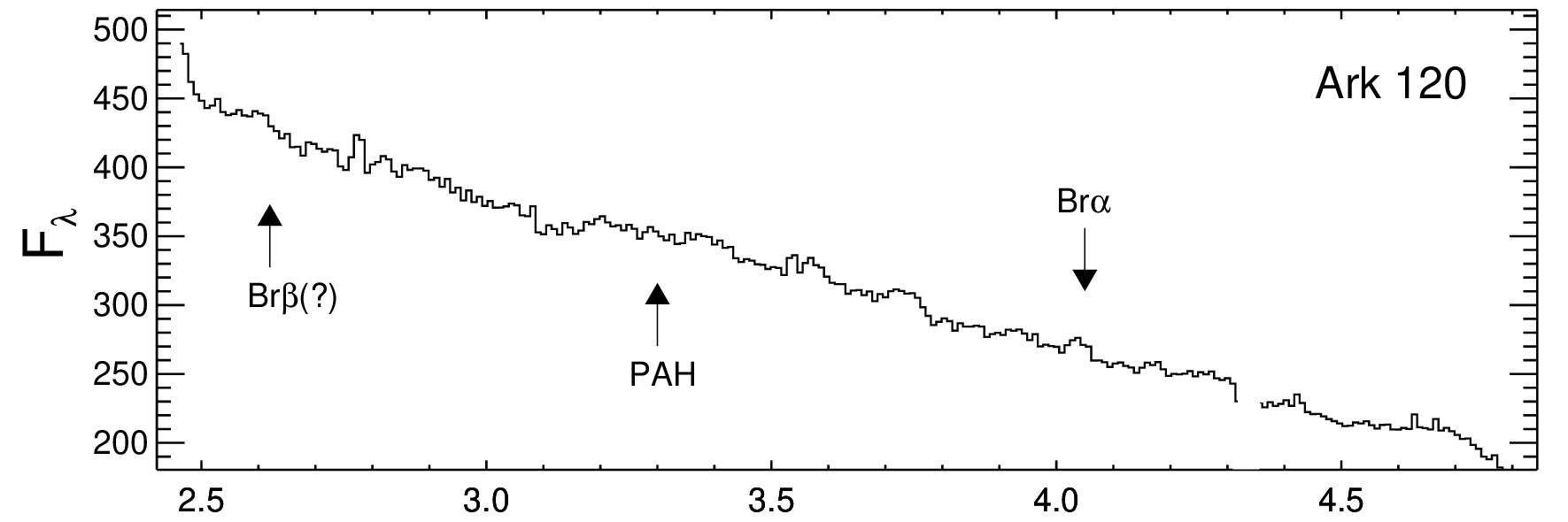}
\includegraphics[scale=0.20]{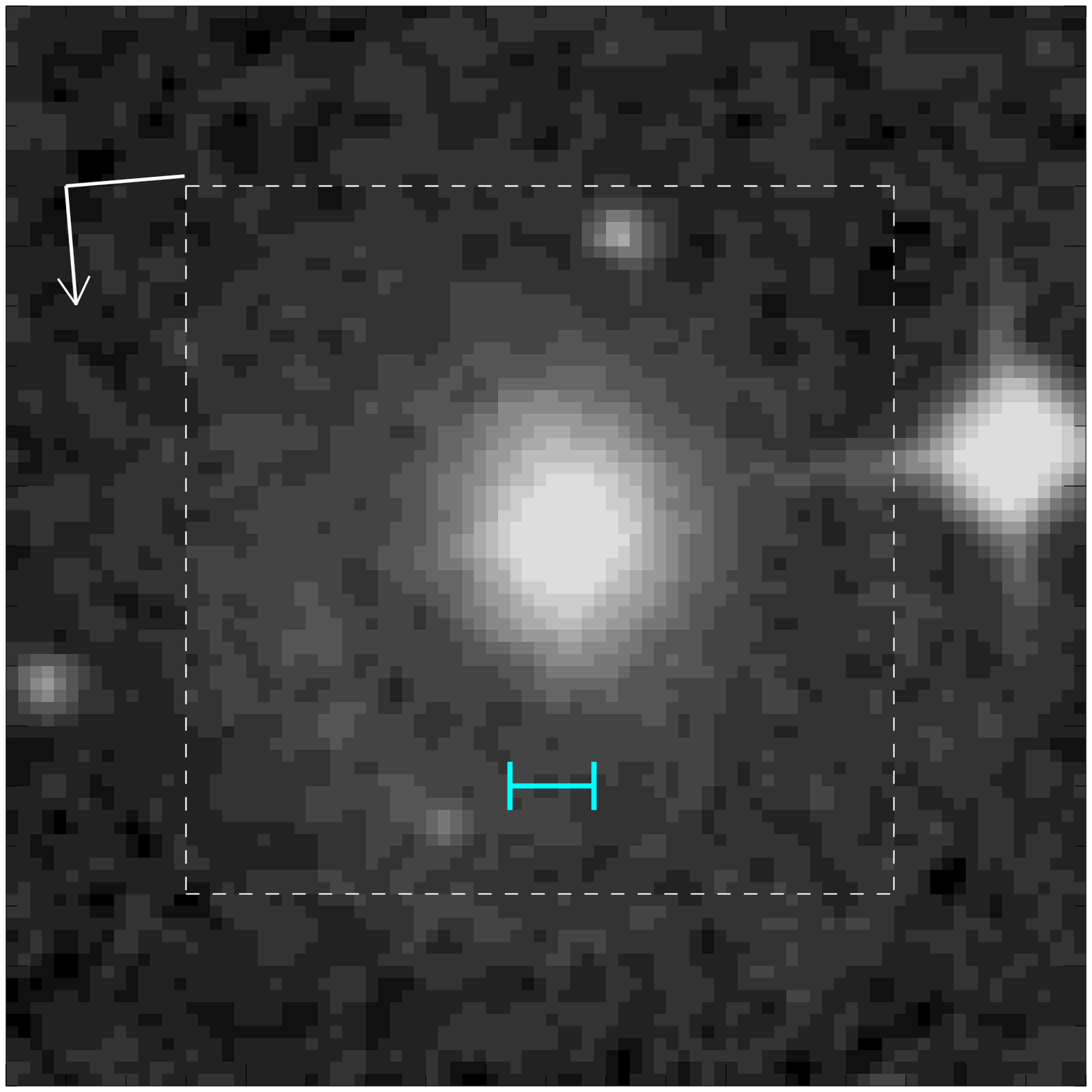}\\
\includegraphics[width=12cm,height=4cm]{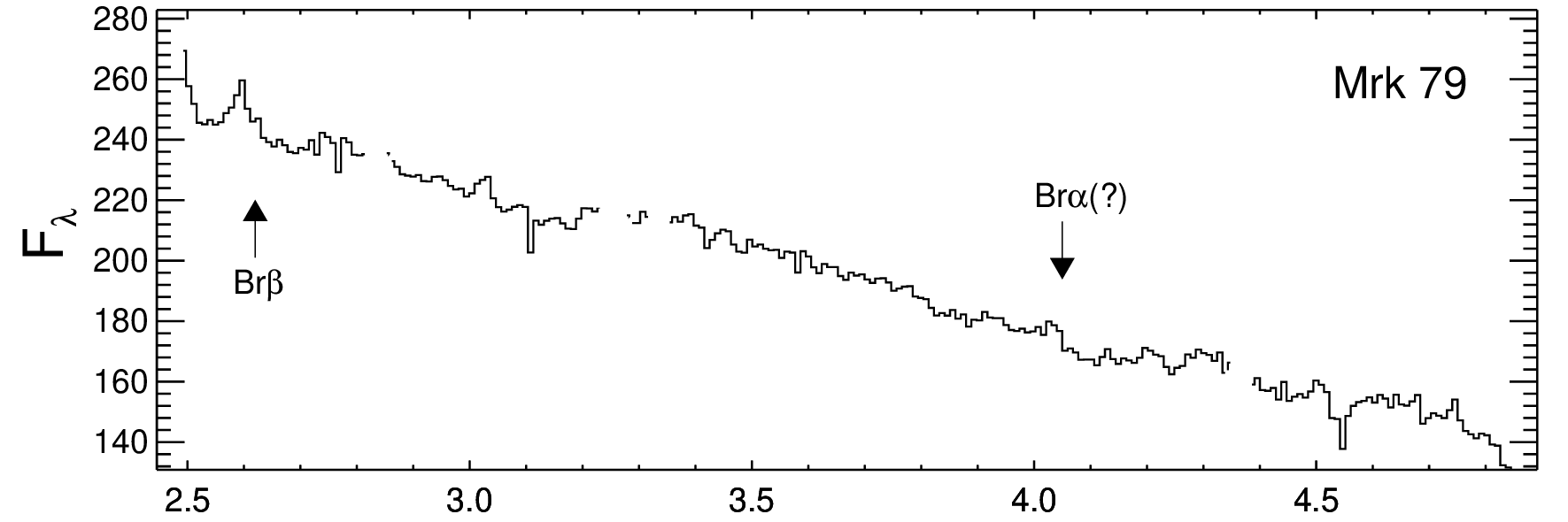}
\includegraphics[scale=0.20]{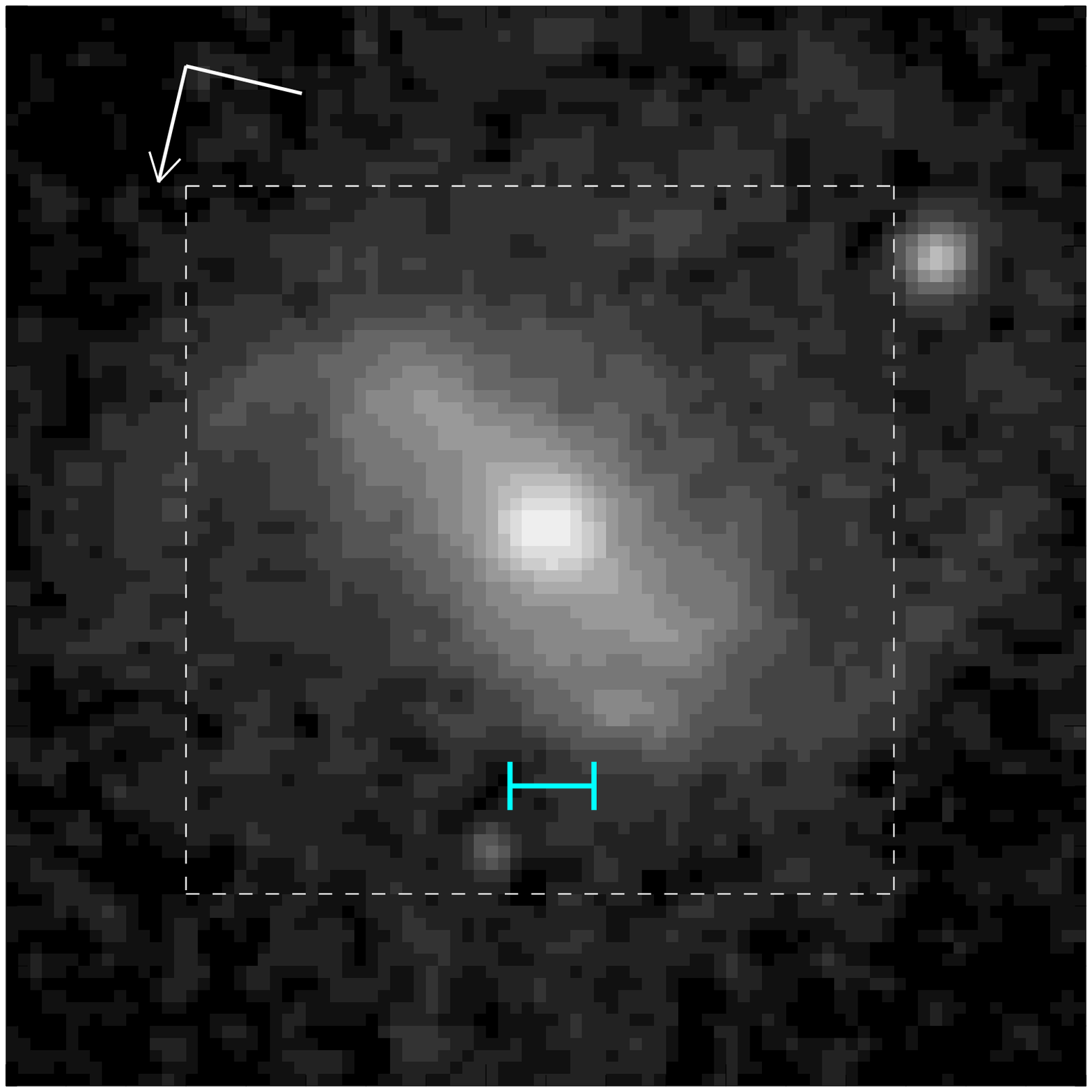}\\
\includegraphics[width=12cm,height=4cm]{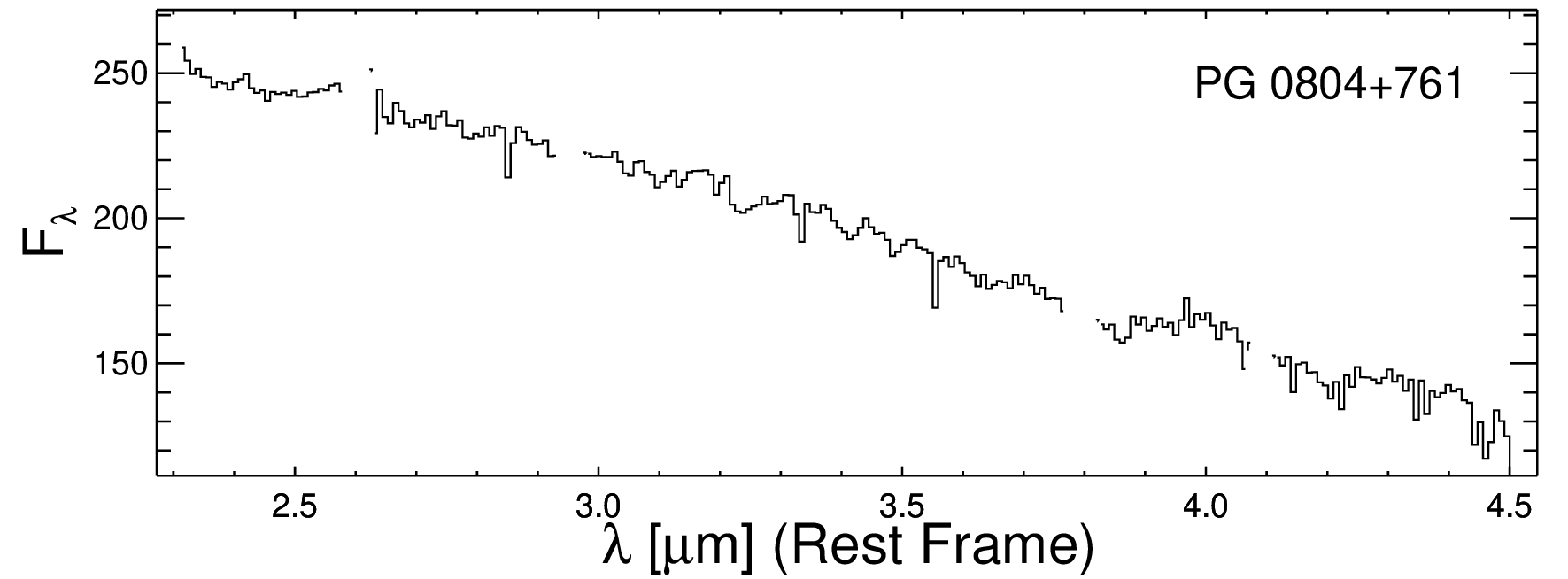}
\includegraphics[scale=0.20]{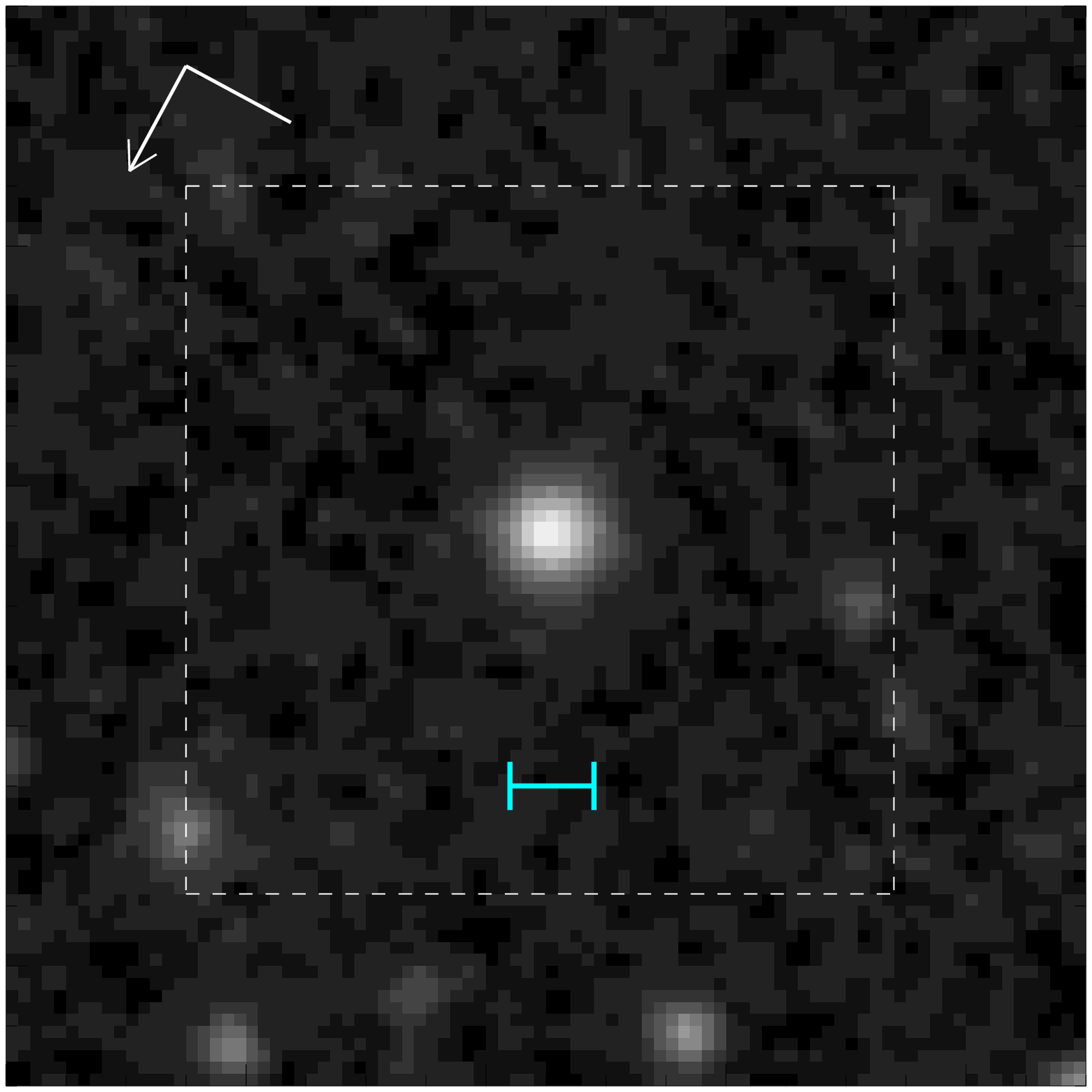}\\

\caption{Continued}
\end{figure}
\clearpage

\begin{figure}
\figurenum{5}
\includegraphics[width=12cm,height=4cm]{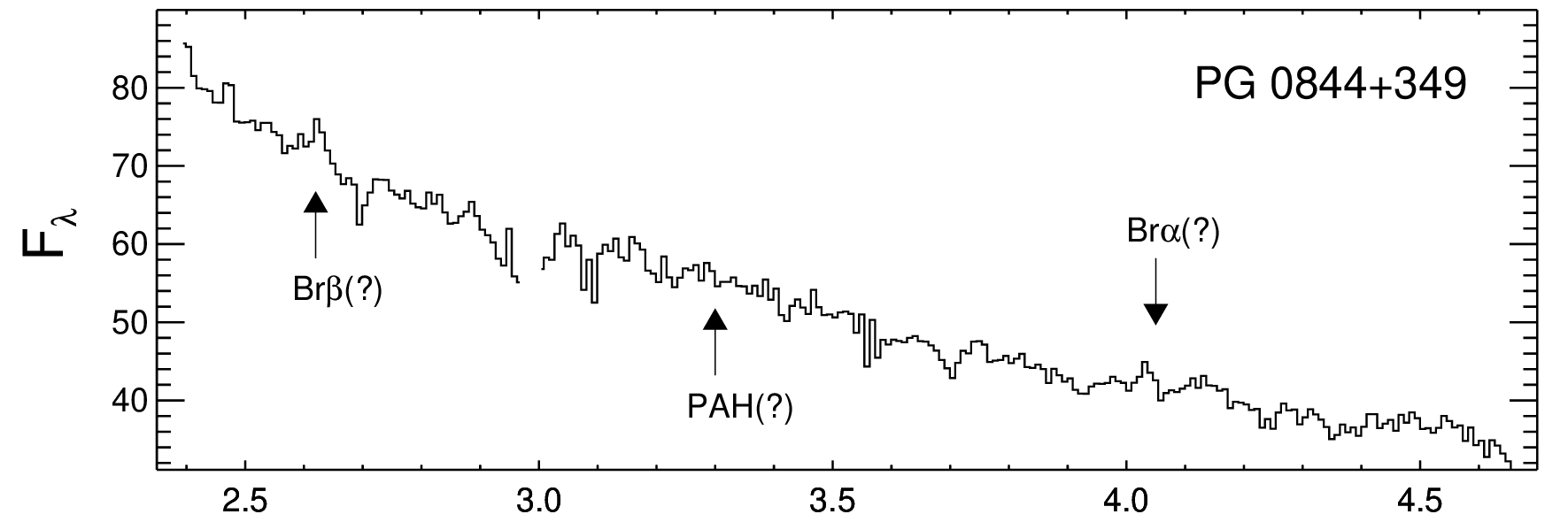}
\includegraphics[scale=0.20]{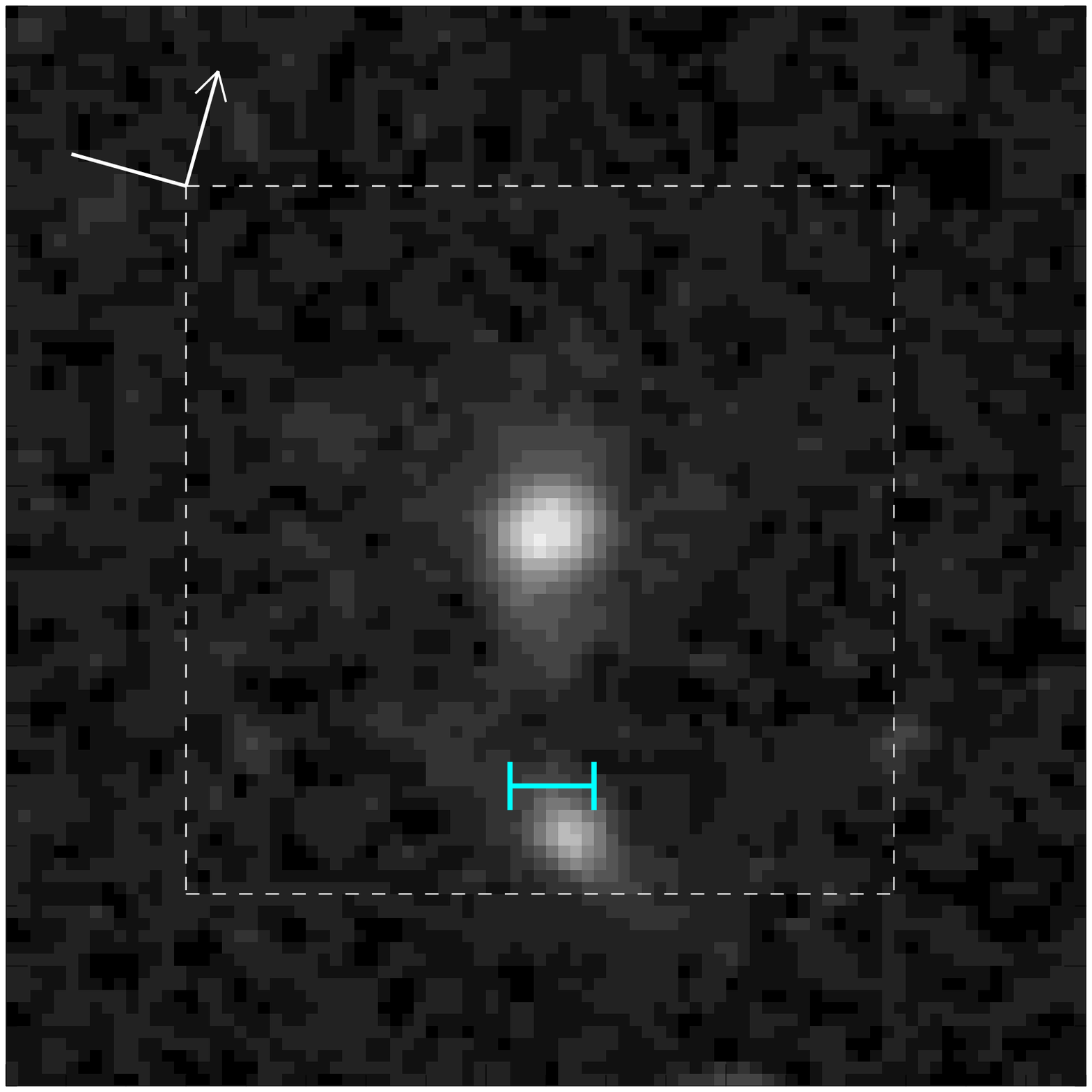}\\
\includegraphics[width=12cm,height=4cm]{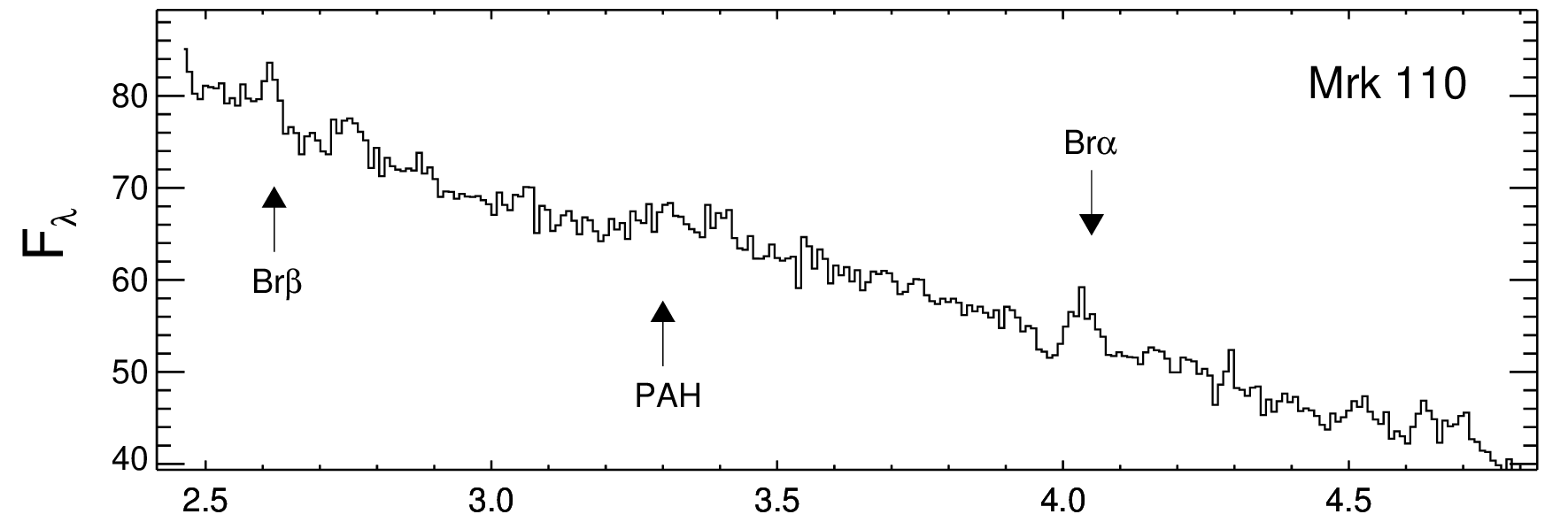}
\includegraphics[scale=0.20]{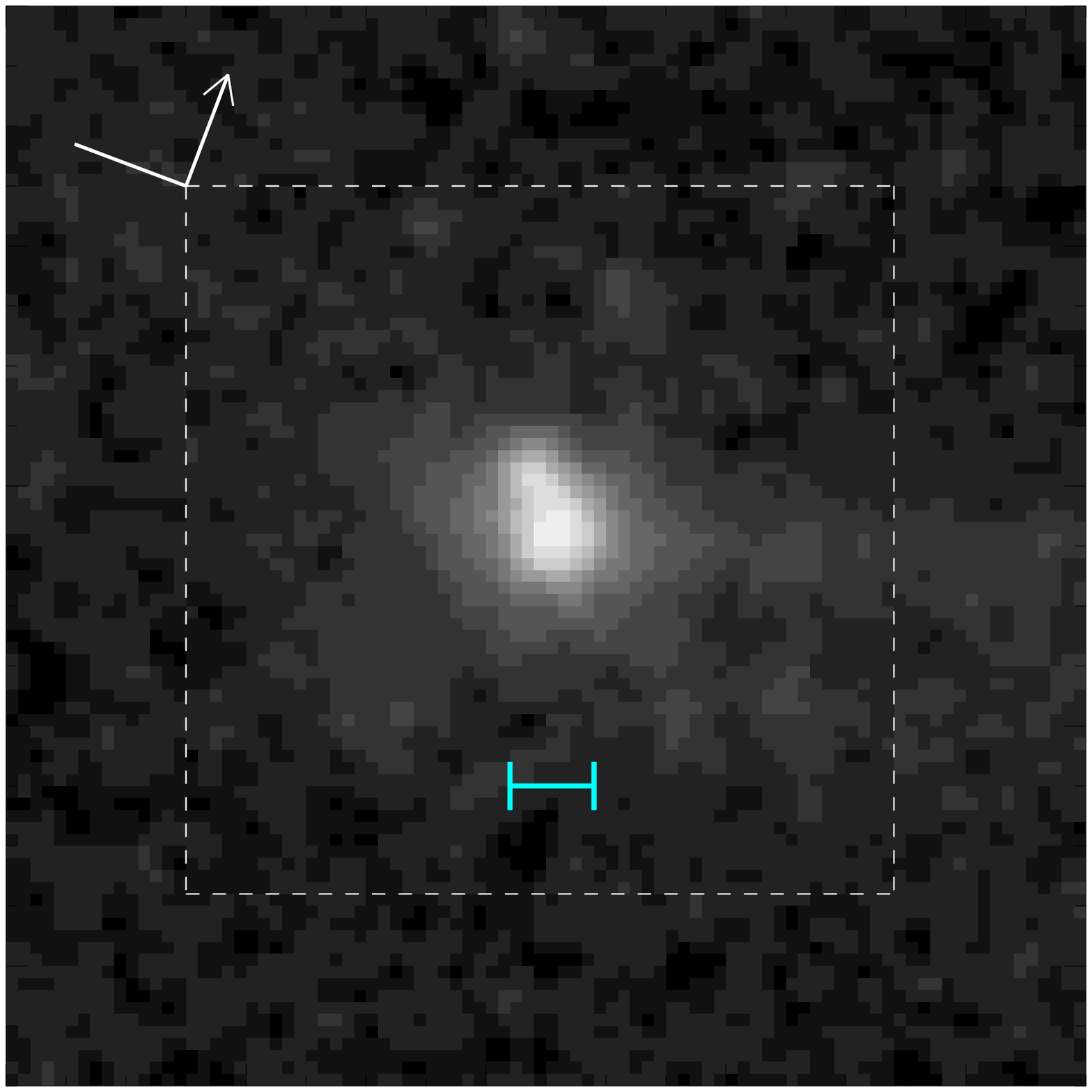}\\
\includegraphics[width=12cm,height=4cm]{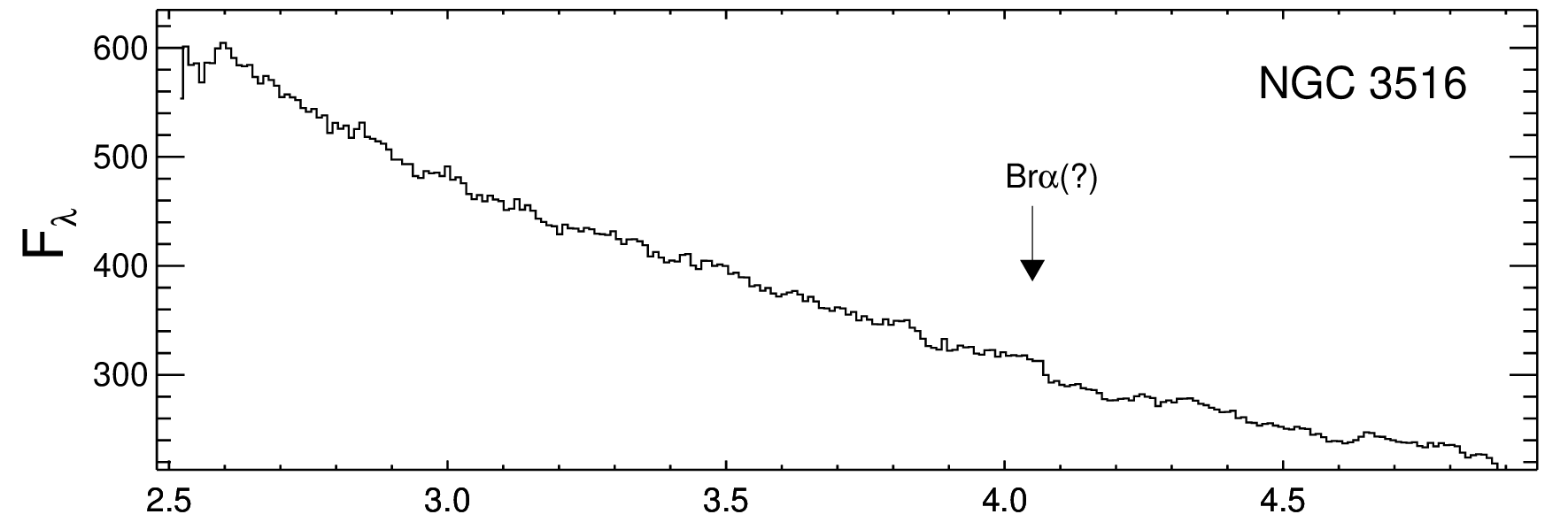}
\includegraphics[scale=0.20]{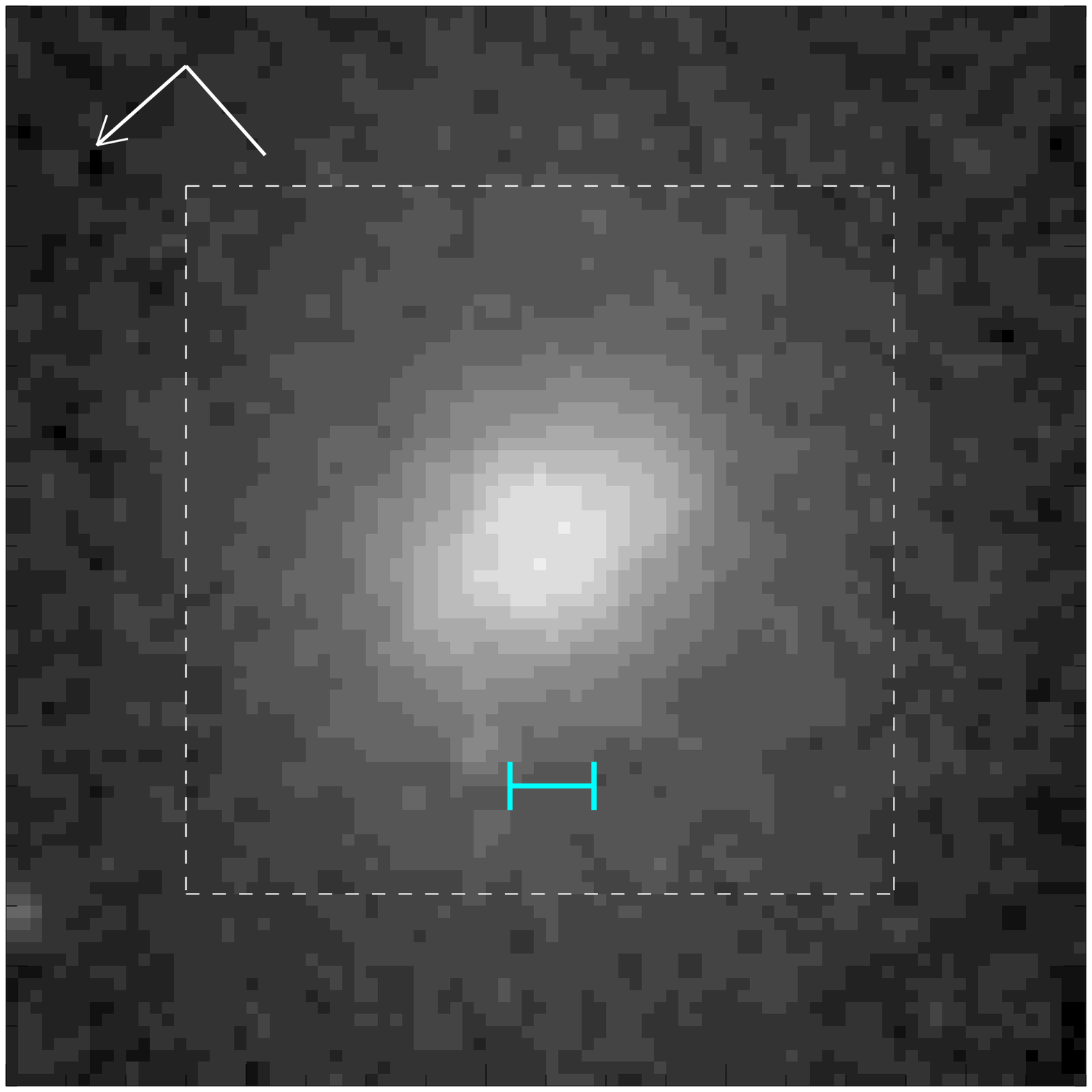}\\
\includegraphics[width=12cm,height=4cm]{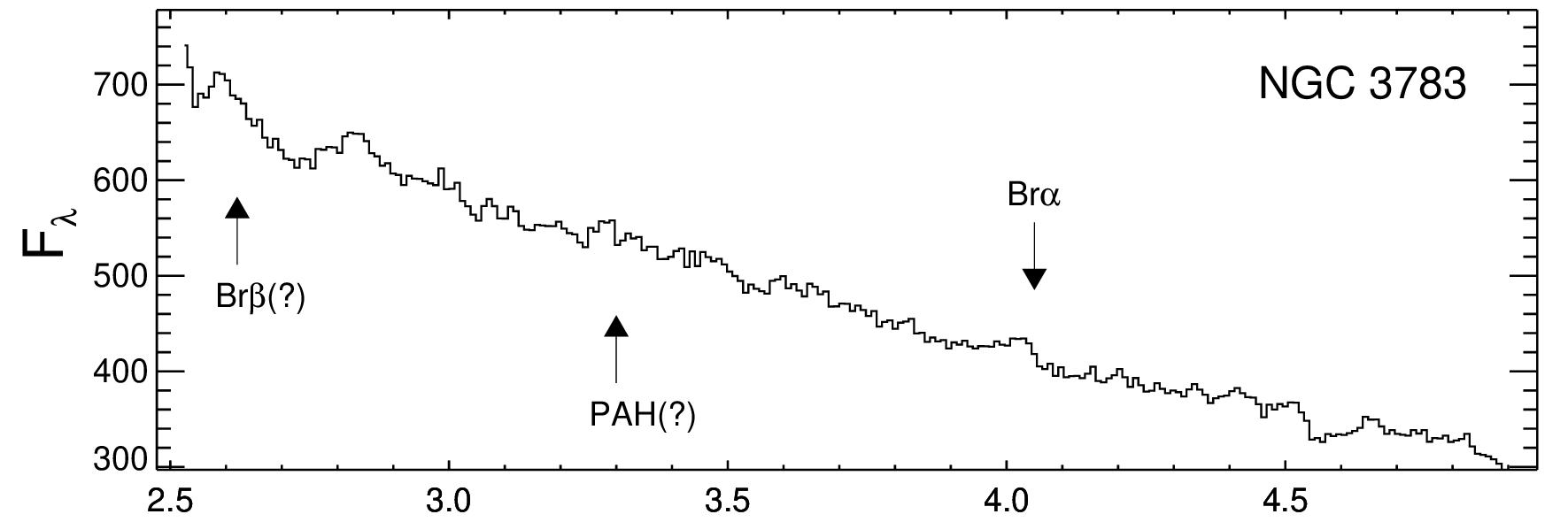}
\includegraphics[scale=0.20]{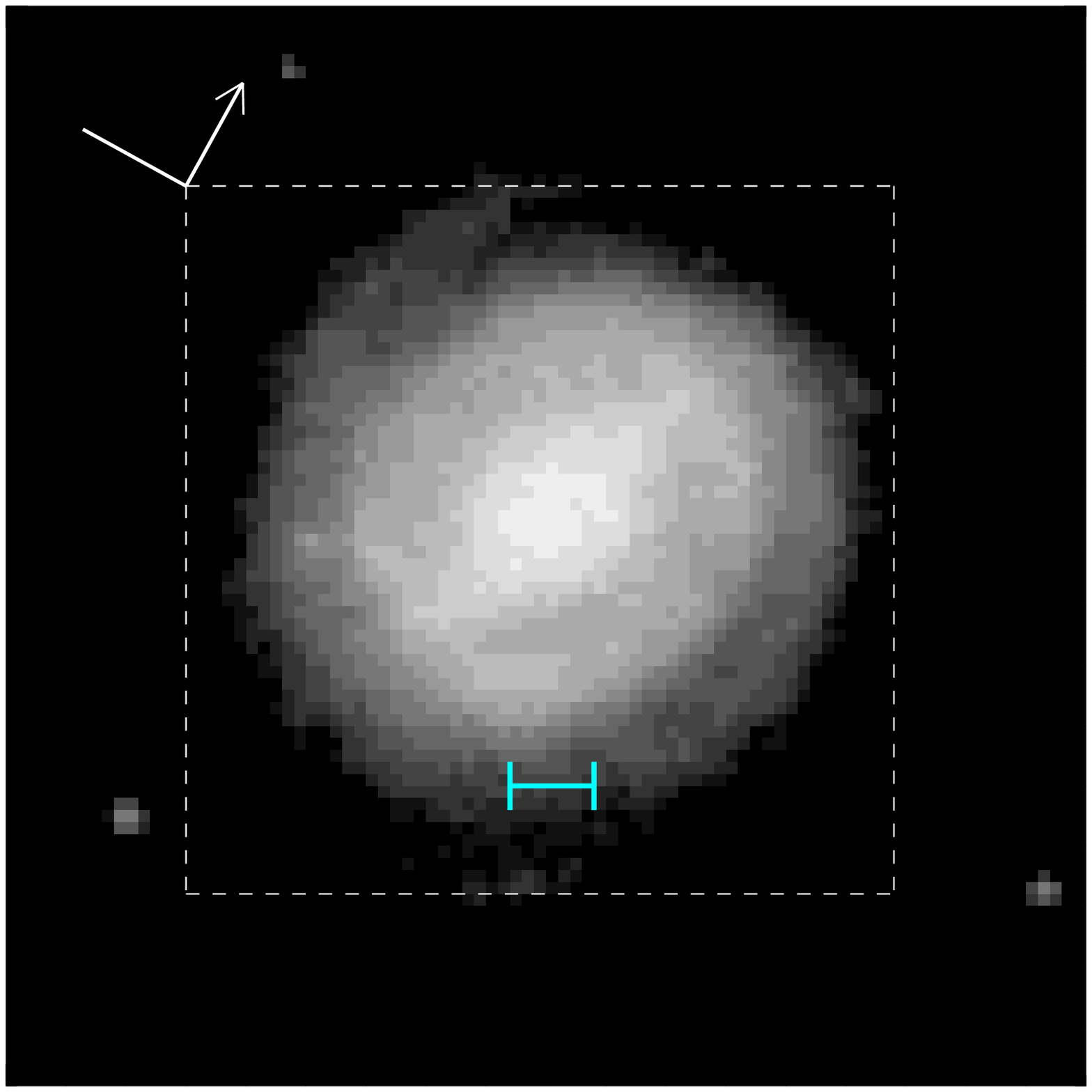}\\
\includegraphics[width=12cm,height=4cm]{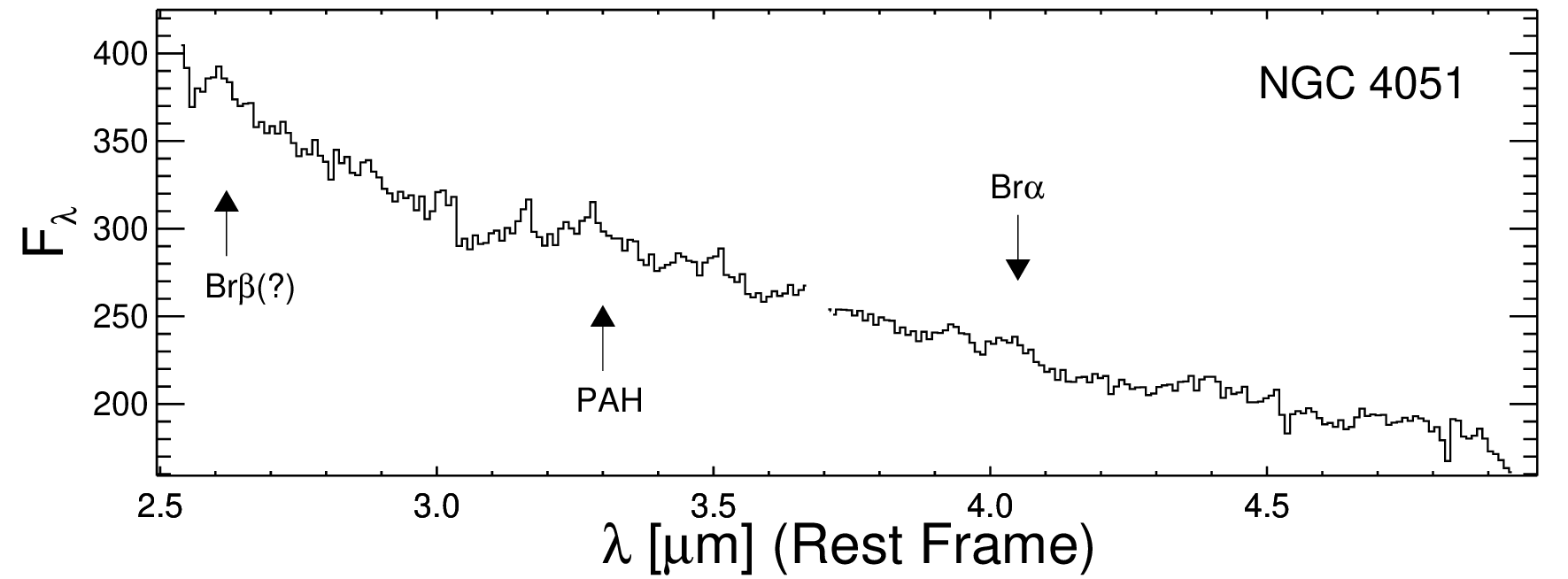}
\includegraphics[scale=0.20]{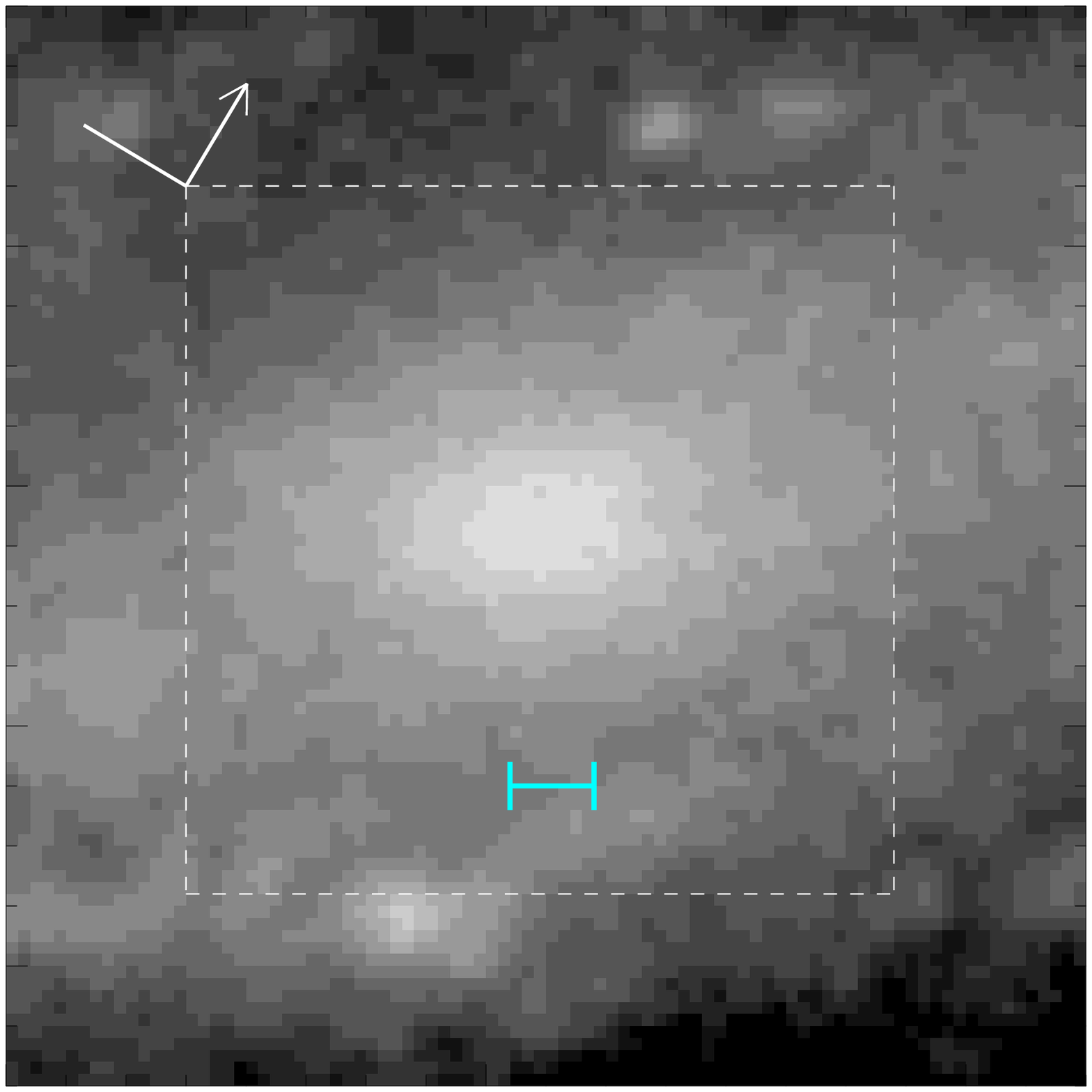}\\
\caption{Continued}
\end{figure}
\clearpage

\begin{figure}
\figurenum{5}
\includegraphics[width=12cm,height=4cm]{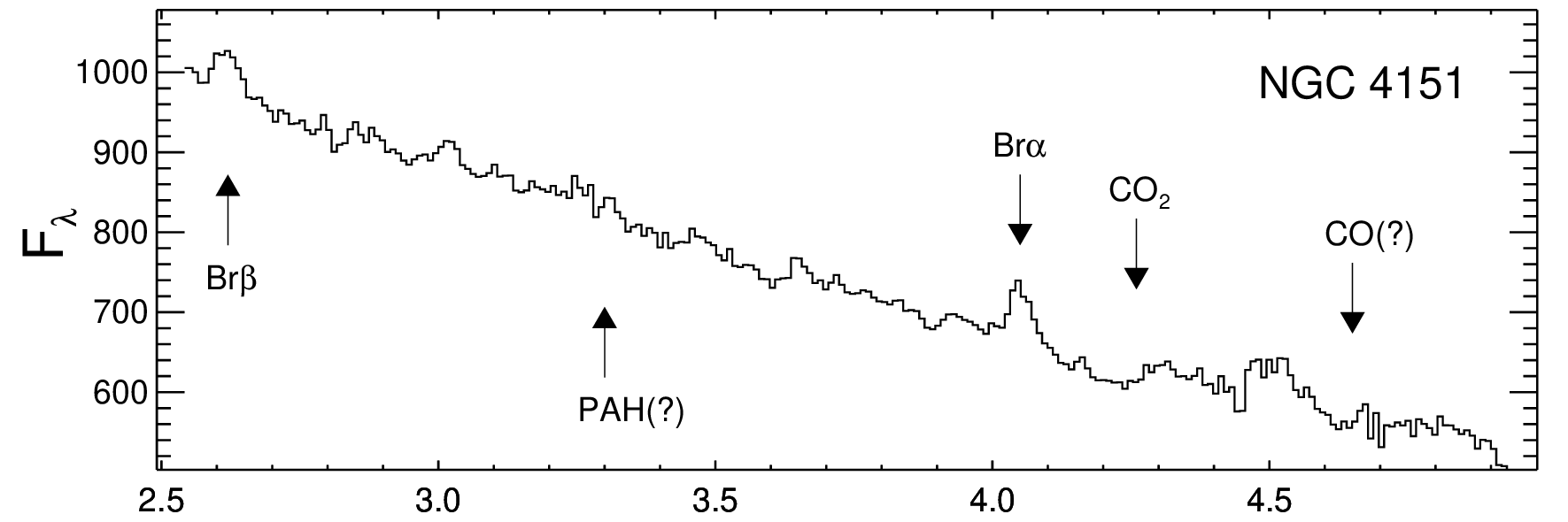}
\includegraphics[scale=0.20]{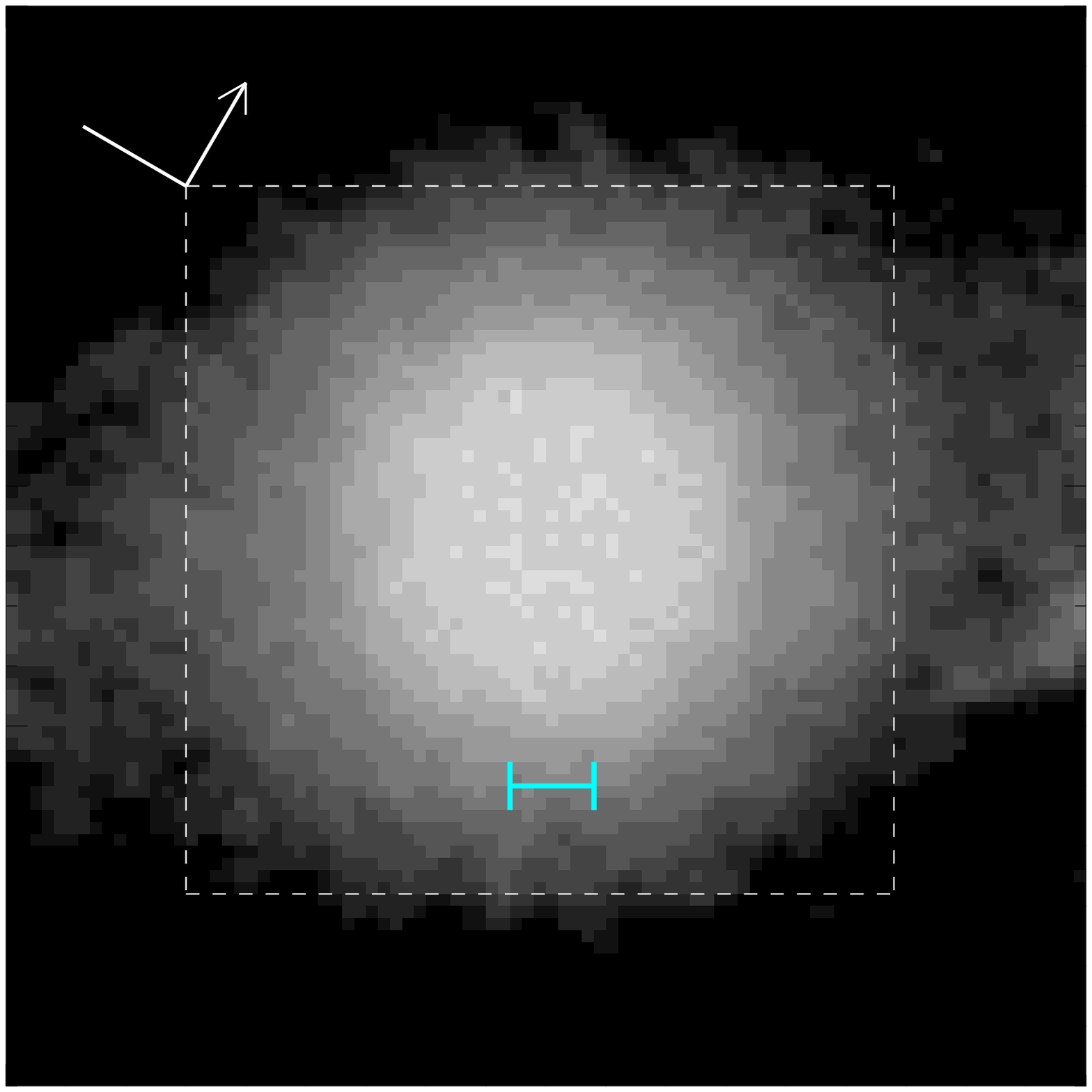}\\
\includegraphics[width=12cm,height=4cm]{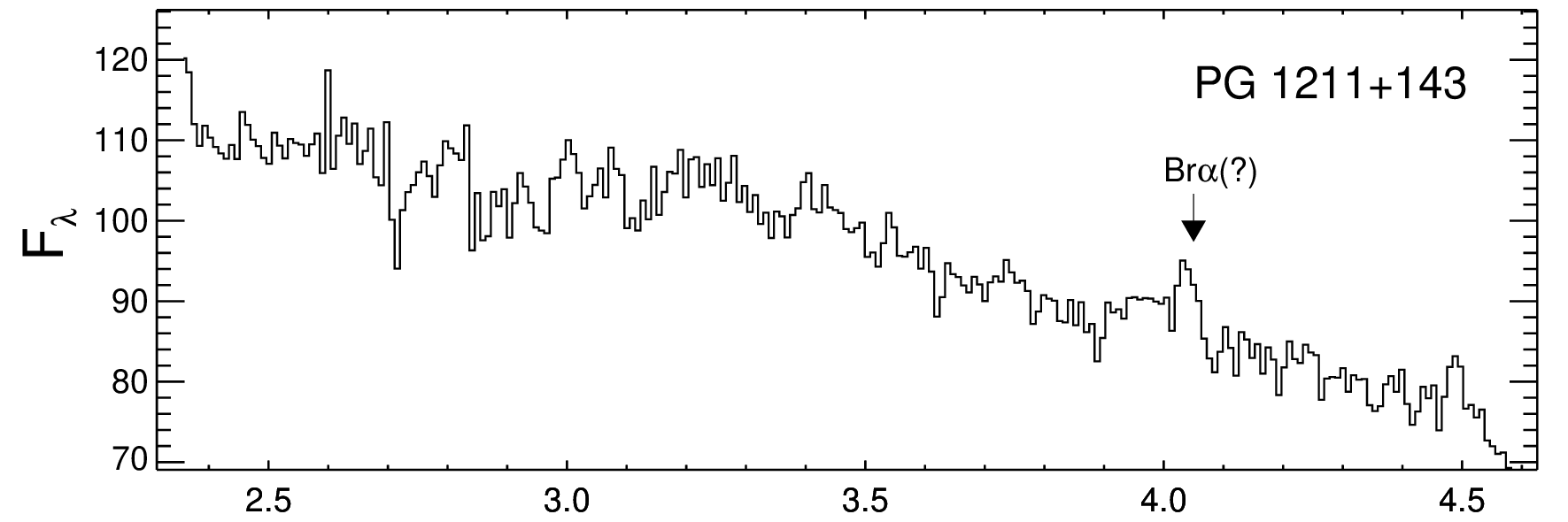}
\includegraphics[scale=0.20]{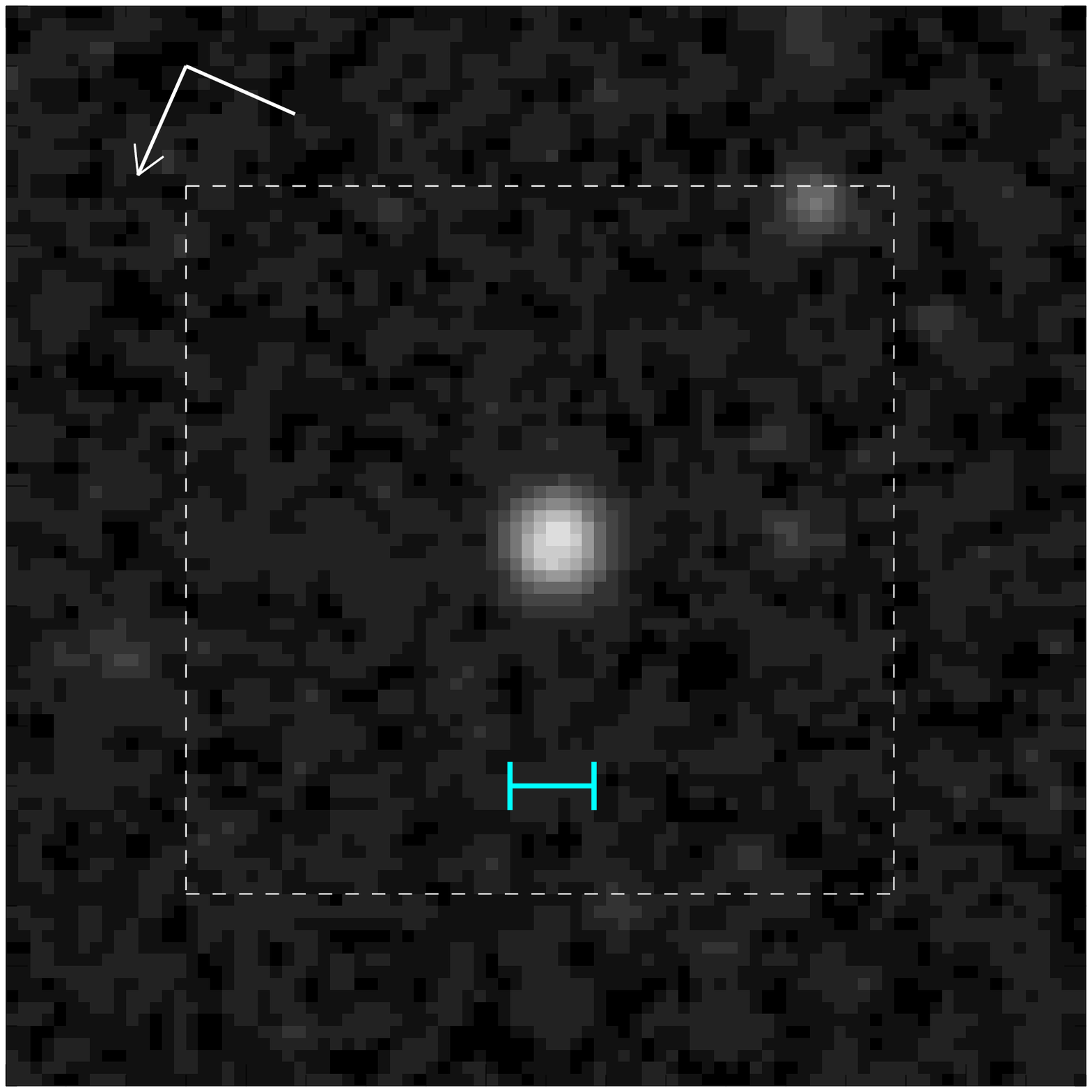}\\
\includegraphics[width=12cm,height=4cm]{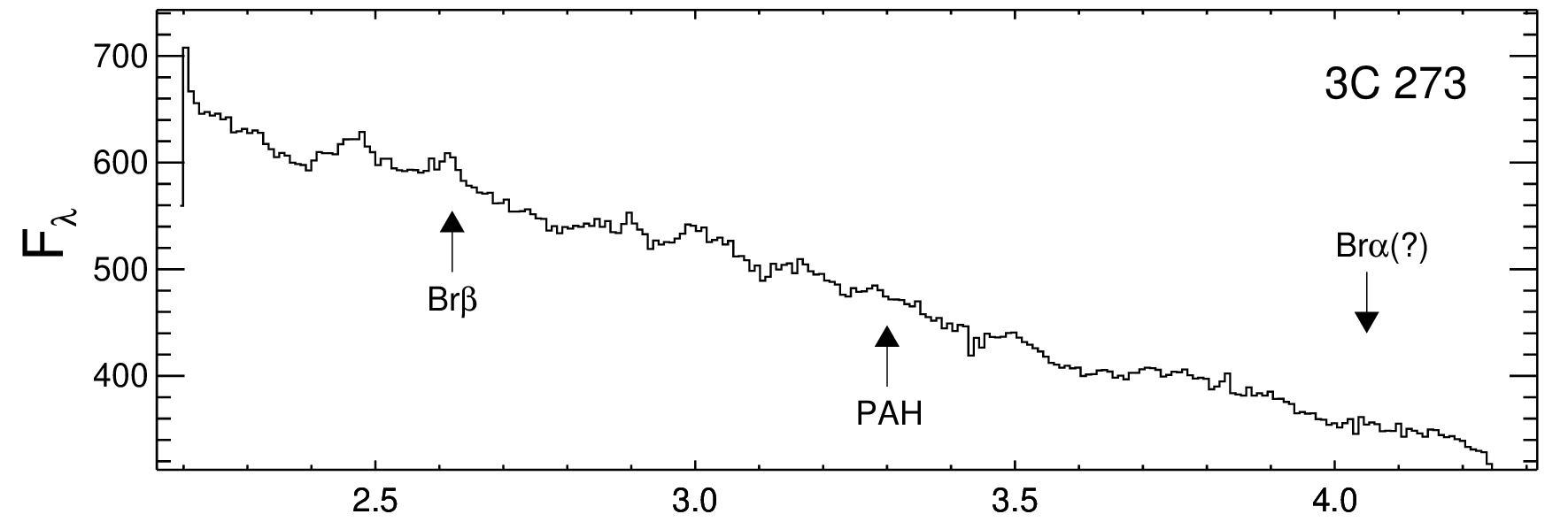}
\includegraphics[scale=0.20]{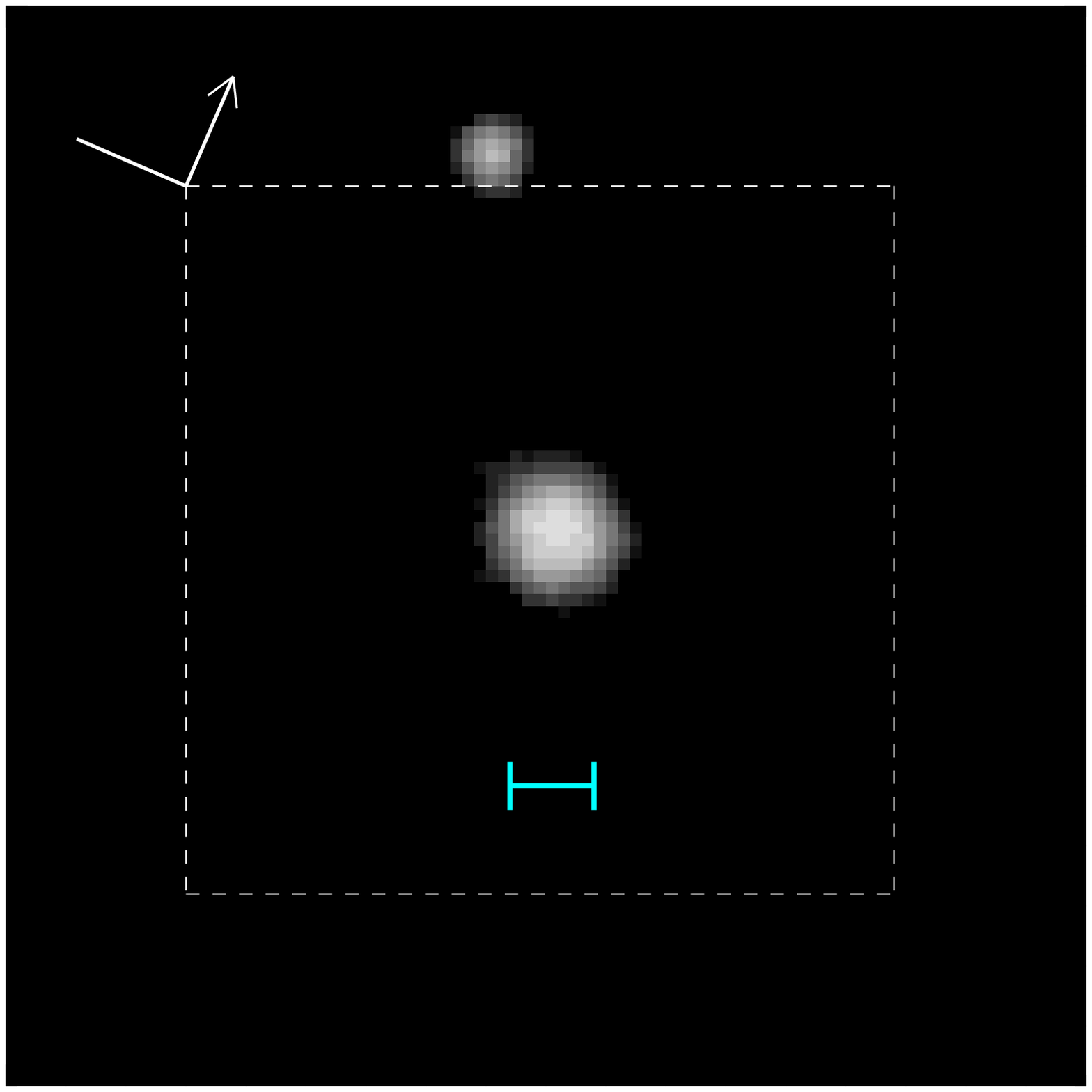}\\
\includegraphics[width=12cm,height=4cm]{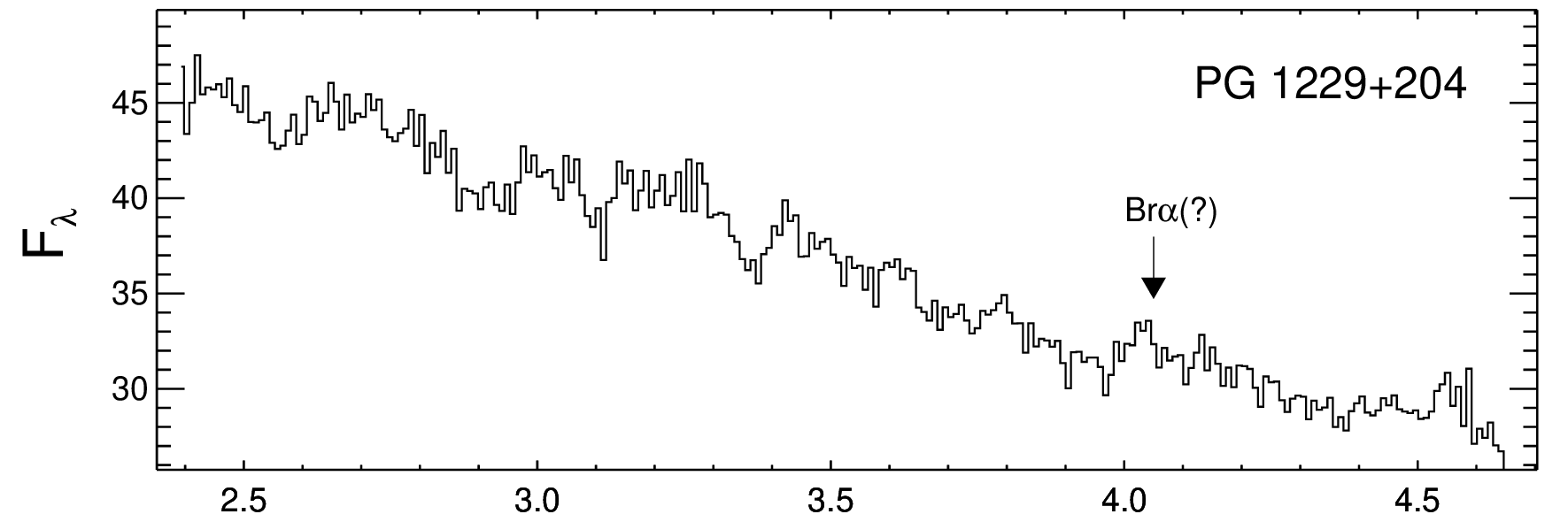}
\includegraphics[scale=0.20]{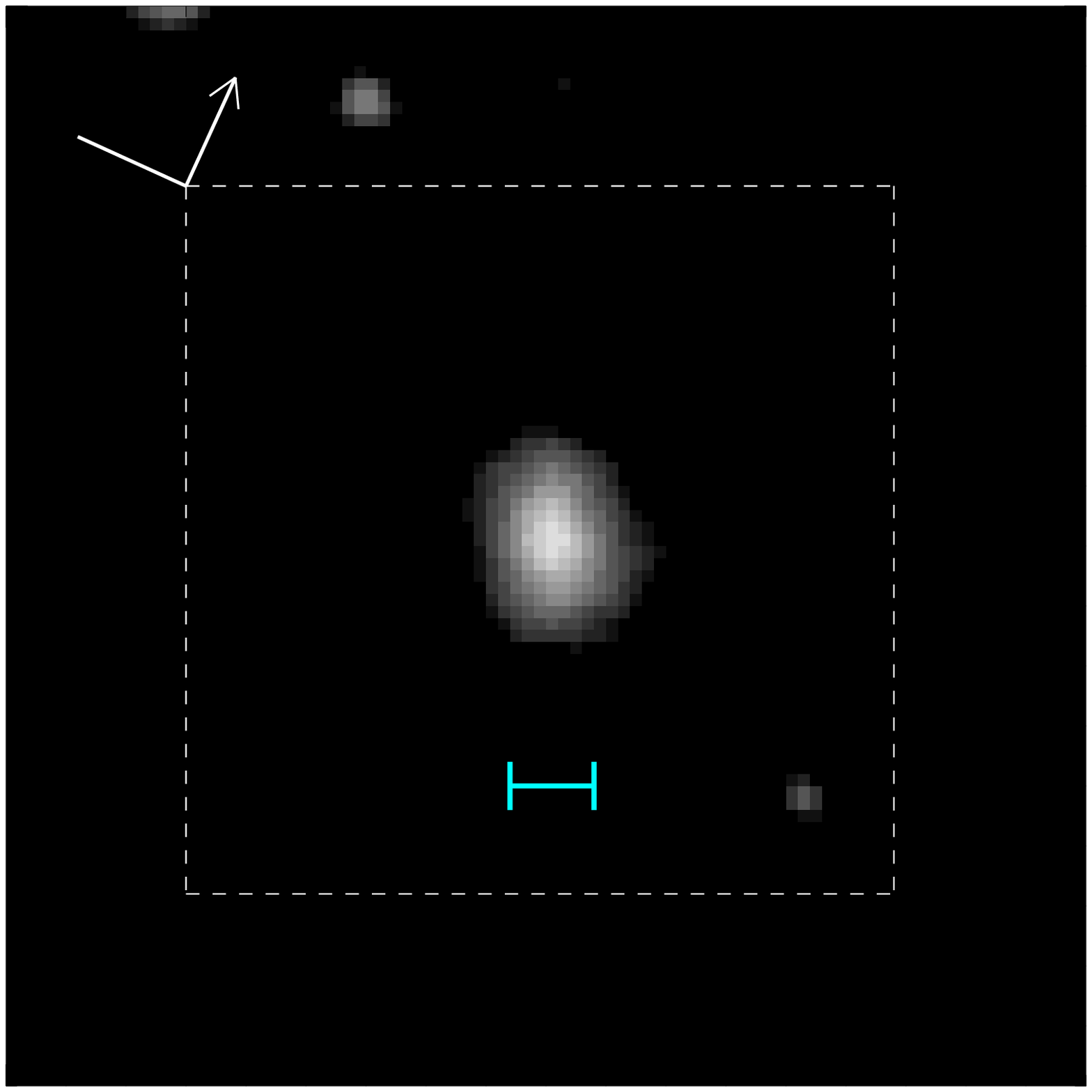}\\
\includegraphics[width=12cm,height=4cm]{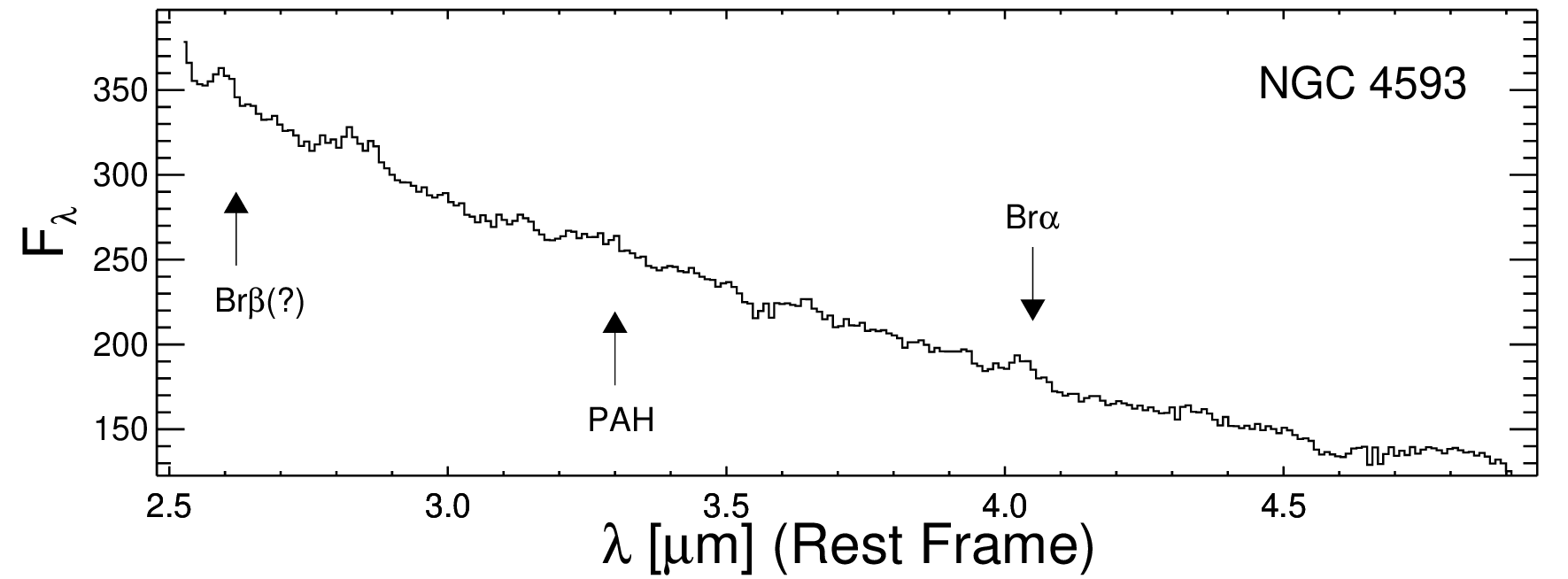}
\includegraphics[scale=0.20]{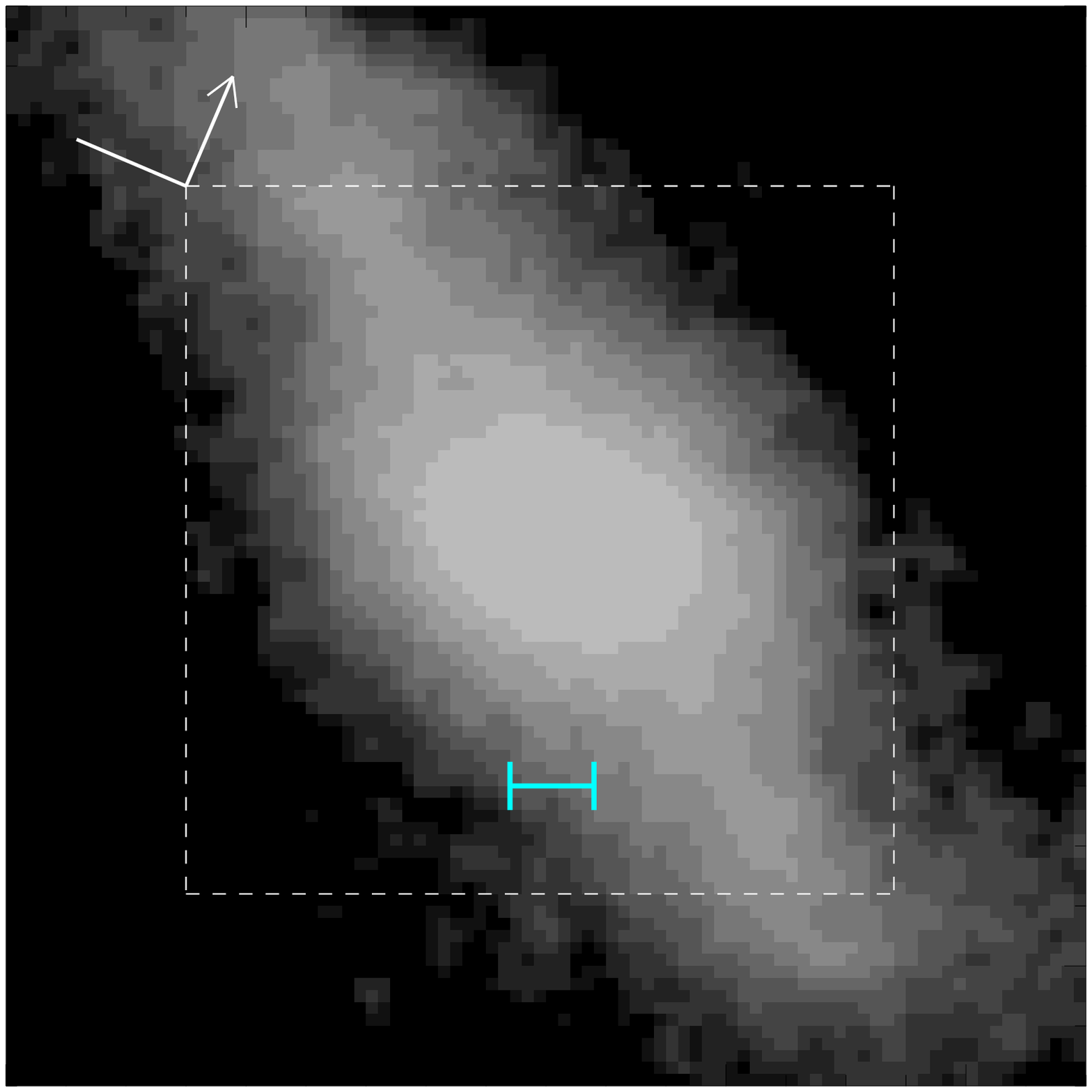}\\
\caption{Continued}
\end{figure}
\clearpage

\begin{figure}
\figurenum{5}
\includegraphics[width=12cm,height=4cm]{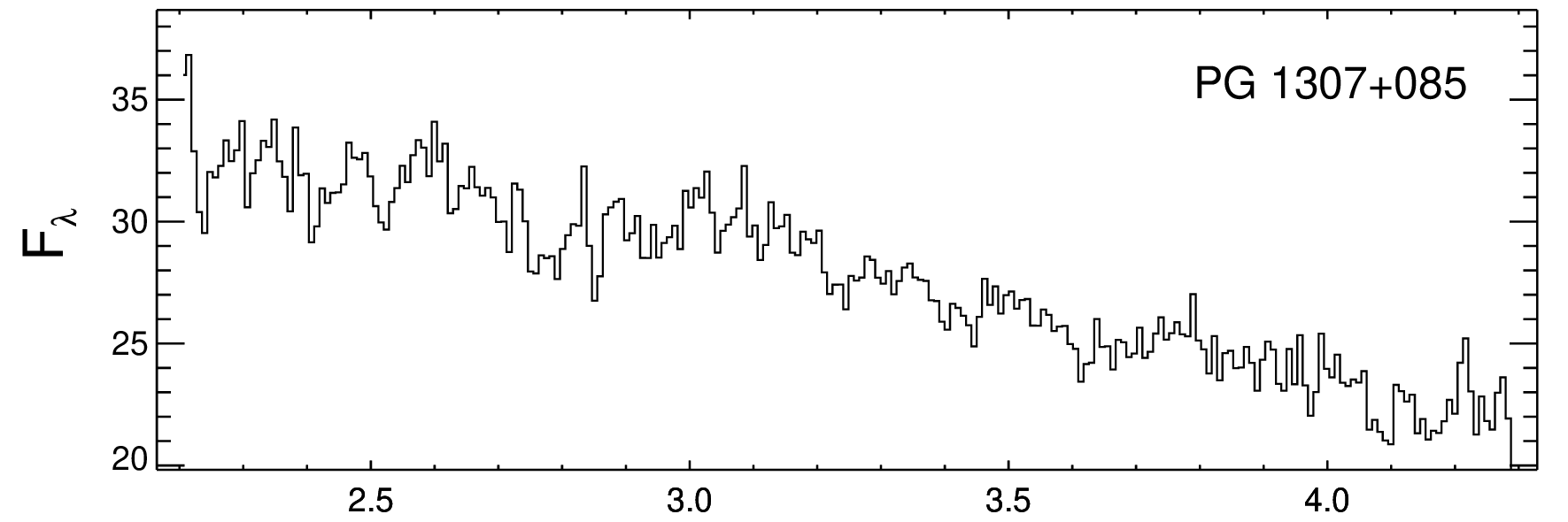}
\includegraphics[scale=0.20]{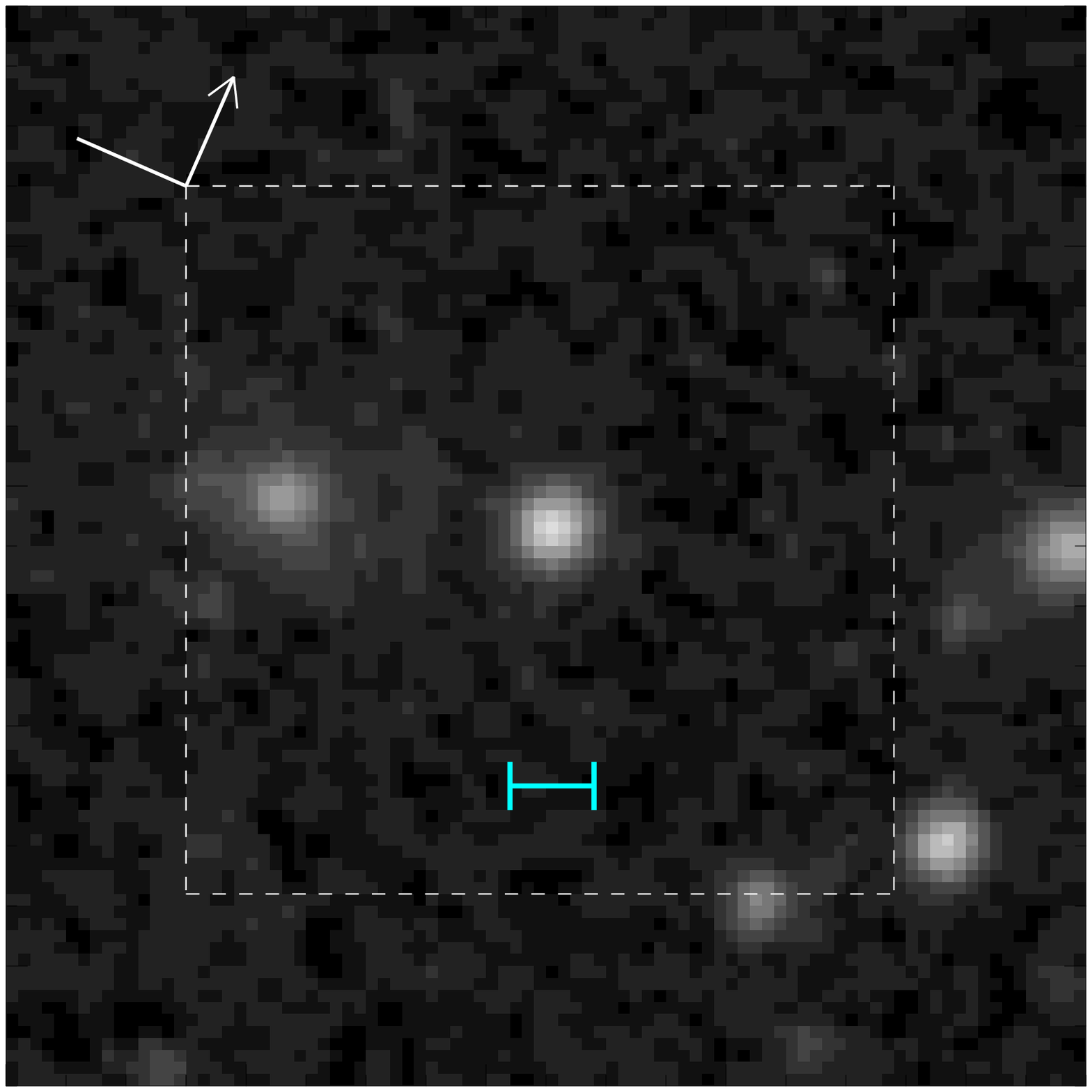}\\
\includegraphics[width=12cm,height=4cm]{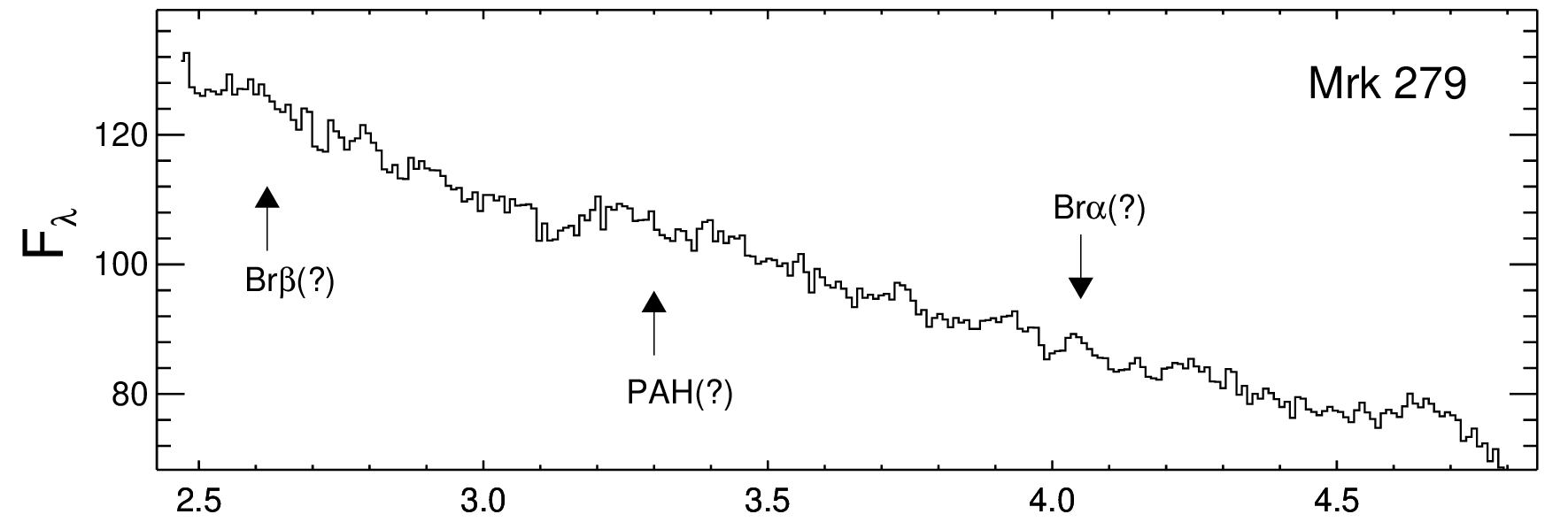}
\includegraphics[scale=0.20]{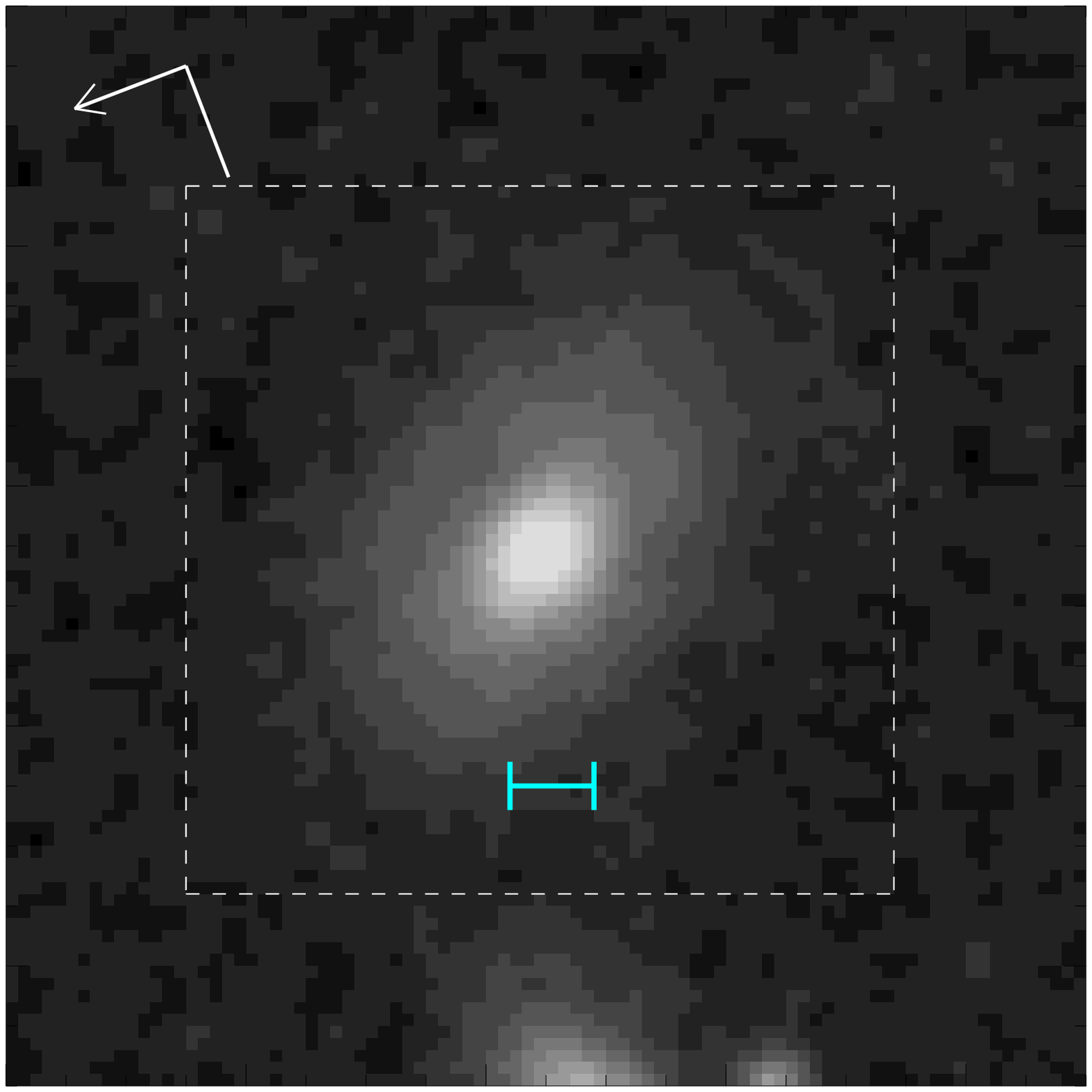}\\
\includegraphics[width=12cm,height=4cm]{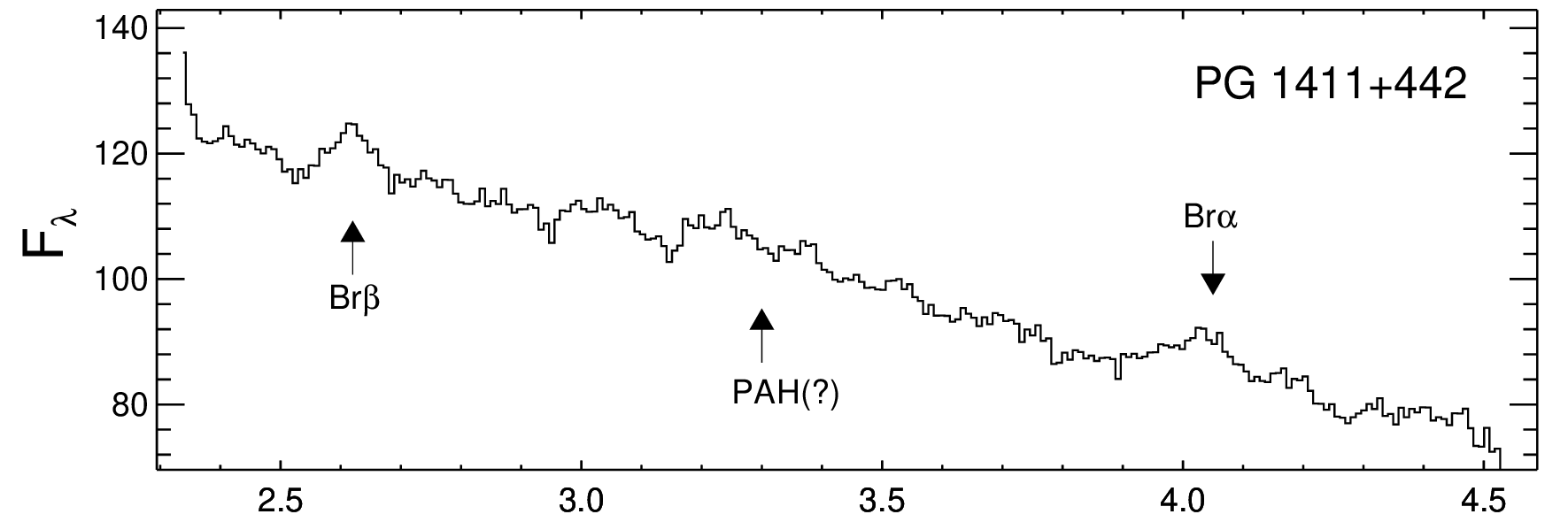}
\includegraphics[scale=0.20]{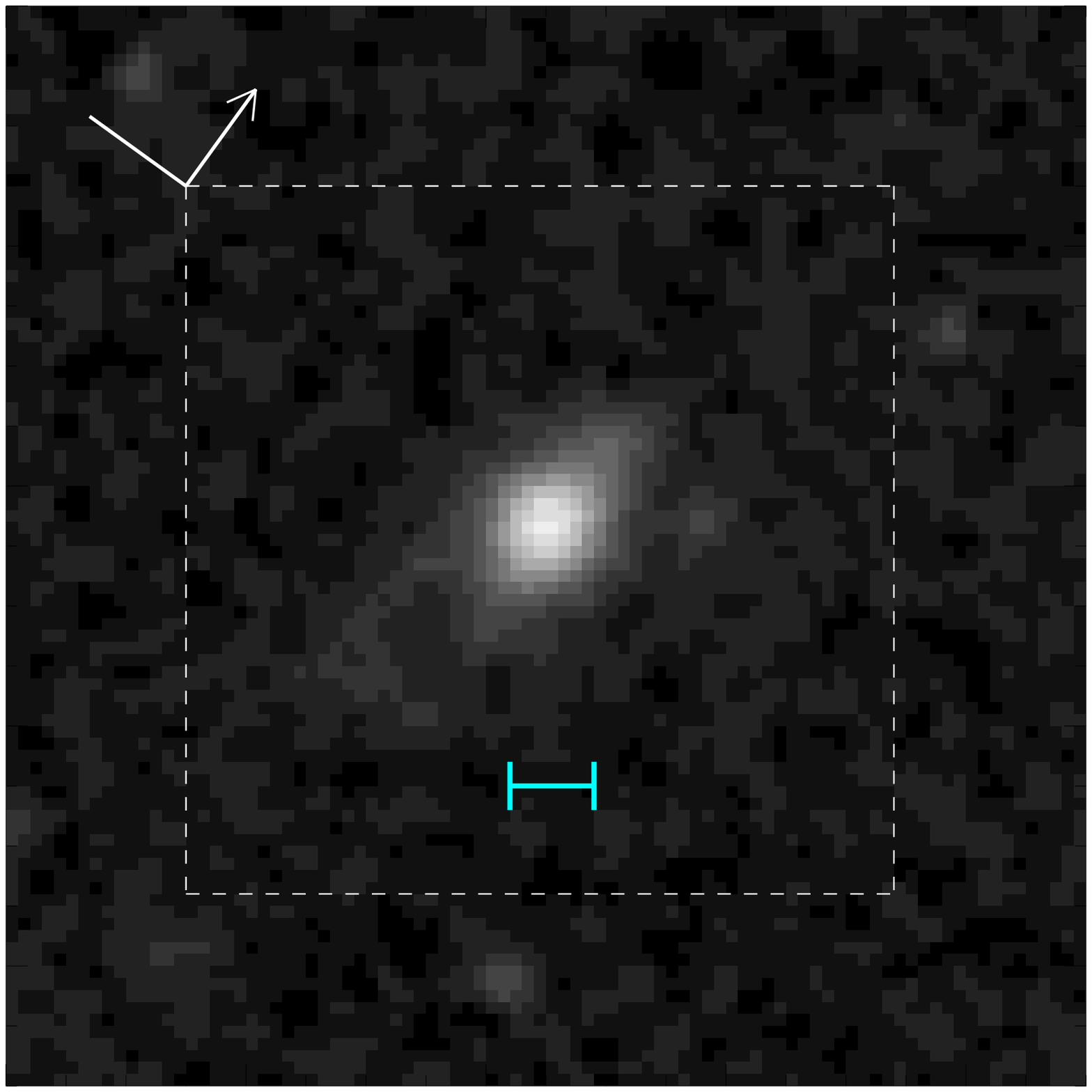}\\
\includegraphics[width=12cm,height=4cm]{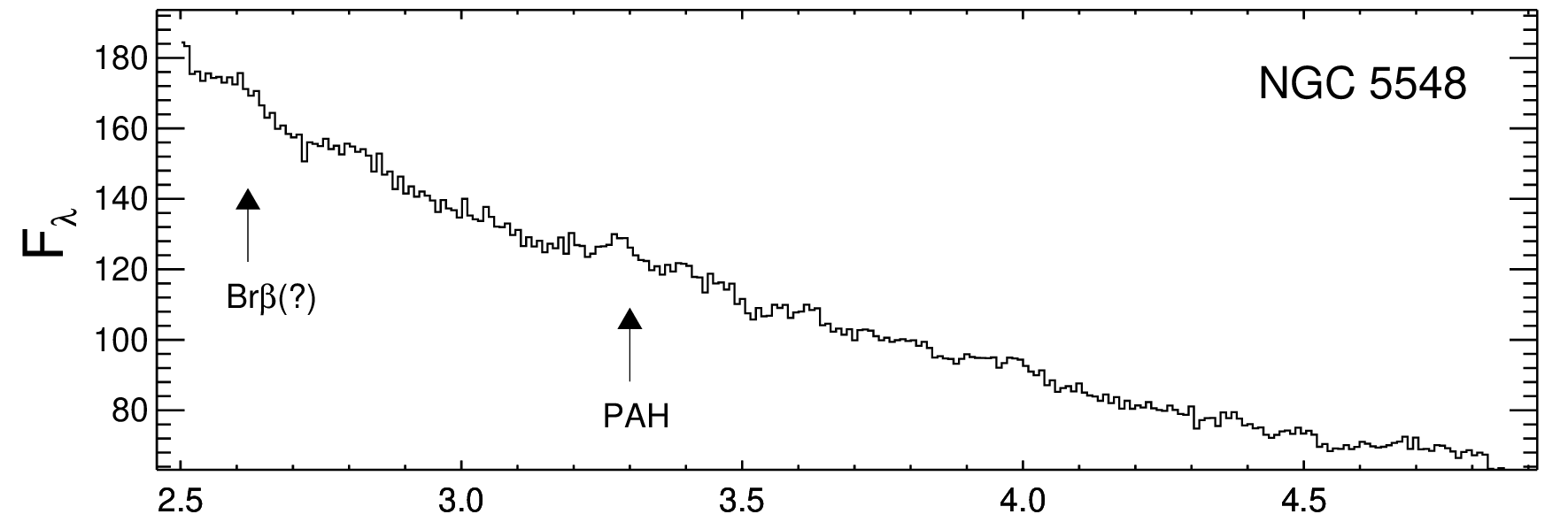}
\includegraphics[scale=0.20]{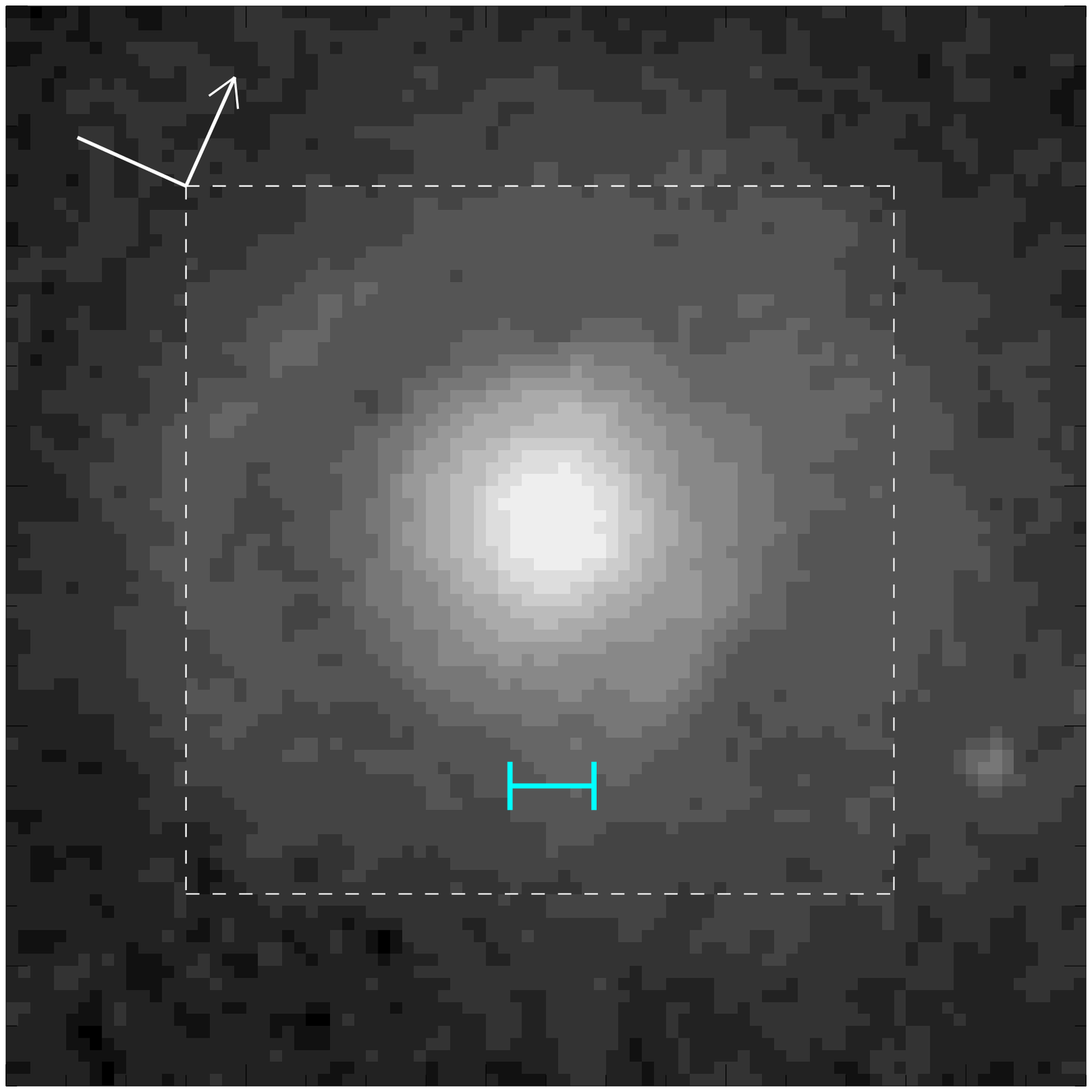}\\
\includegraphics[width=12cm,height=4cm]{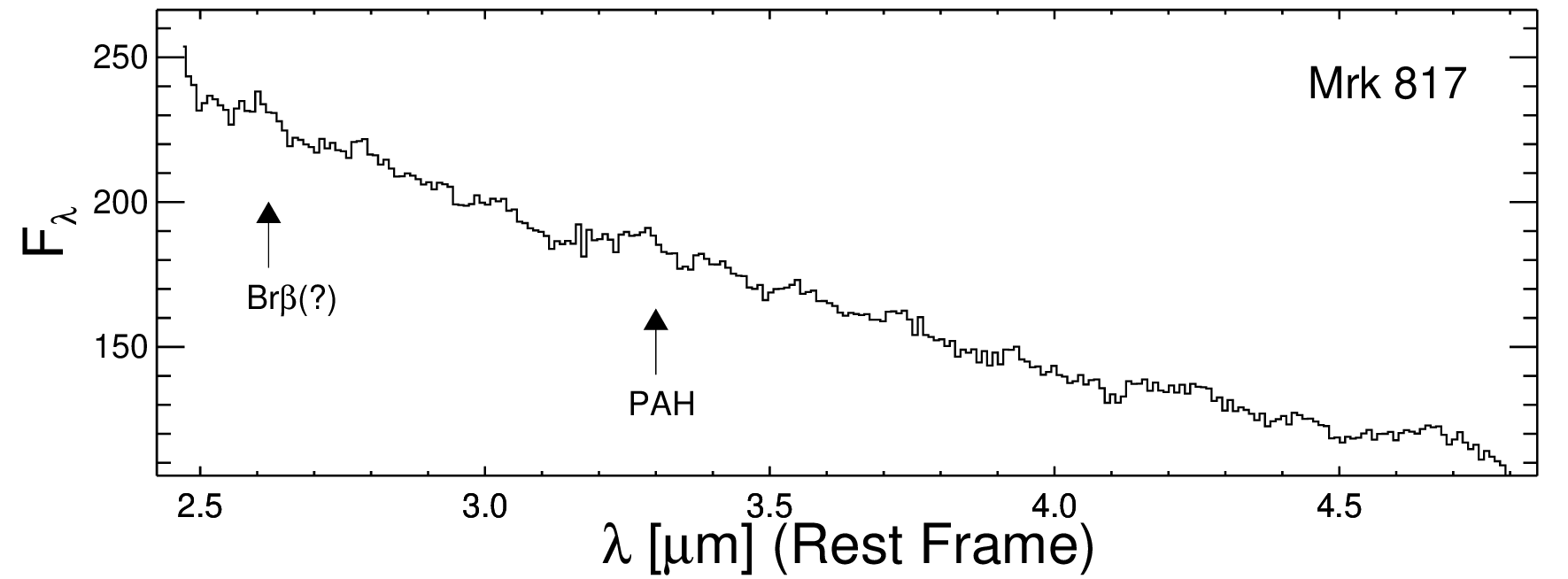}
\includegraphics[scale=0.20]{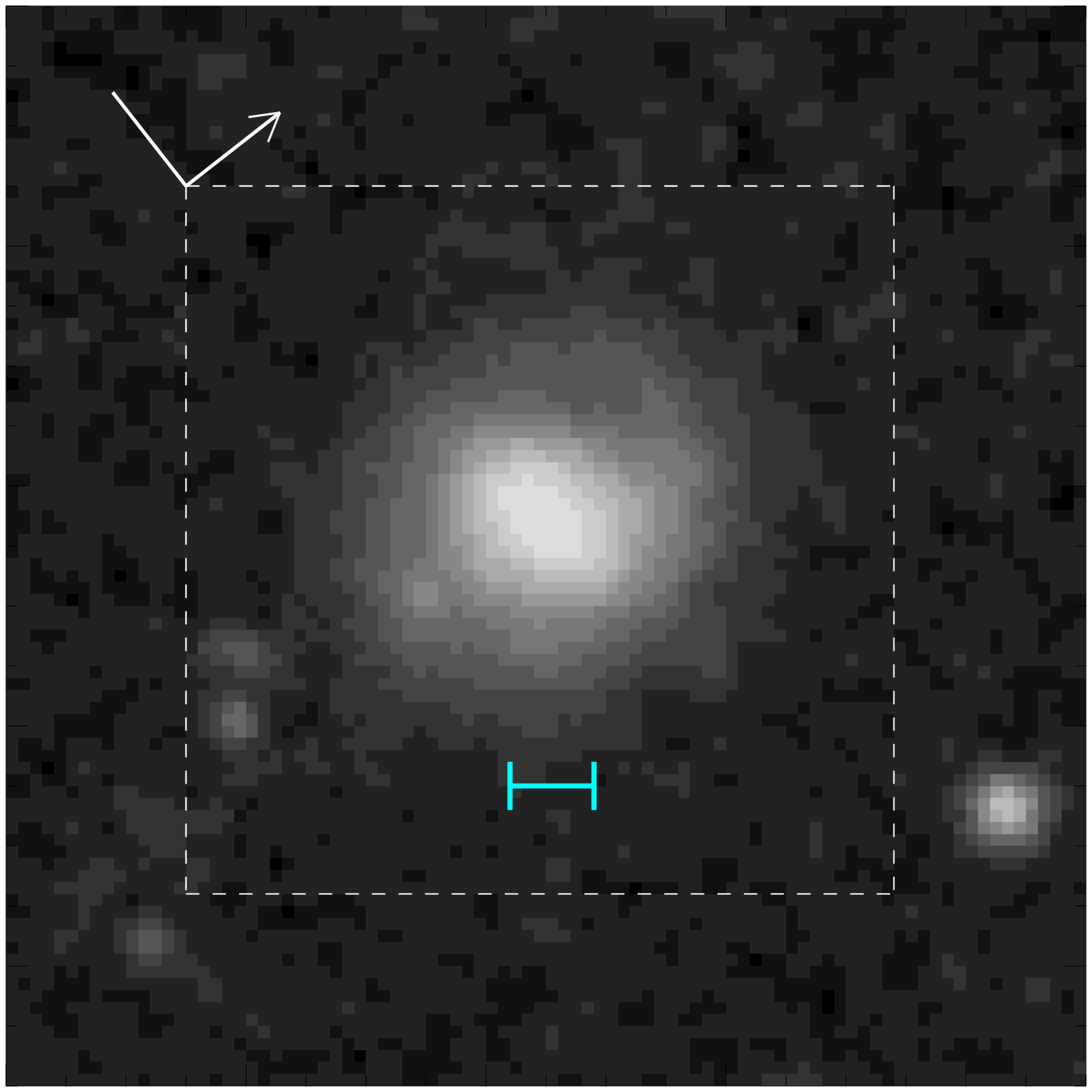}\\
\caption{Continued}
\end{figure}
\clearpage

\begin{figure}
\figurenum{5}
\includegraphics[width=12cm,height=4cm]{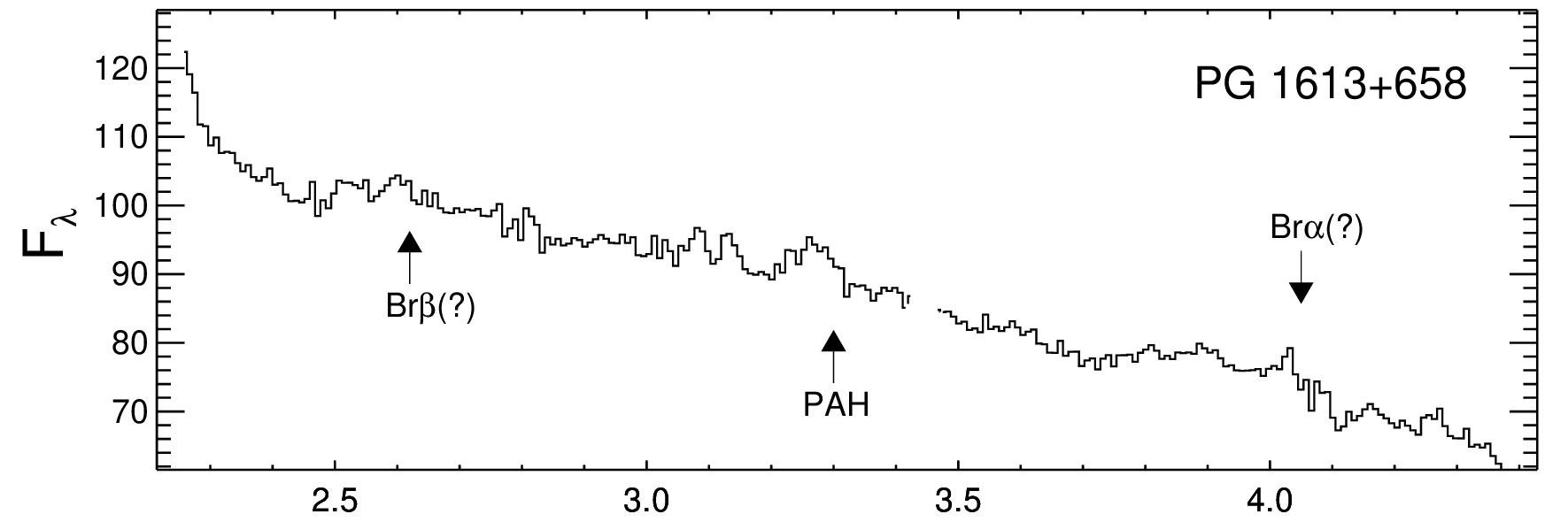}
\includegraphics[scale=0.20]{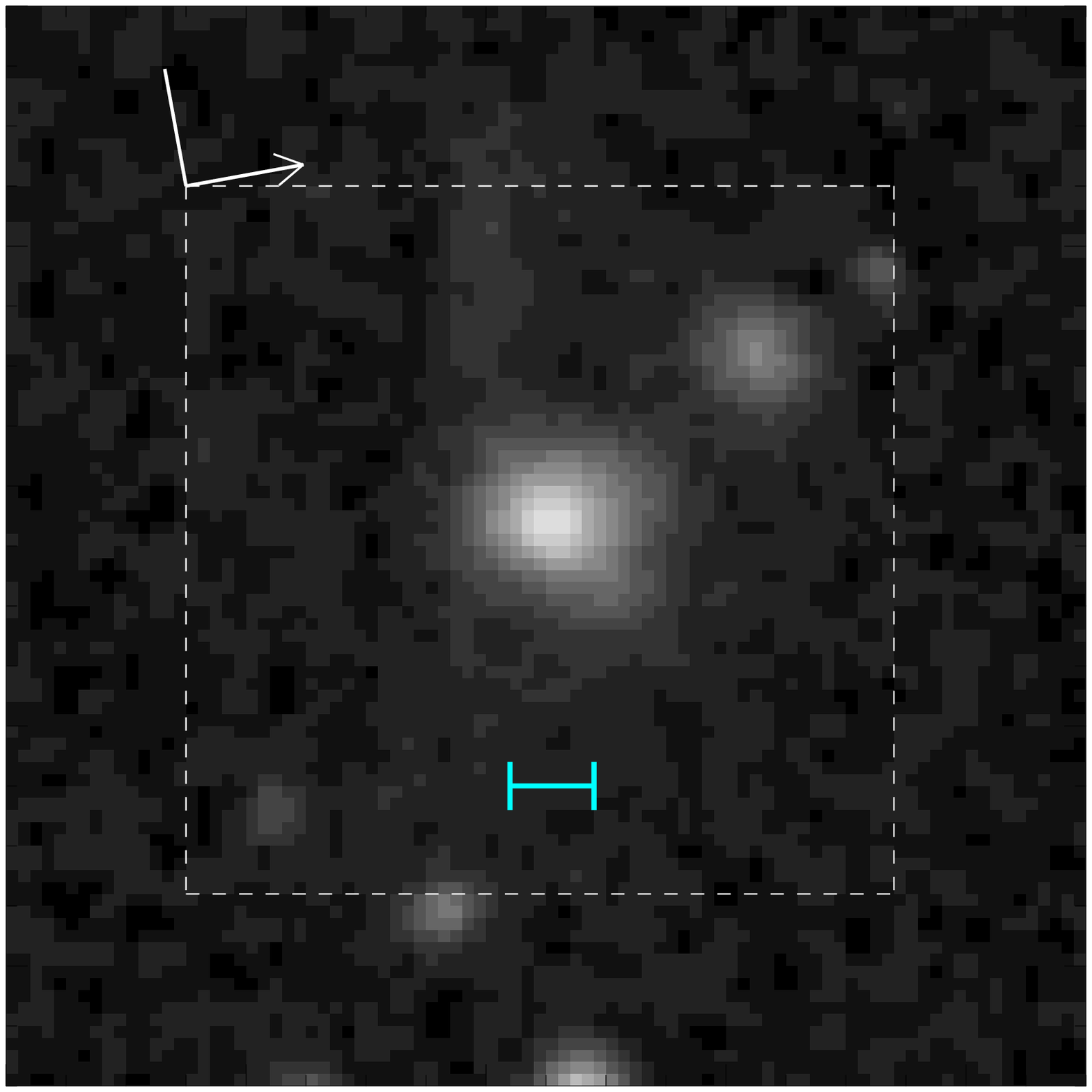}\\
\includegraphics[width=12cm,height=4cm]{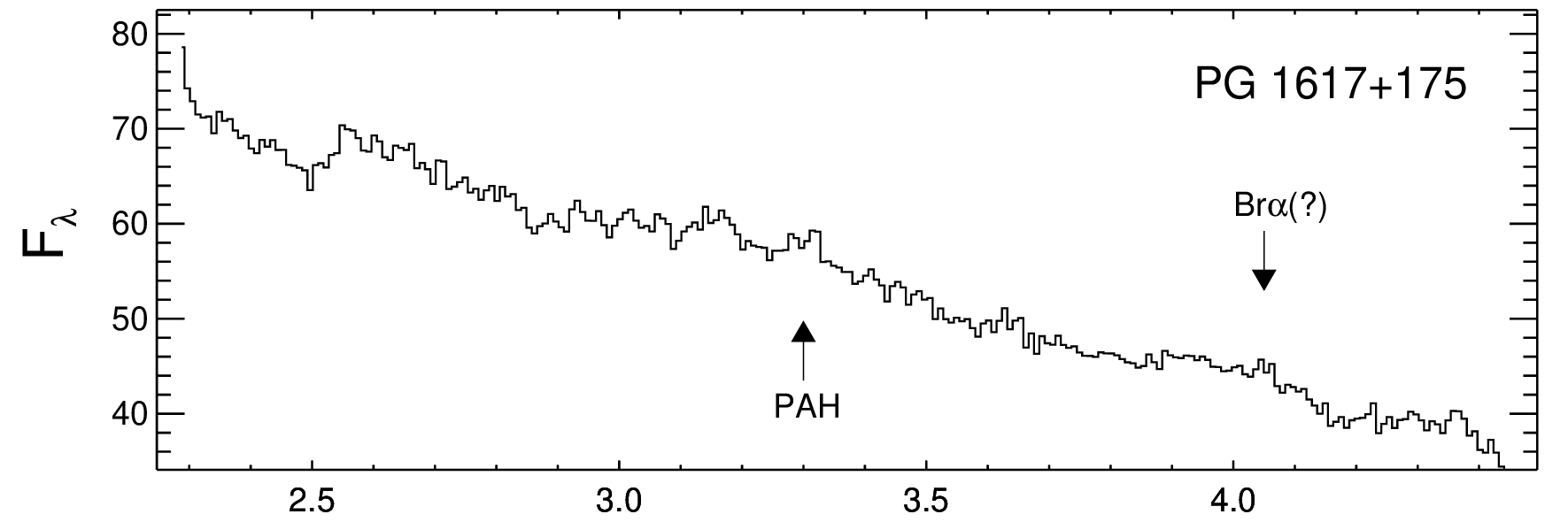}
\includegraphics[scale=0.20]{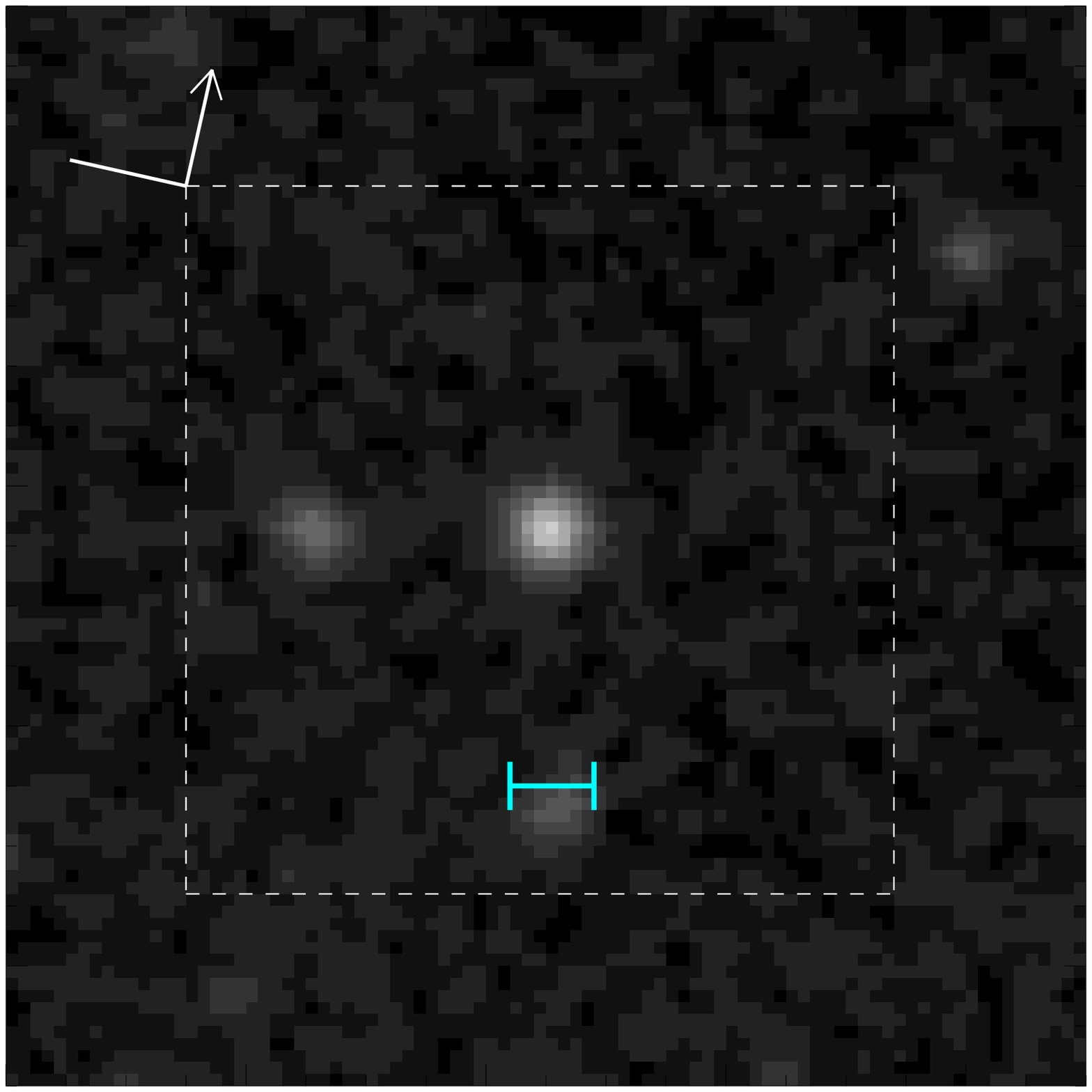}\\
\includegraphics[width=12cm,height=4cm]{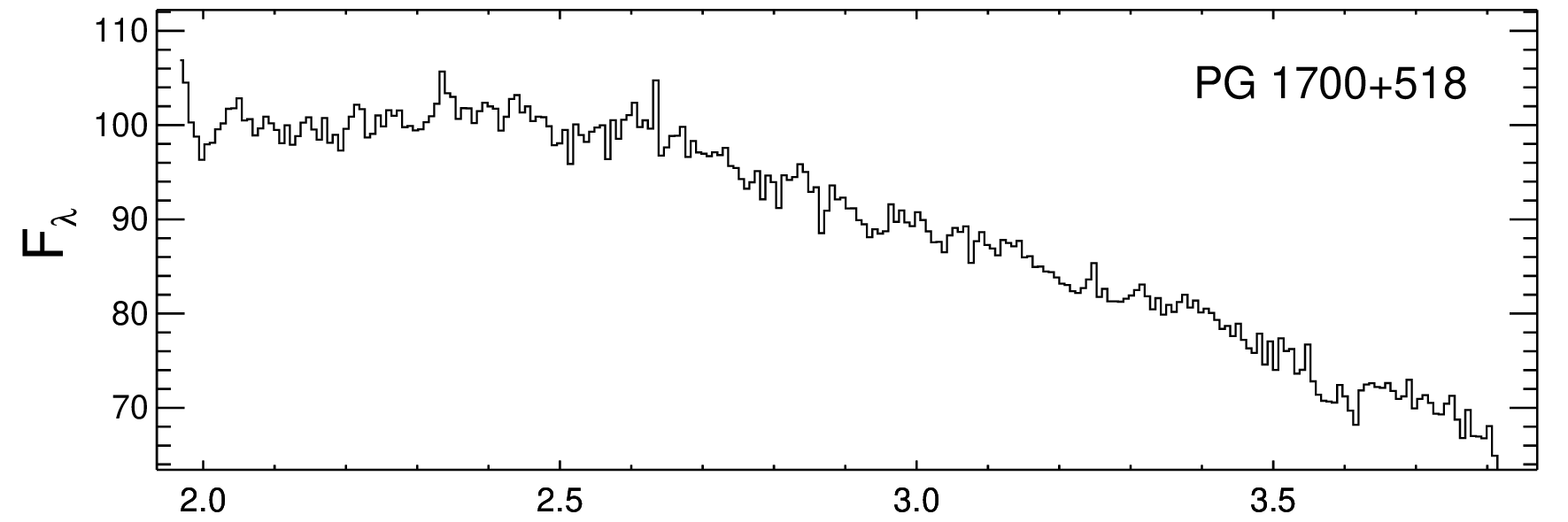}
\includegraphics[scale=0.20]{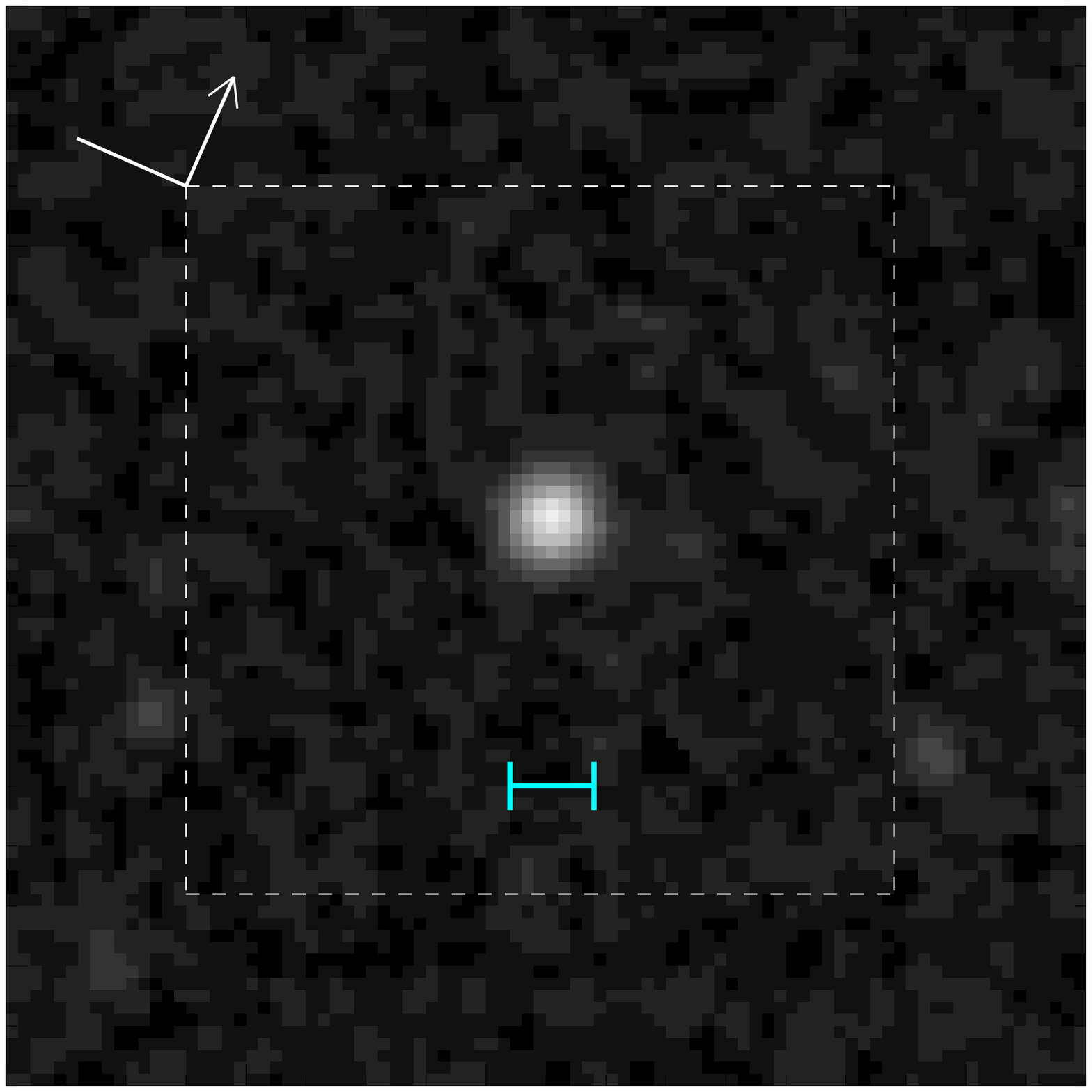}\\
\includegraphics[width=12cm,height=4cm]{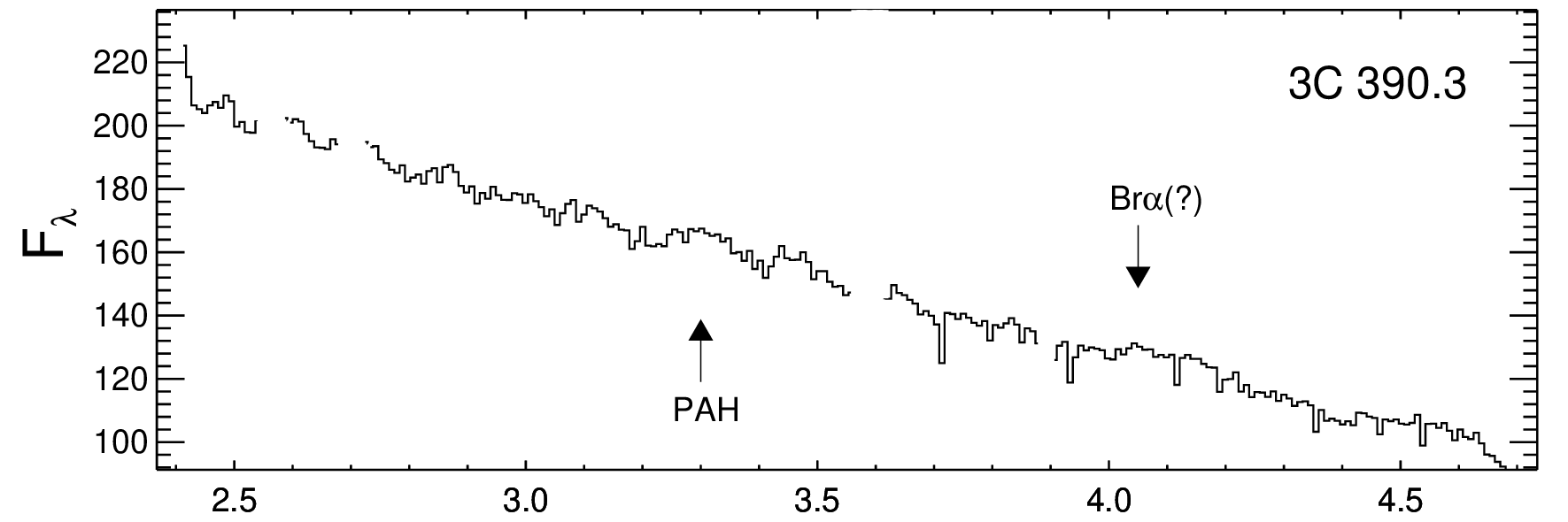}
\includegraphics[scale=0.20]{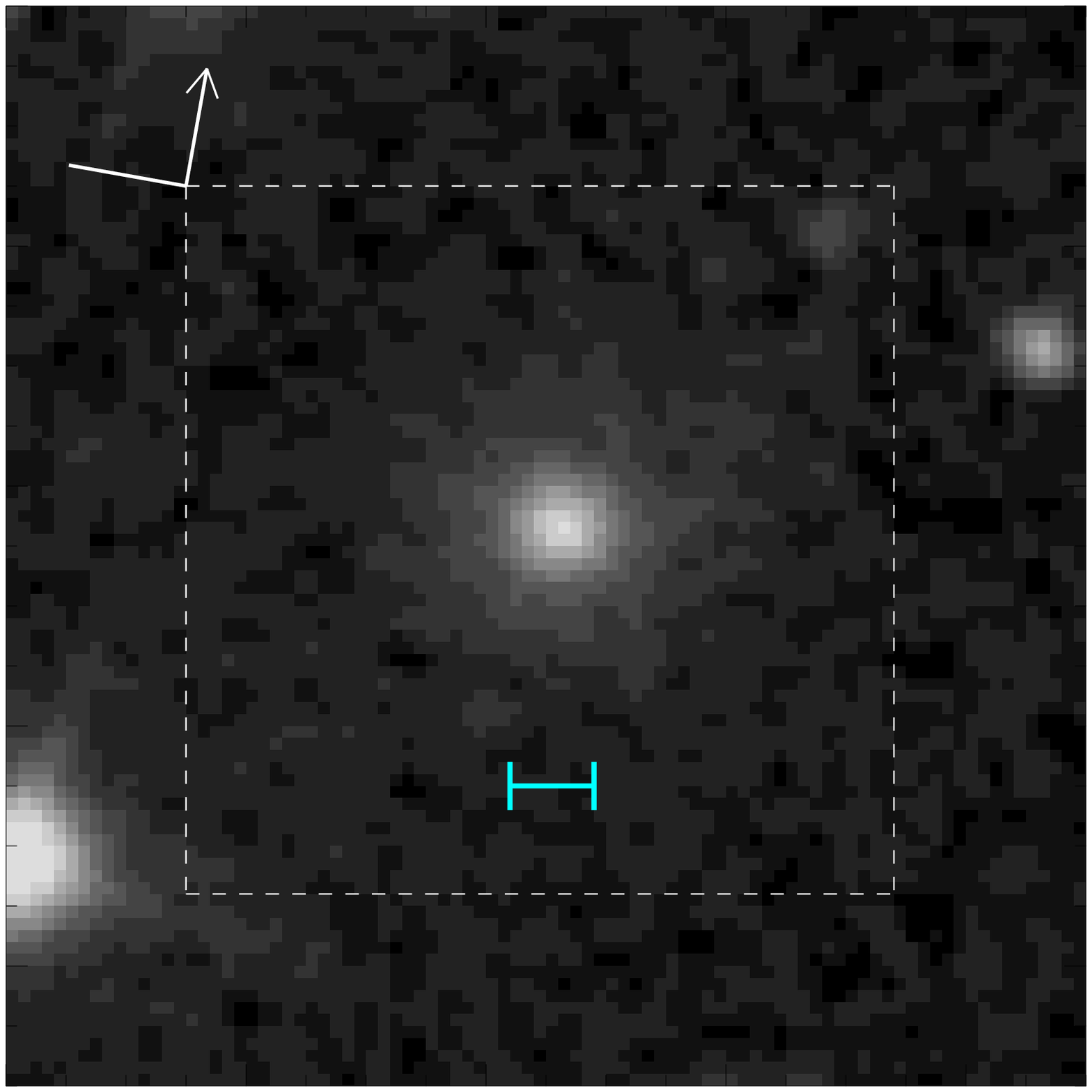}\\
\includegraphics[width=12cm,height=4cm]{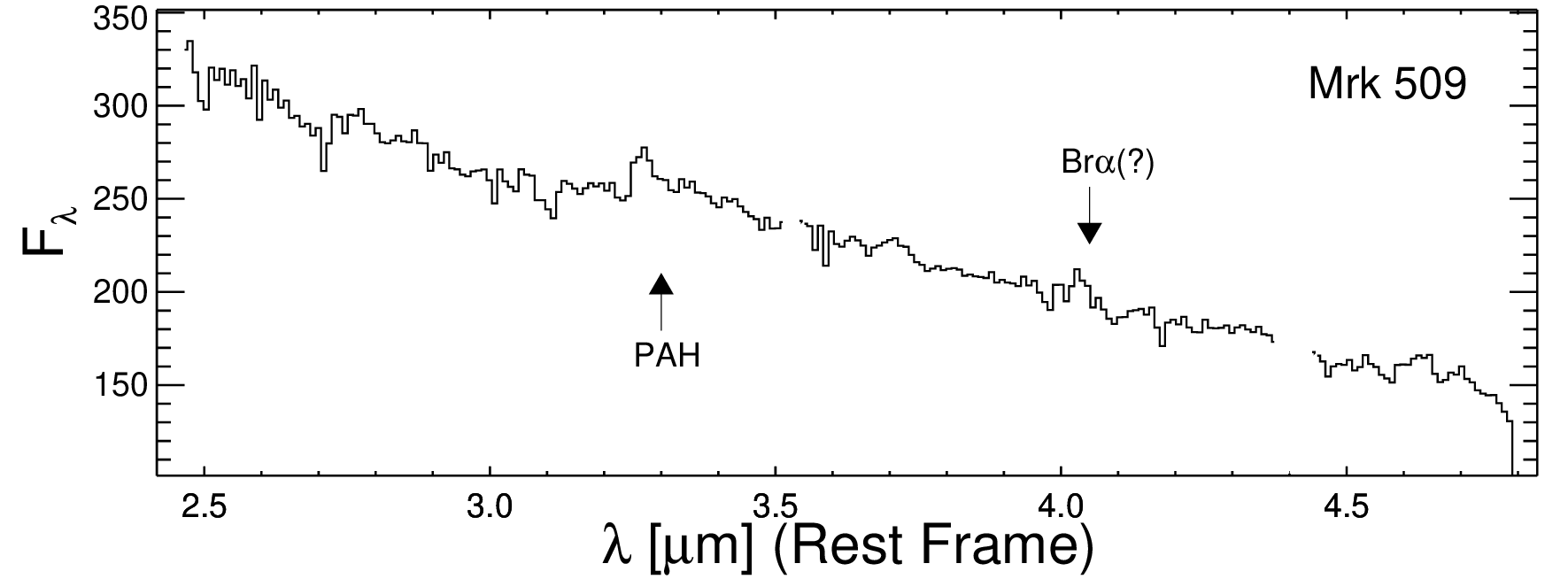}
\includegraphics[scale=0.20]{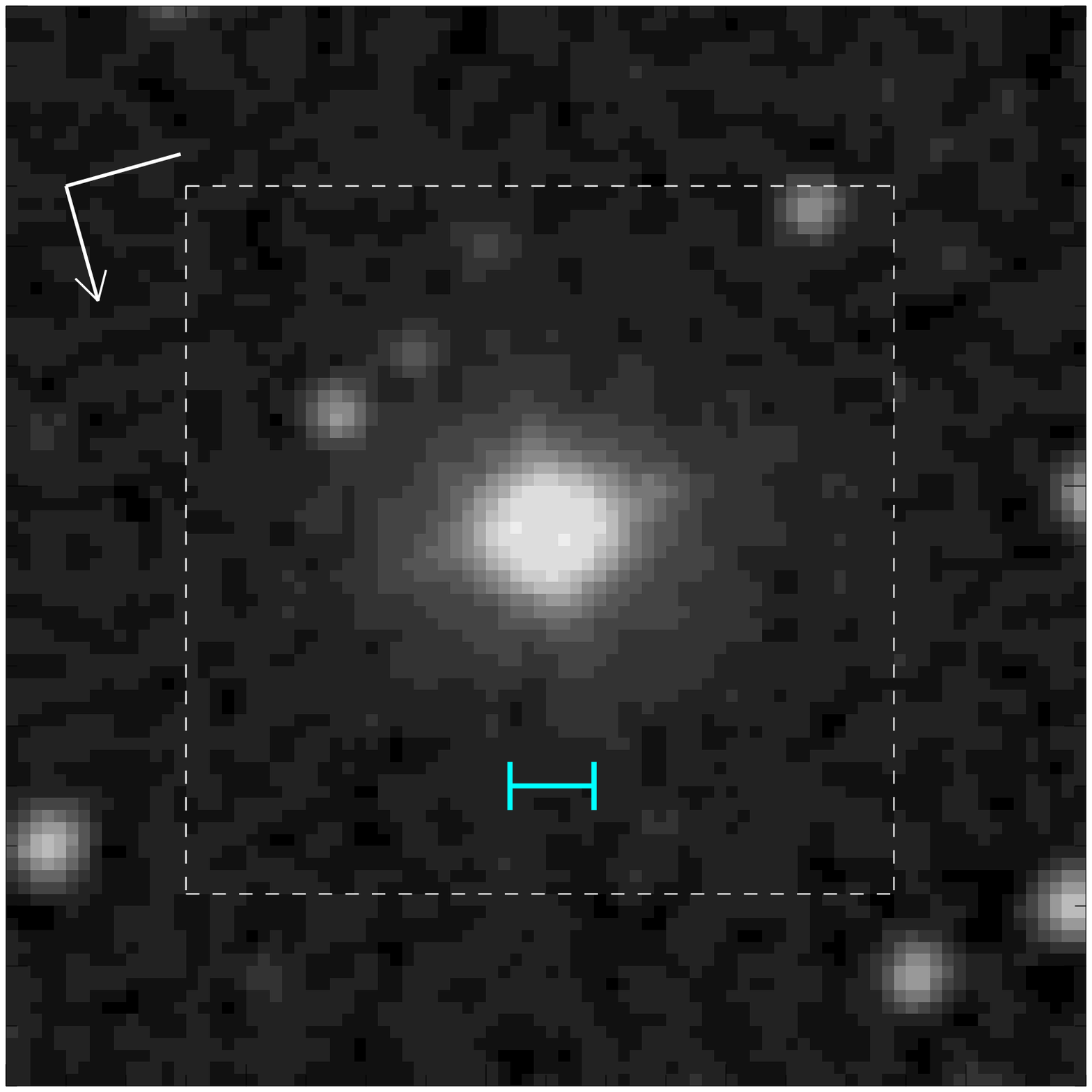}\\
\caption{Continued}
\end{figure}
\clearpage

\begin{figure}
\figurenum{5}
\includegraphics[width=12cm,height=4cm]{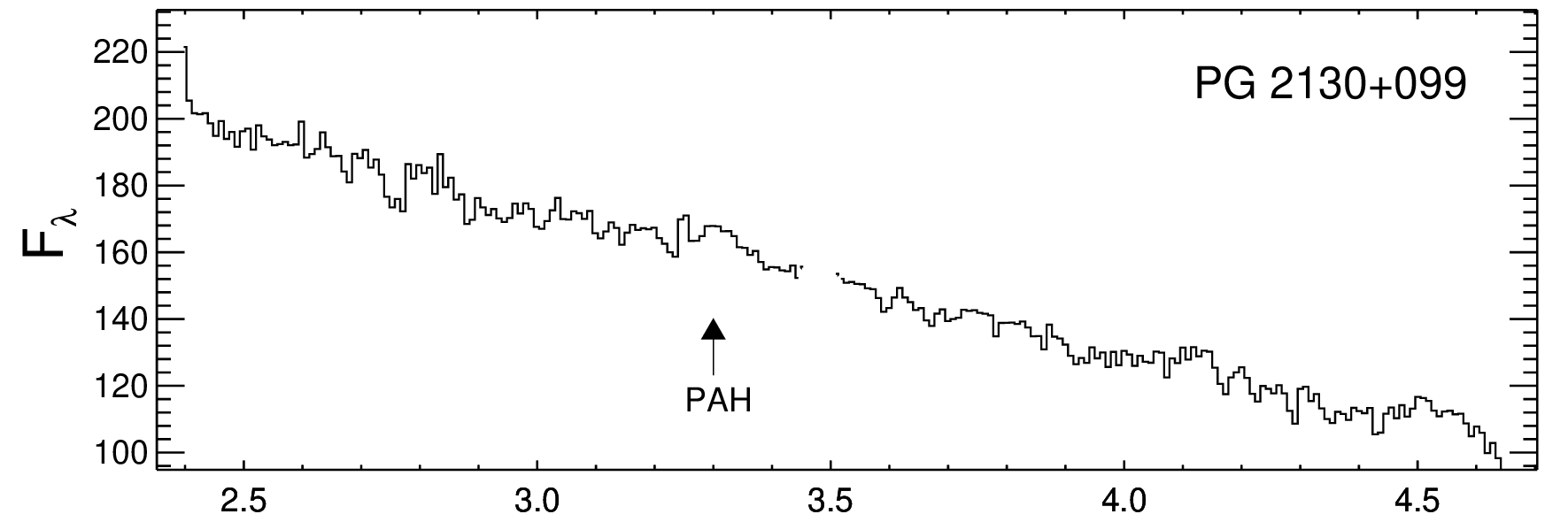}
\includegraphics[scale=0.20]{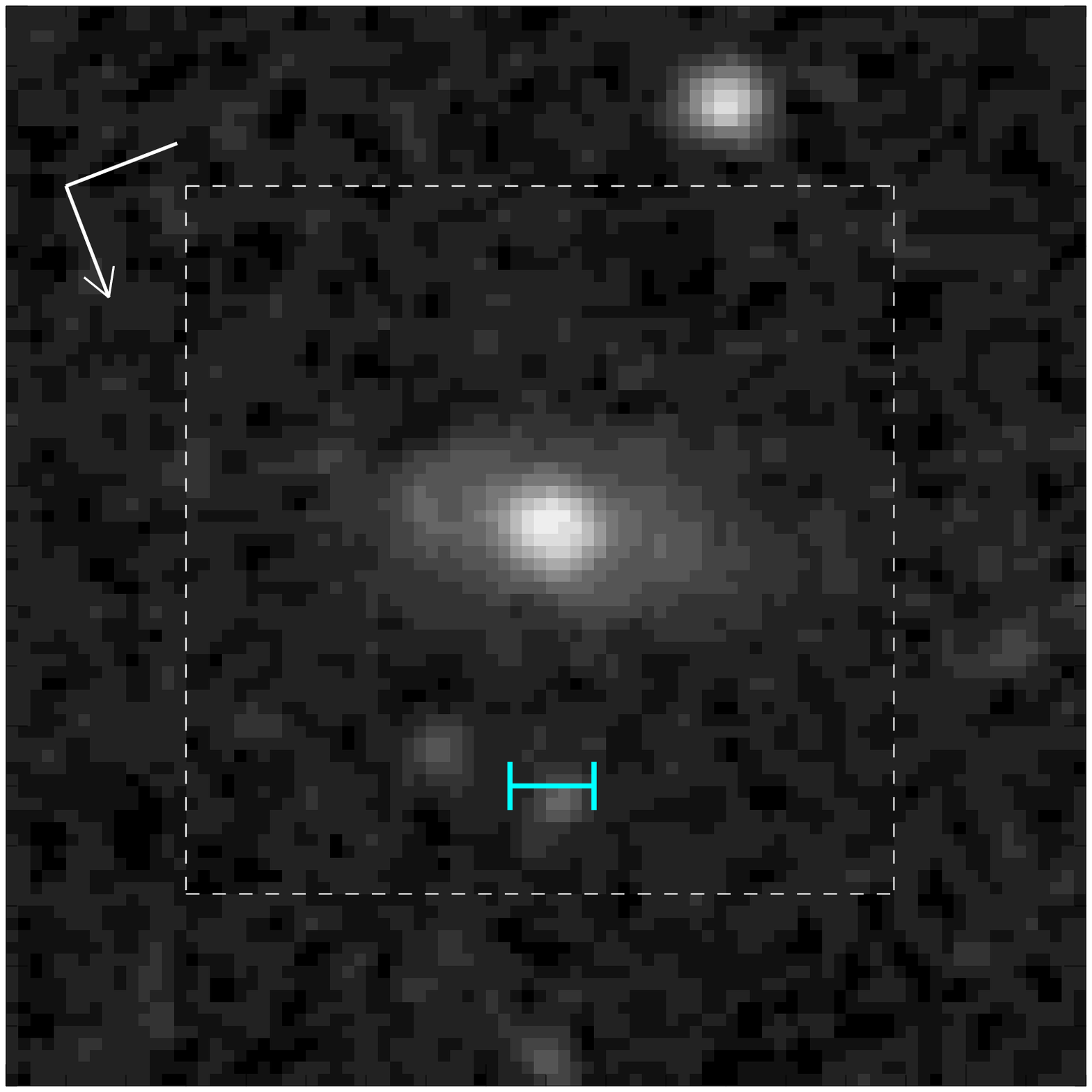}\\
\includegraphics[width=12cm,height=4cm]{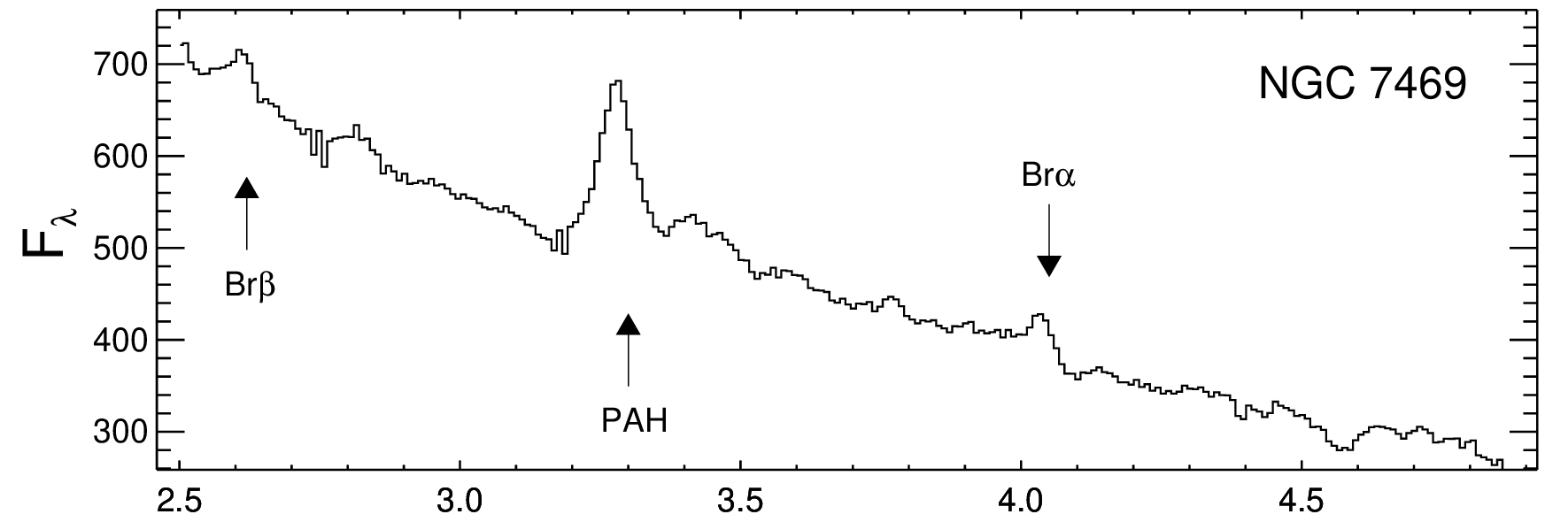}
\includegraphics[scale=0.20]{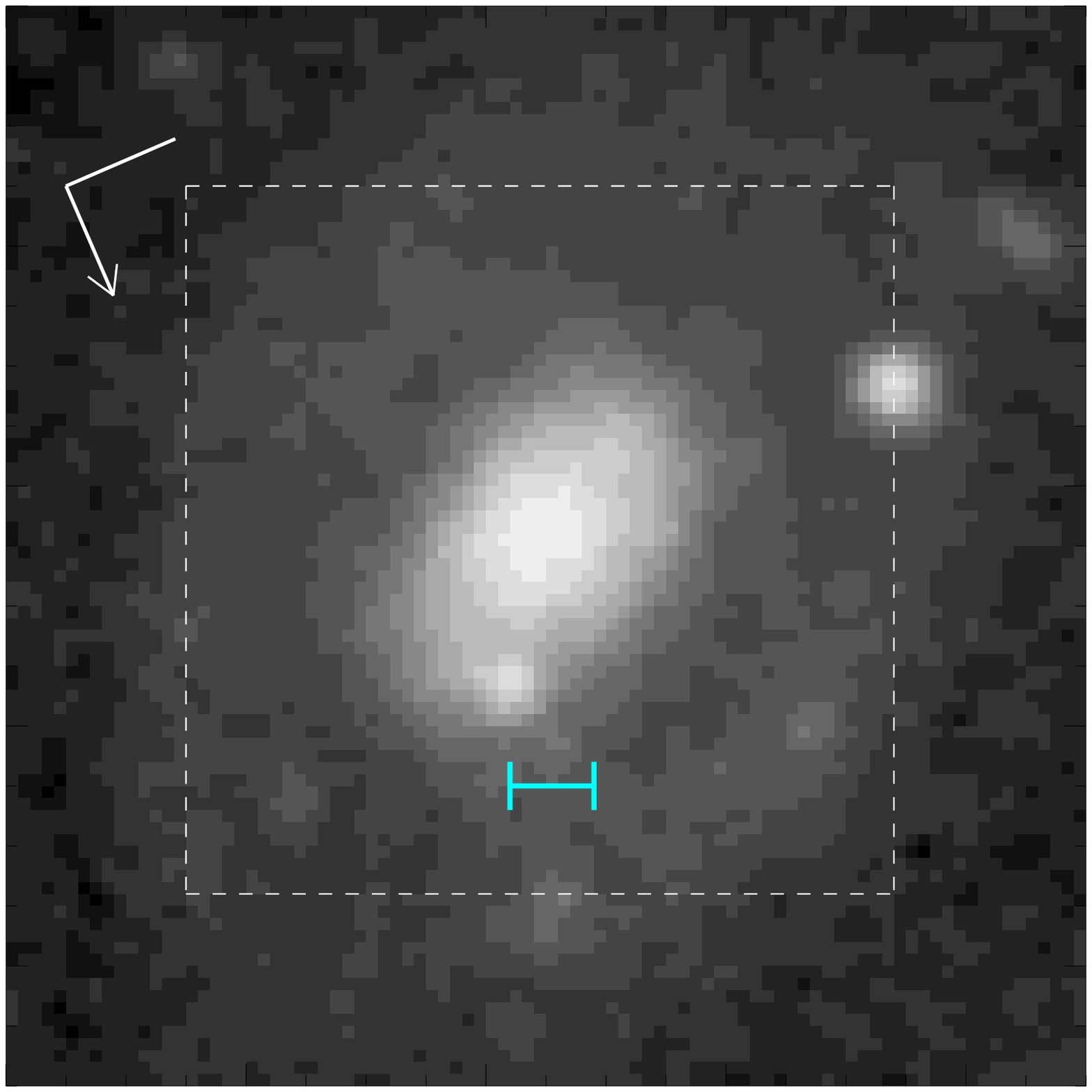}\\
\includegraphics[width=12cm,height=4cm]{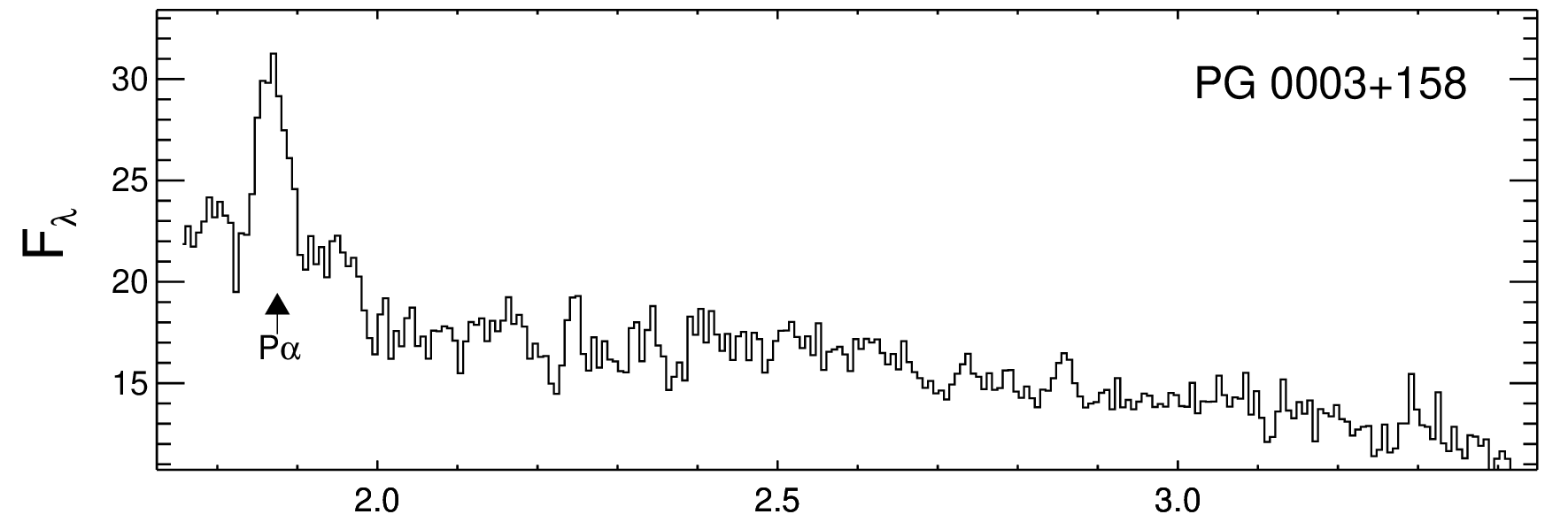}
\includegraphics[scale=0.20]{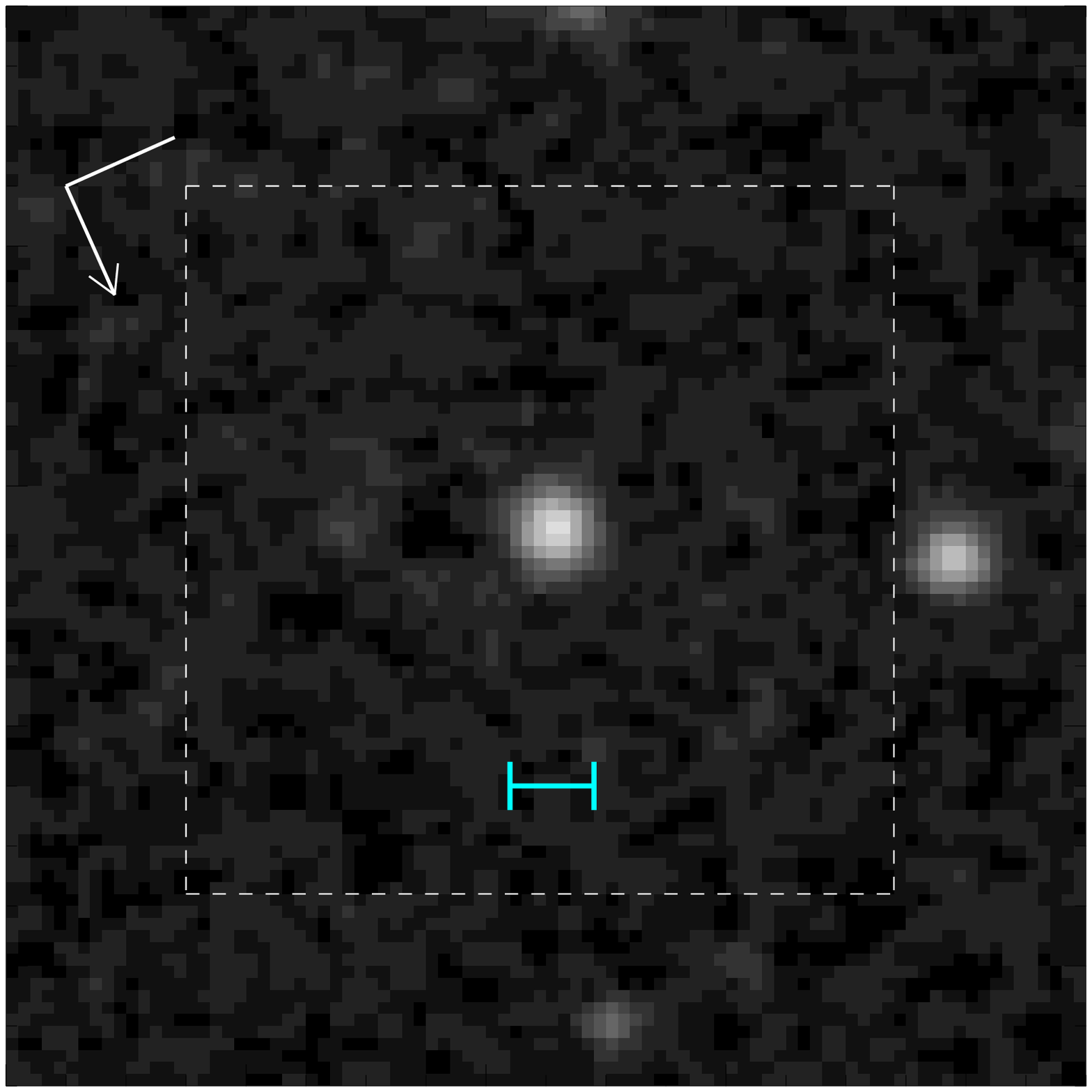}\\
\includegraphics[width=12cm,height=4cm]{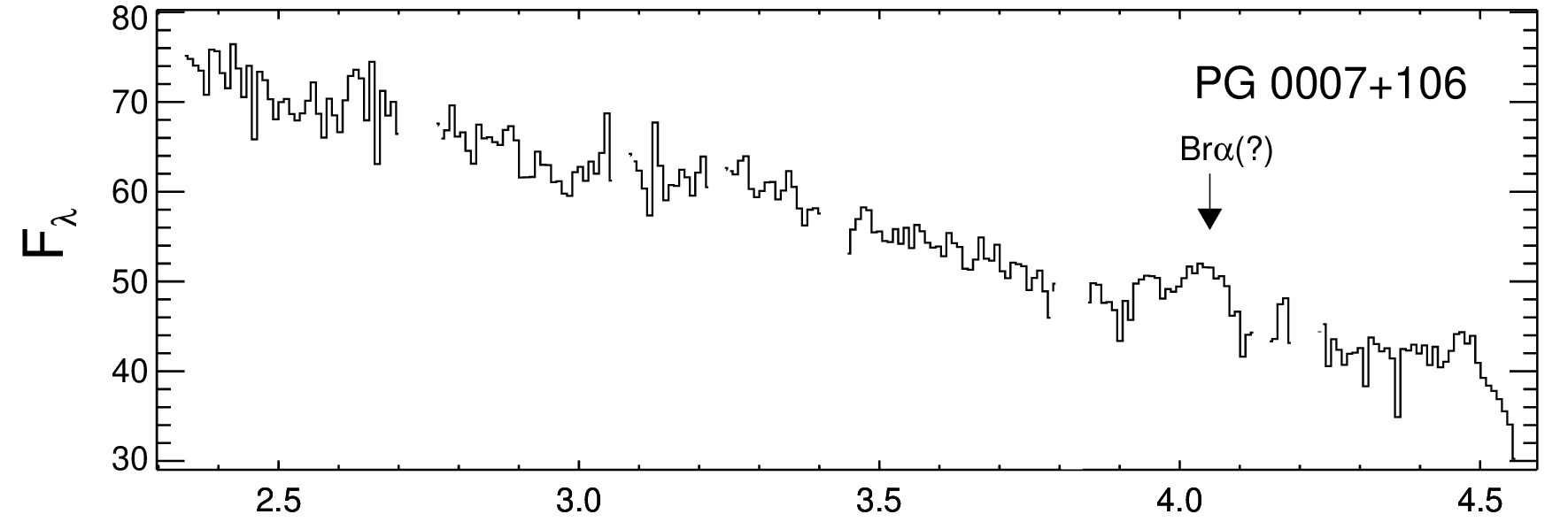}
\includegraphics[scale=0.20]{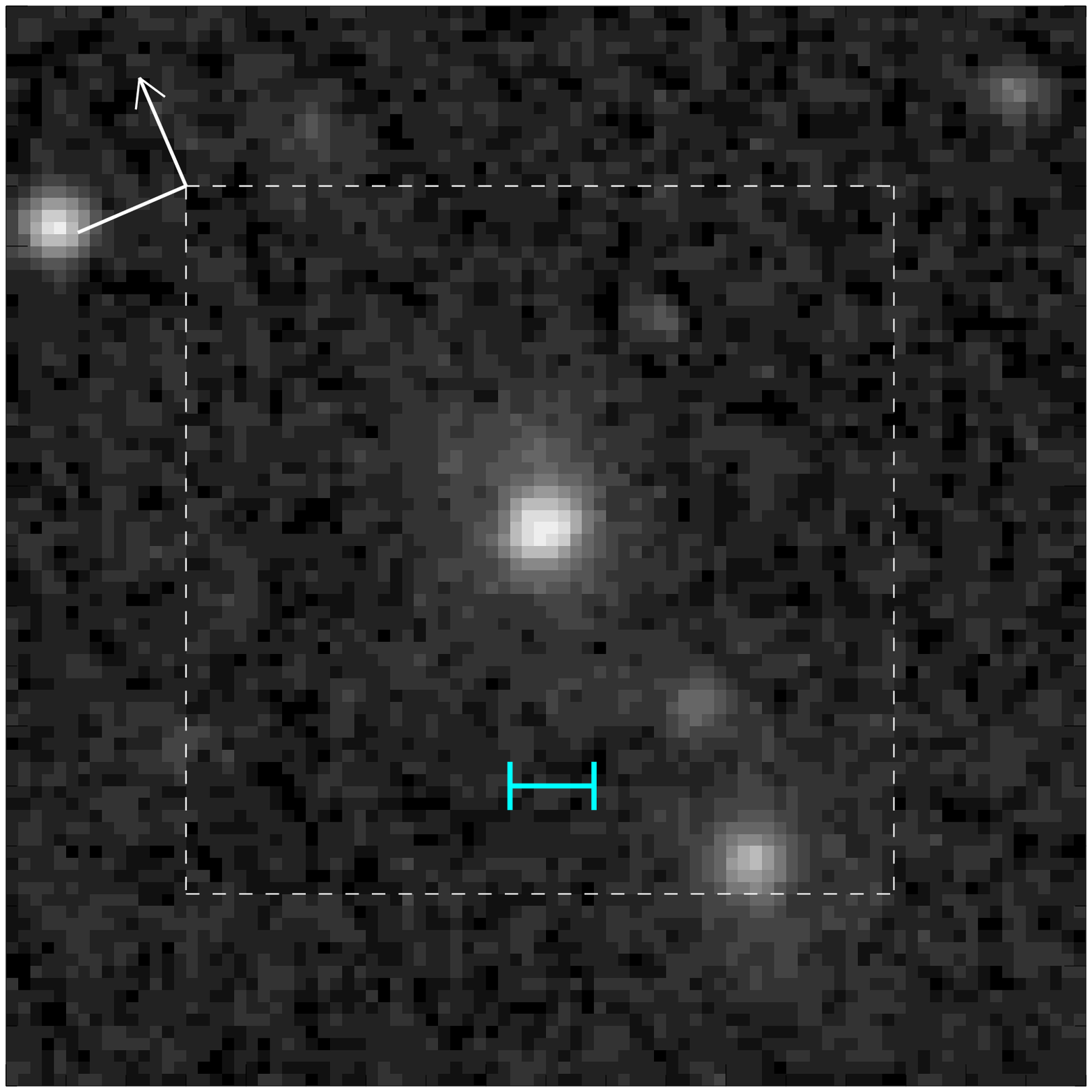}\\
\includegraphics[width=12cm,height=4cm]{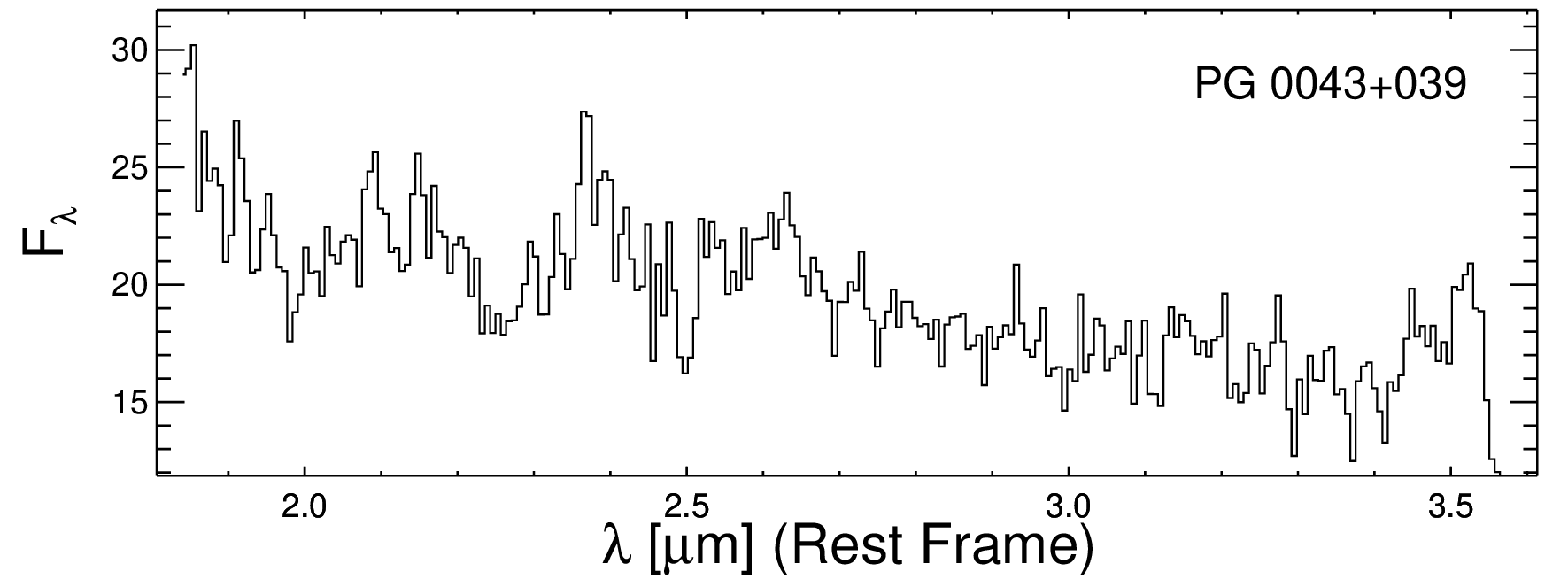}
\includegraphics[scale=0.20]{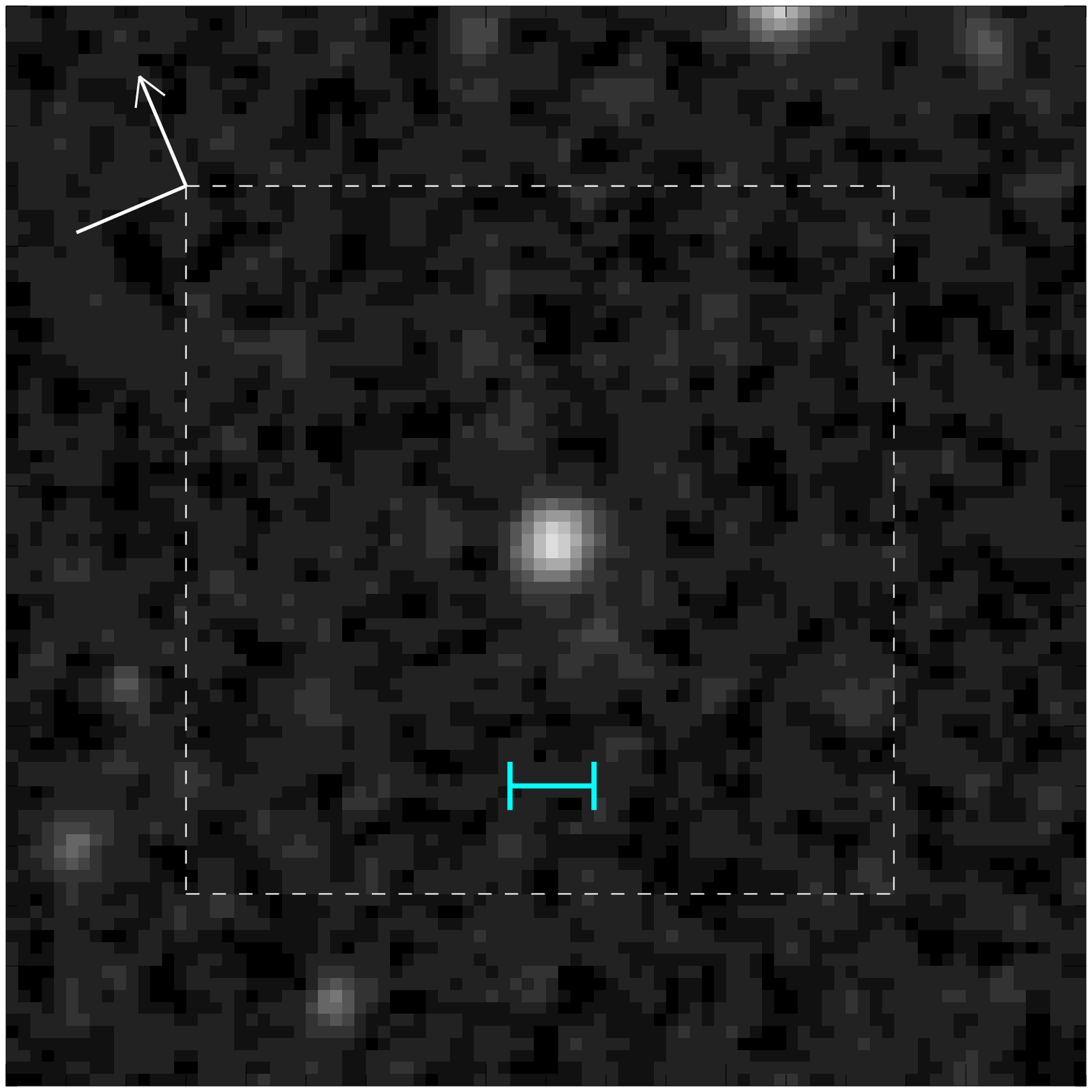}\\
\caption{Continued}
\end{figure}
\clearpage

\begin{figure}
\figurenum{5}
\includegraphics[width=12cm,height=4cm]{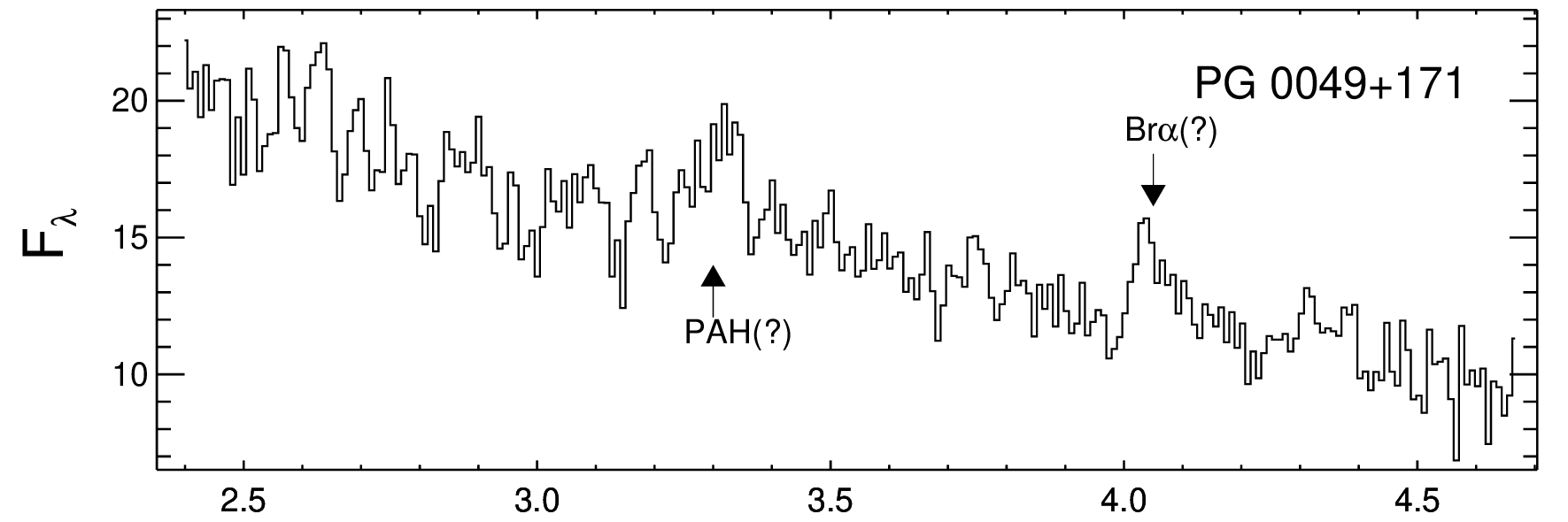}
\includegraphics[scale=0.20]{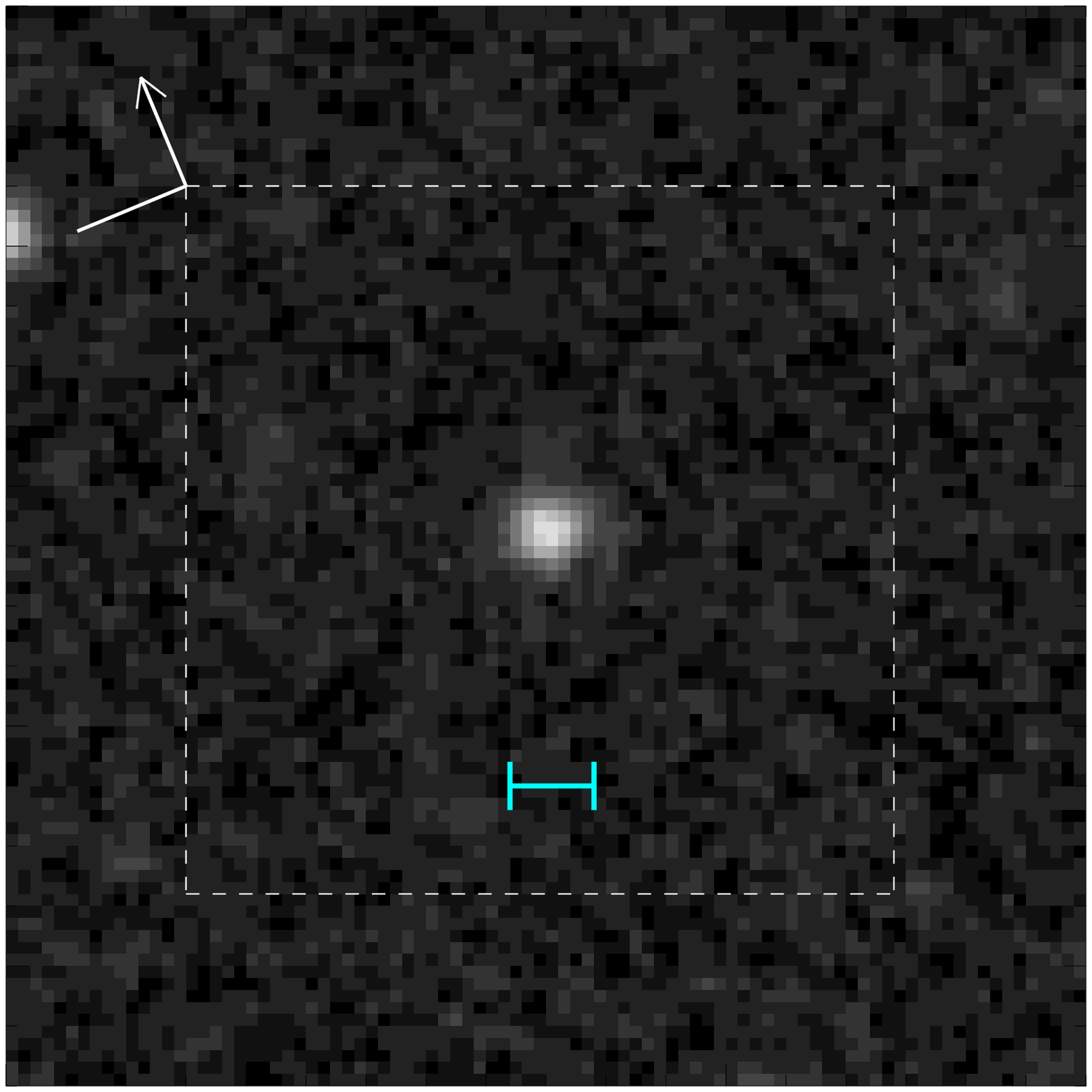}\\
\includegraphics[width=12cm,height=4cm]{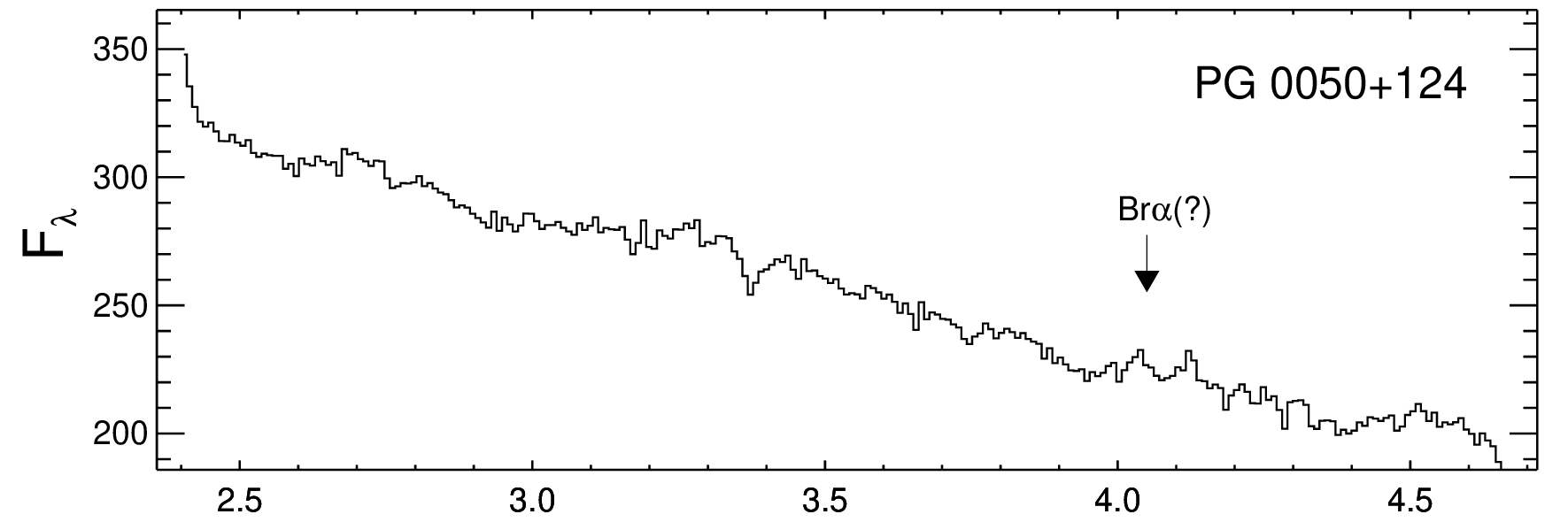}
\includegraphics[scale=0.20]{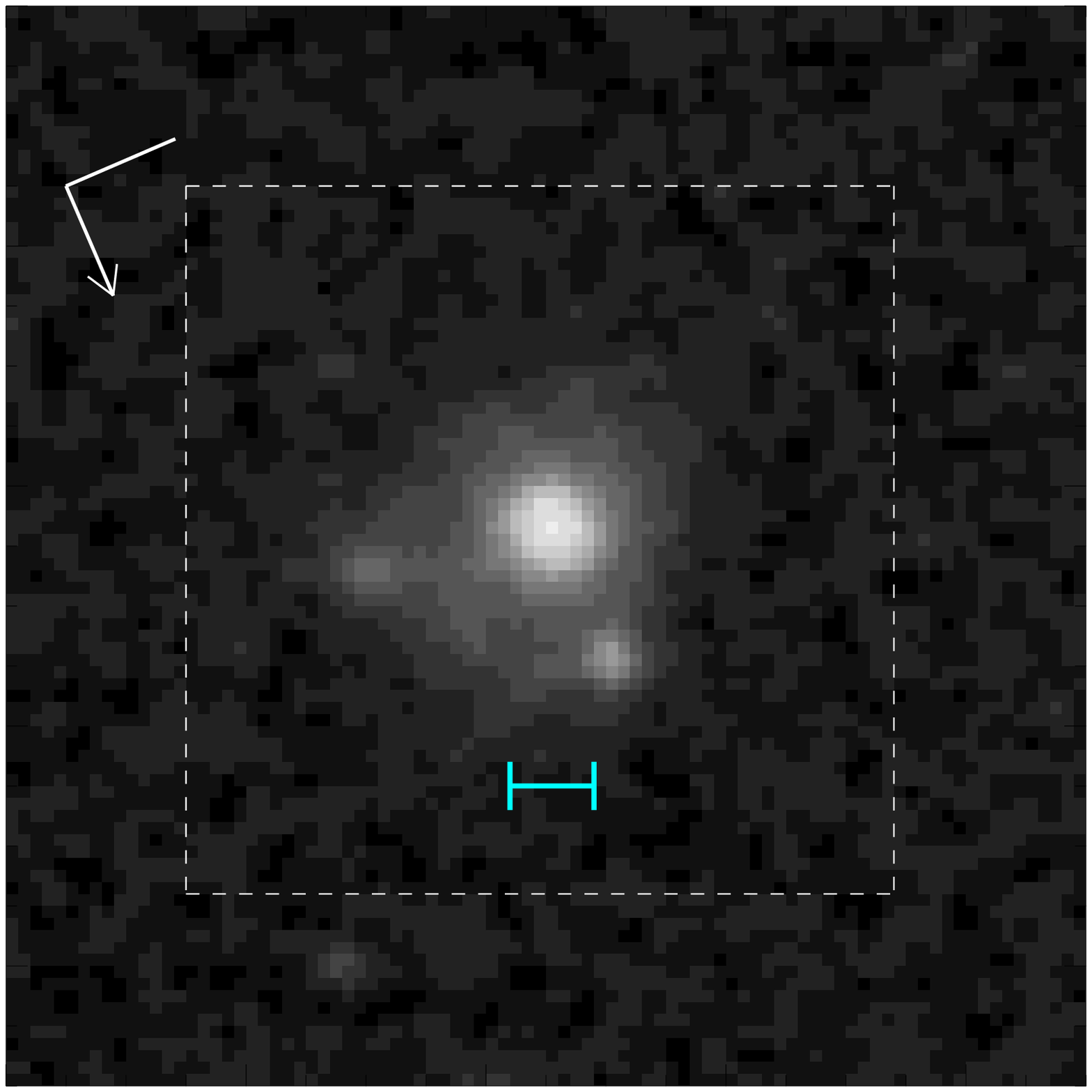}\\
\includegraphics[width=12cm,height=4cm]{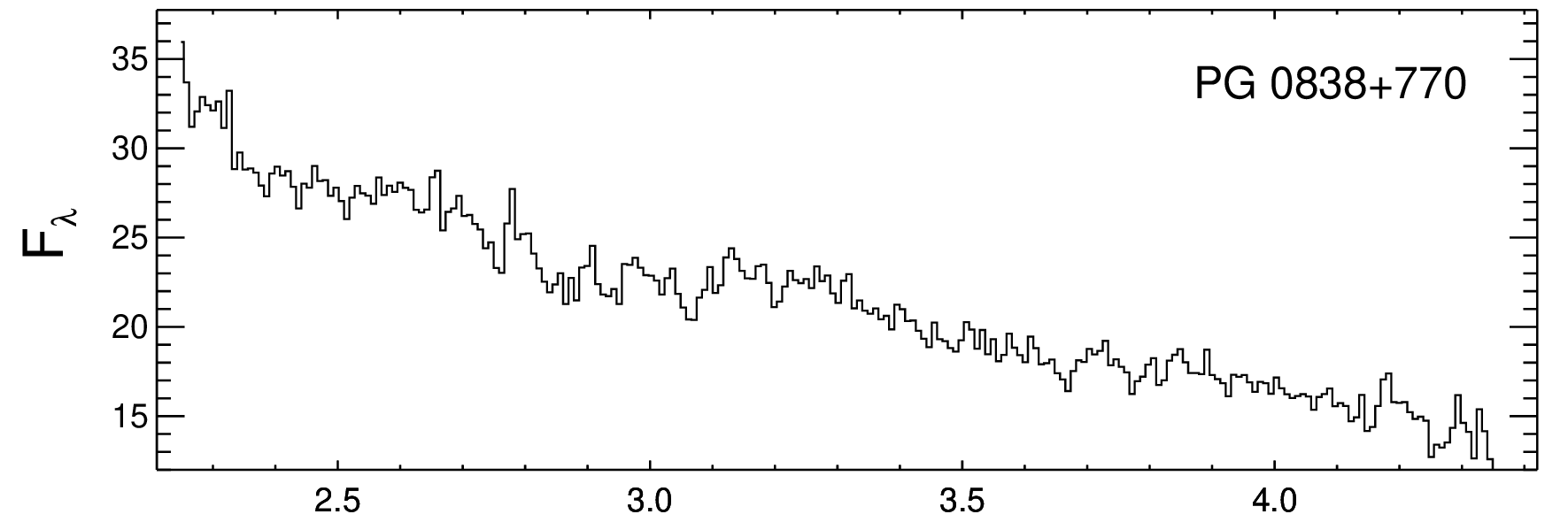}
\includegraphics[scale=0.20]{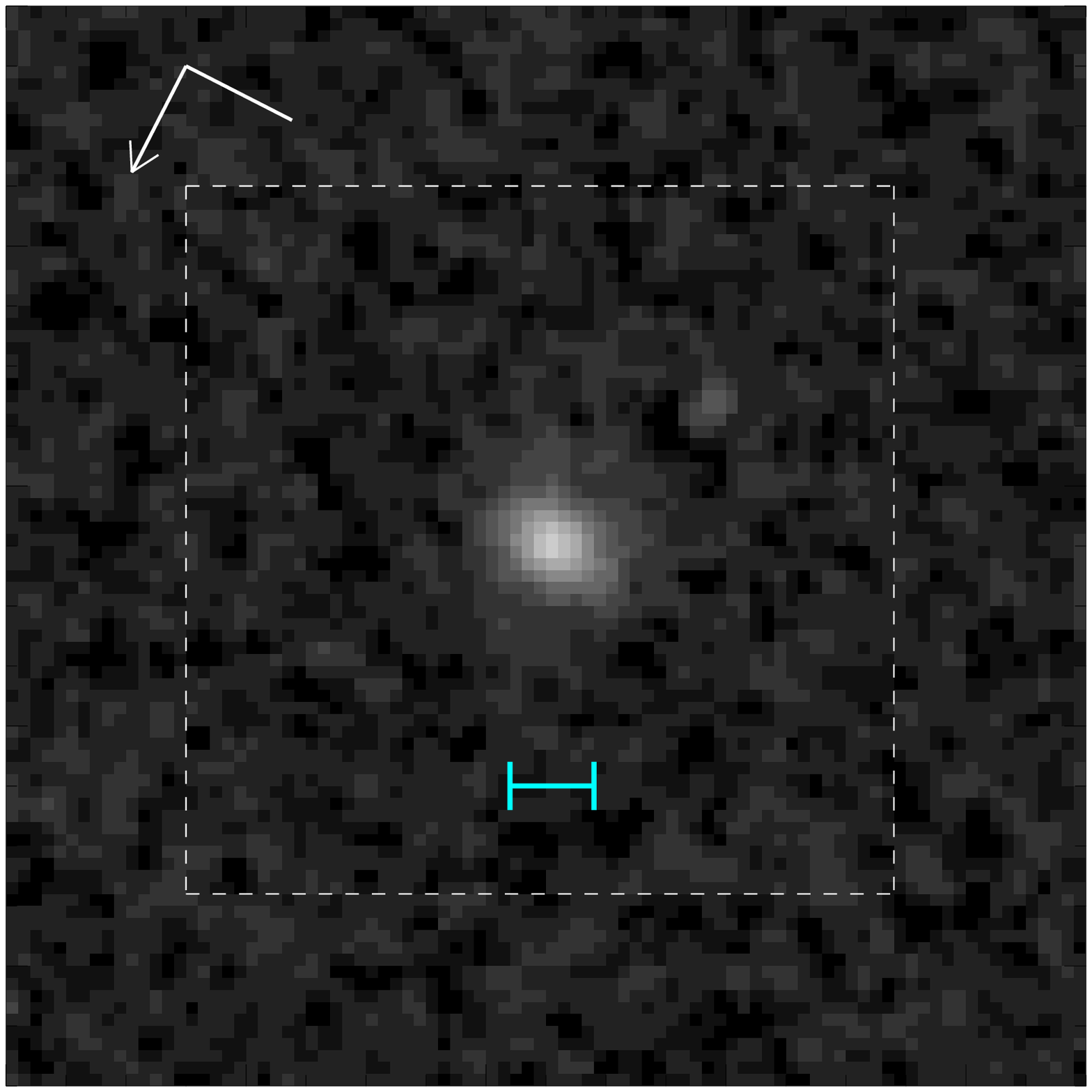}\\
\includegraphics[width=12cm,height=4cm]{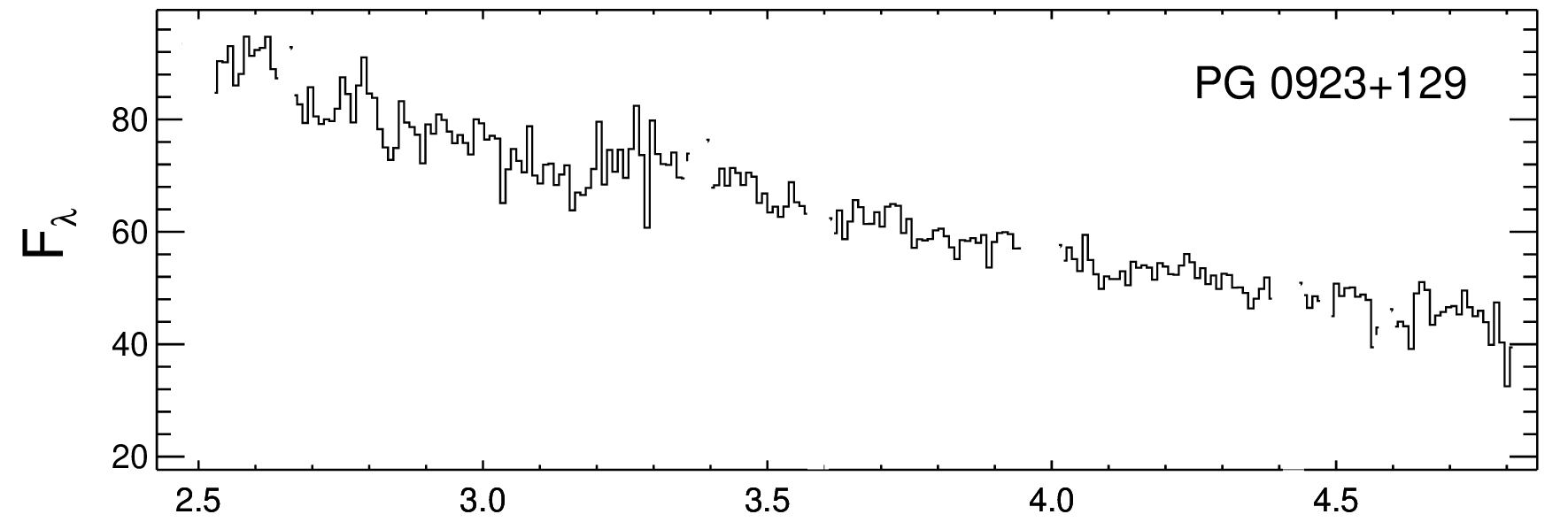}
\includegraphics[scale=0.20]{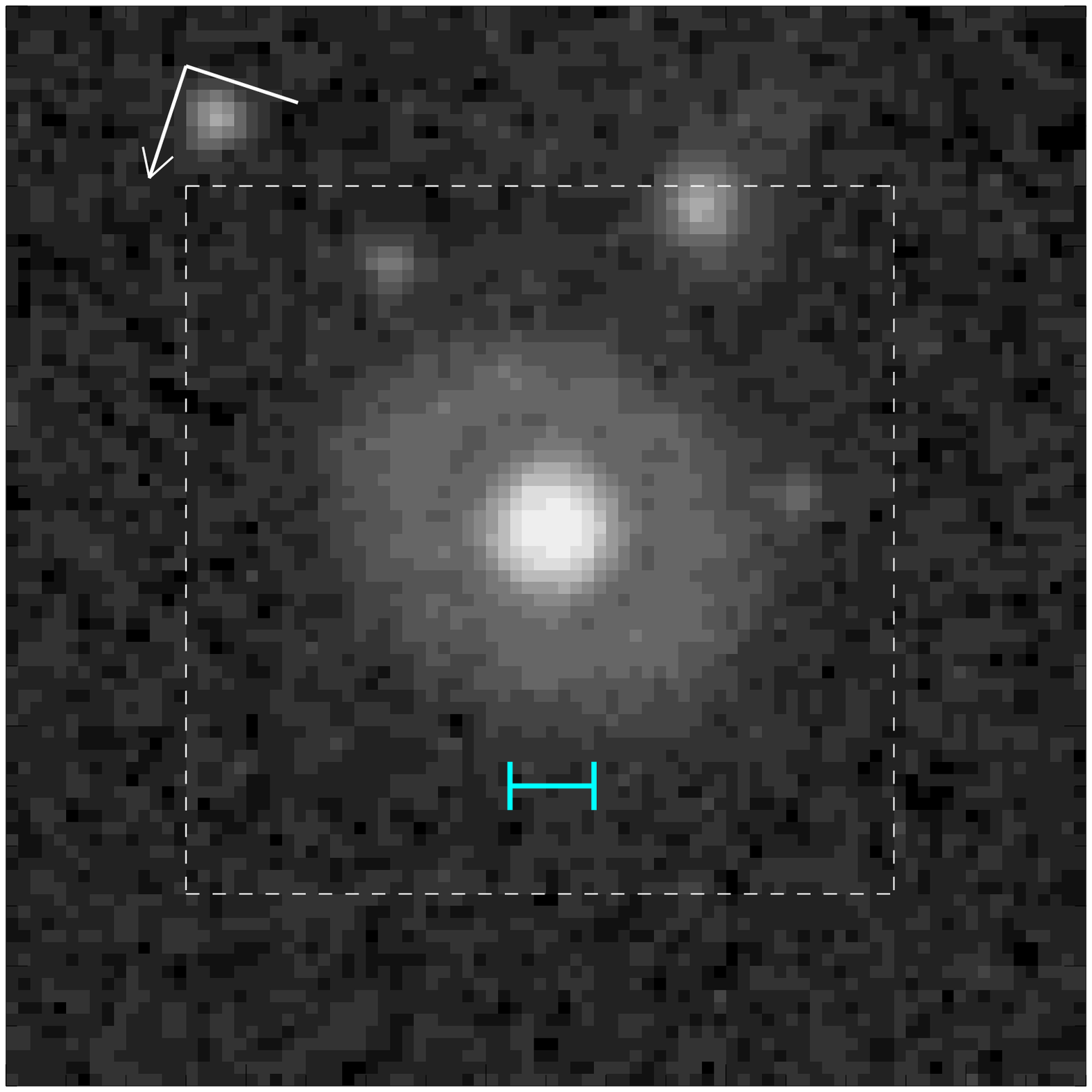}\\
\includegraphics[width=12cm,height=4cm]{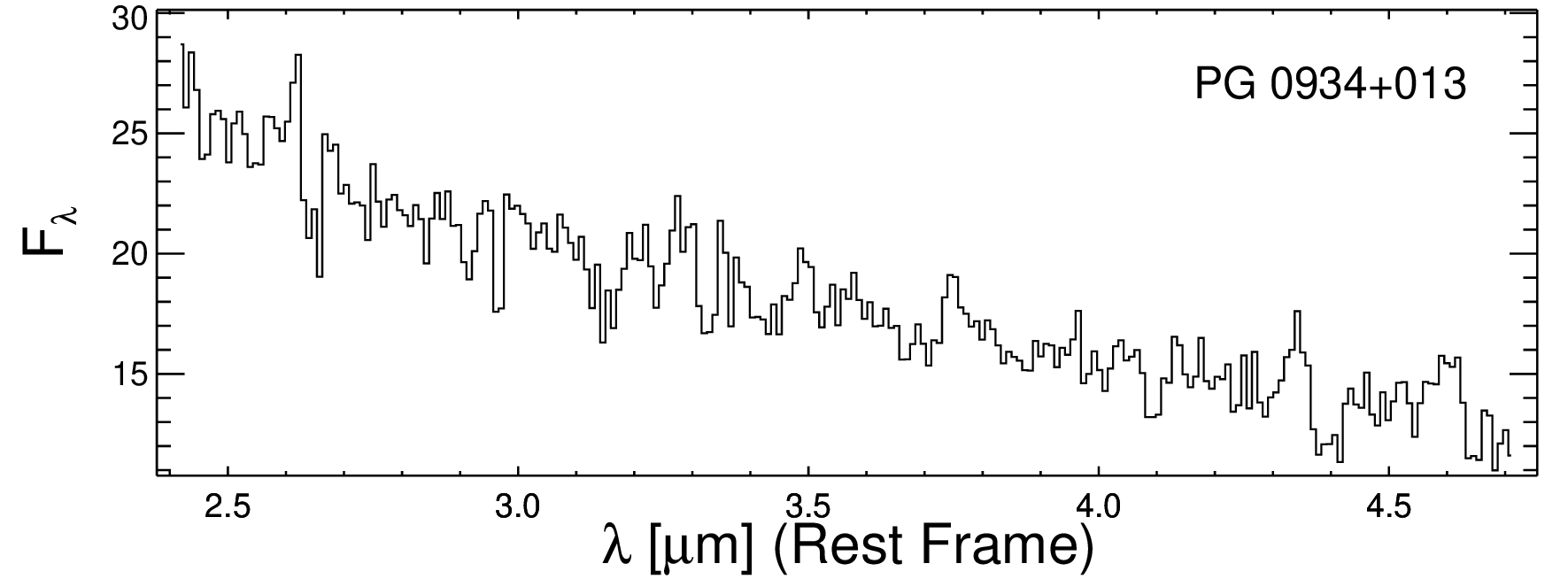}
\includegraphics[scale=0.20]{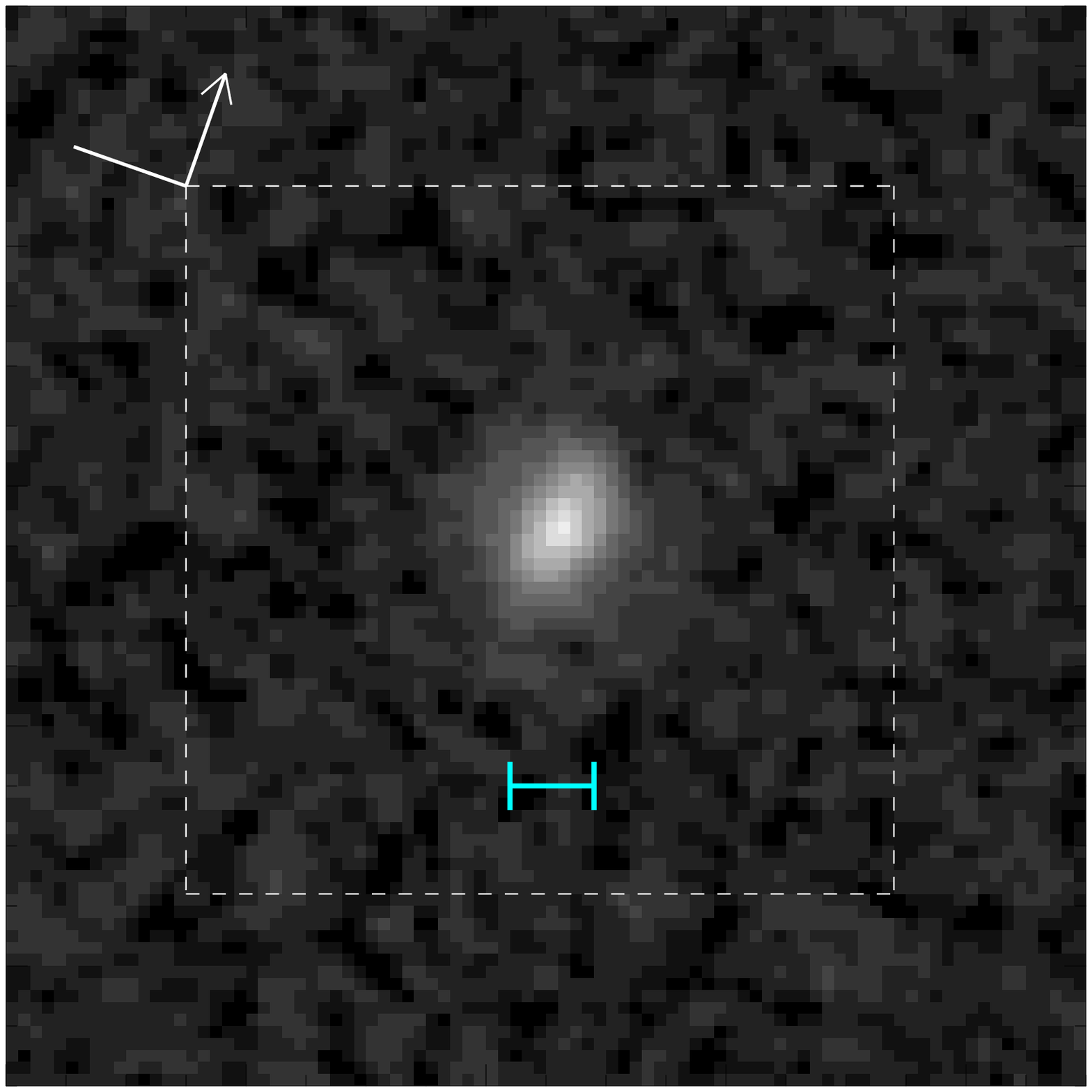}\\
\caption{Continued}
\end{figure}
\clearpage

\begin{figure}
\figurenum{5}
\includegraphics[width=12cm,height=4cm]{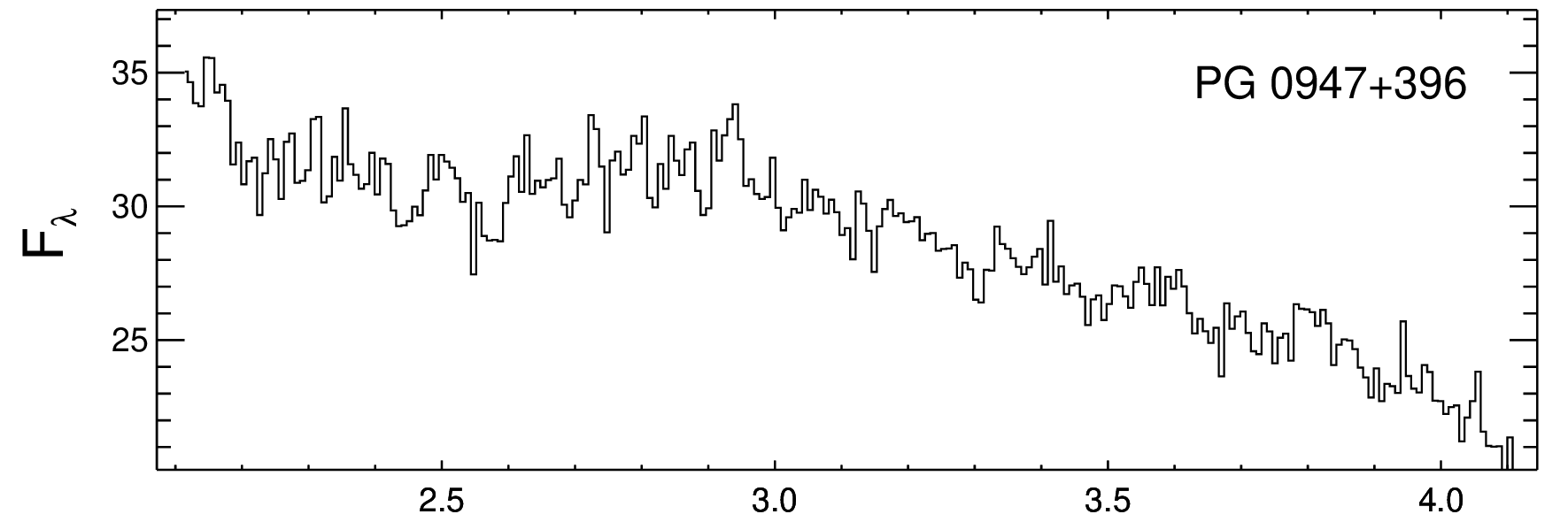}
\includegraphics[scale=0.20]{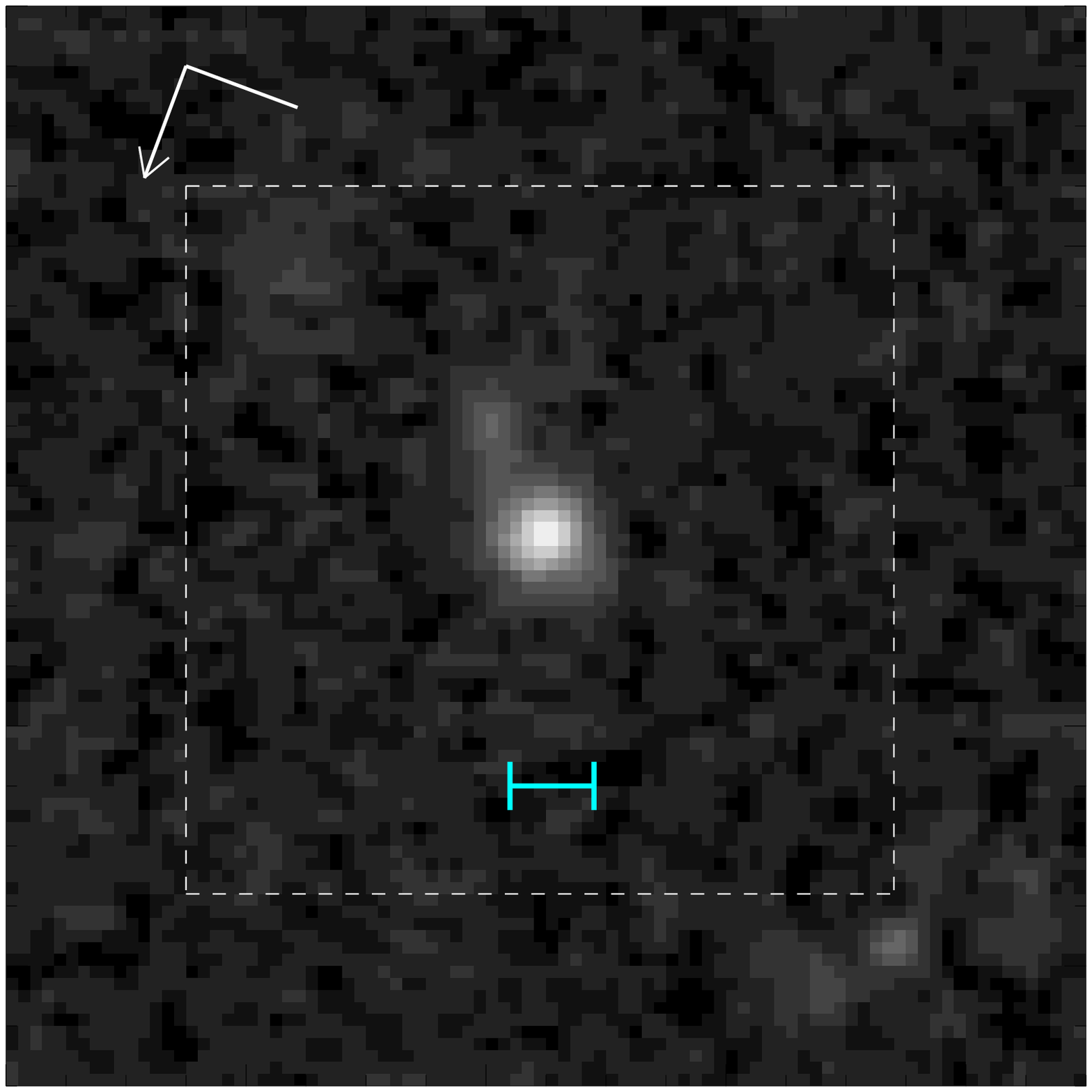}\\
\includegraphics[width=12cm,height=4cm]{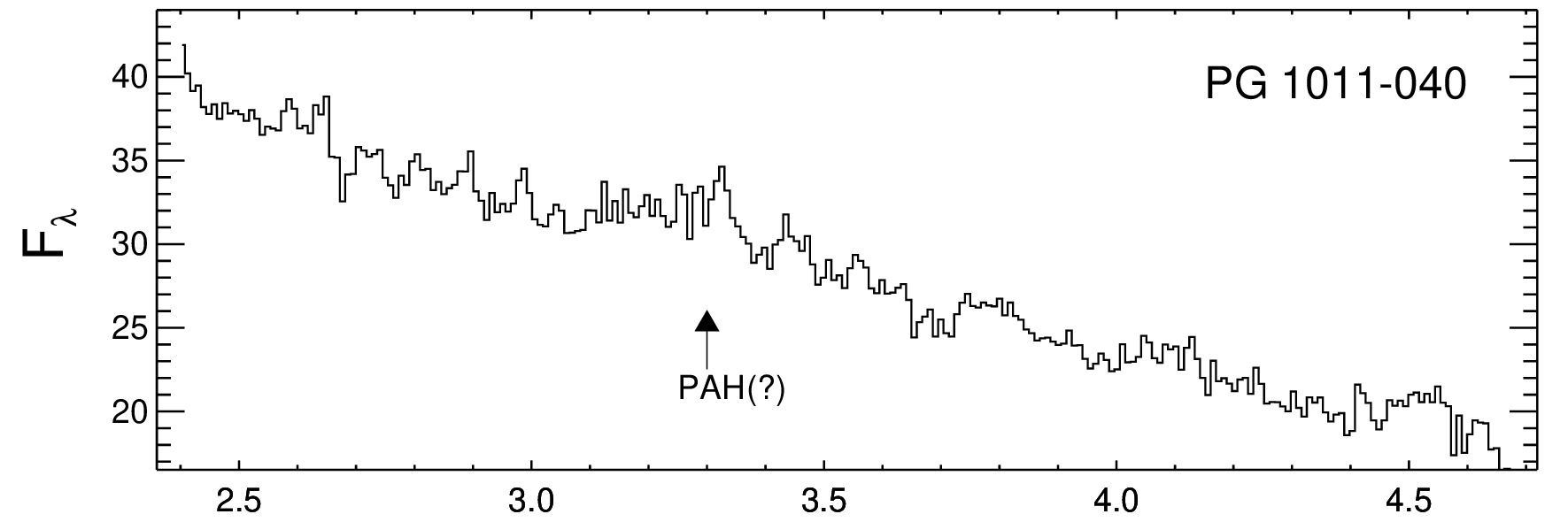}
\includegraphics[scale=0.20]{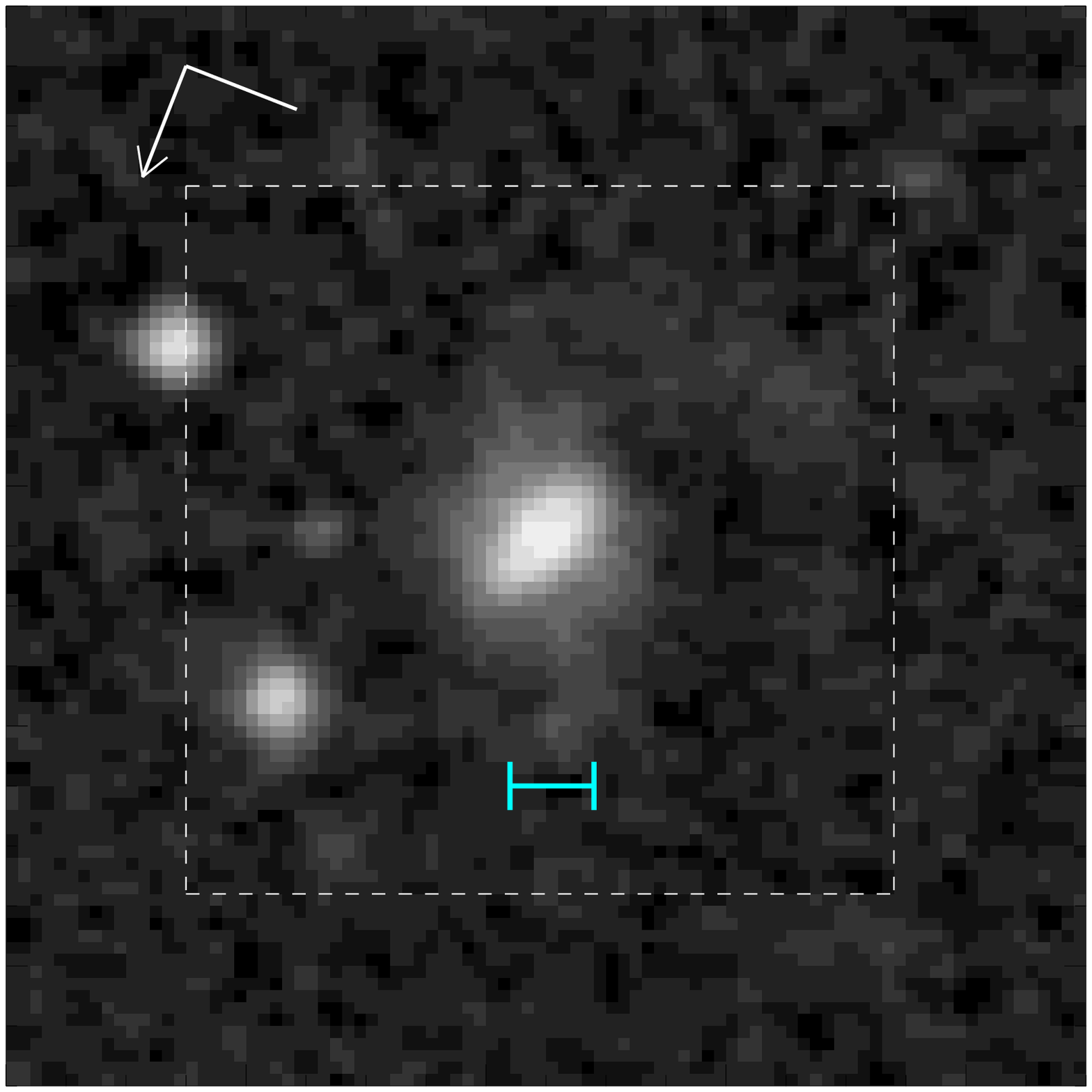}\\
\includegraphics[width=12cm,height=4cm]{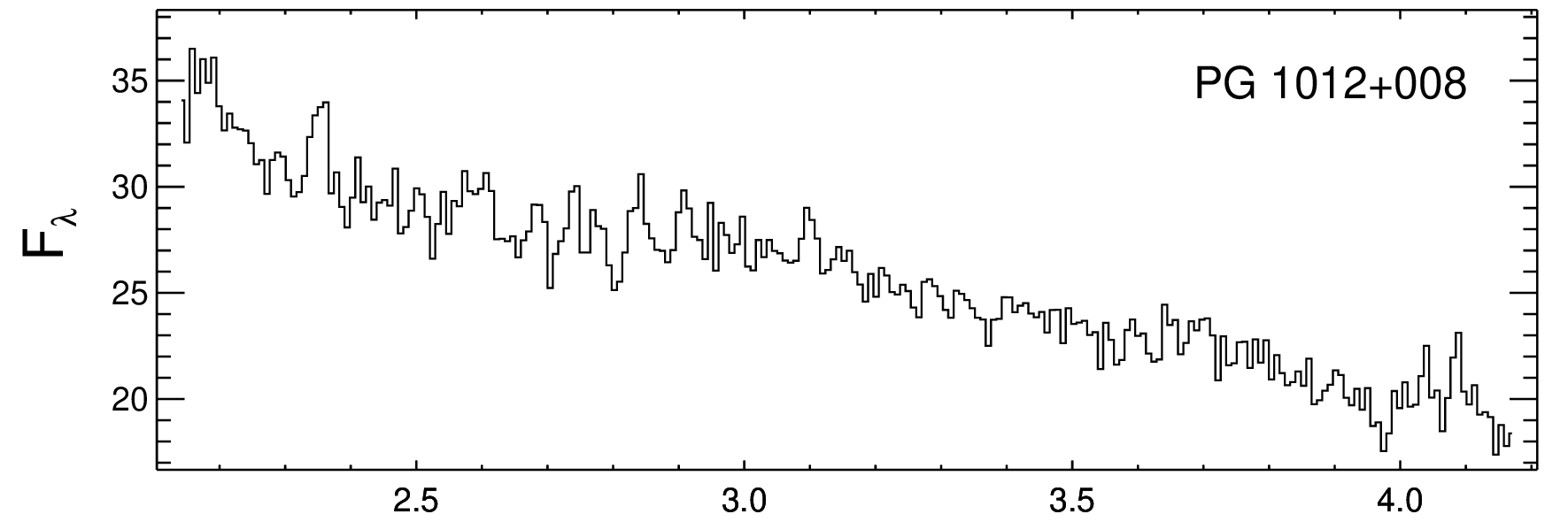}
\includegraphics[scale=0.20]{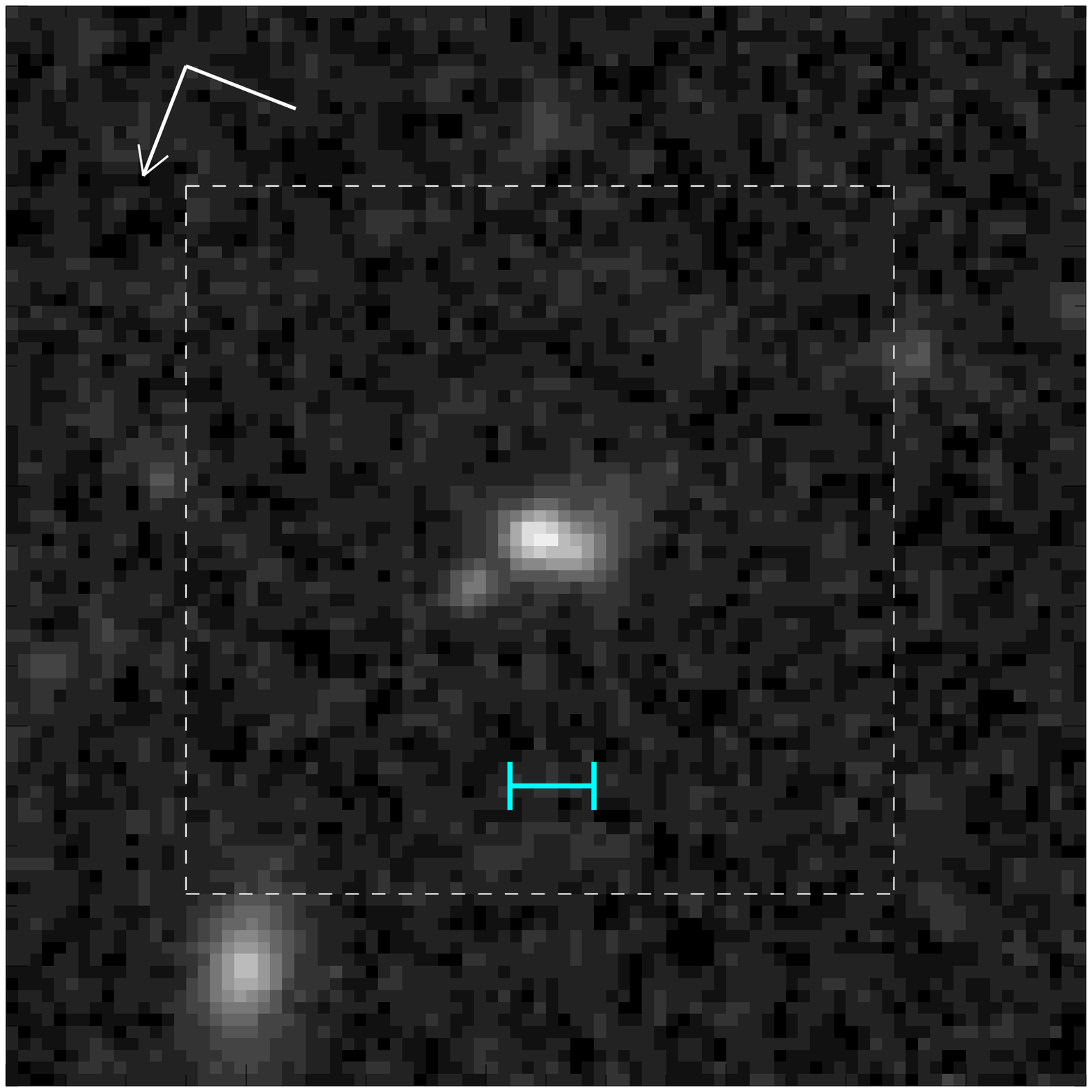}\\
\includegraphics[width=12cm,height=4cm]{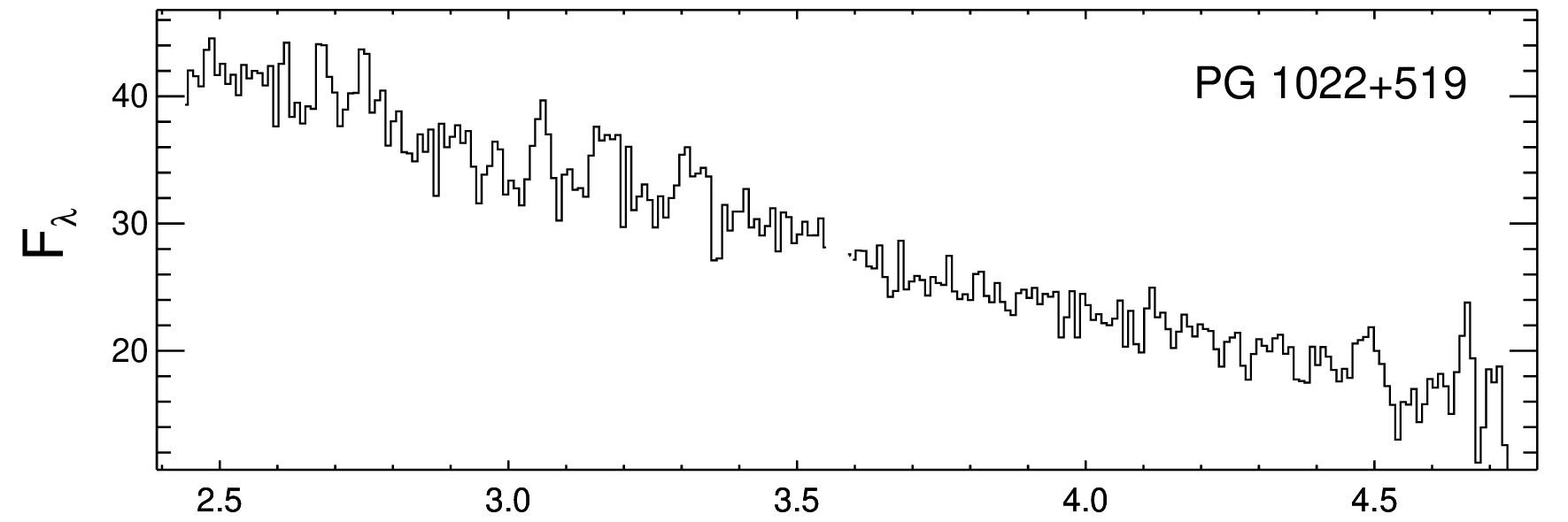}
\includegraphics[scale=0.20]{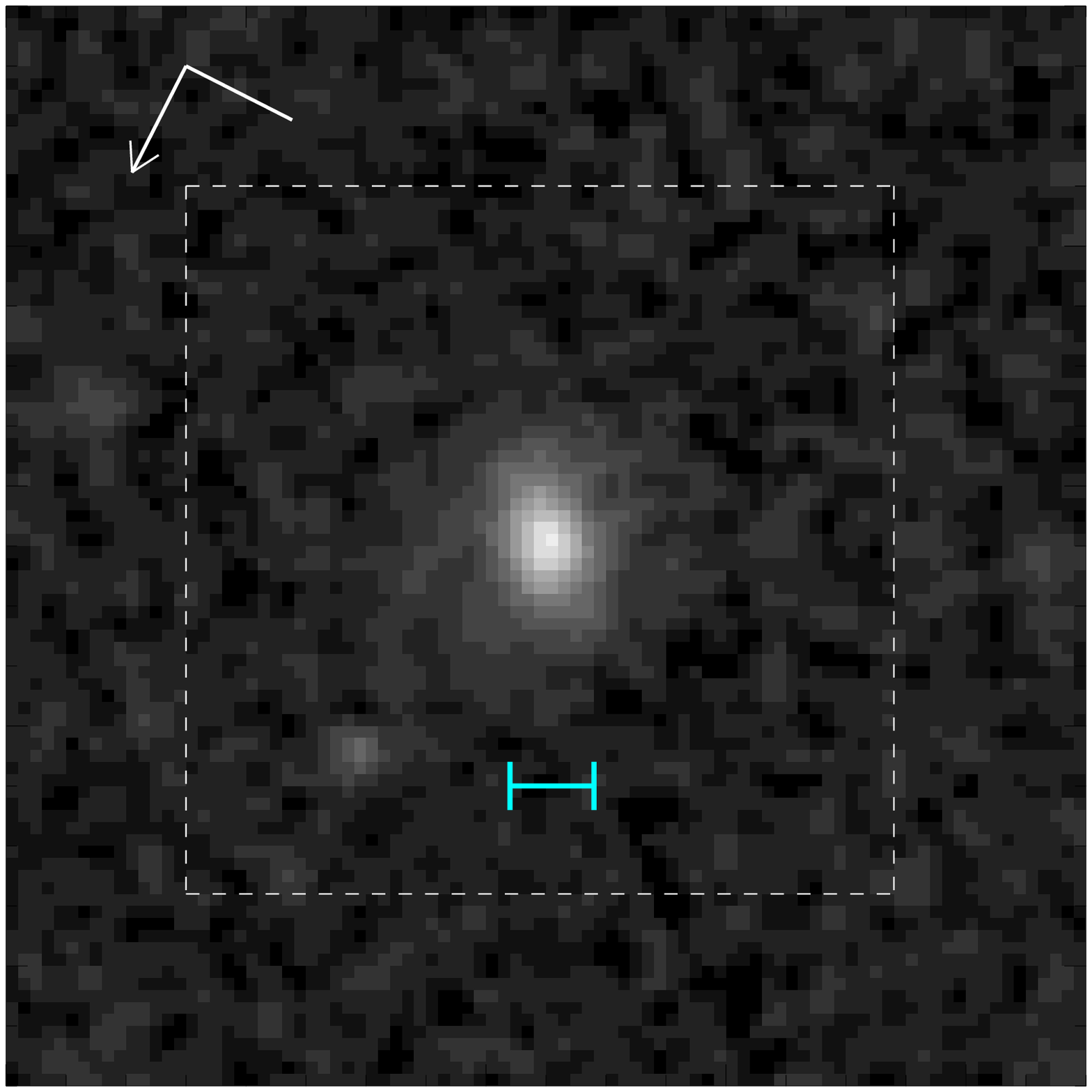}\\
\includegraphics[width=12cm,height=4cm]{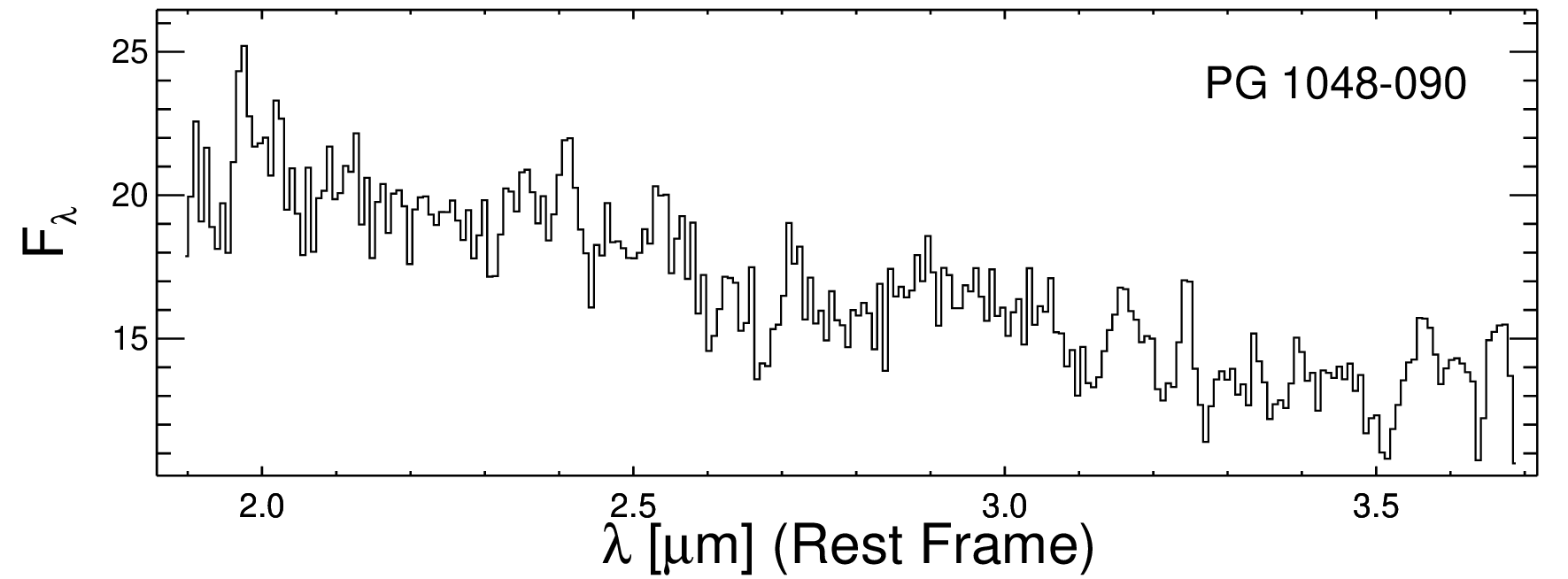}
\includegraphics[scale=0.20]{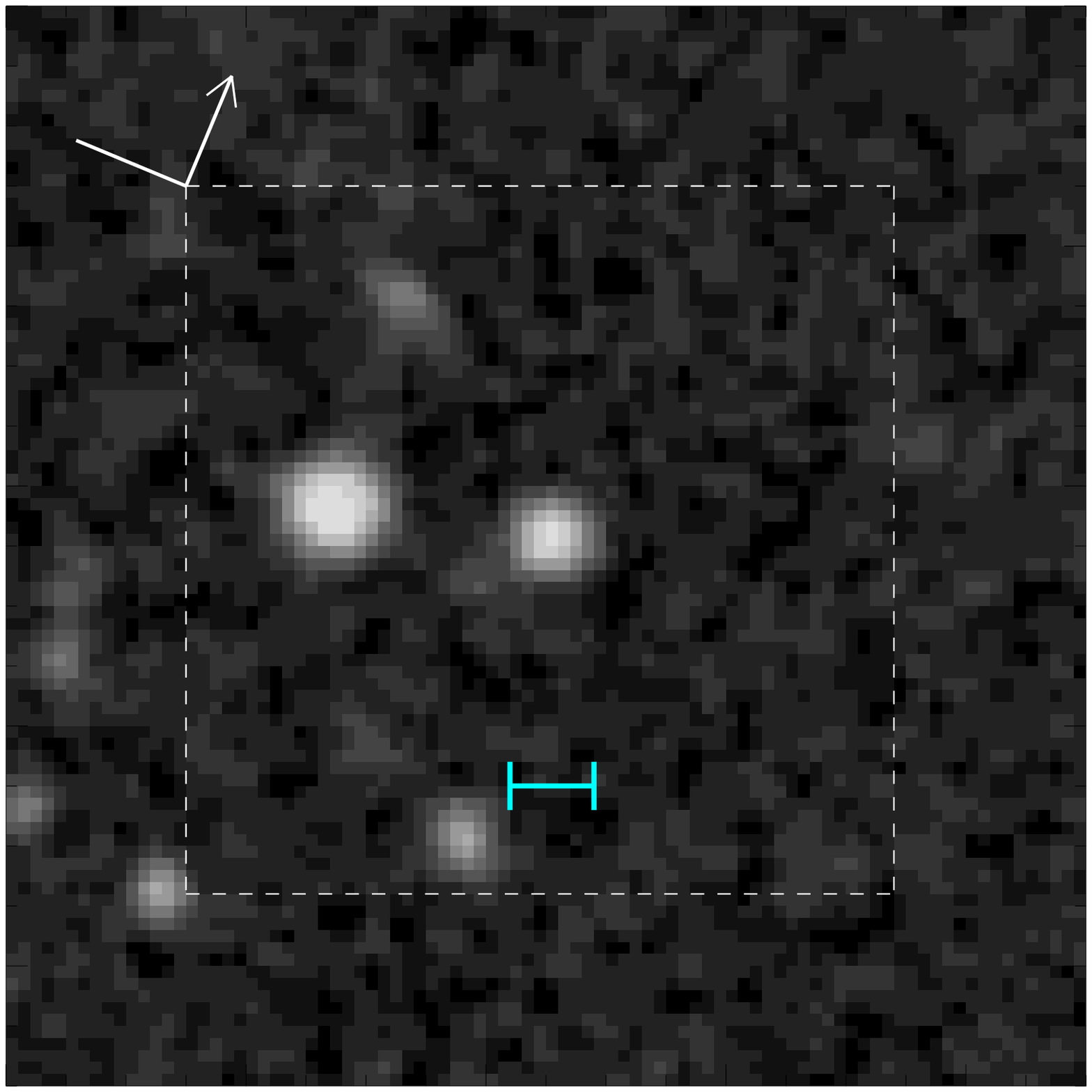}\\
\caption{Continued}
\end{figure}
\clearpage

\begin{figure}
\figurenum{5}
\includegraphics[width=12cm,height=4cm]{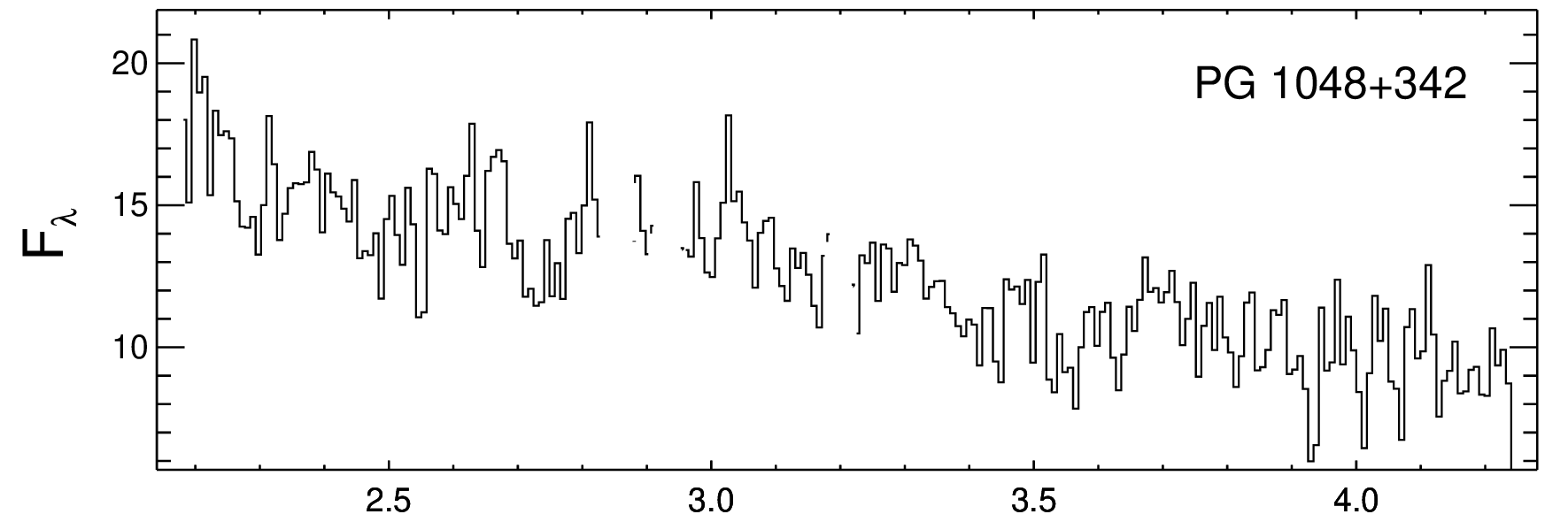}
\includegraphics[scale=0.20]{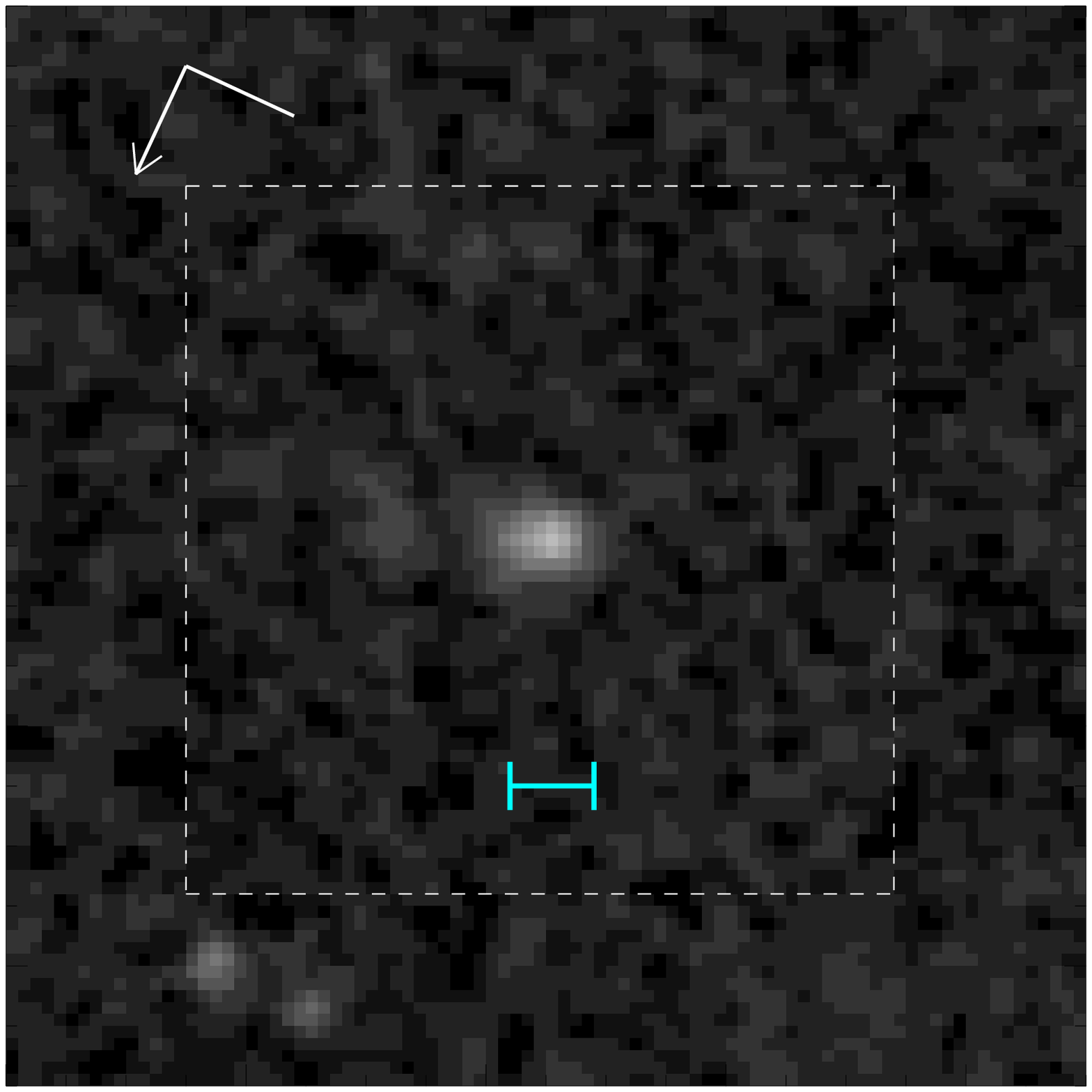}\\
\includegraphics[width=12cm,height=4cm]{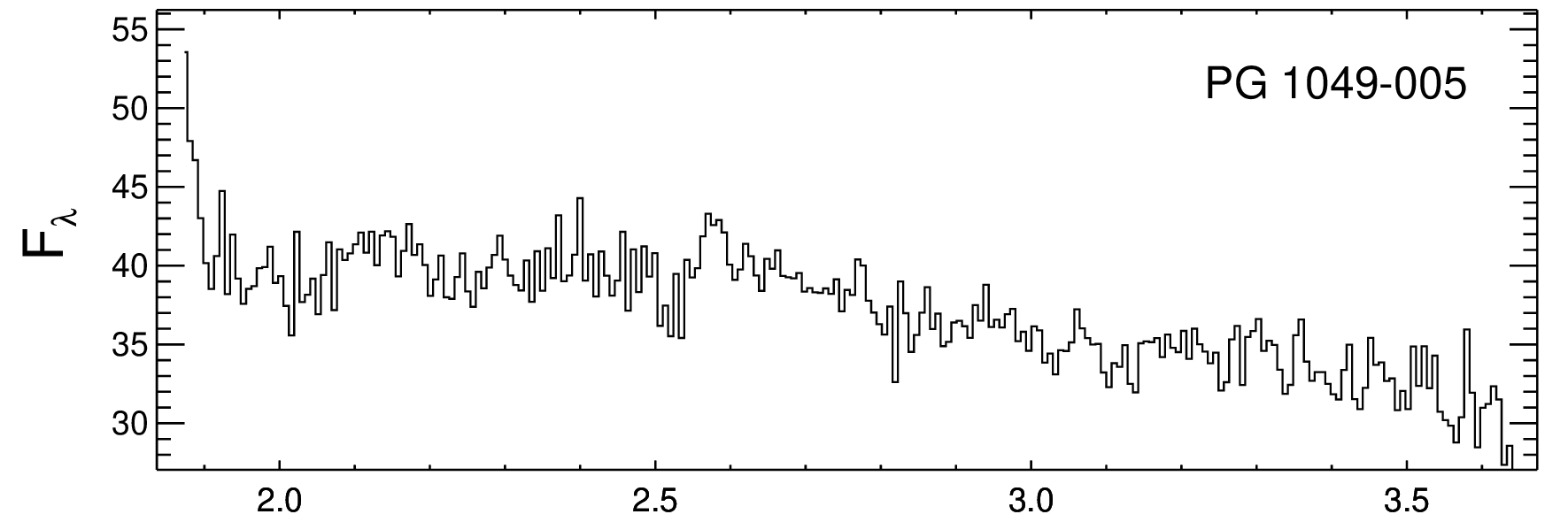}
\includegraphics[scale=0.20]{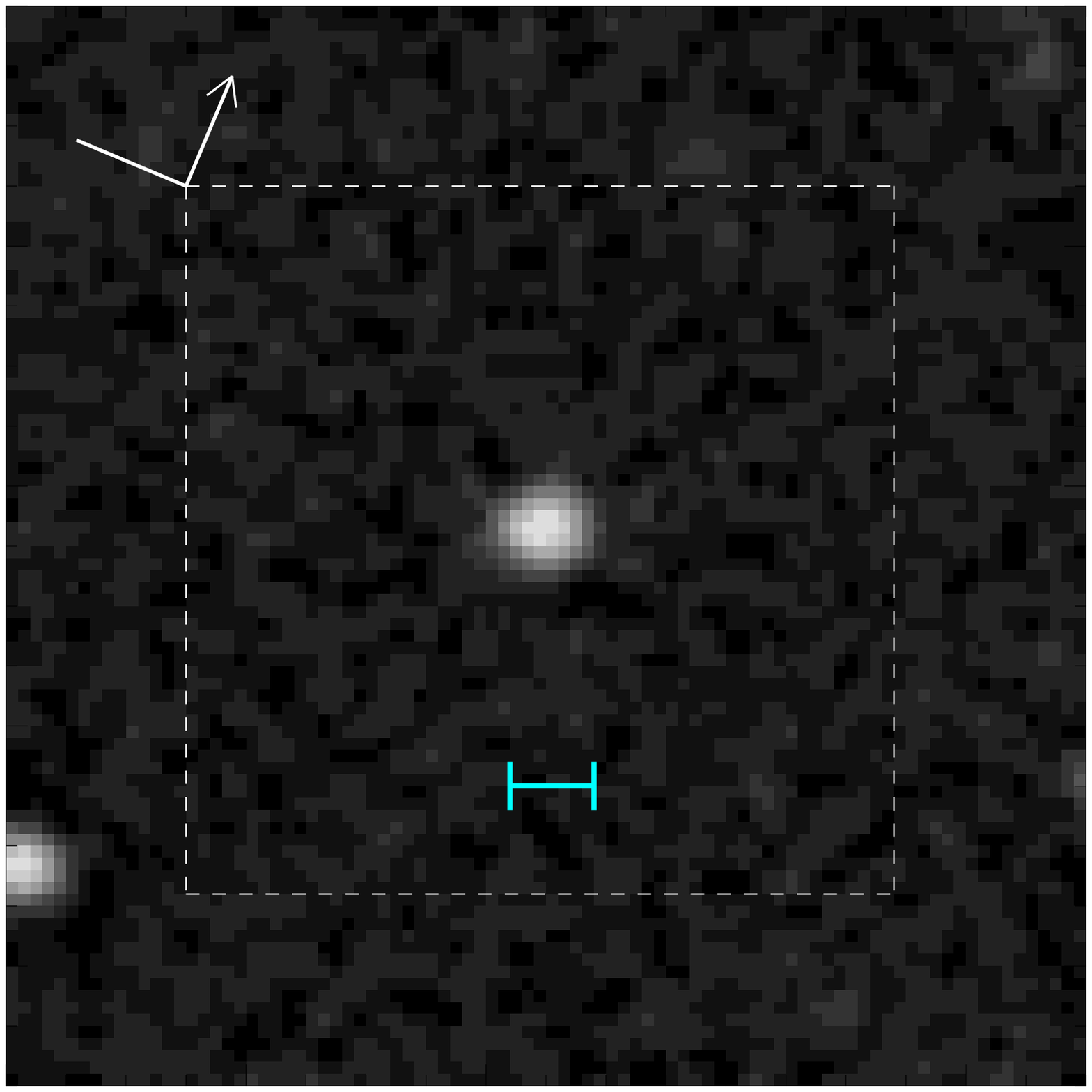}\\
\includegraphics[width=12cm,height=4cm]{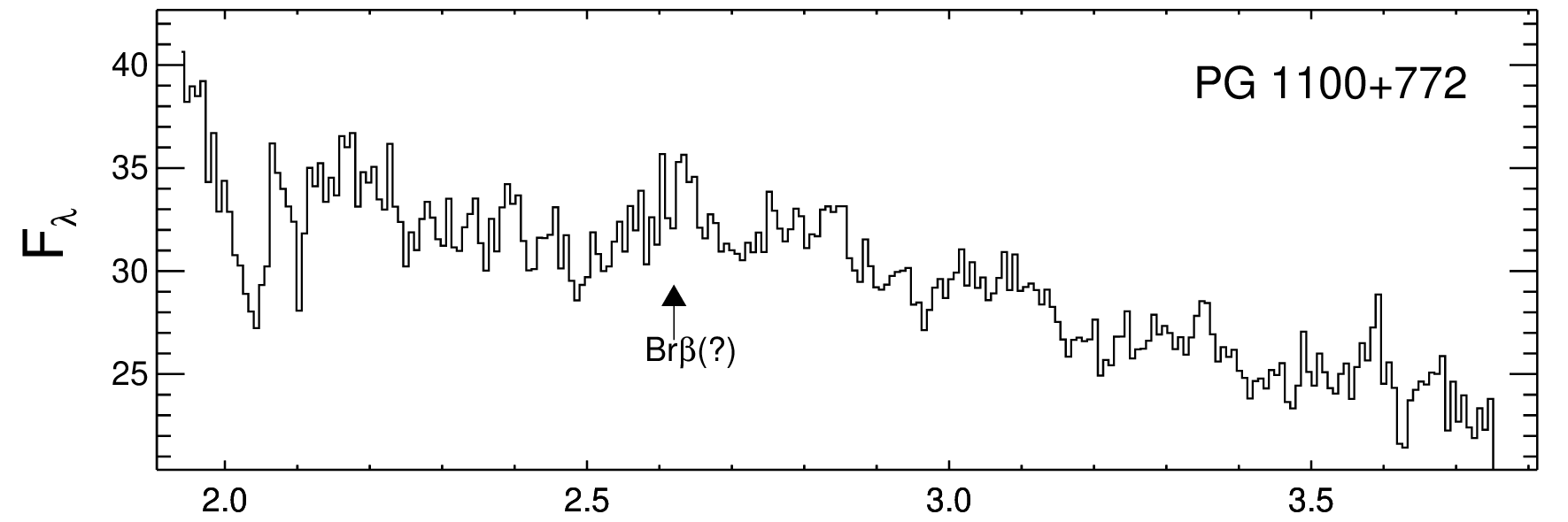}
\includegraphics[scale=0.20]{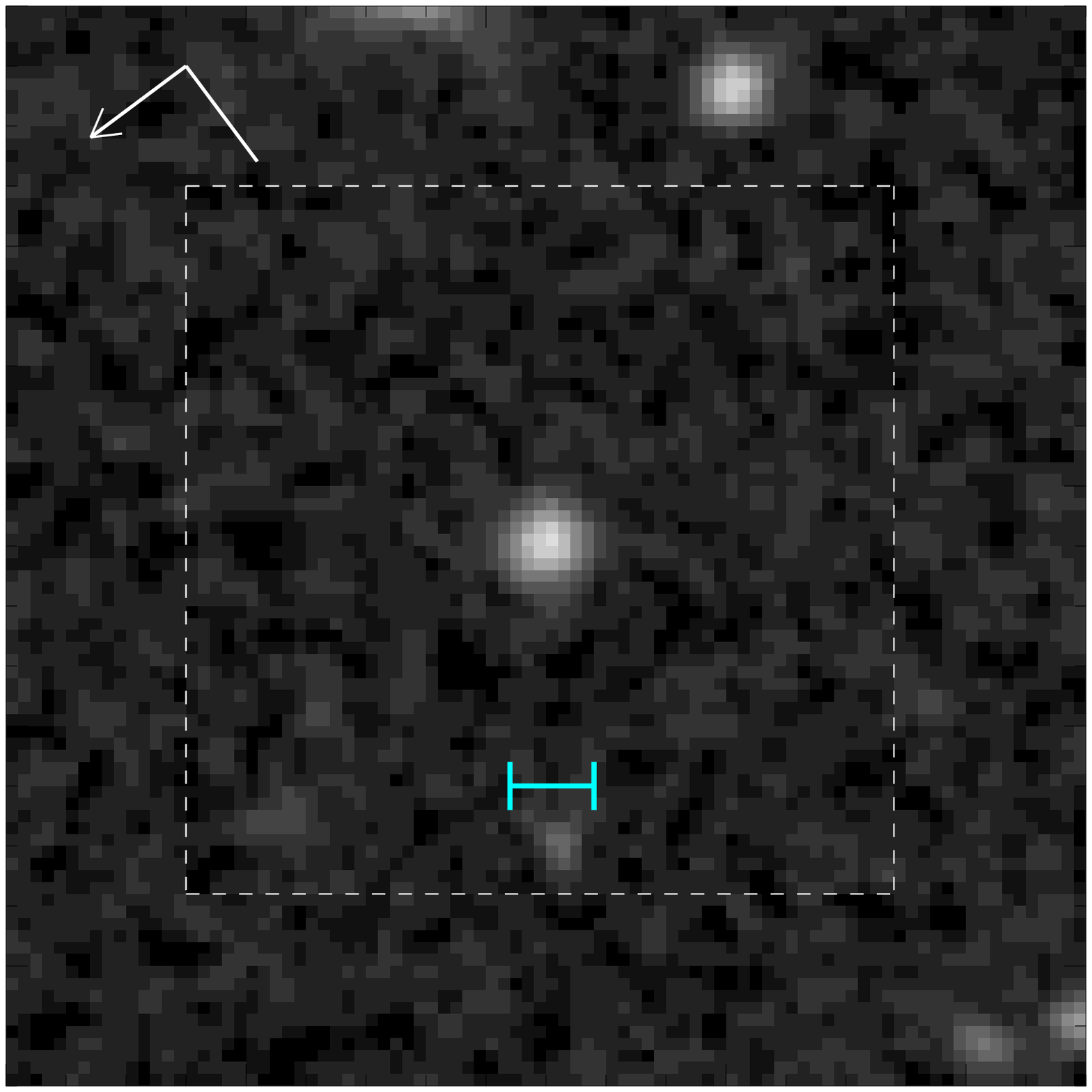}\\
\includegraphics[width=12cm,height=4cm]{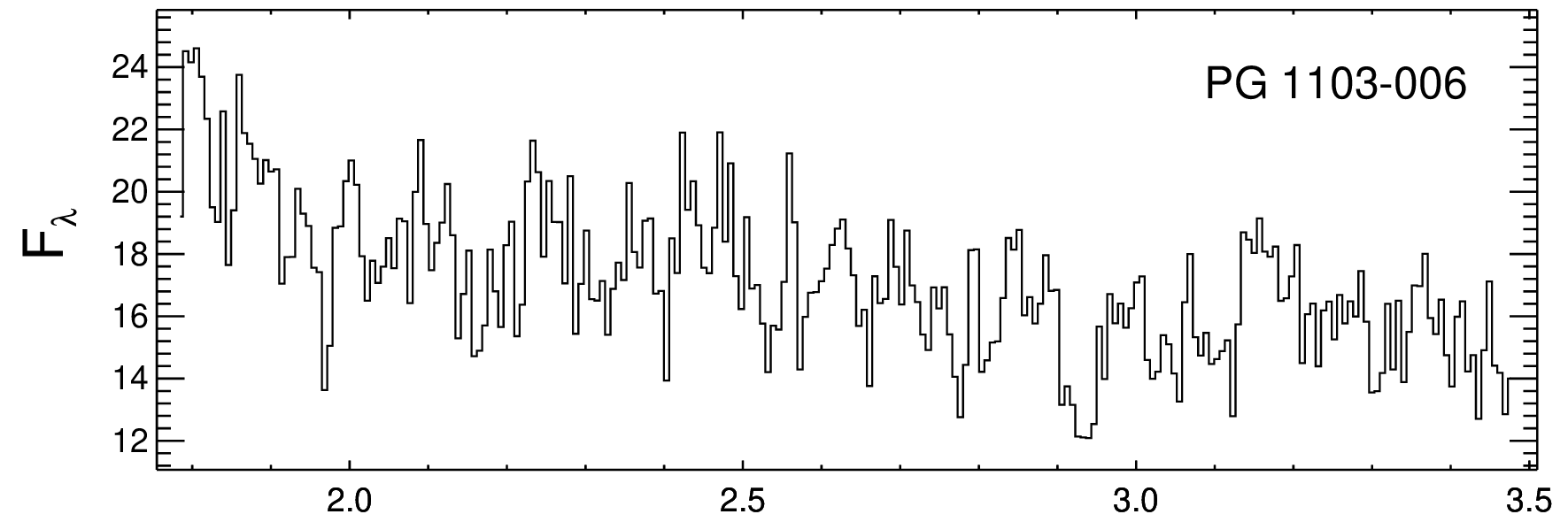}
\includegraphics[scale=0.20]{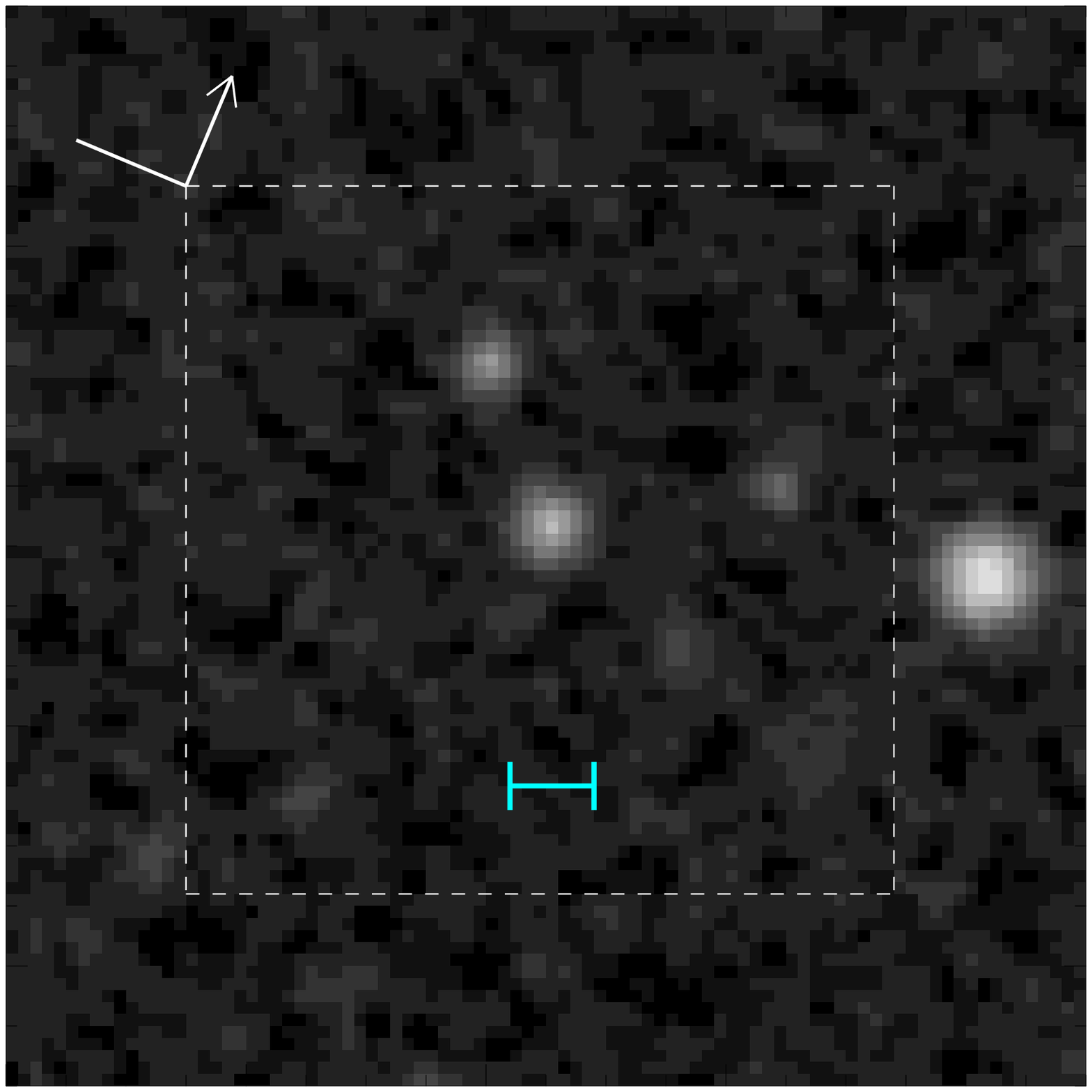}\\
\includegraphics[width=12cm,height=4cm]{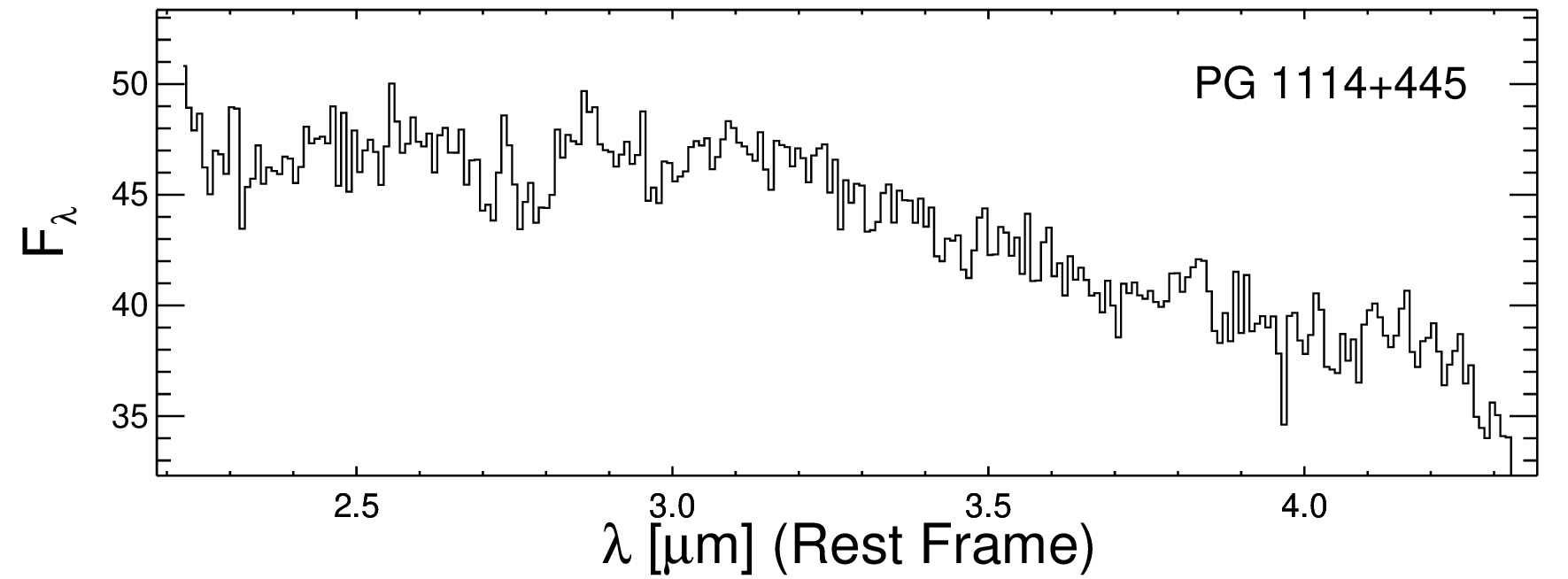}
\includegraphics[scale=0.20]{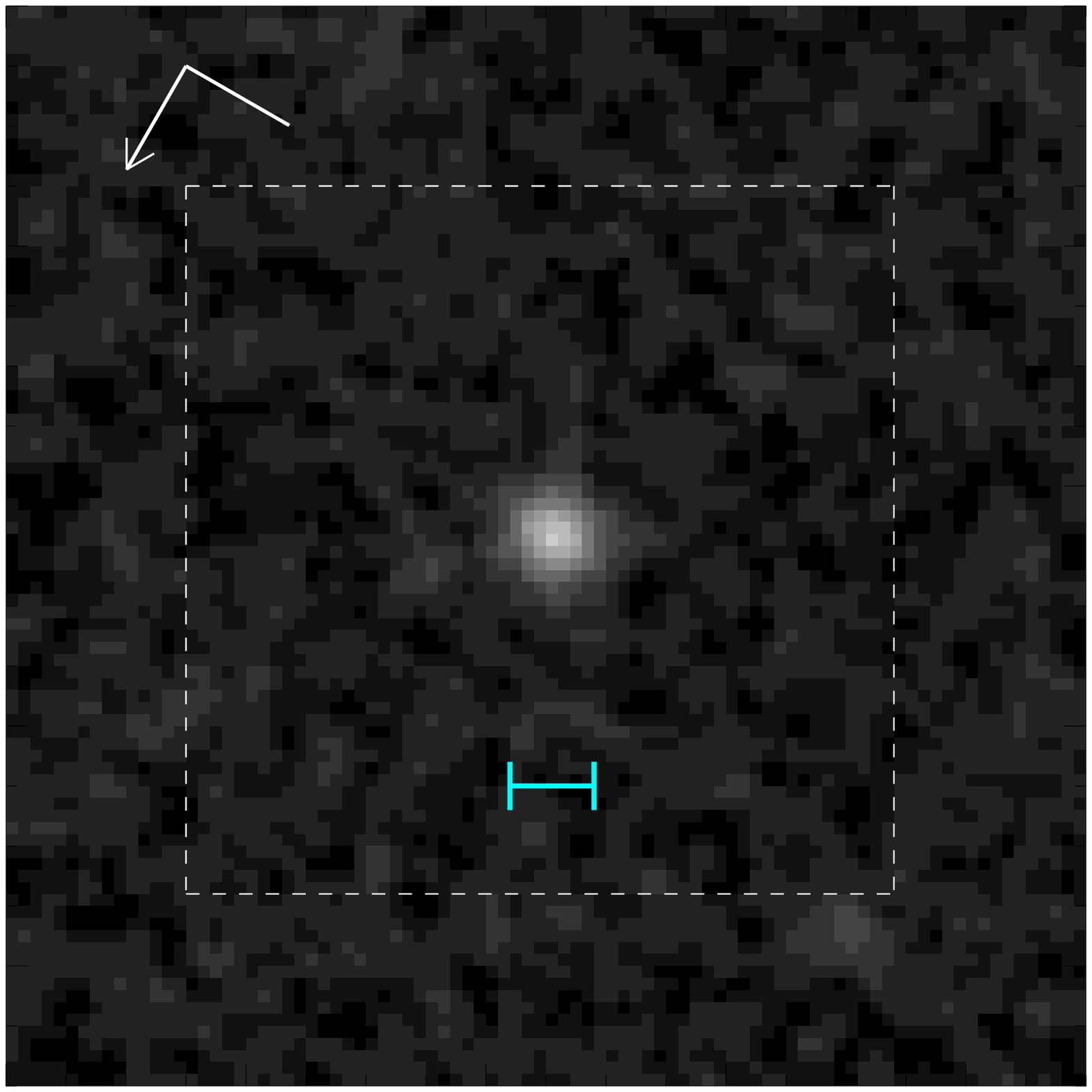}\\
\caption{Continued}
\end{figure}
\clearpage

\begin{figure}
\figurenum{5}
\includegraphics[width=12cm,height=4cm]{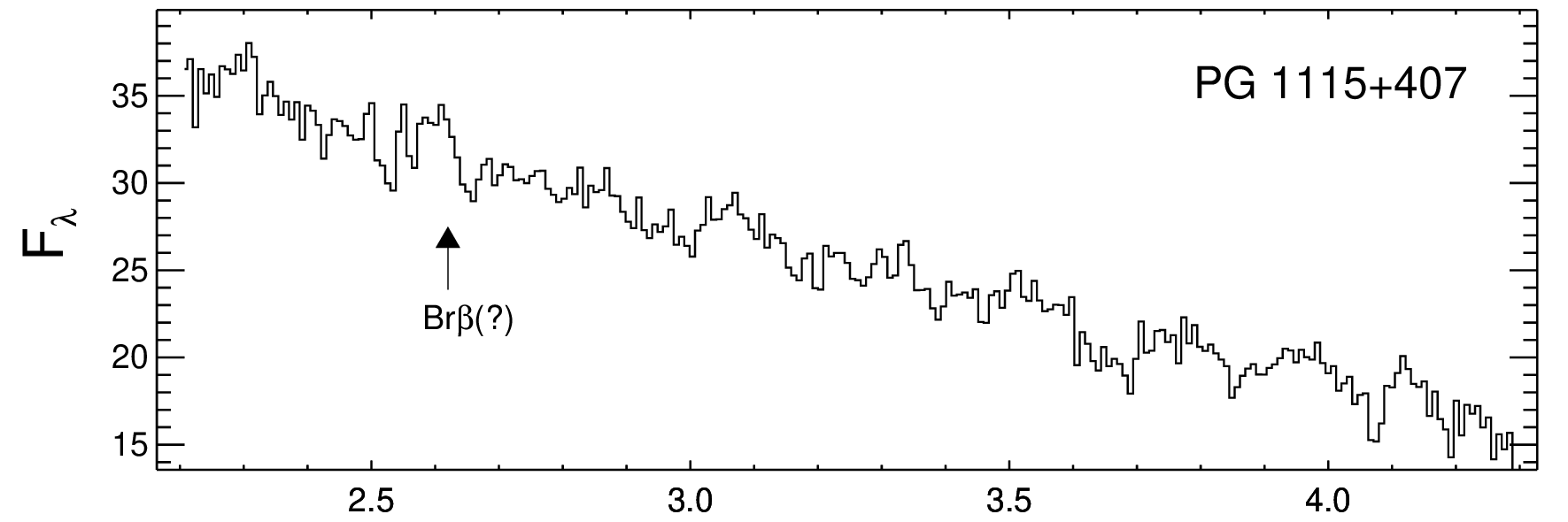}
\includegraphics[scale=0.20]{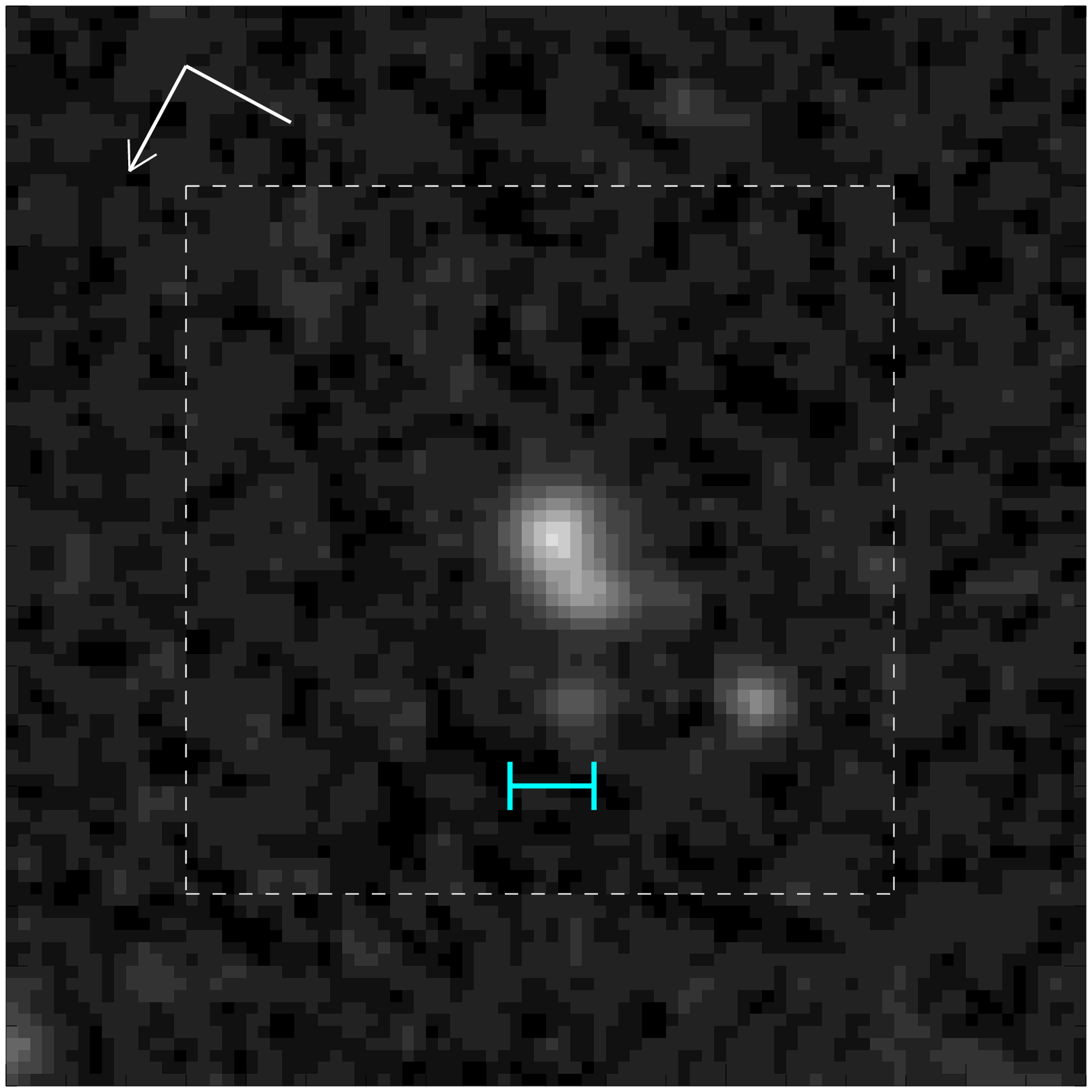}\\
\includegraphics[width=12cm,height=4cm]{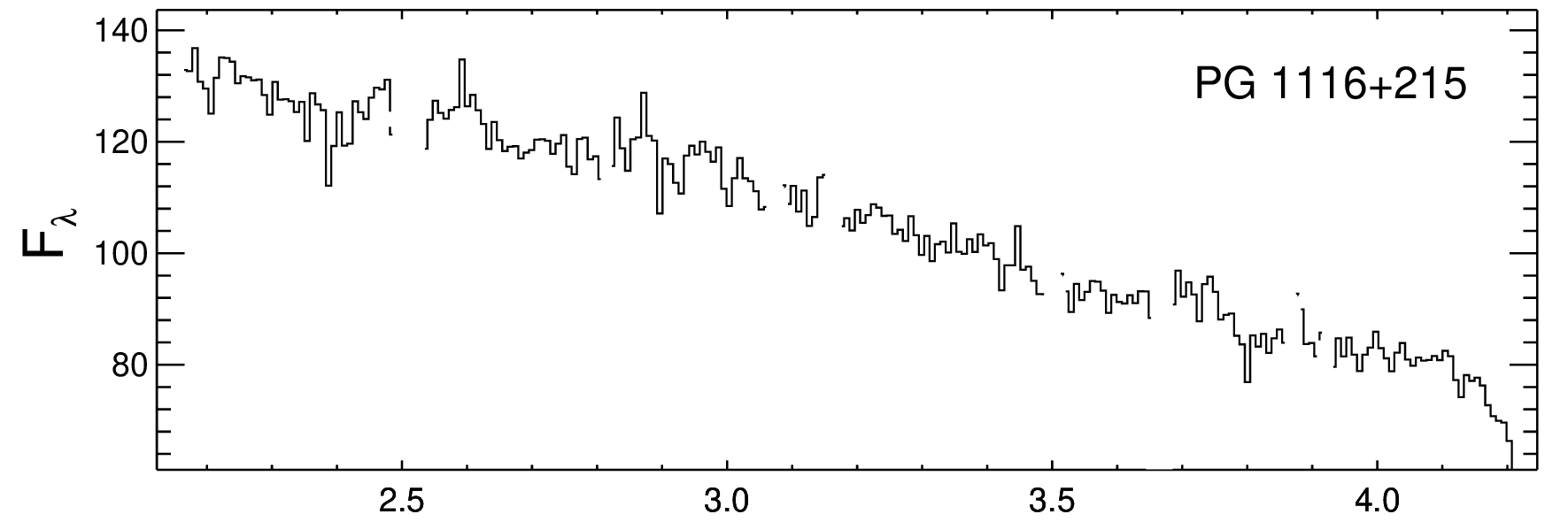}
\includegraphics[scale=0.20]{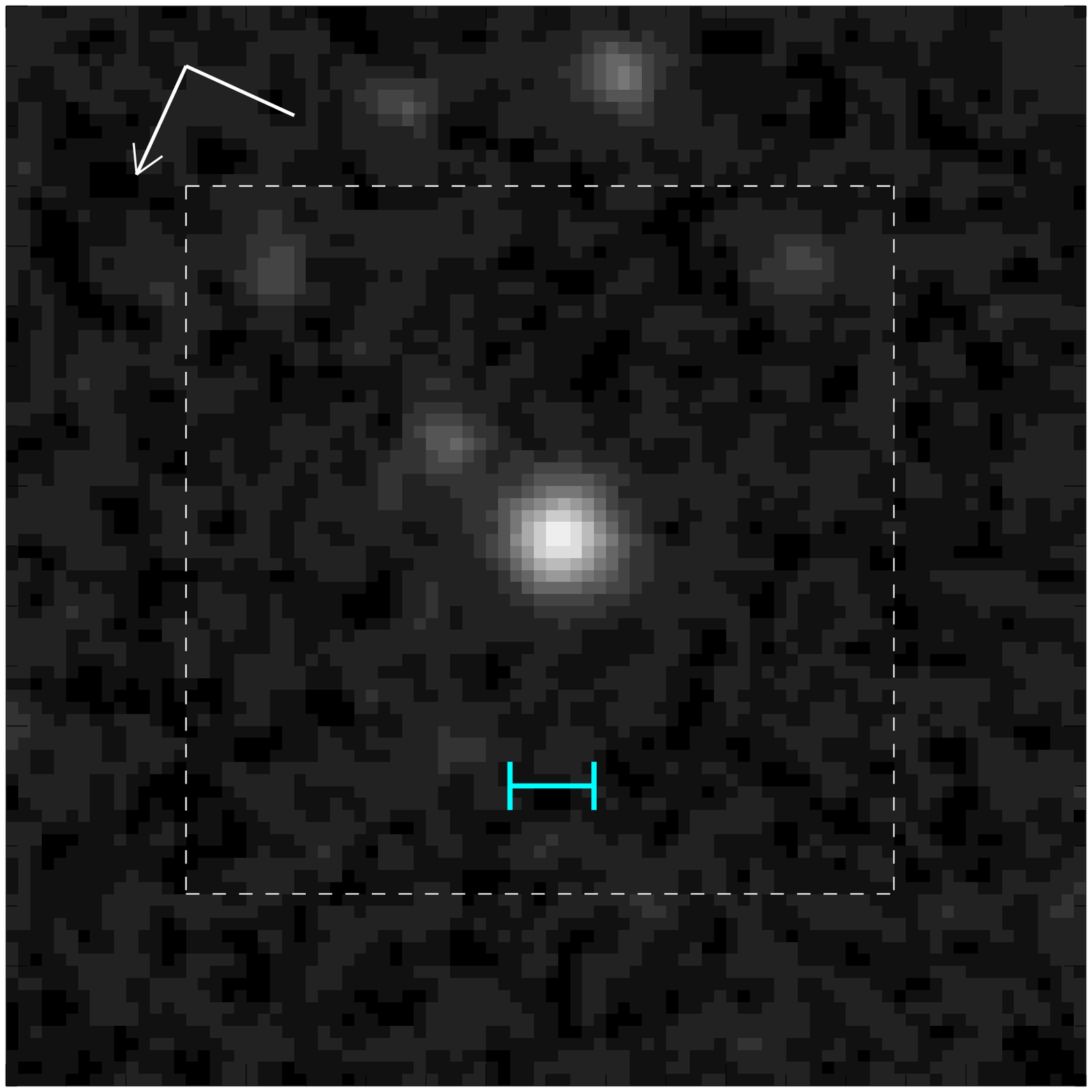}\\
\includegraphics[width=12cm,height=4cm]{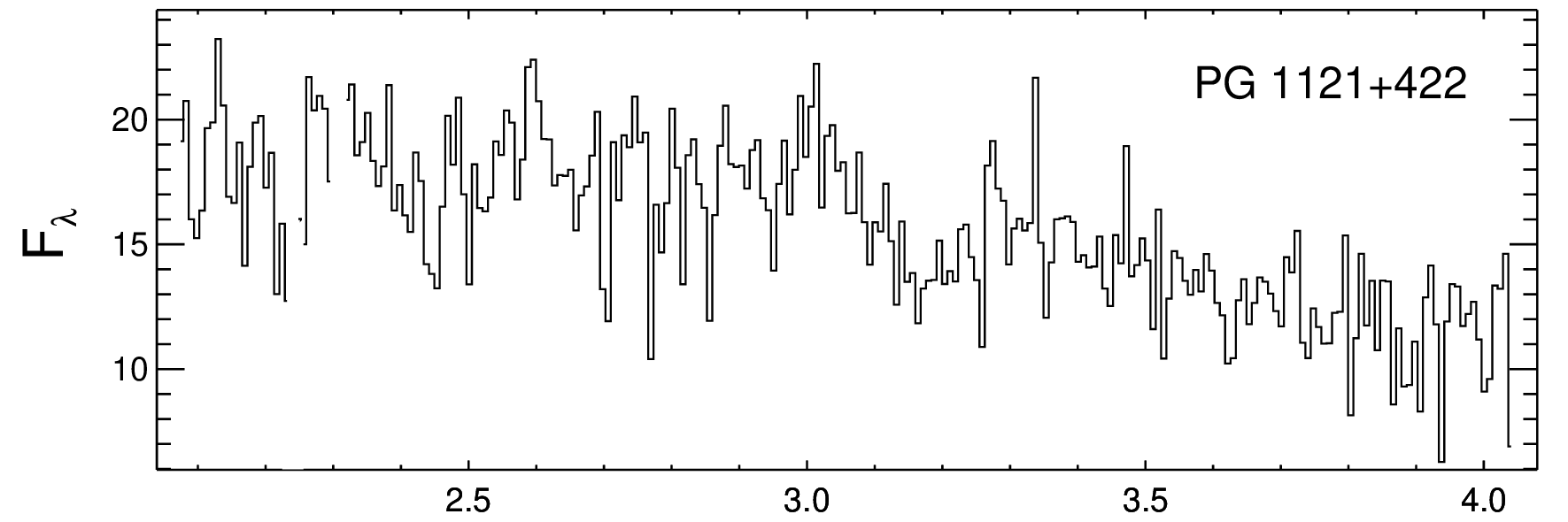}
\includegraphics[scale=0.20]{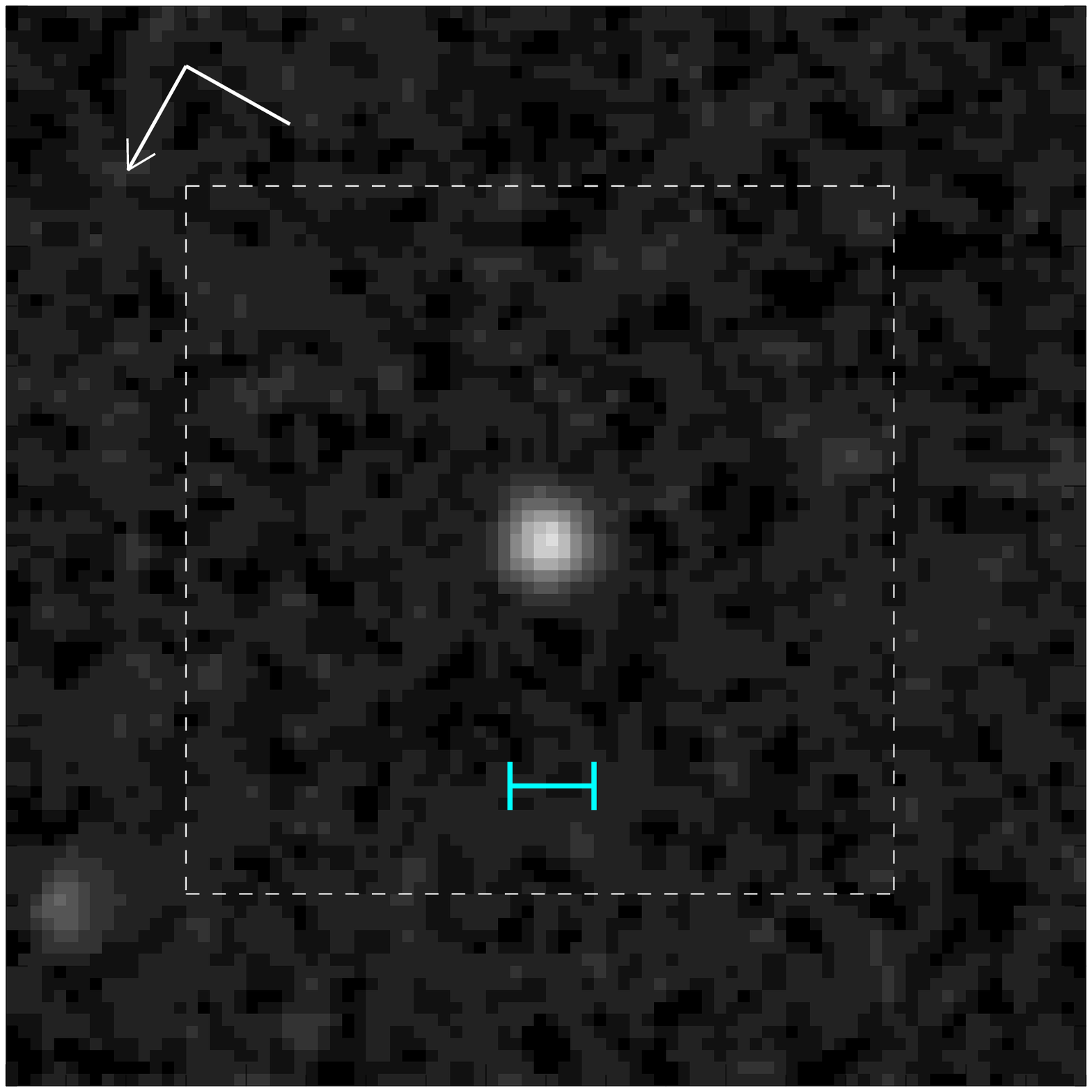}\\
\includegraphics[width=12cm,height=4cm]{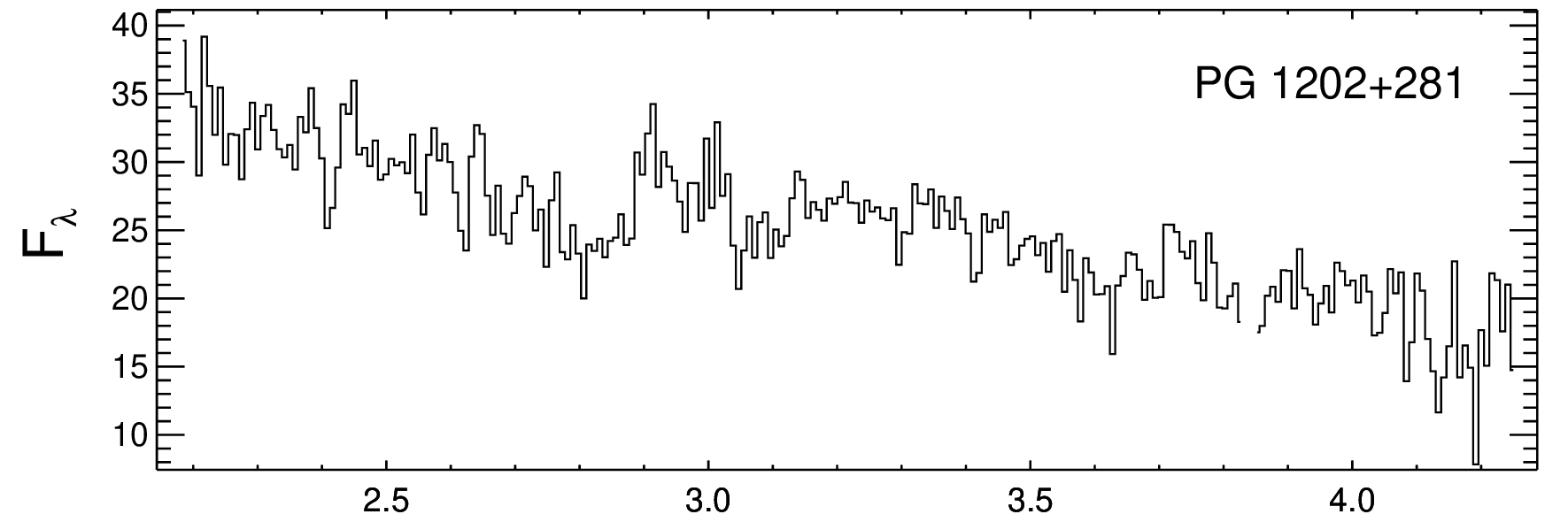}
\includegraphics[scale=0.20]{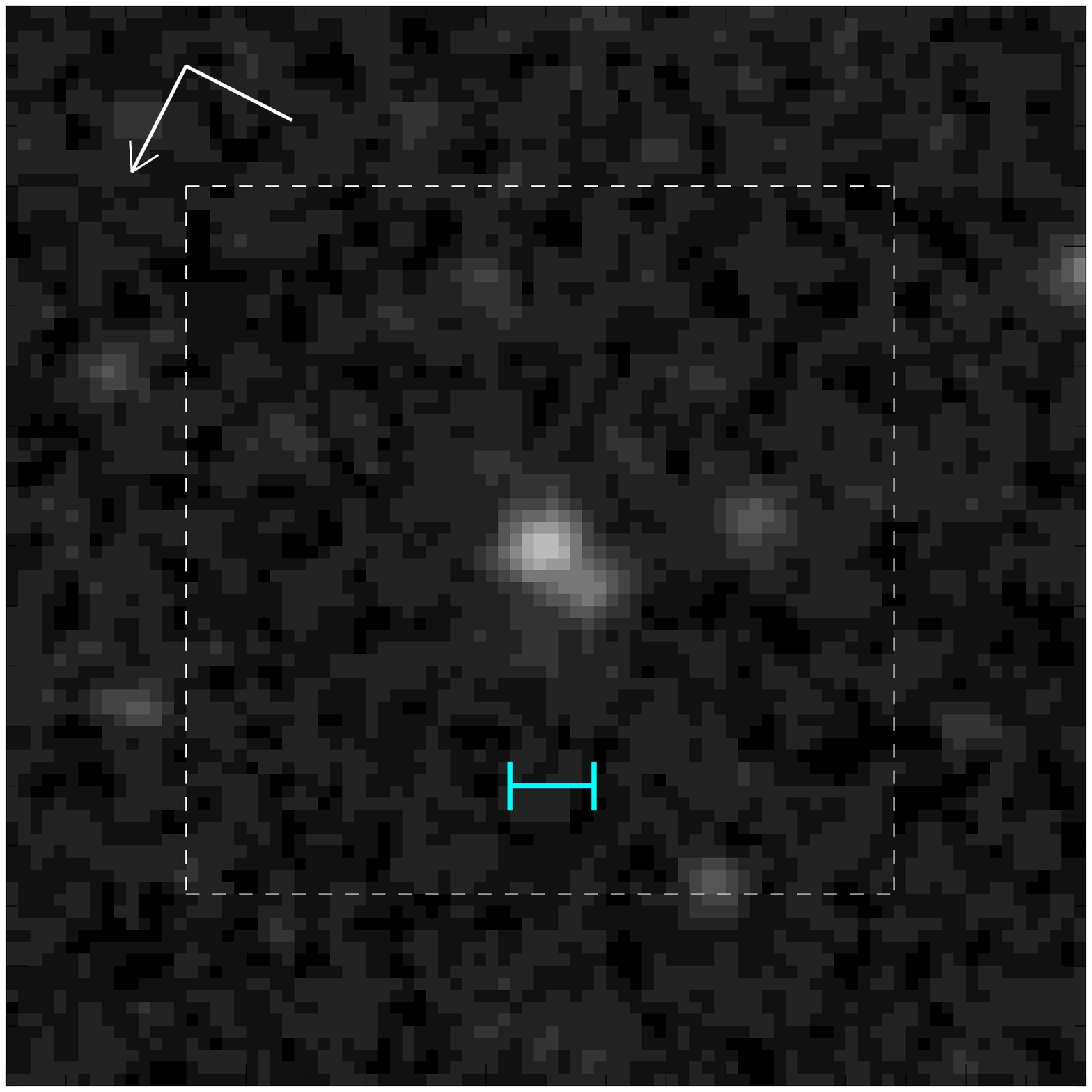}\\
\includegraphics[width=12cm,height=4cm]{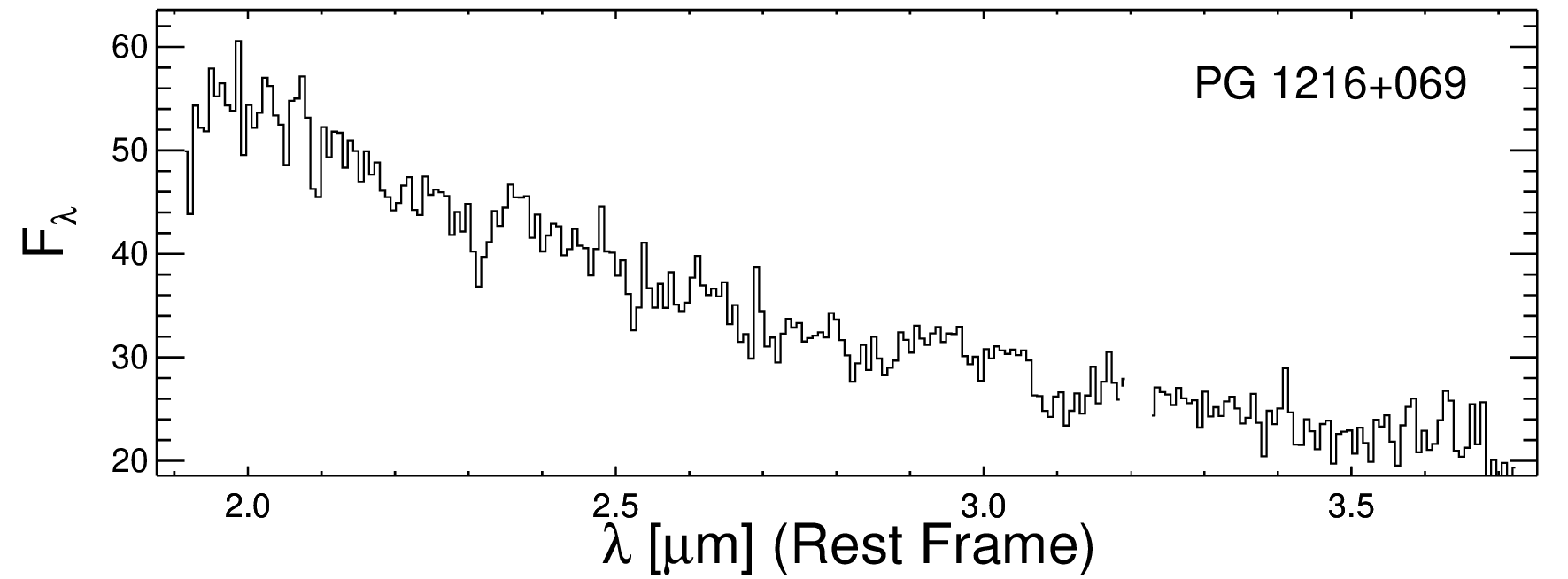}
\includegraphics[scale=0.20]{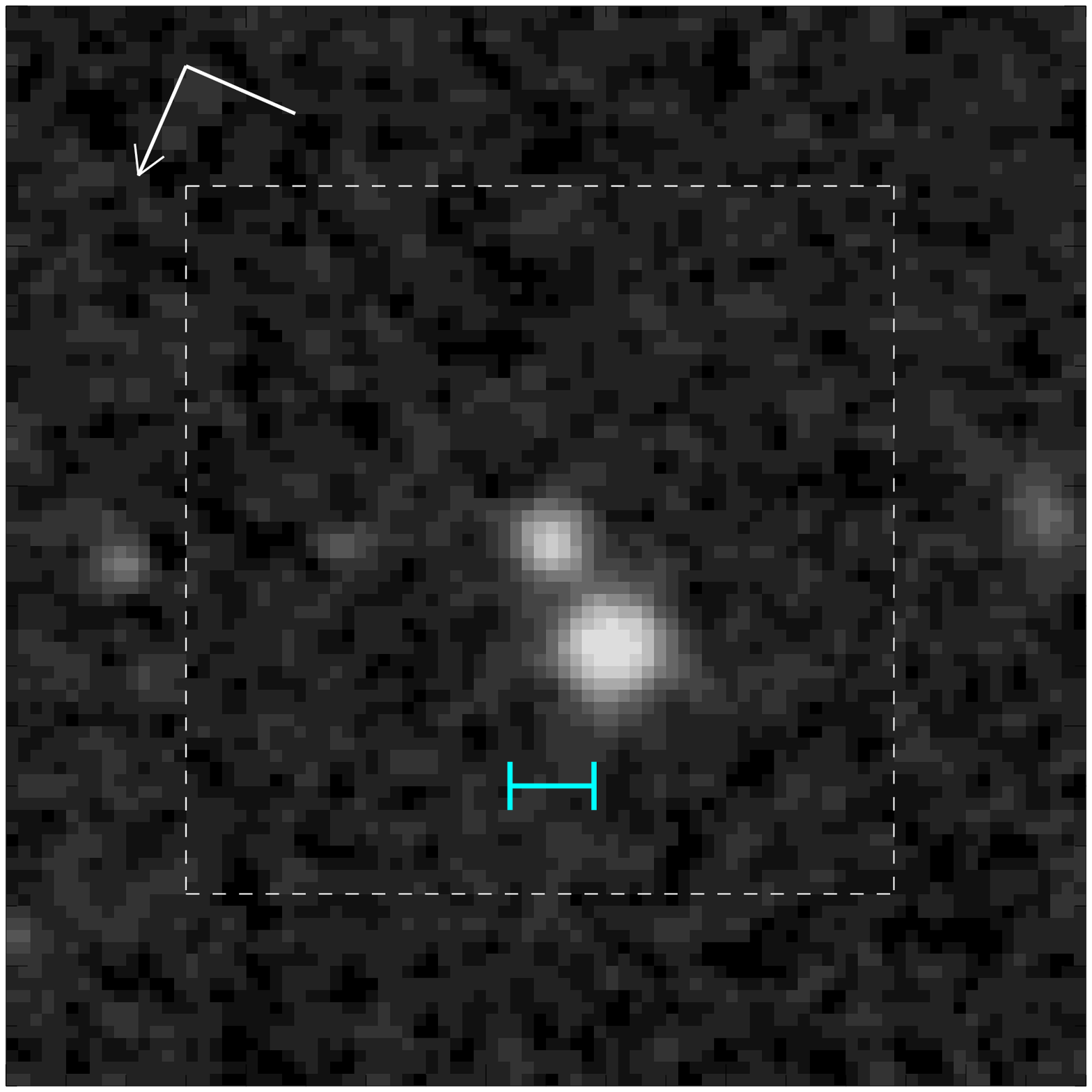}\\
\caption{Continued}
\end{figure}
\clearpage

\begin{figure}
\figurenum{5}

\includegraphics[width=12cm,height=4cm]{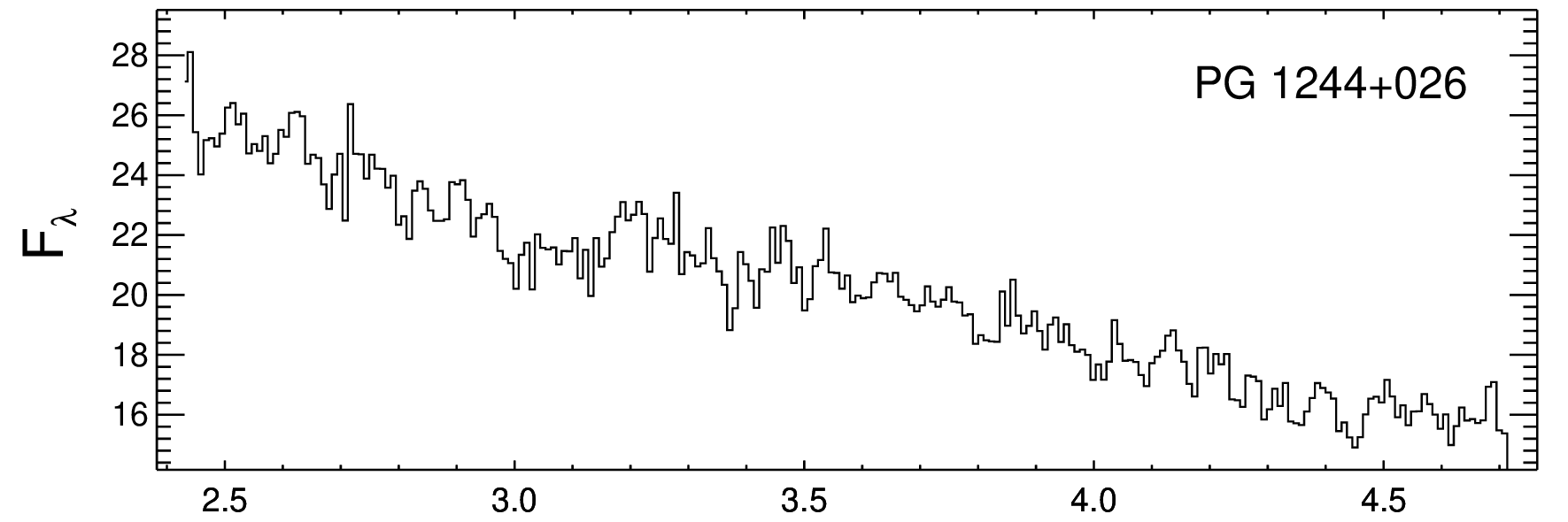}
\includegraphics[scale=0.20]{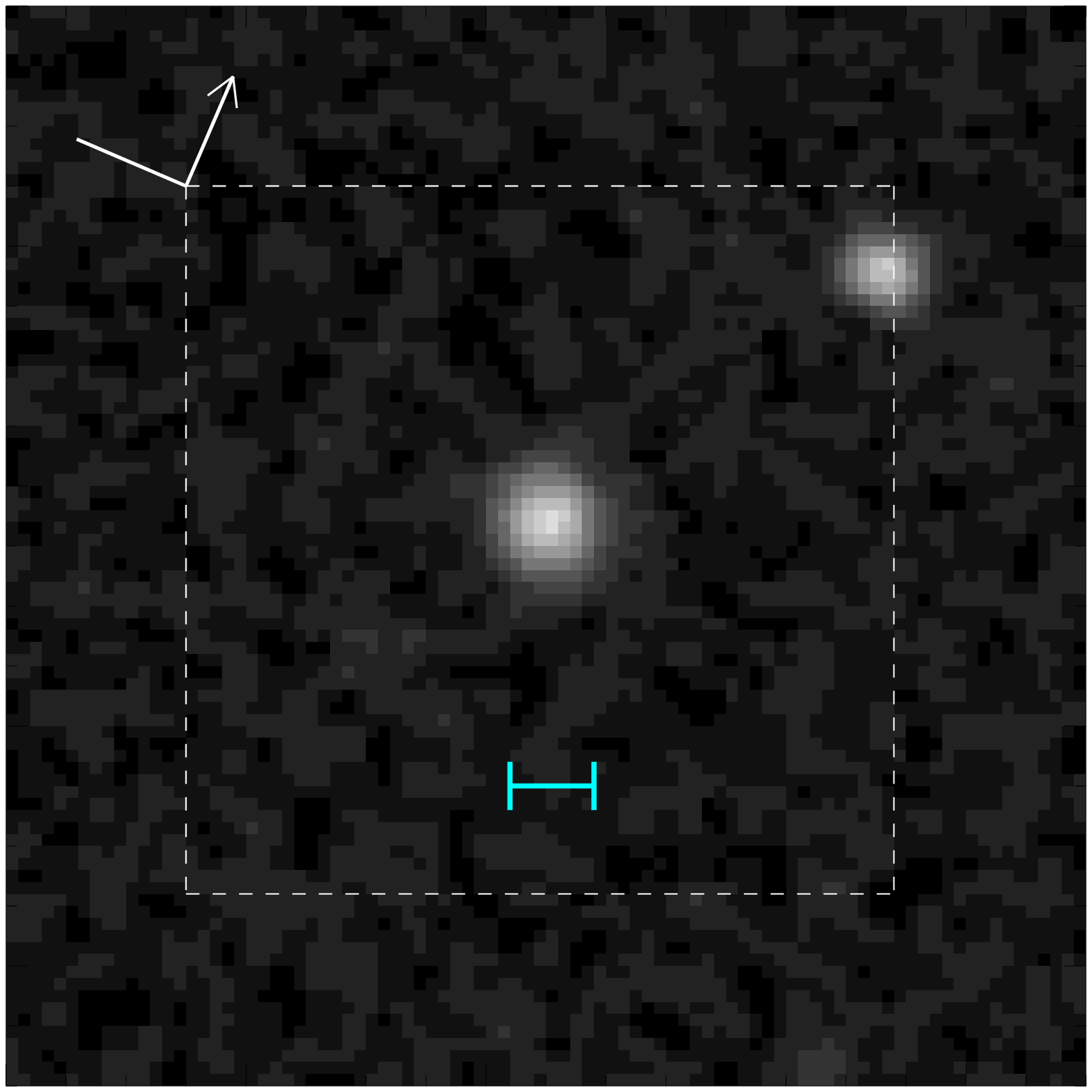}\\
\includegraphics[width=12cm,height=4cm]{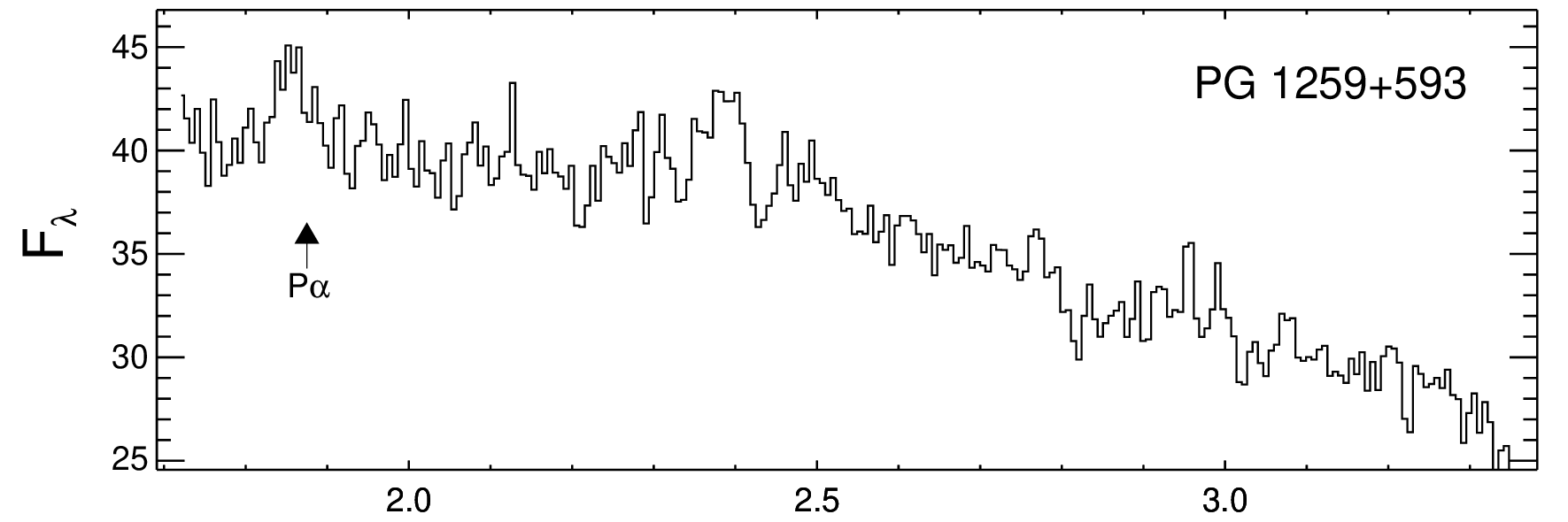}
\includegraphics[scale=0.20]{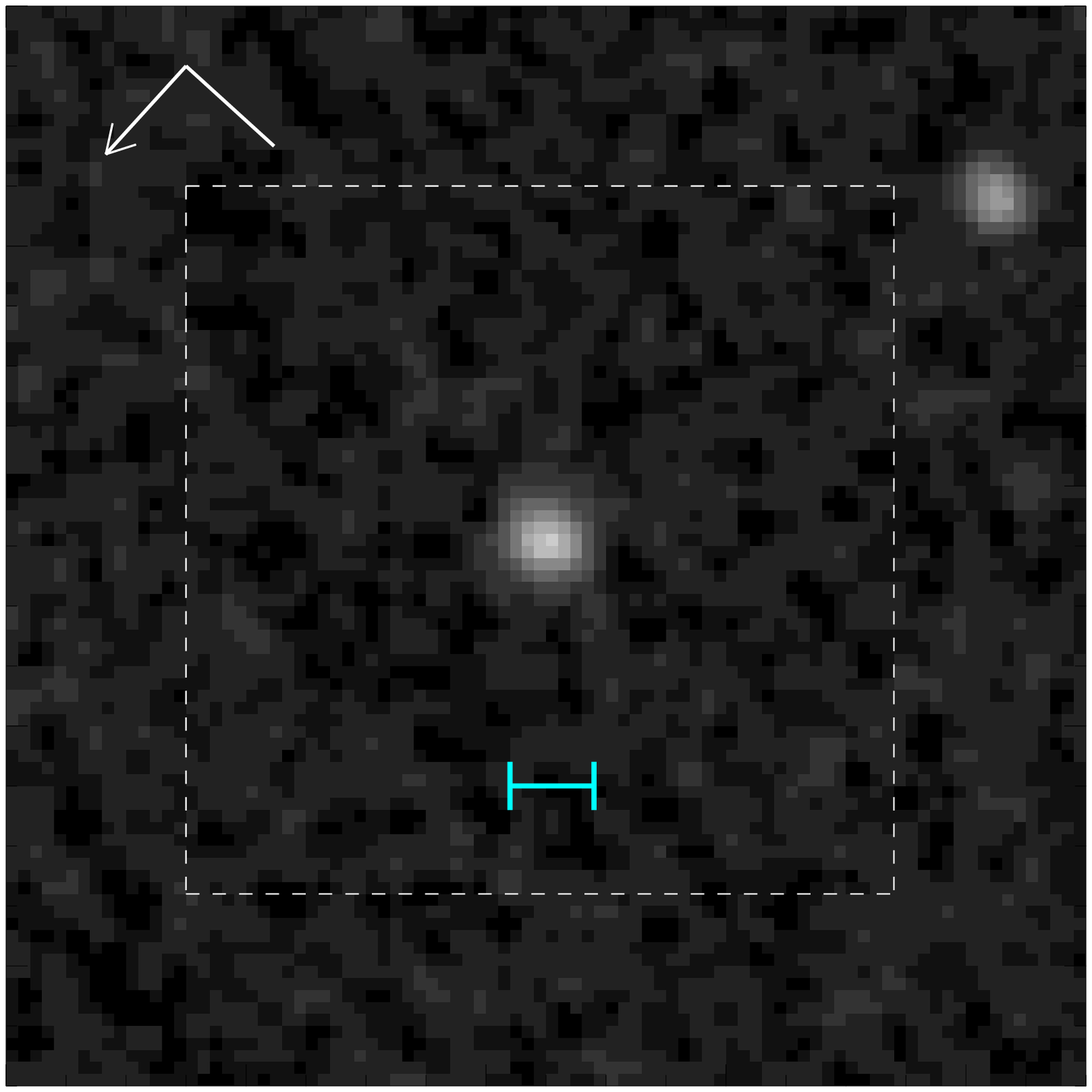}\\
\includegraphics[width=12cm,height=4cm]{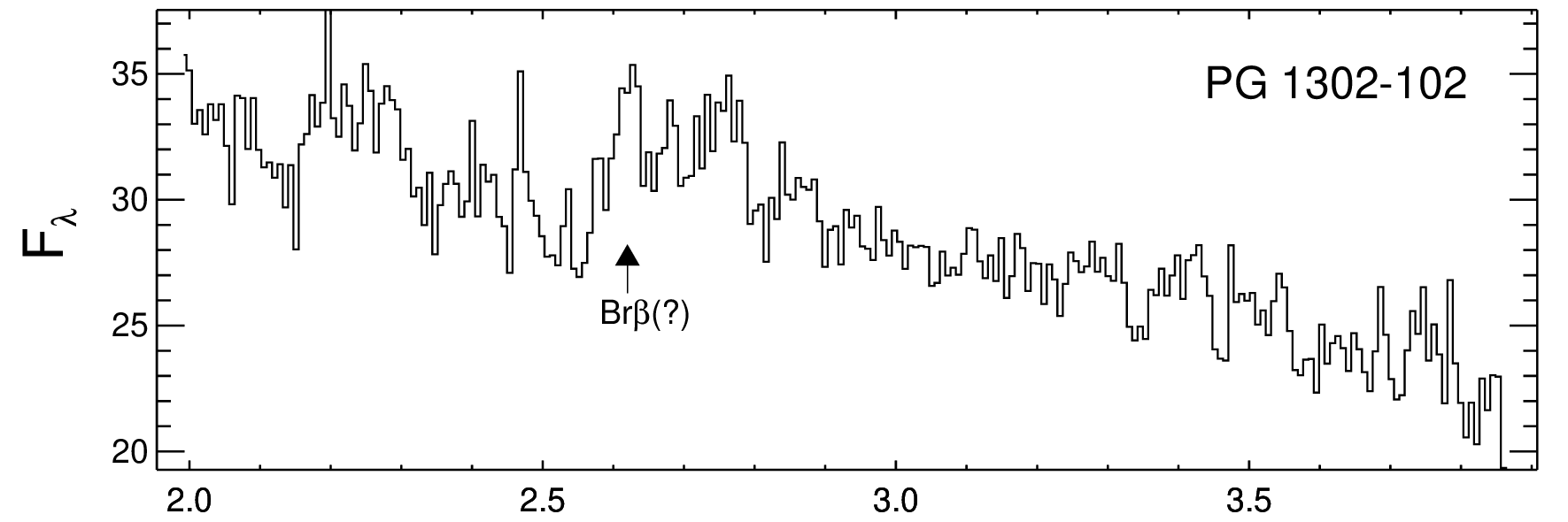}
\includegraphics[scale=0.20]{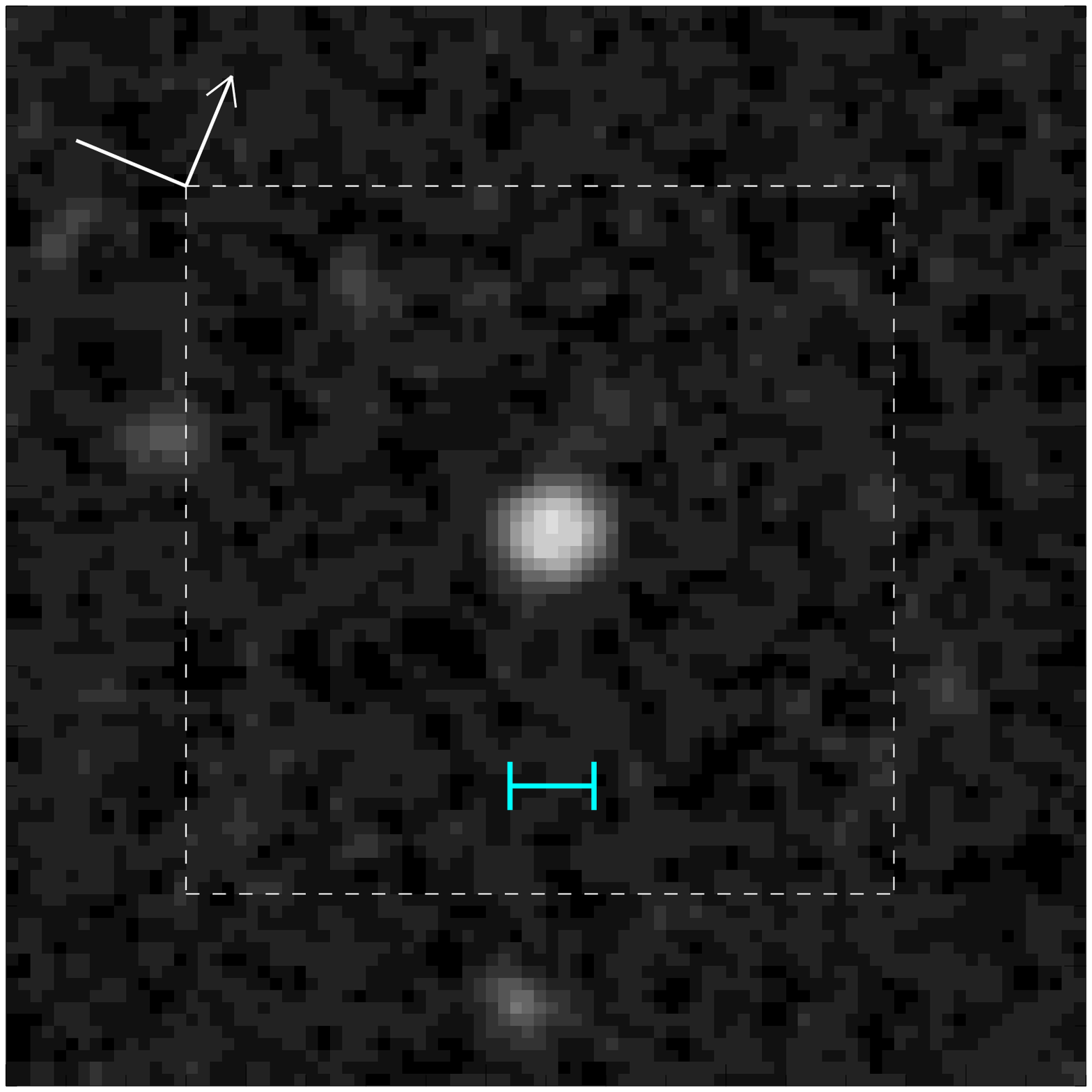}\\
\includegraphics[width=12cm,height=4cm]{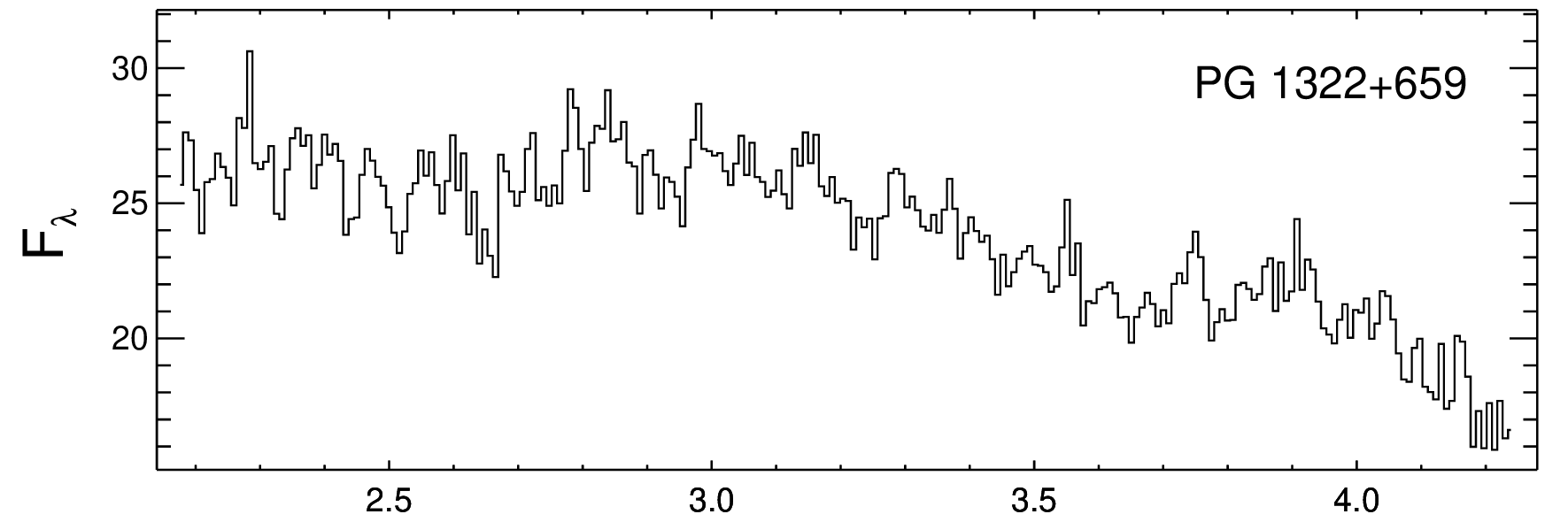}
\includegraphics[scale=0.20]{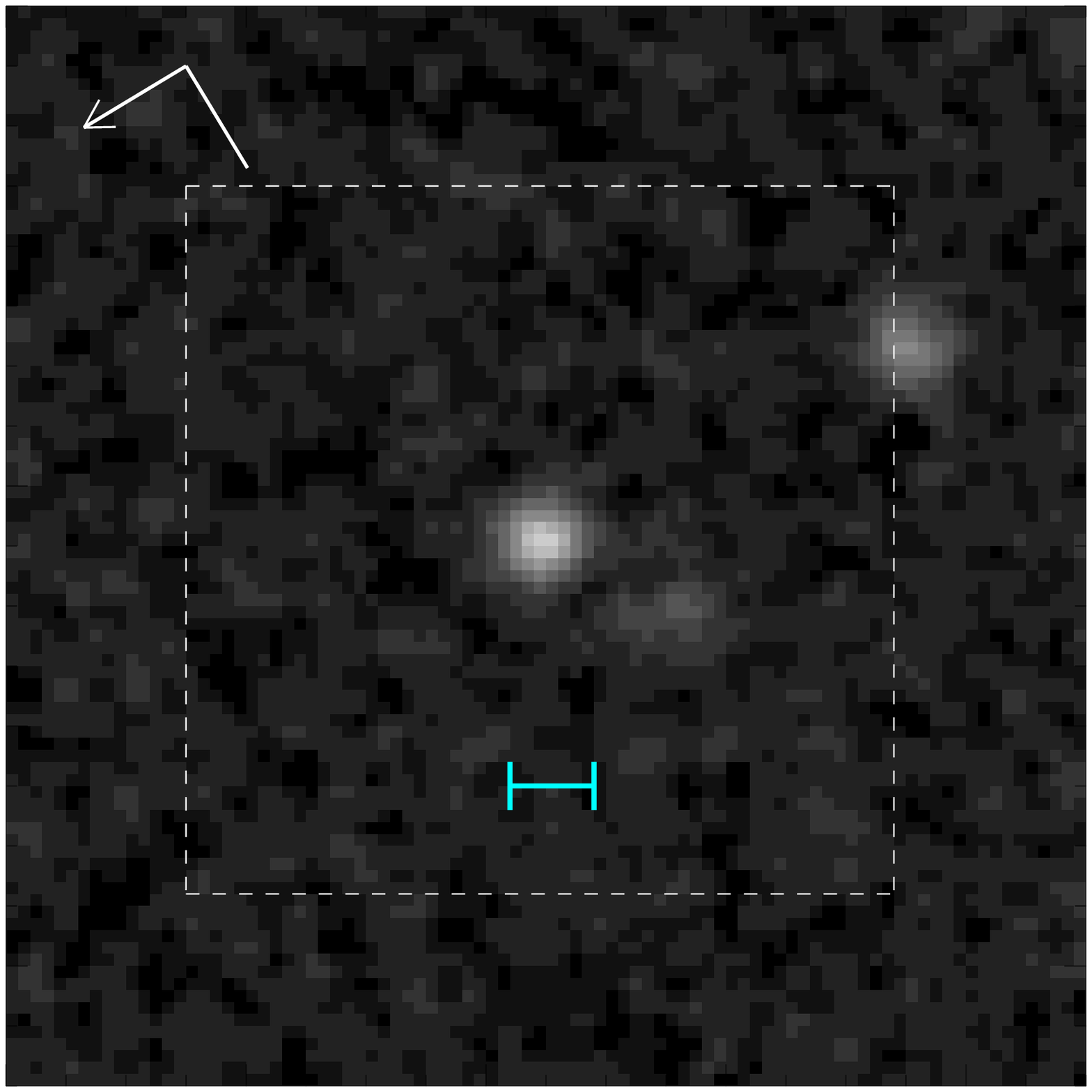}\\
\includegraphics[width=12cm,height=4cm]{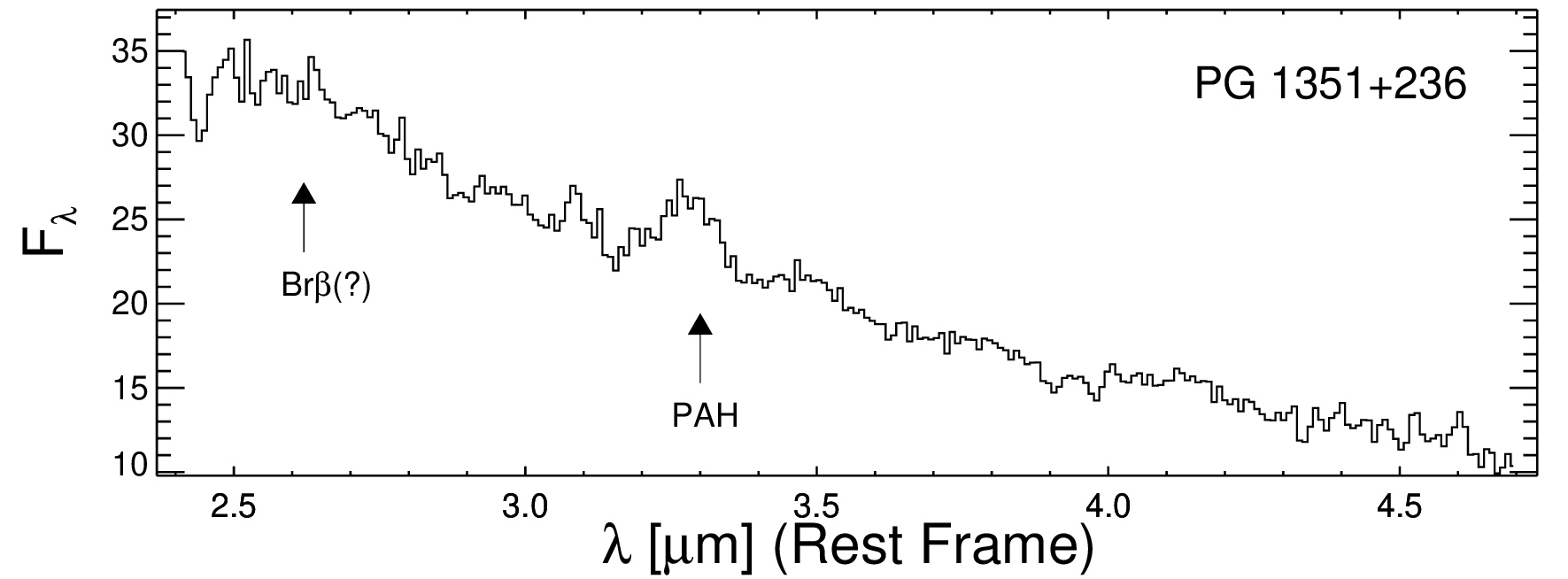}
\includegraphics[scale=0.20]{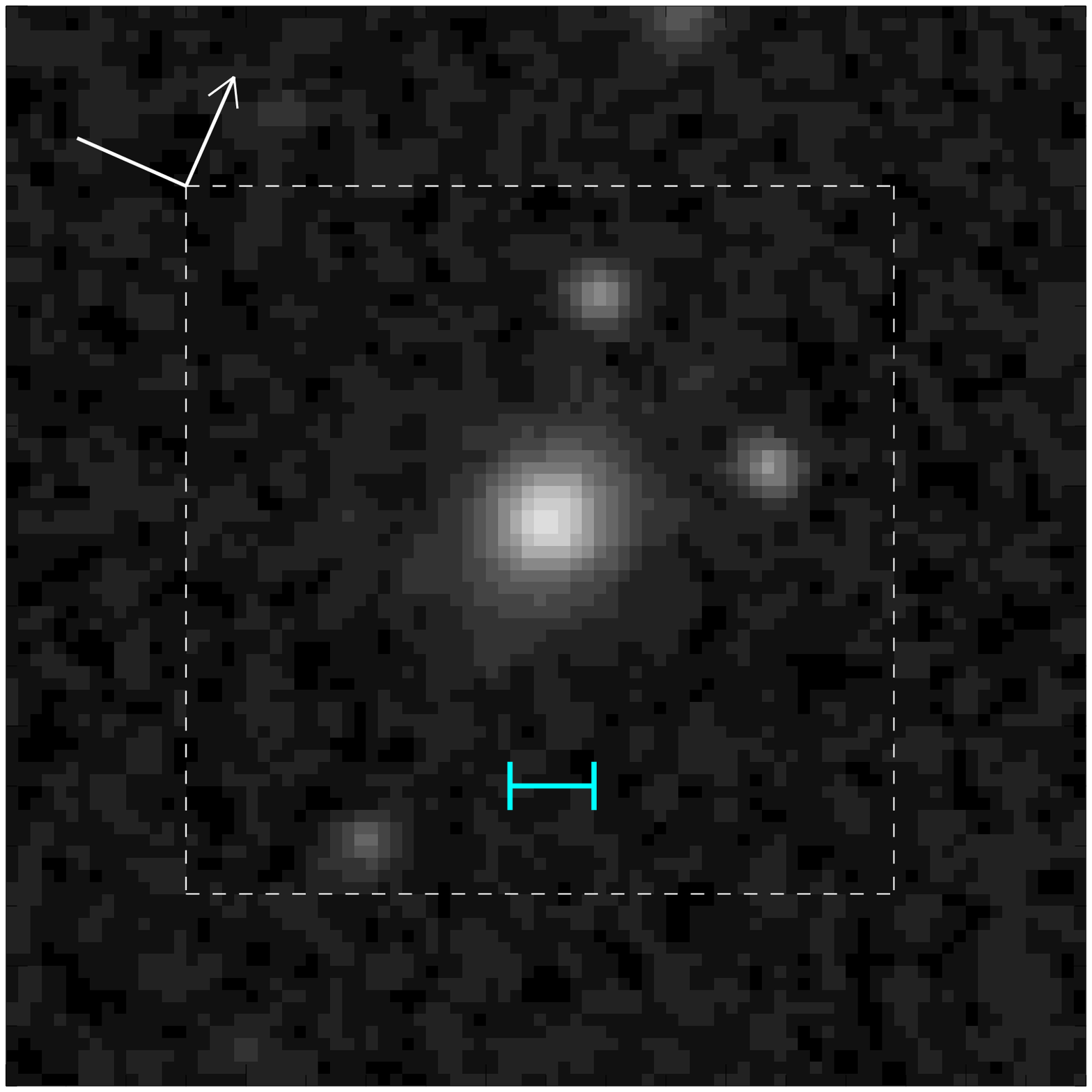}\\
\caption{Continued}
\end{figure}
\clearpage

\begin{figure}
\figurenum{5}
\includegraphics[width=12cm,height=4cm]{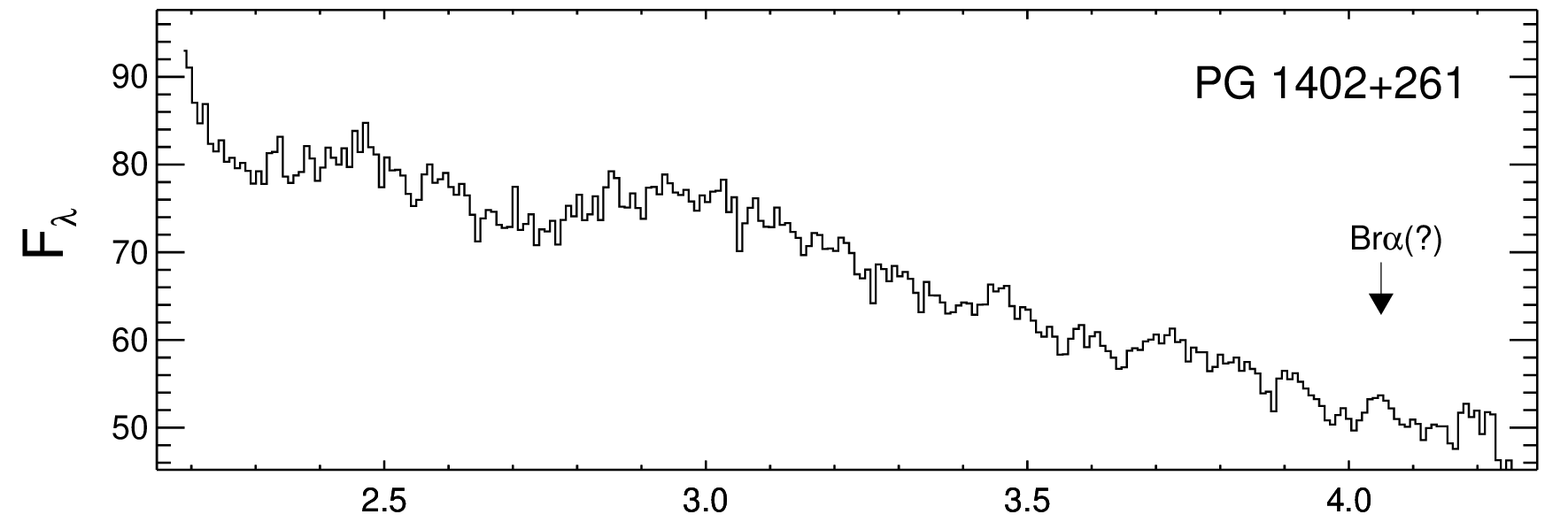}
\includegraphics[scale=0.20]{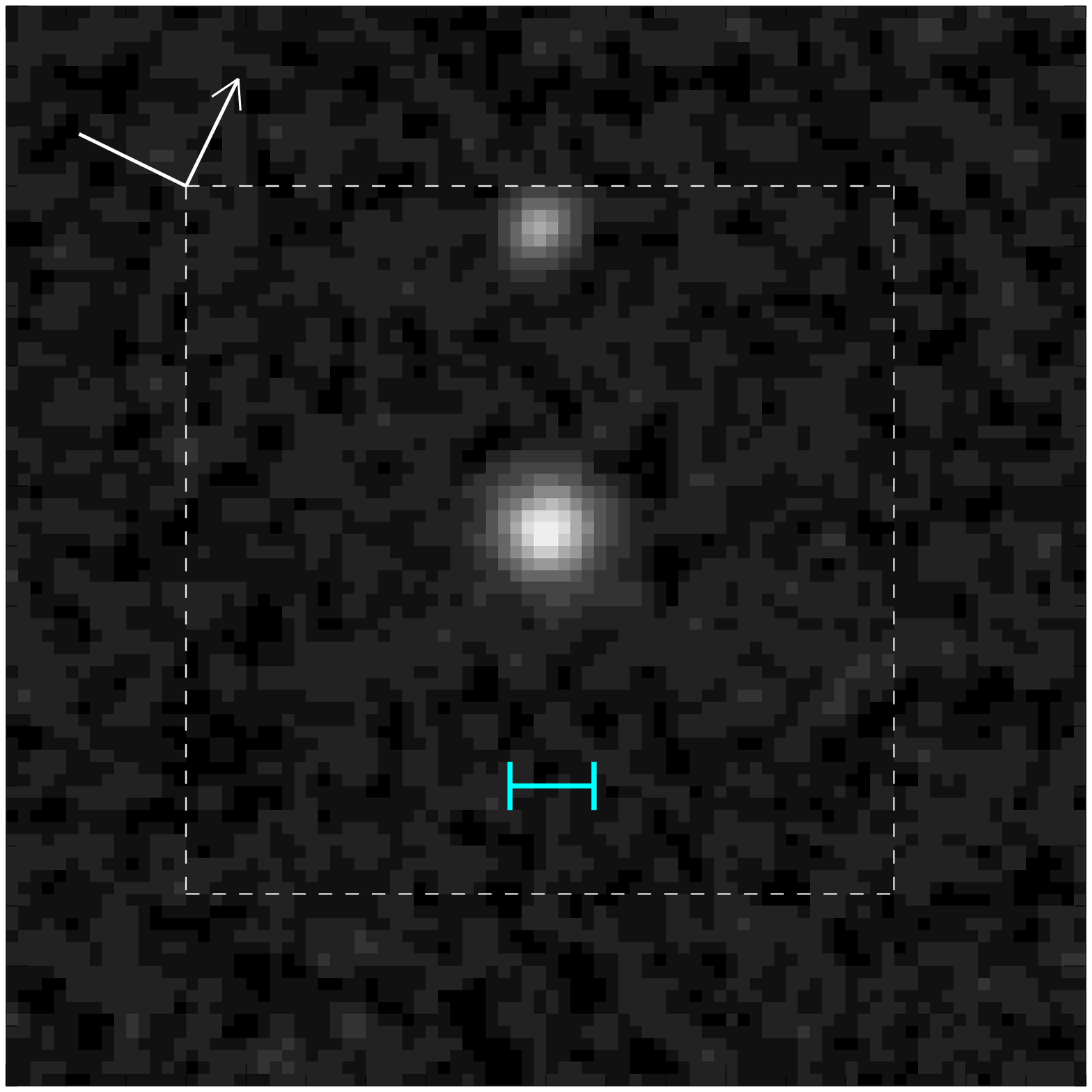}\\
\includegraphics[width=12cm,height=4cm]{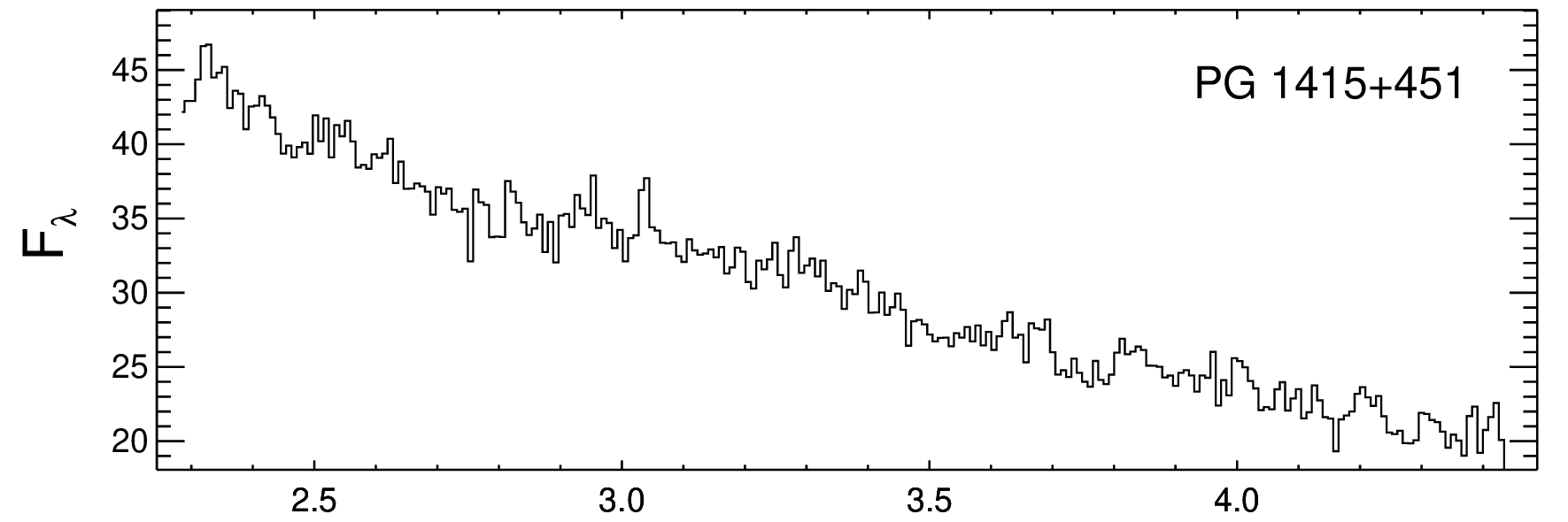}
\includegraphics[scale=0.20]{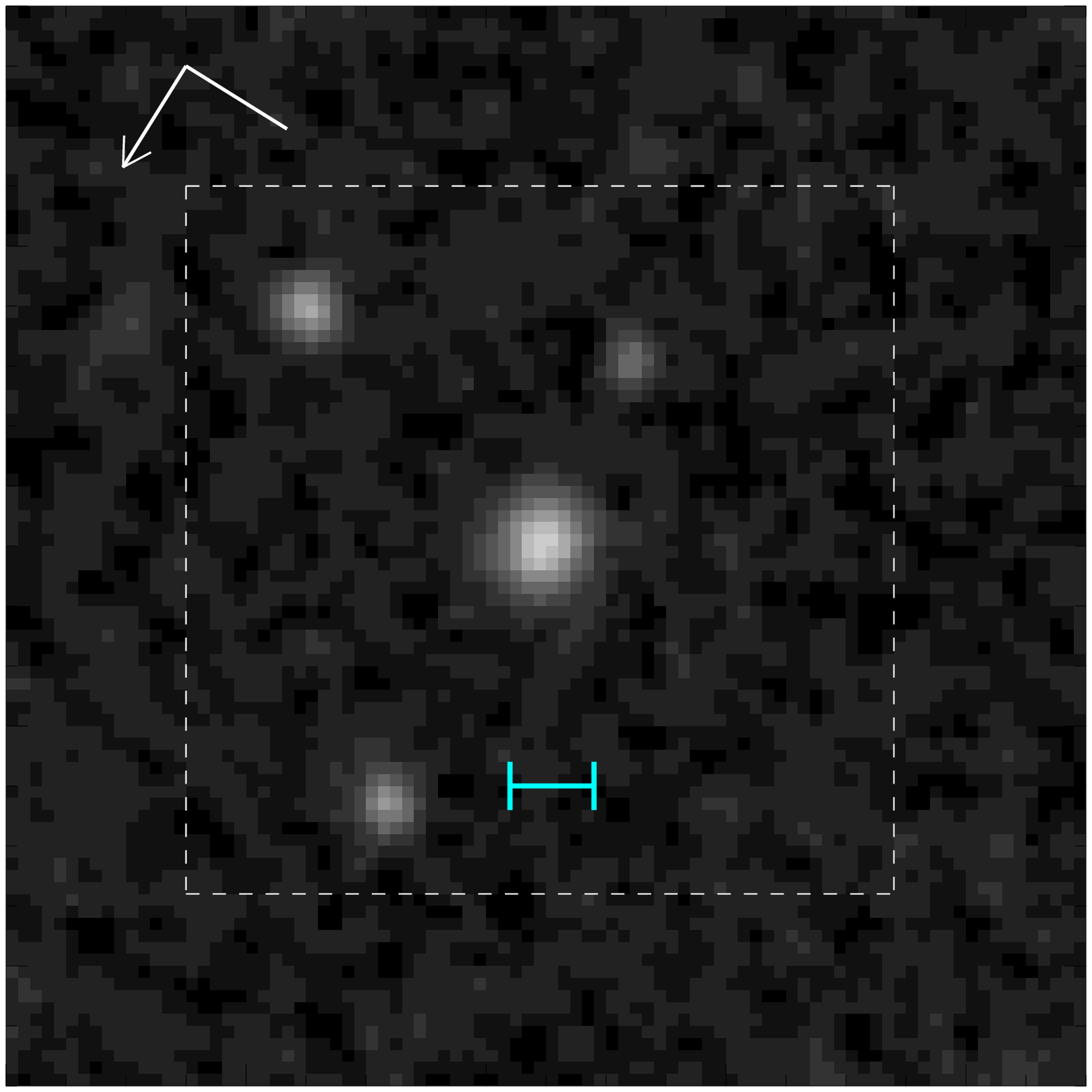}\\
\includegraphics[width=12cm,height=4cm]{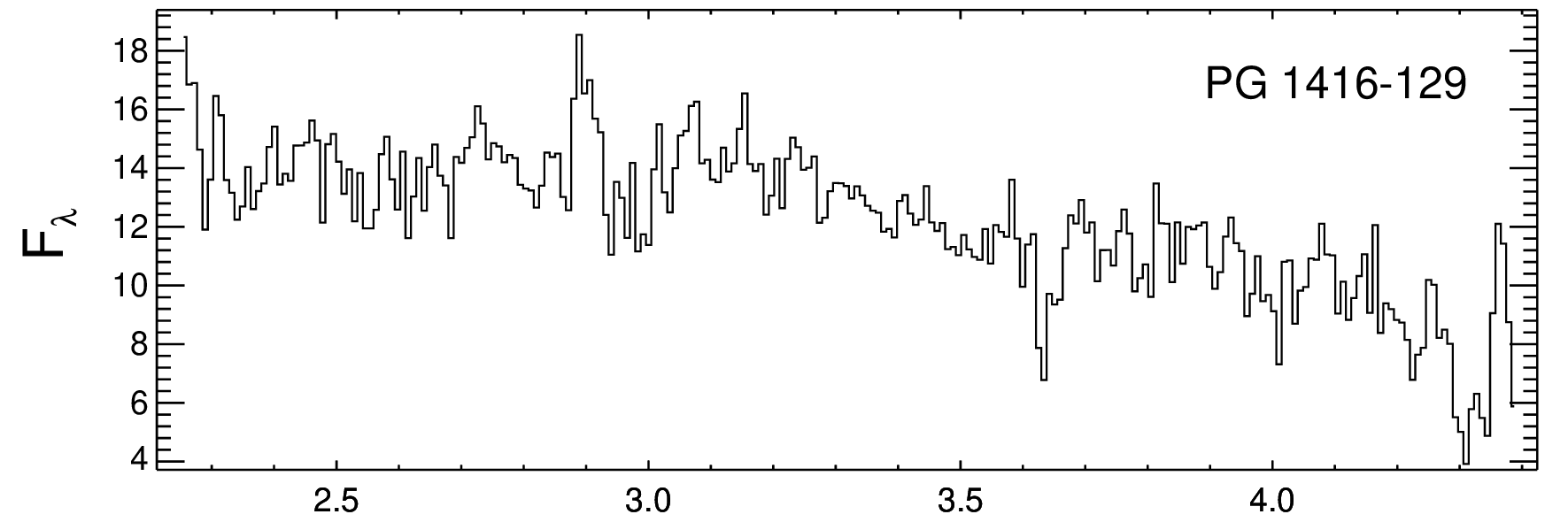}
\includegraphics[scale=0.20]{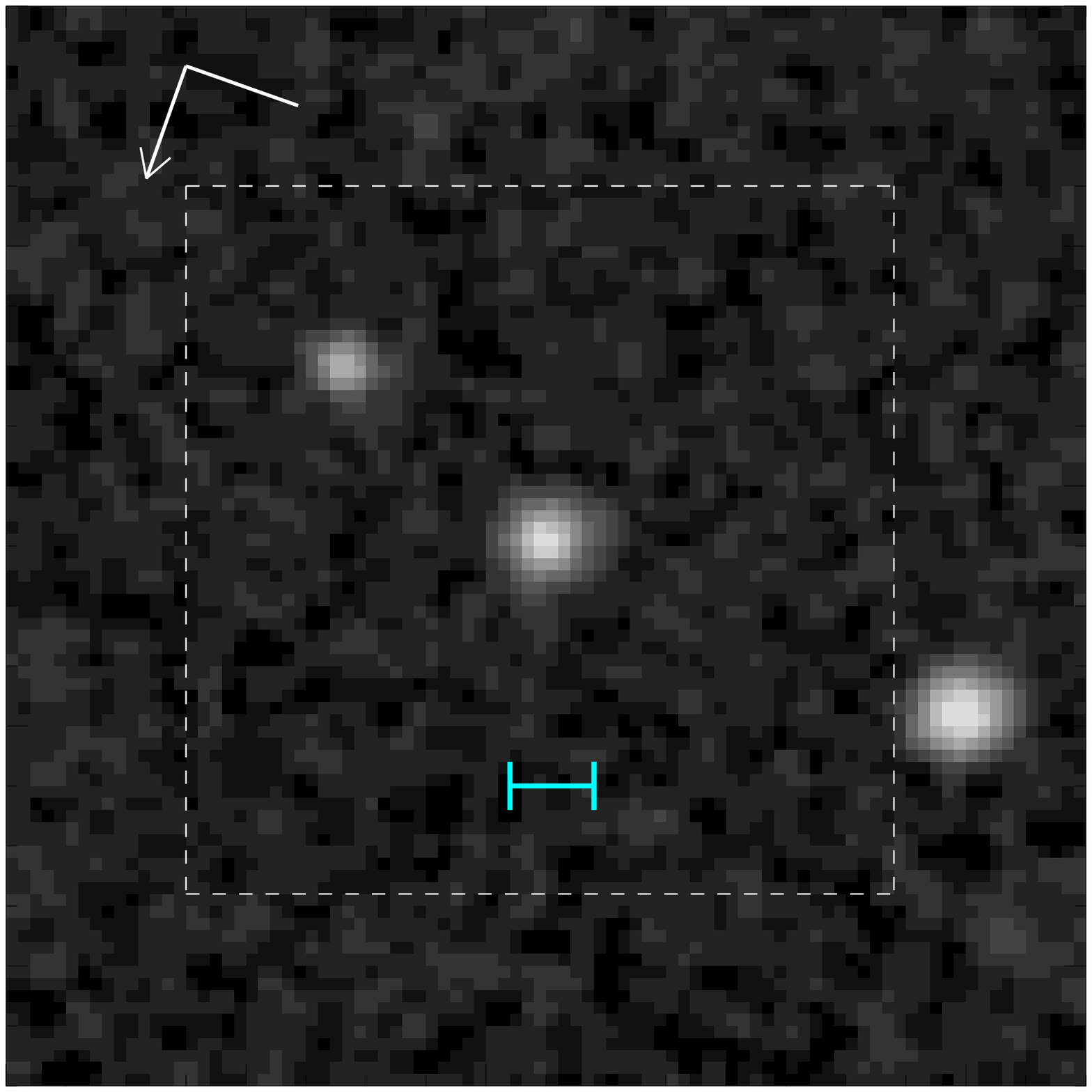}\\
\includegraphics[width=12cm,height=4cm]{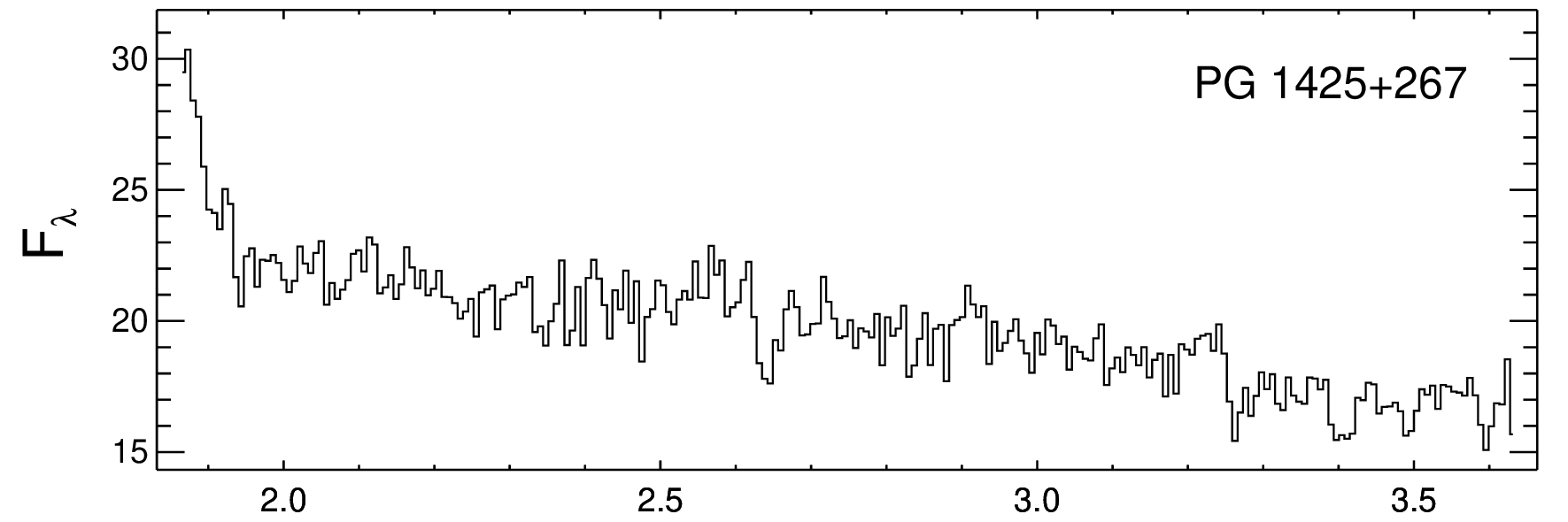}
\includegraphics[scale=0.20]{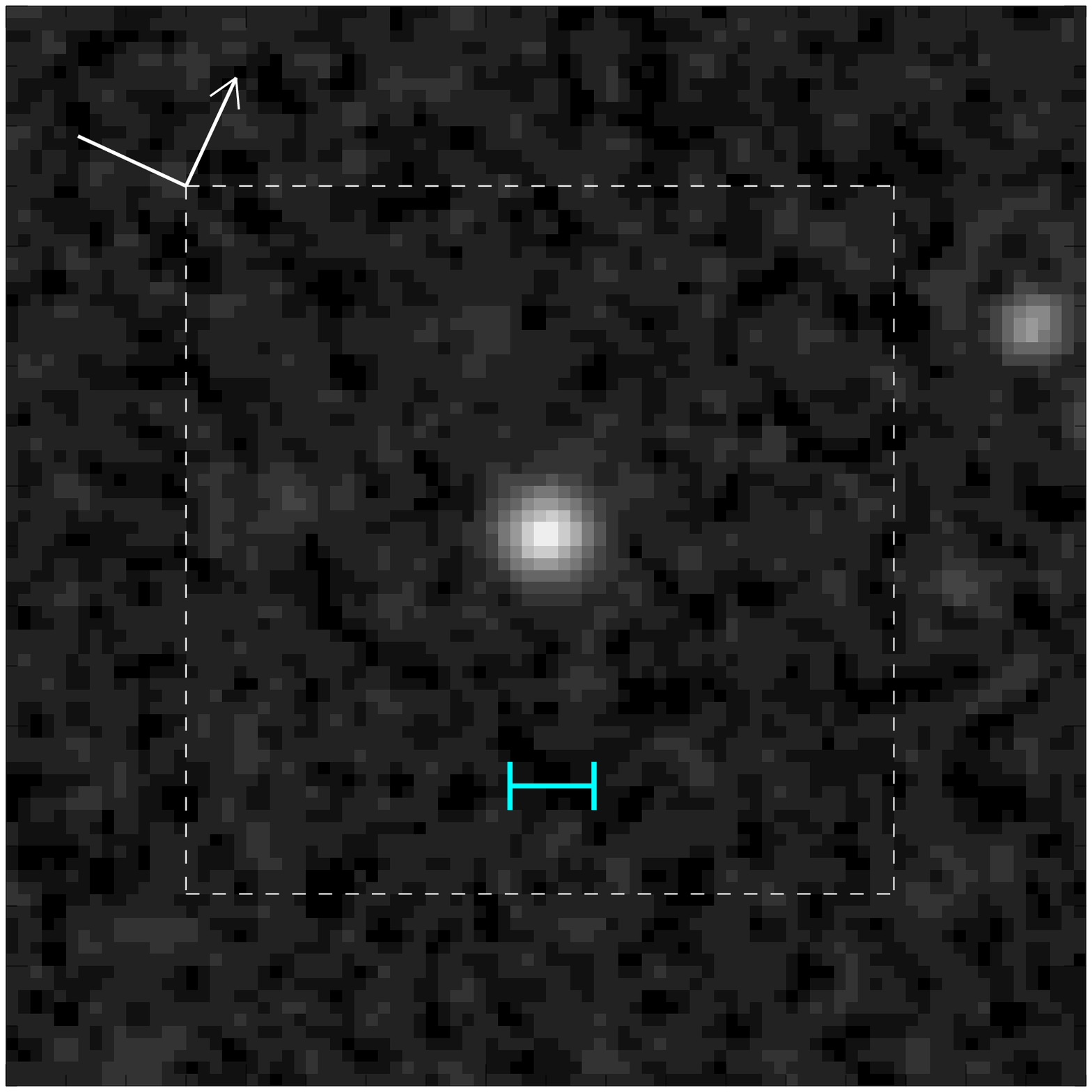}\\
\includegraphics[width=12cm,height=4cm]{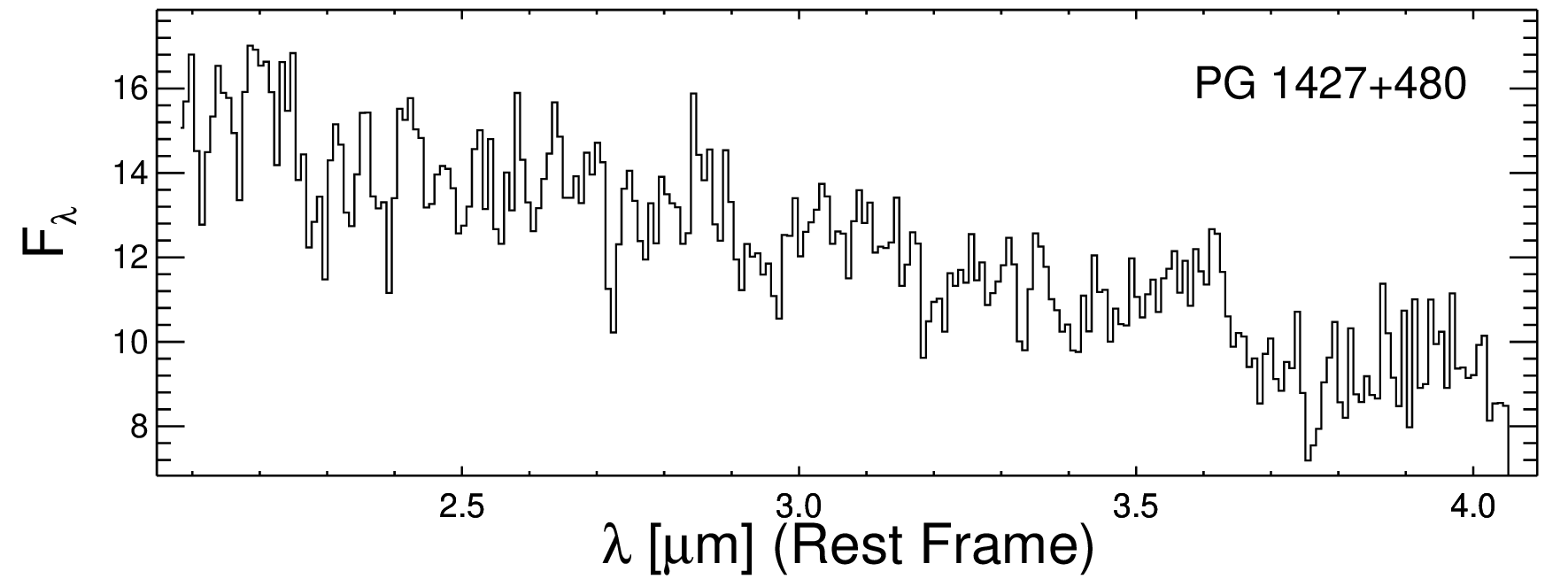}
\includegraphics[scale=0.20]{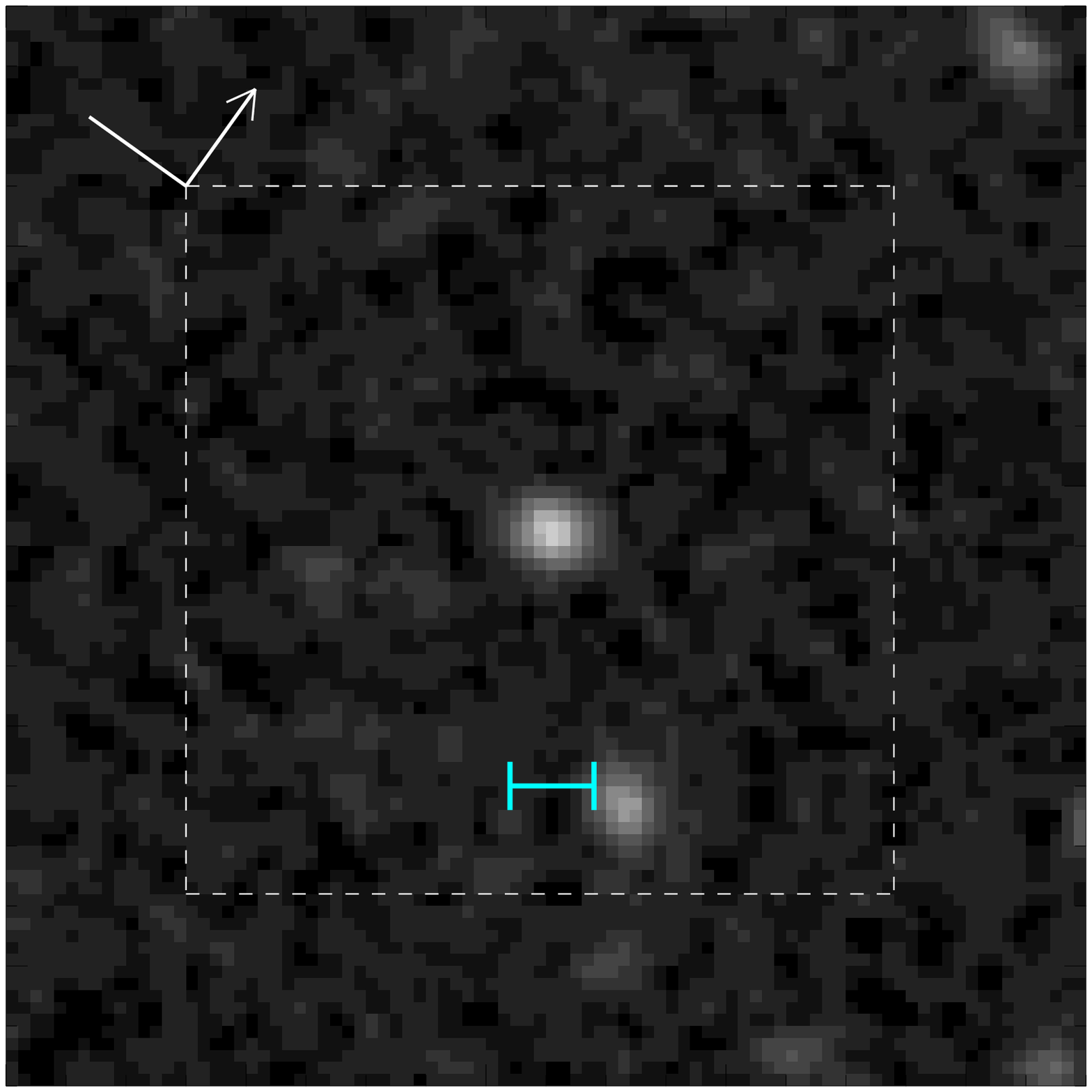}\\
\caption{Continued}
\end{figure}
\clearpage

\begin{figure}
\figurenum{5}
\includegraphics[width=12cm,height=4cm]{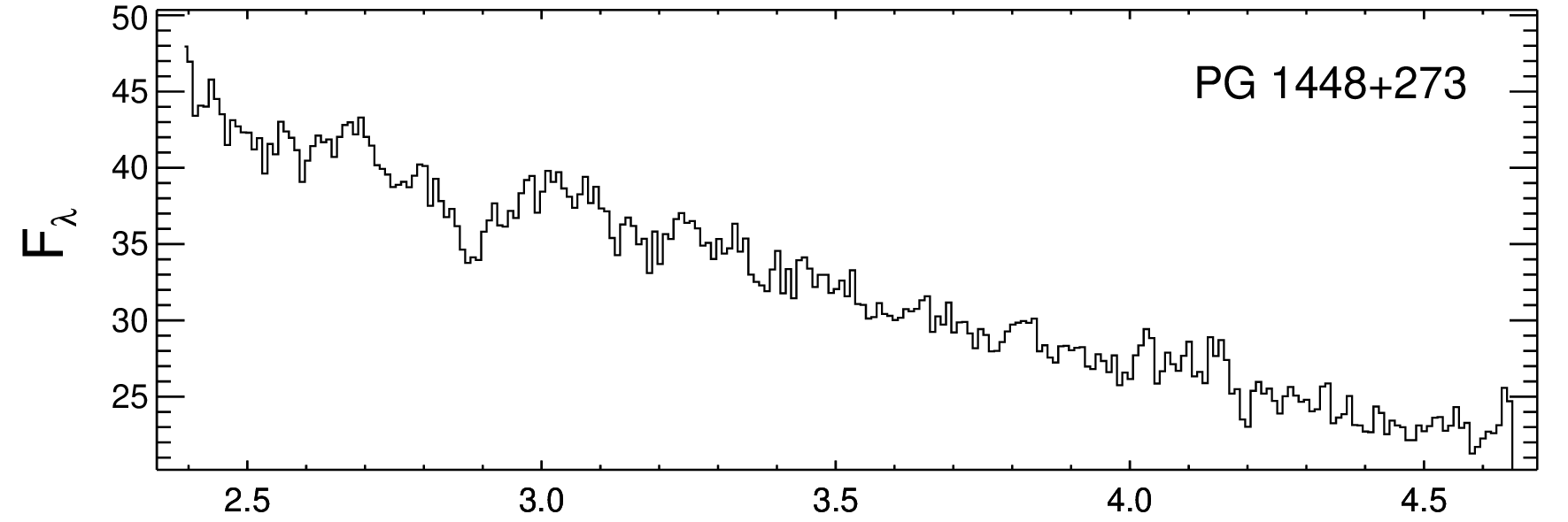}
\includegraphics[scale=0.20]{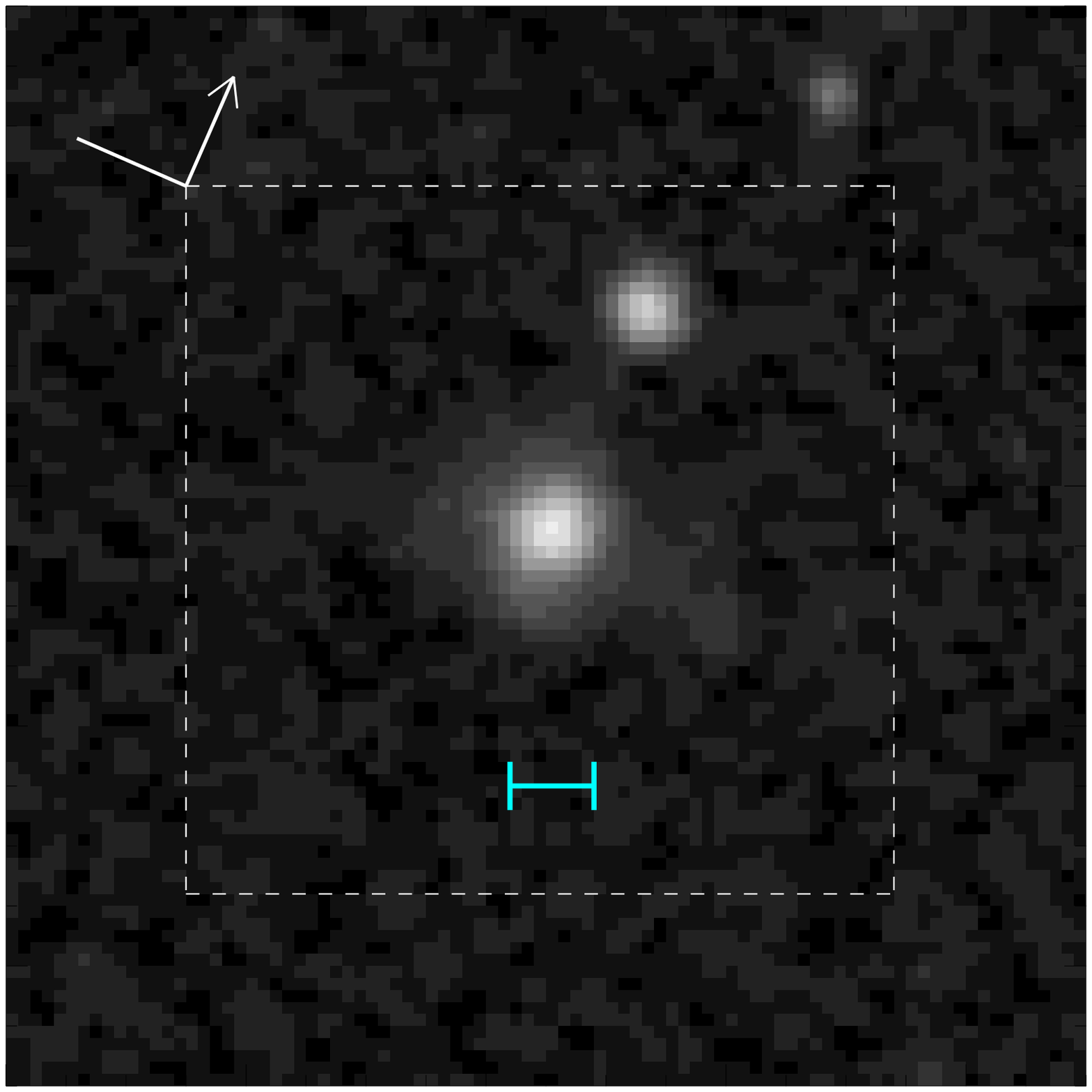}\\
\includegraphics[width=12cm,height=4cm]{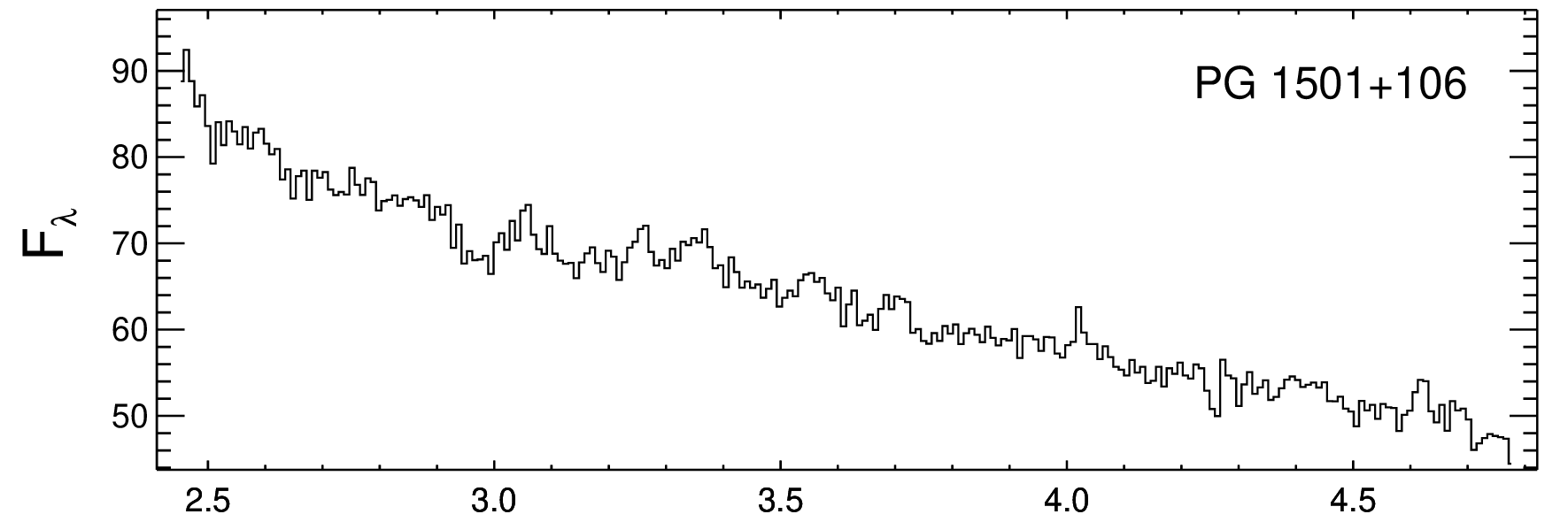}
\includegraphics[scale=0.20]{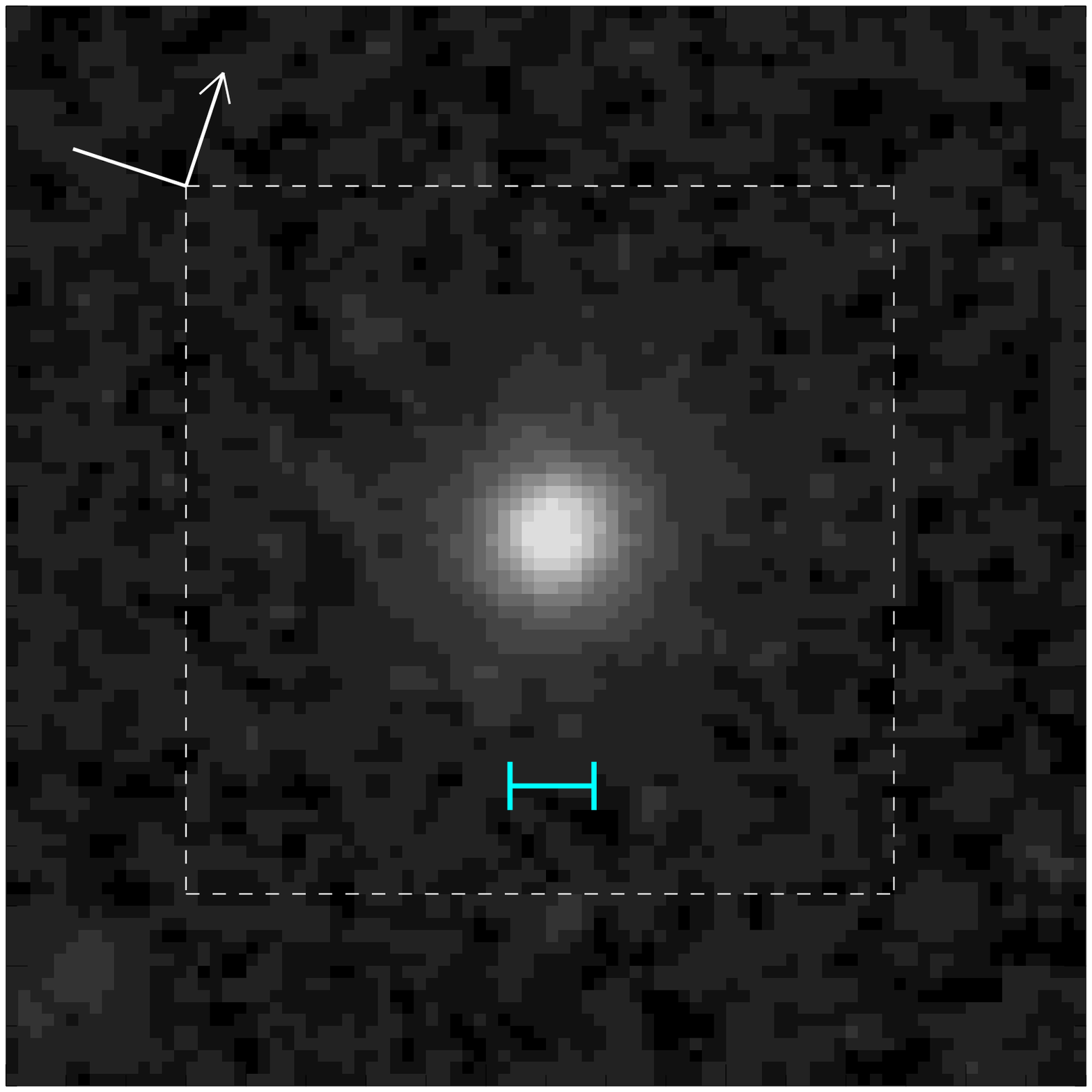}\\
\includegraphics[width=12cm,height=4cm]{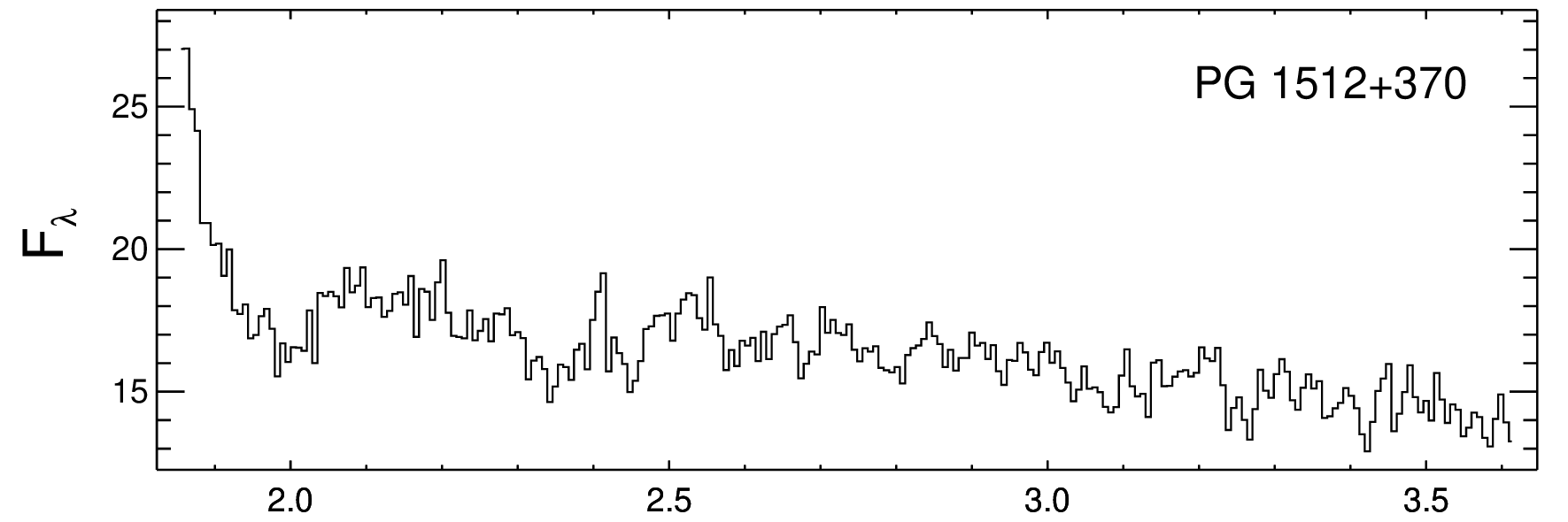}
\includegraphics[scale=0.20]{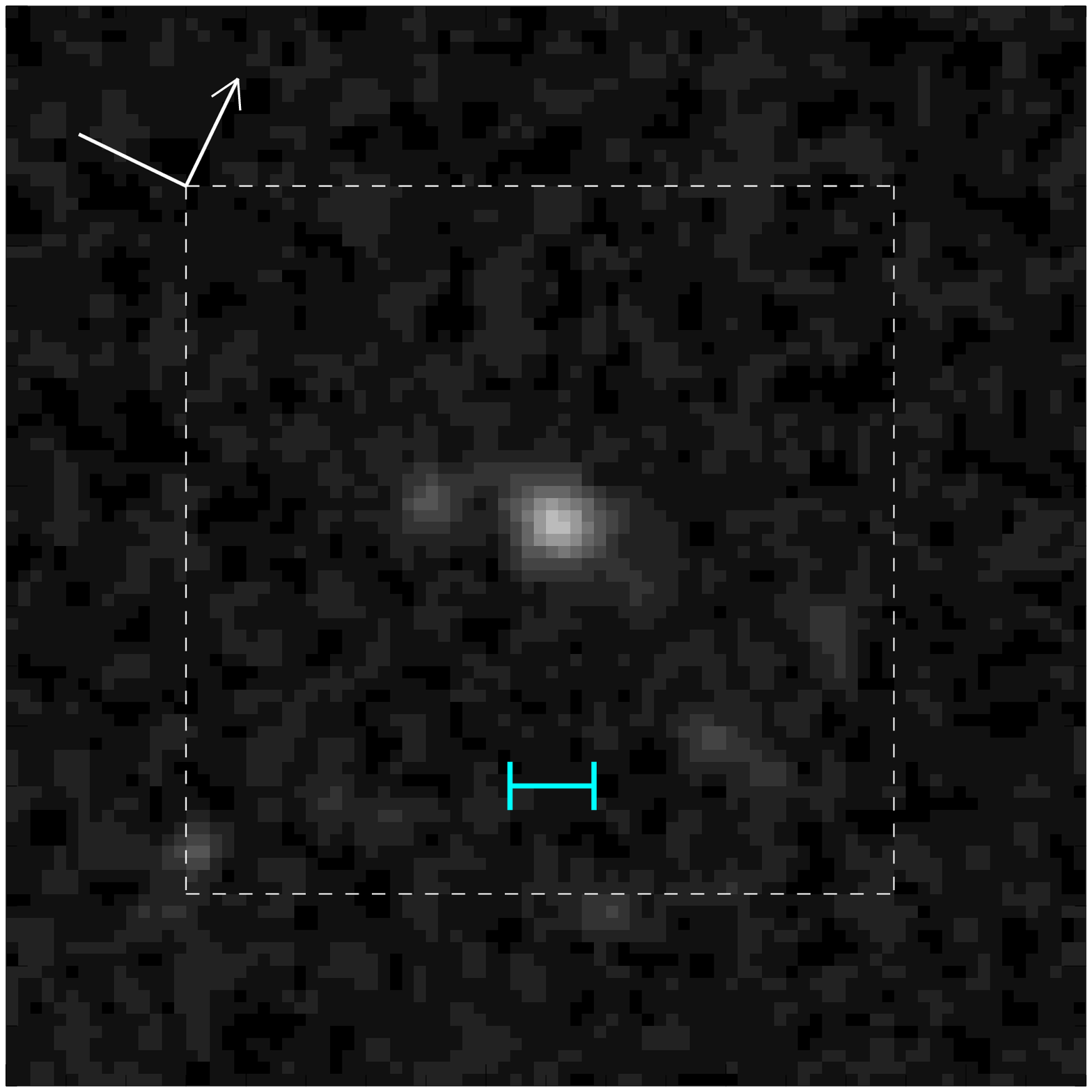}\\
\includegraphics[width=12cm,height=4cm]{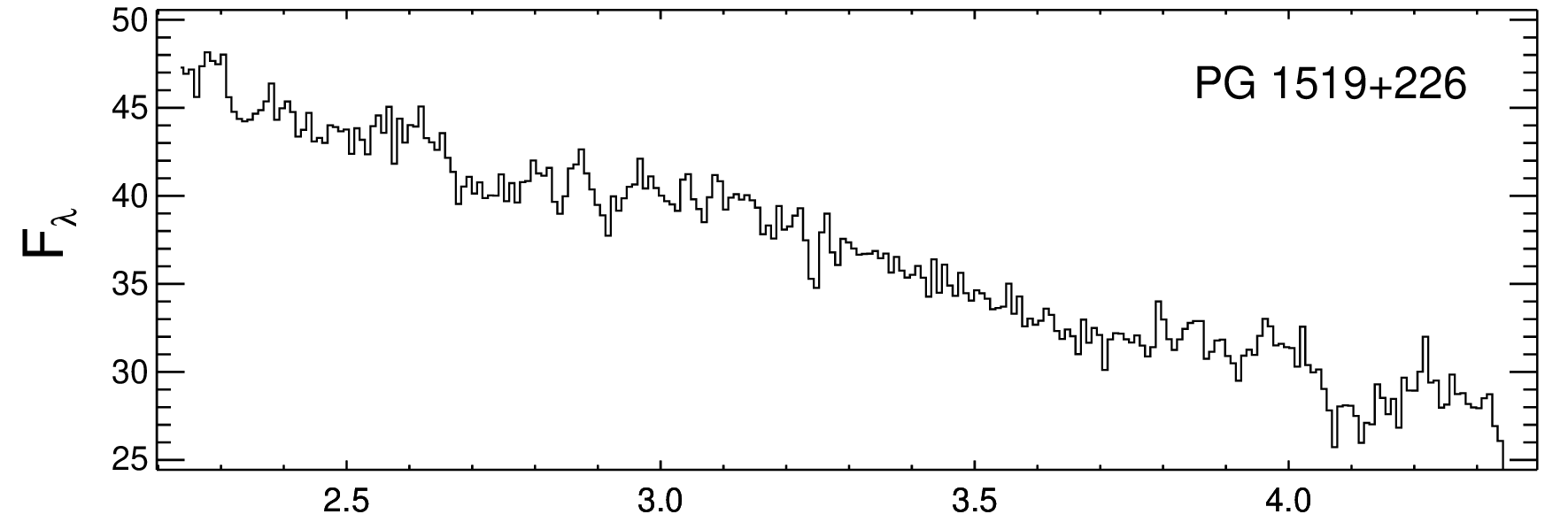}
\includegraphics[scale=0.20]{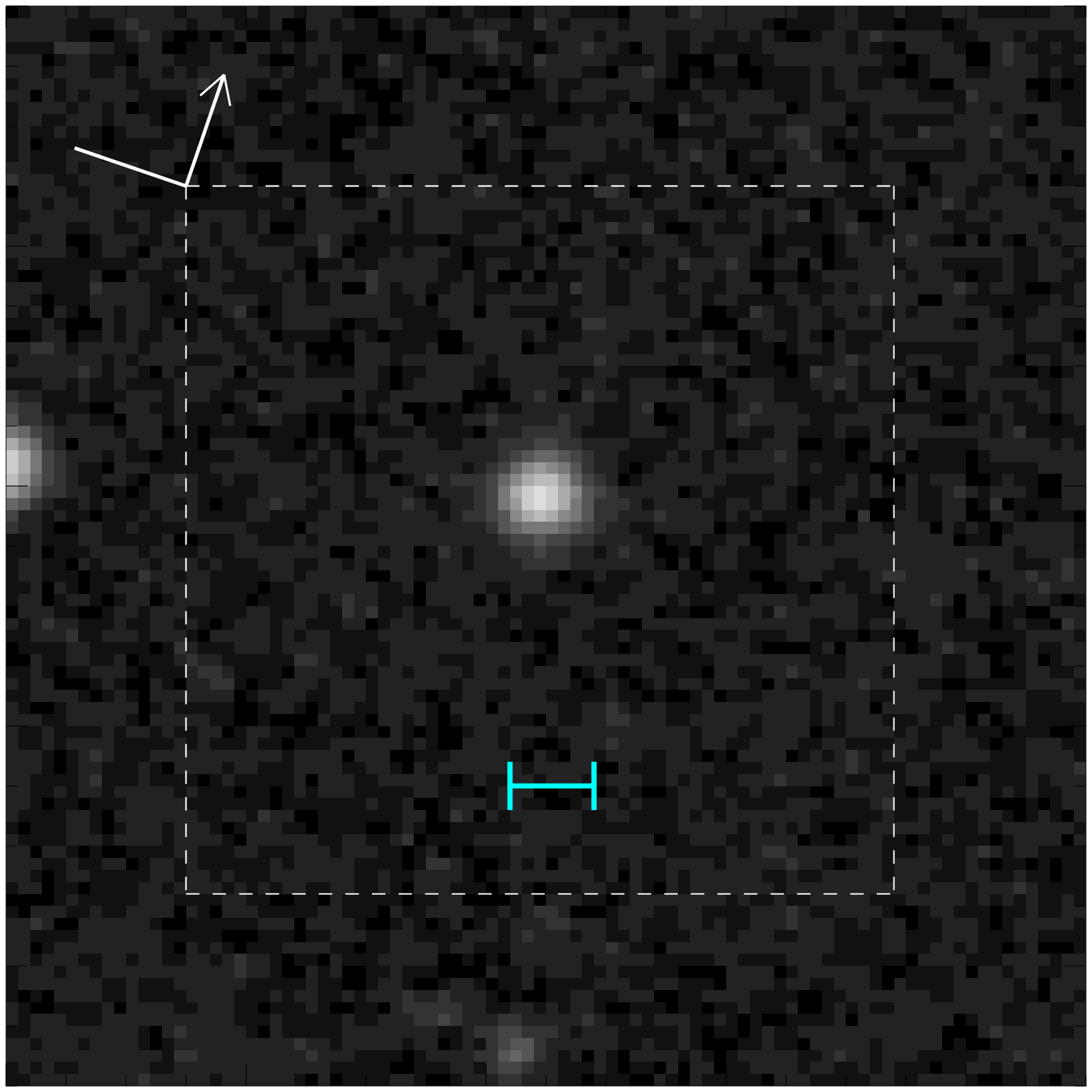}\\
\includegraphics[width=12cm,height=4cm]{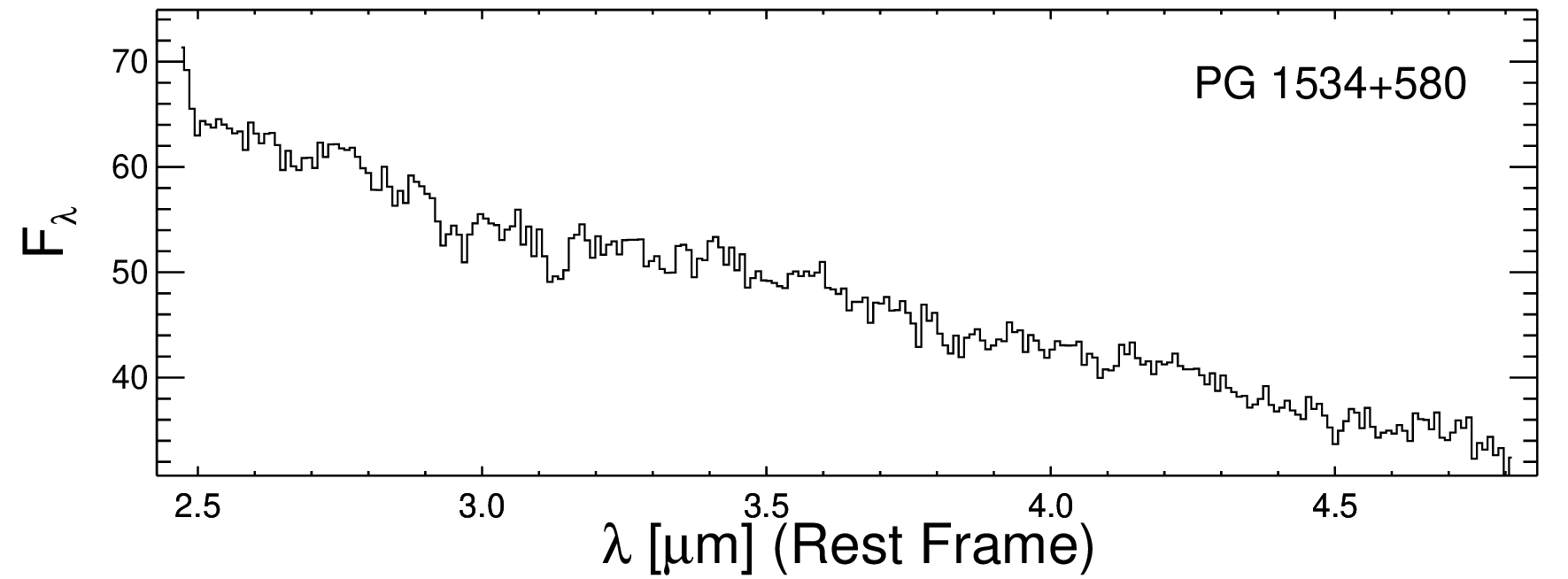}
\includegraphics[scale=0.20]{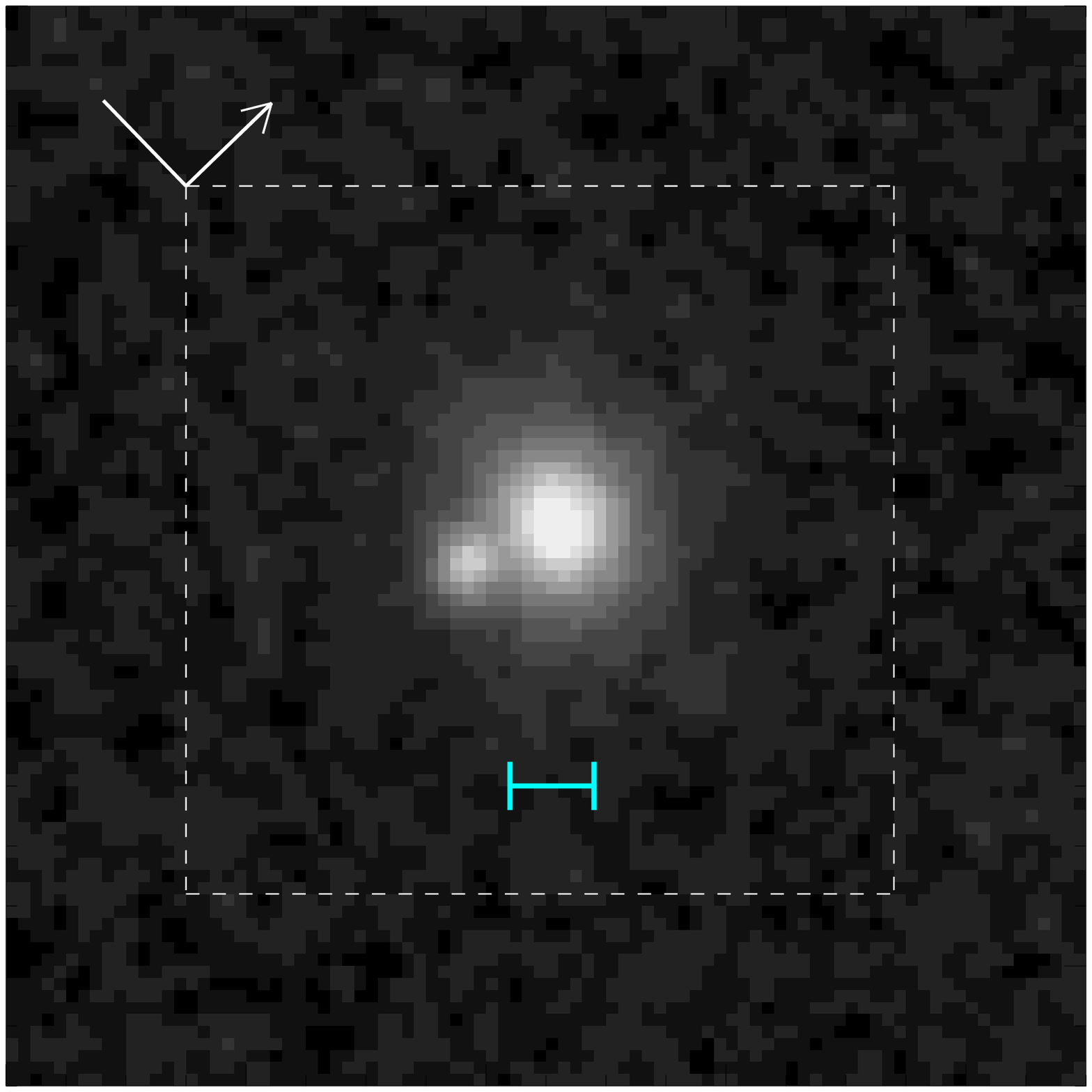}\\
\caption{Continued}
\end{figure}
\clearpage

\begin{figure}
\figurenum{5}
\includegraphics[width=12cm,height=4cm]{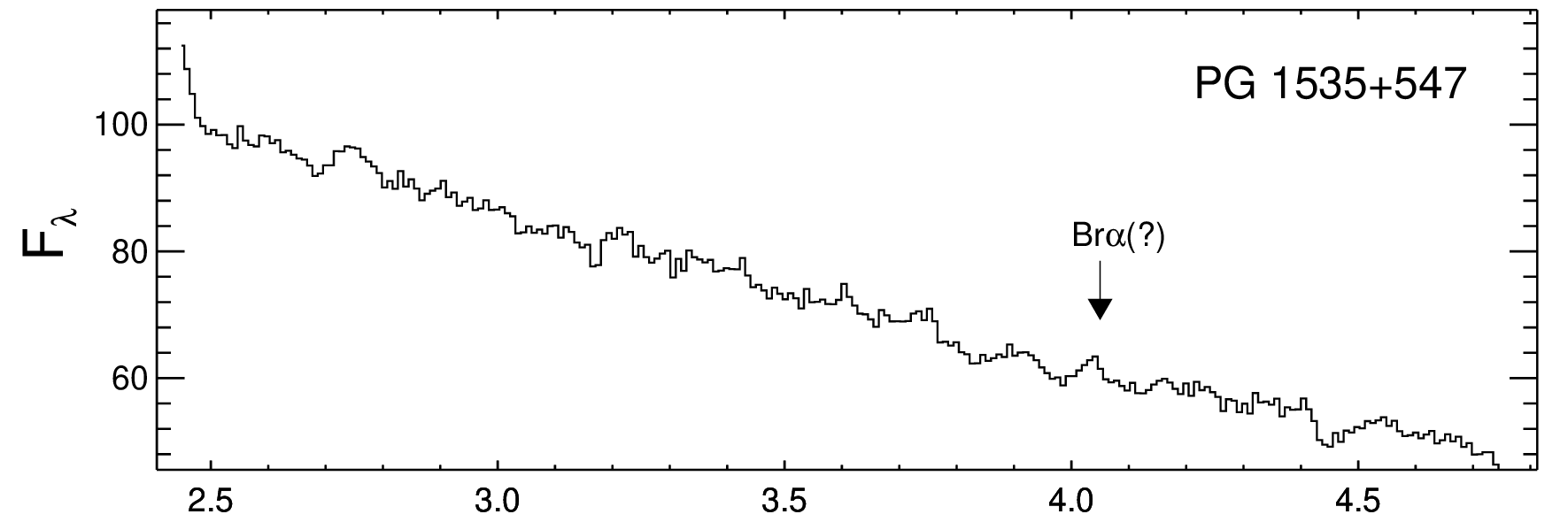}
\includegraphics[scale=0.20]{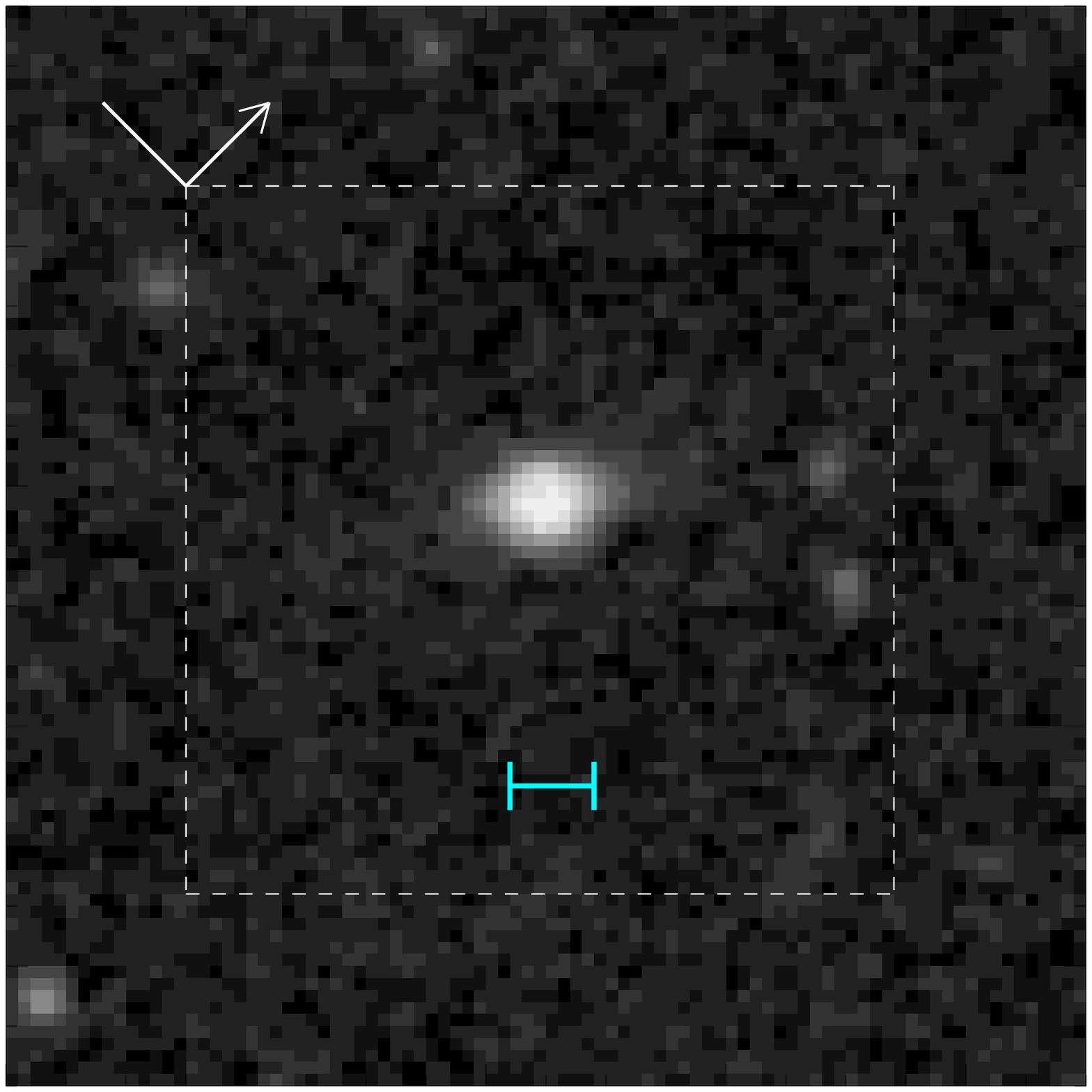}\\
\includegraphics[width=12cm,height=4cm]{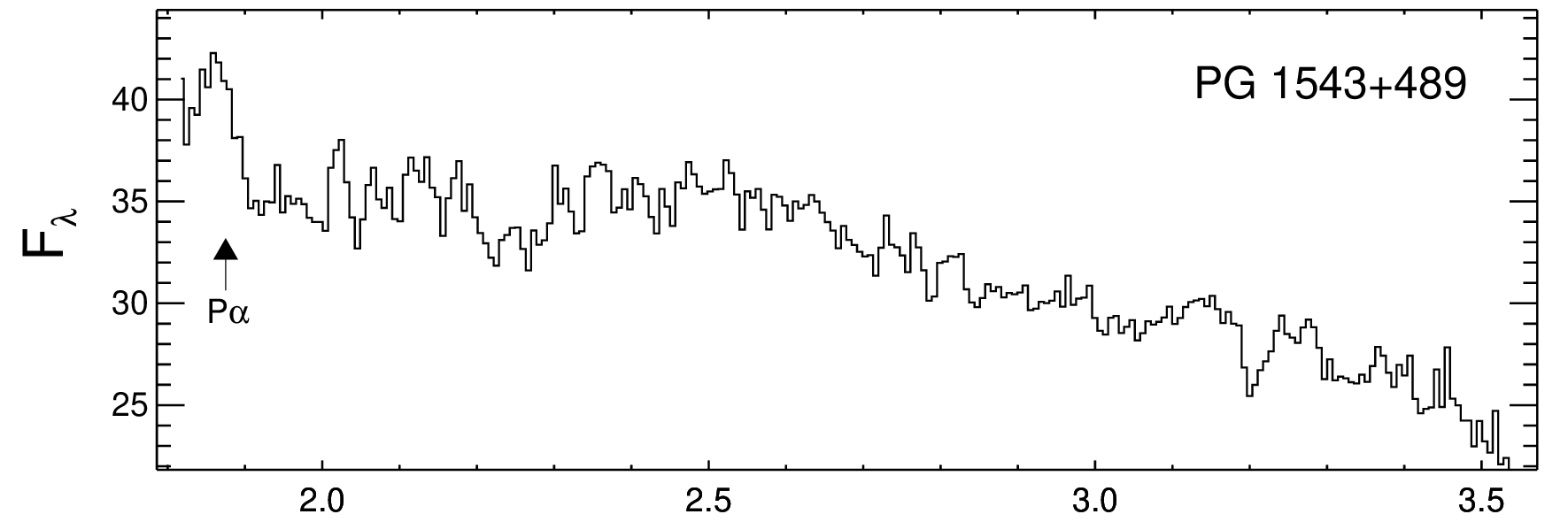}
\includegraphics[scale=0.20]{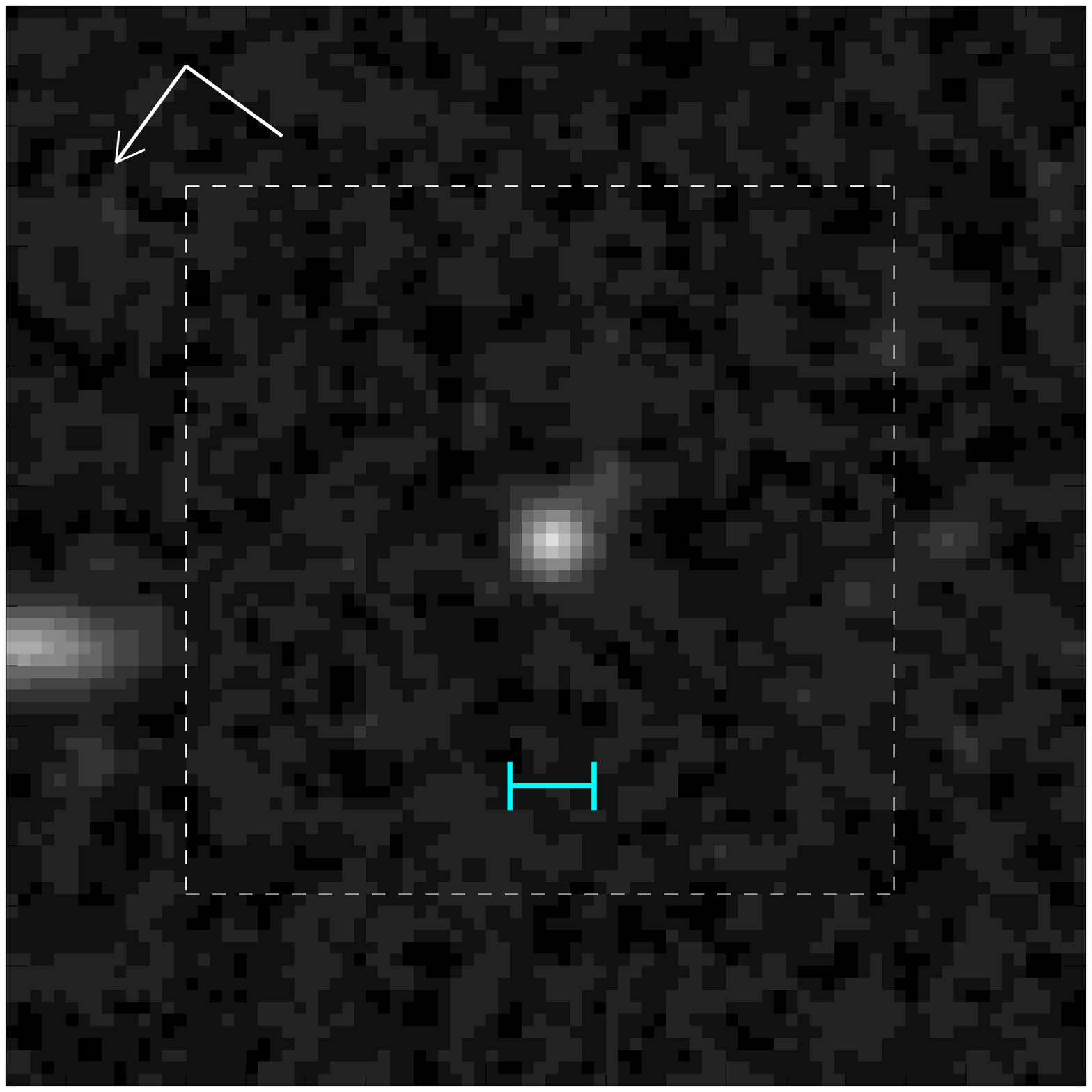}\\
\includegraphics[width=12cm,height=4cm]{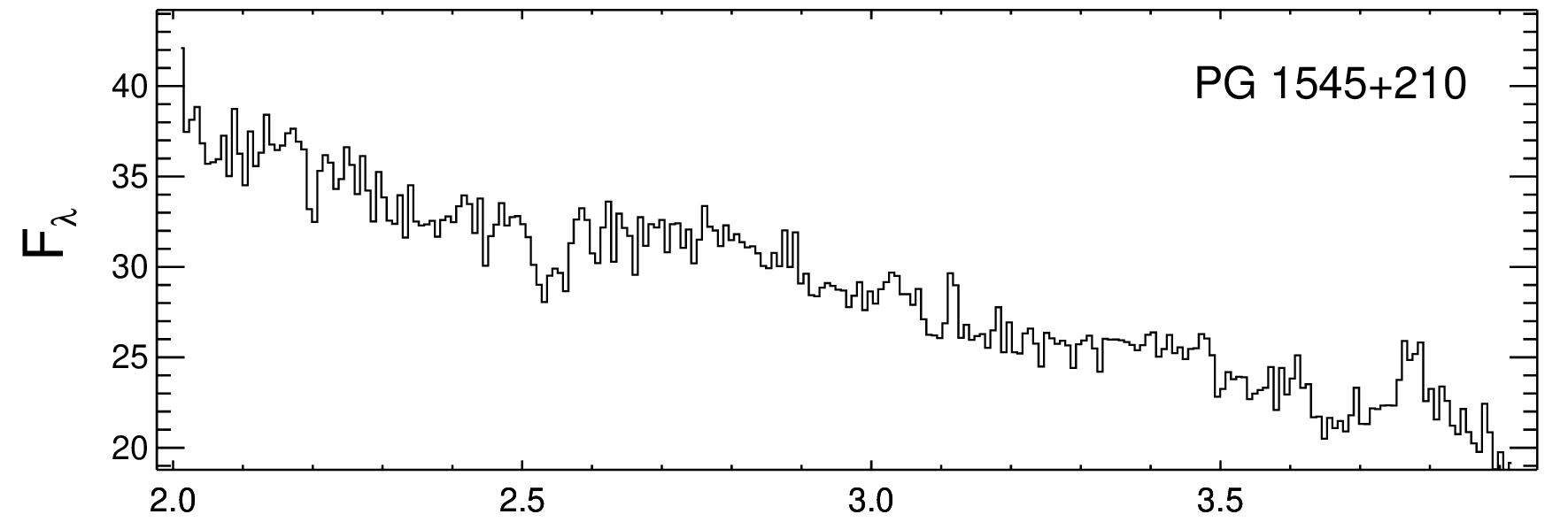}
\includegraphics[scale=0.20]{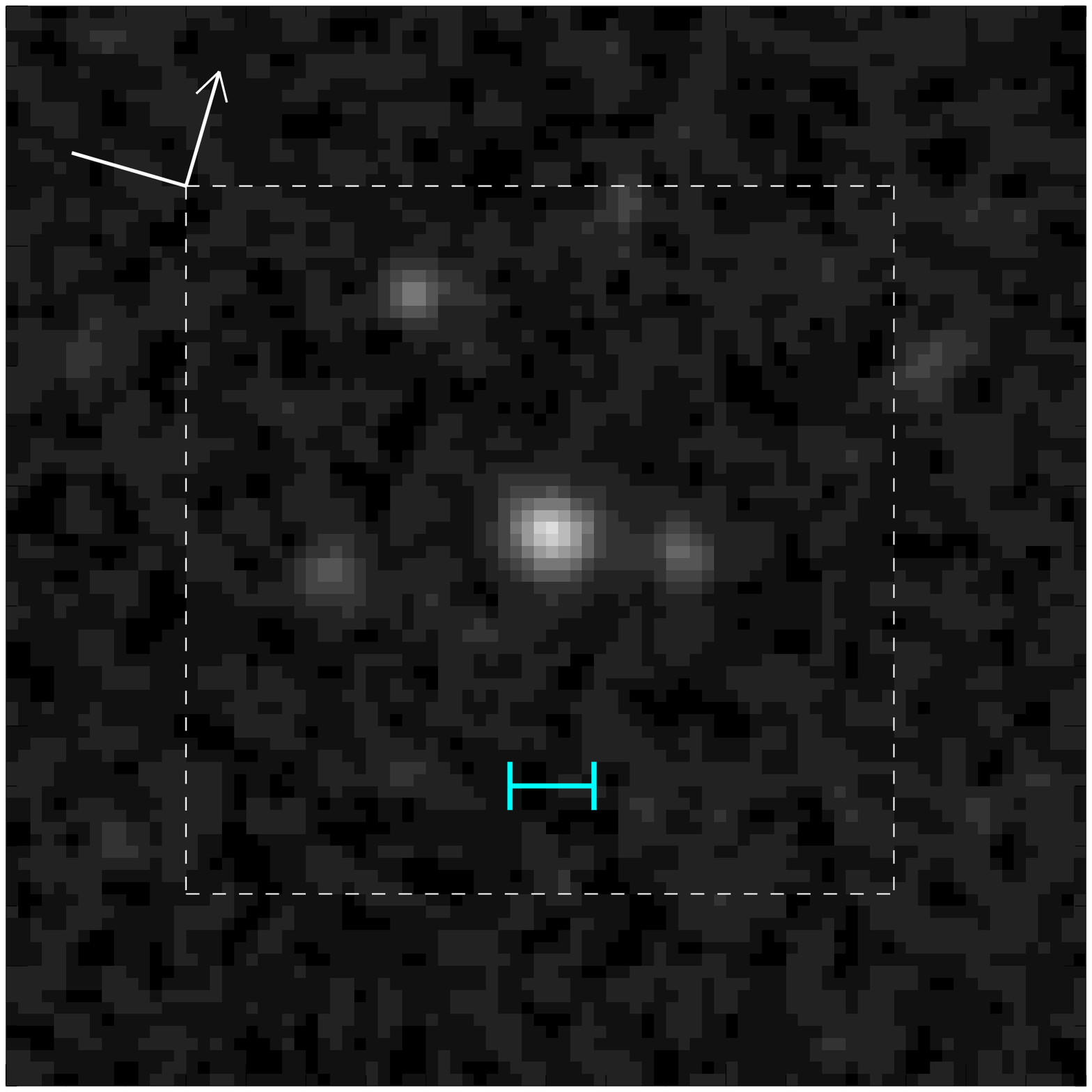}\\
\includegraphics[width=12cm,height=4cm]{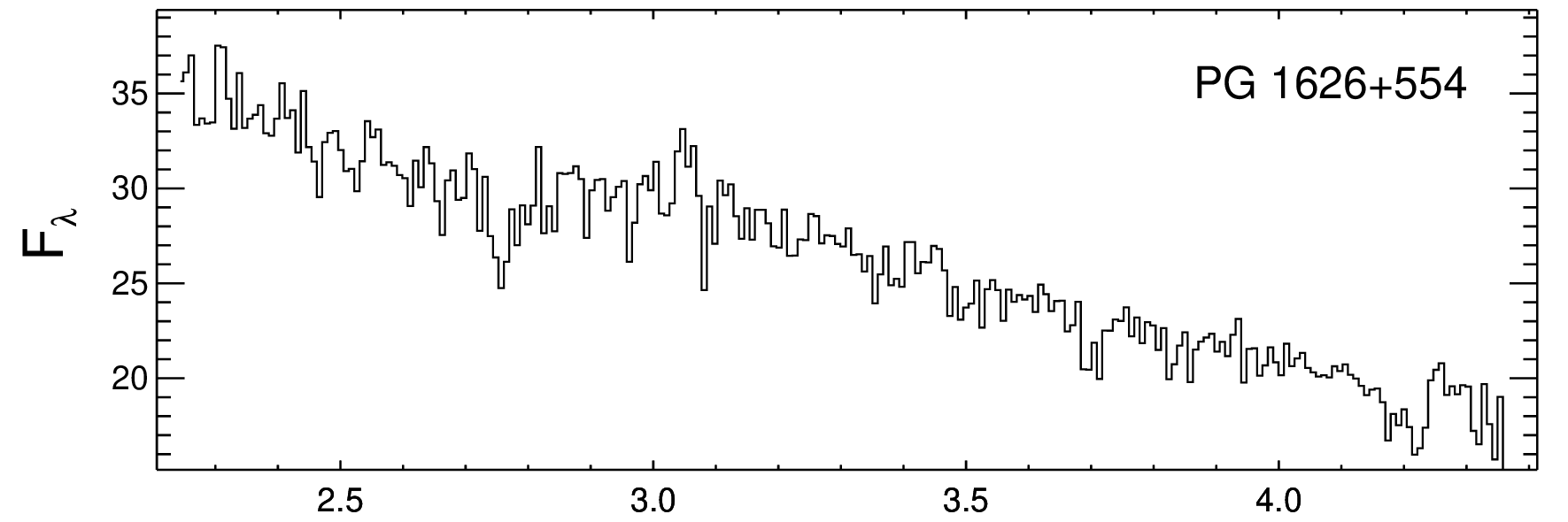}
\includegraphics[scale=0.20]{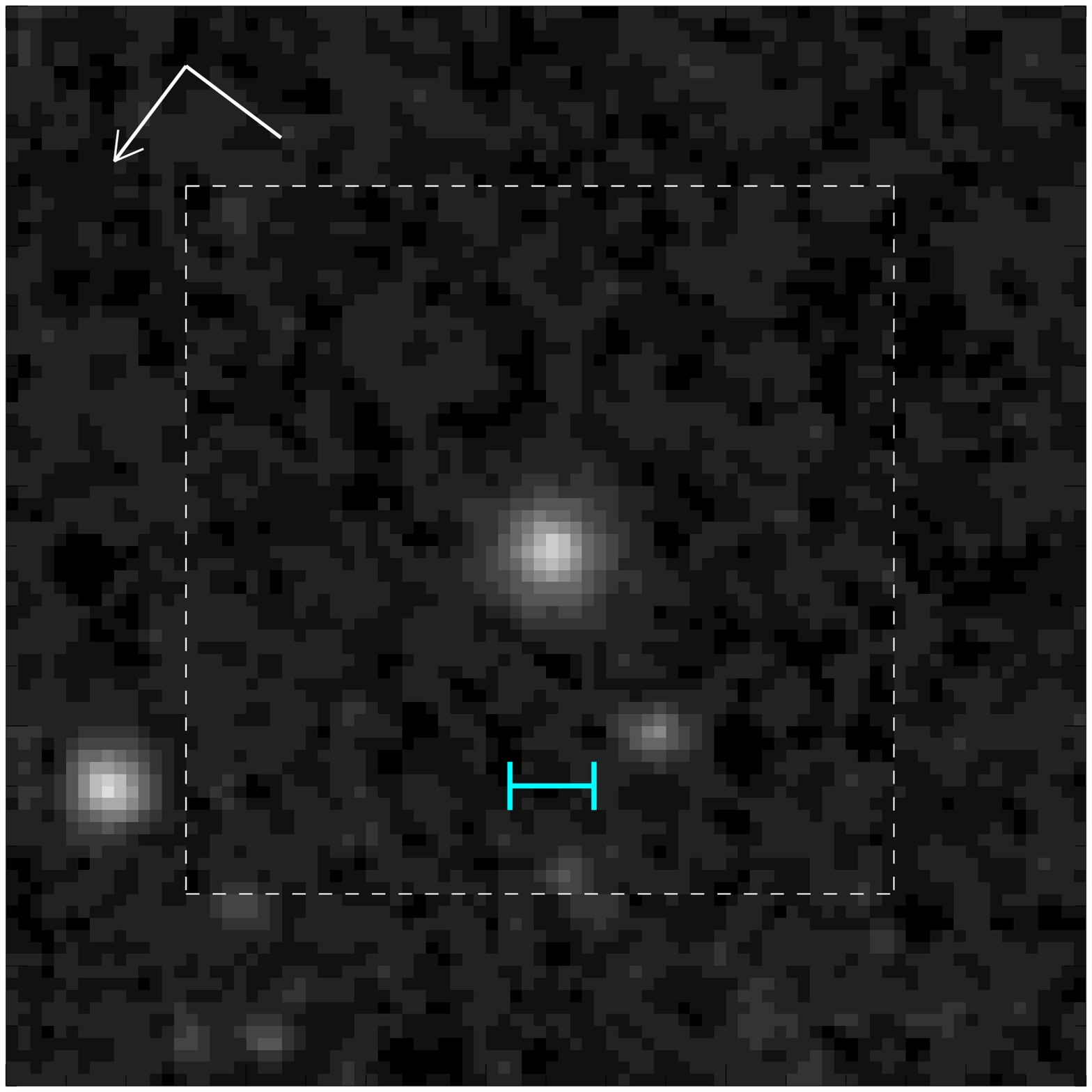}\\
\includegraphics[width=12cm,height=4cm]{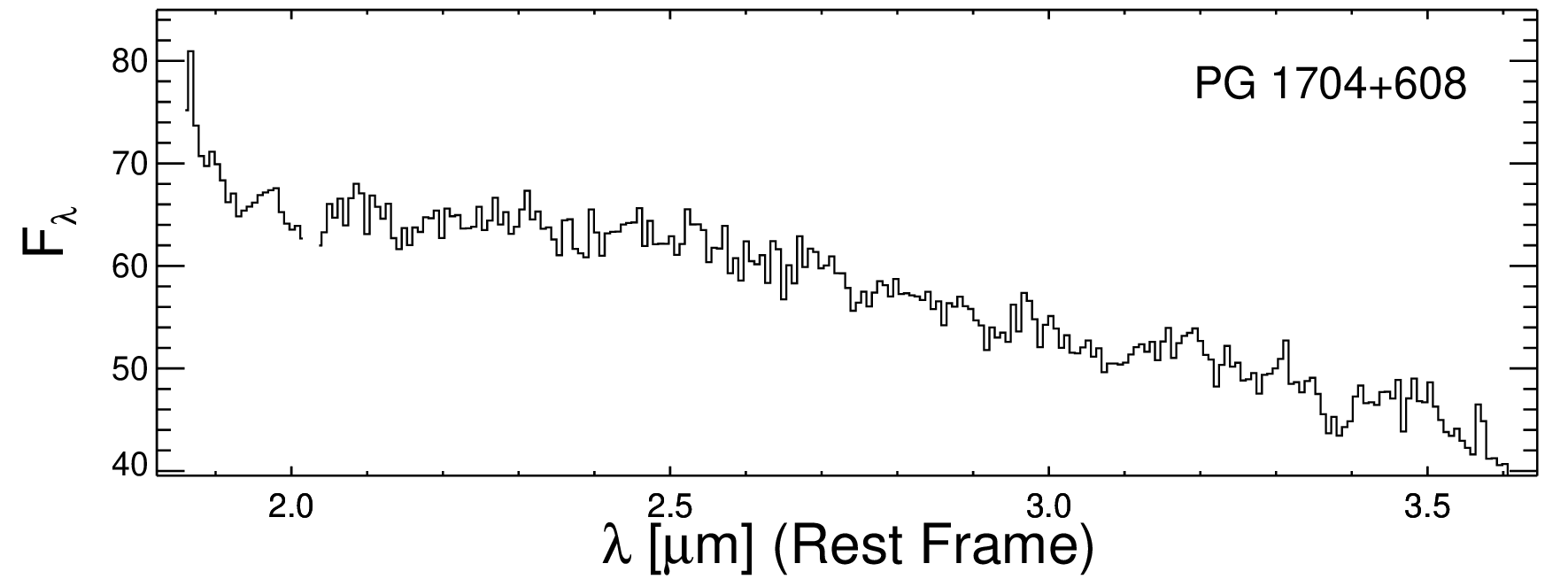}
\includegraphics[scale=0.20]{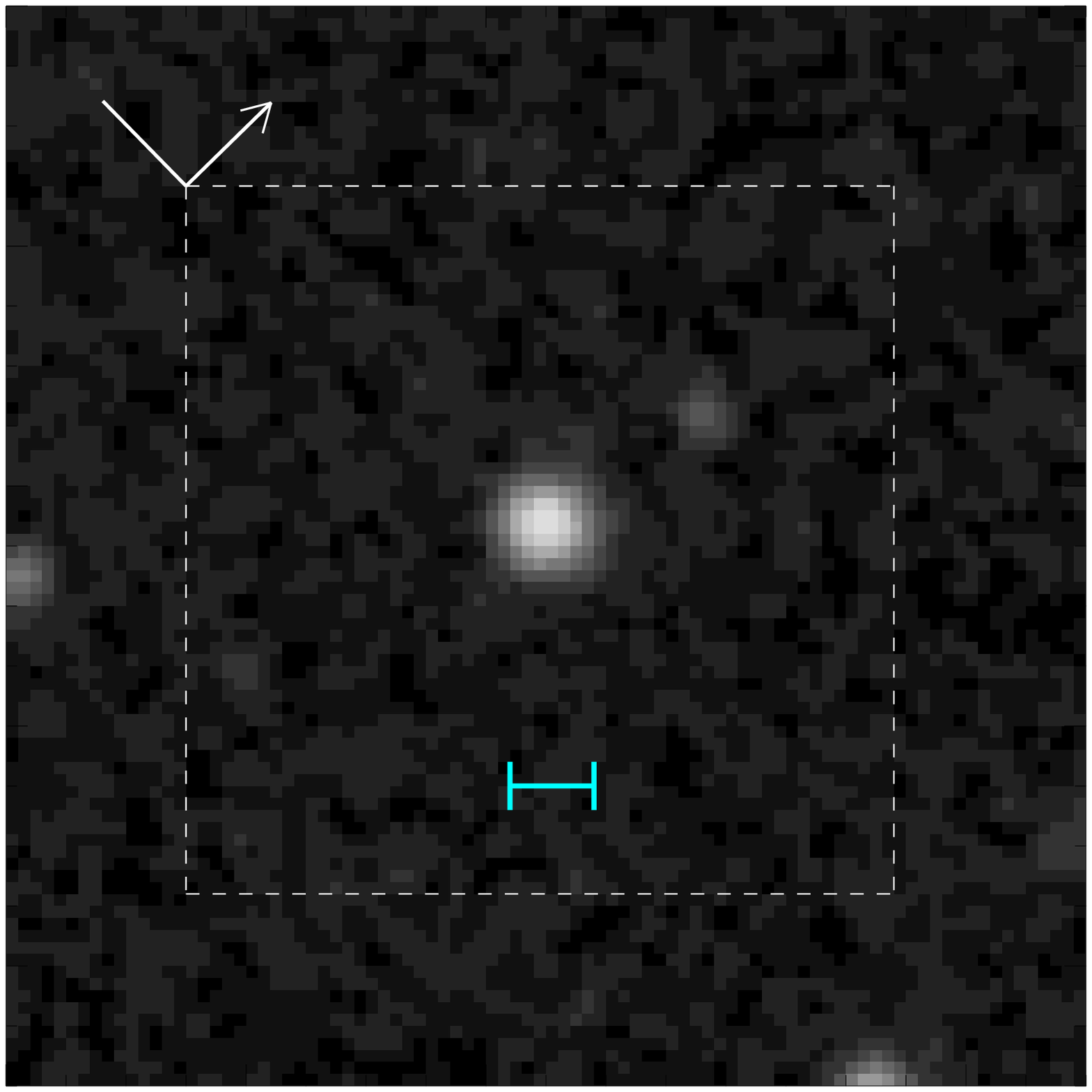}\\
\caption{Continued}
\end{figure}
\clearpage

\begin{figure}
\figurenum{5}
\includegraphics[width=12cm,height=4cm]{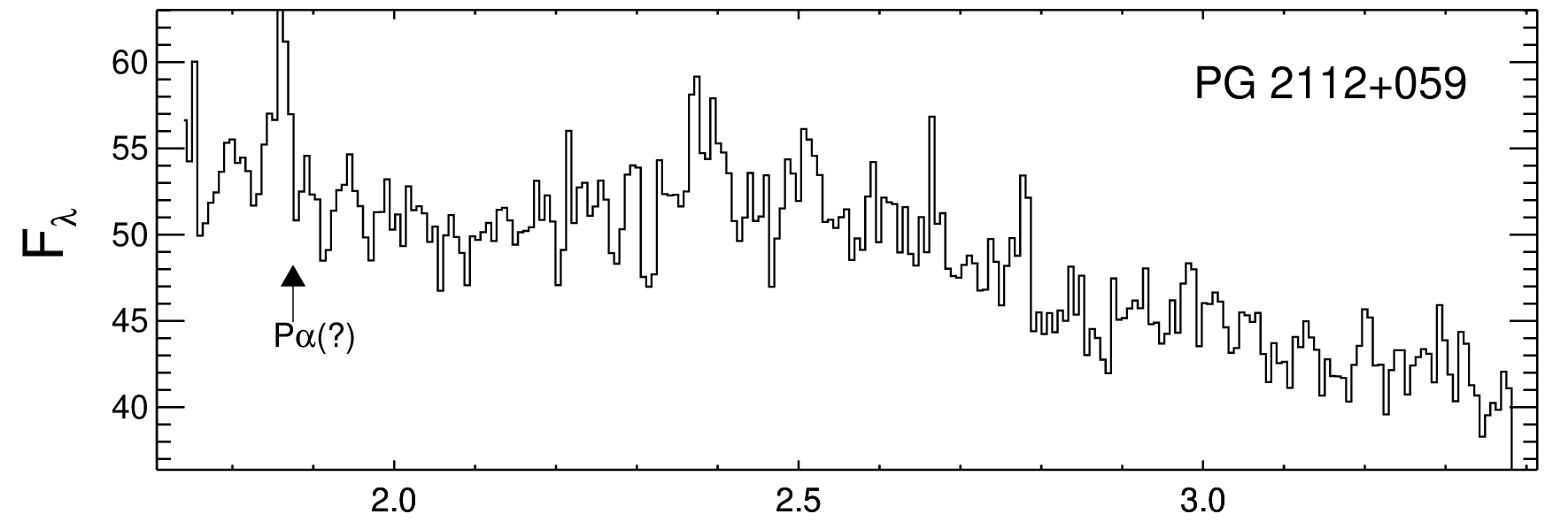}
\includegraphics[scale=0.20]{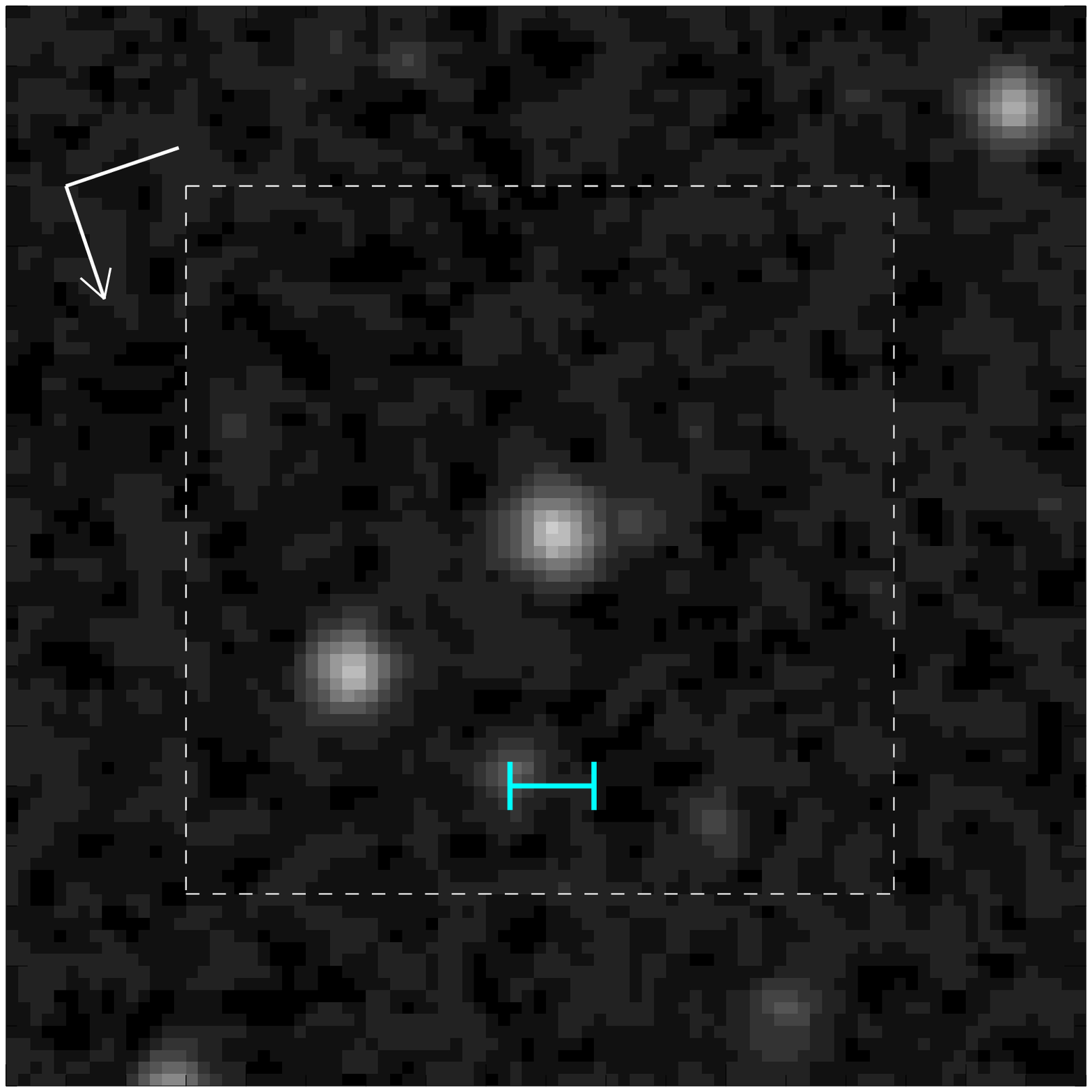}\\
\includegraphics[width=12cm,height=4cm]{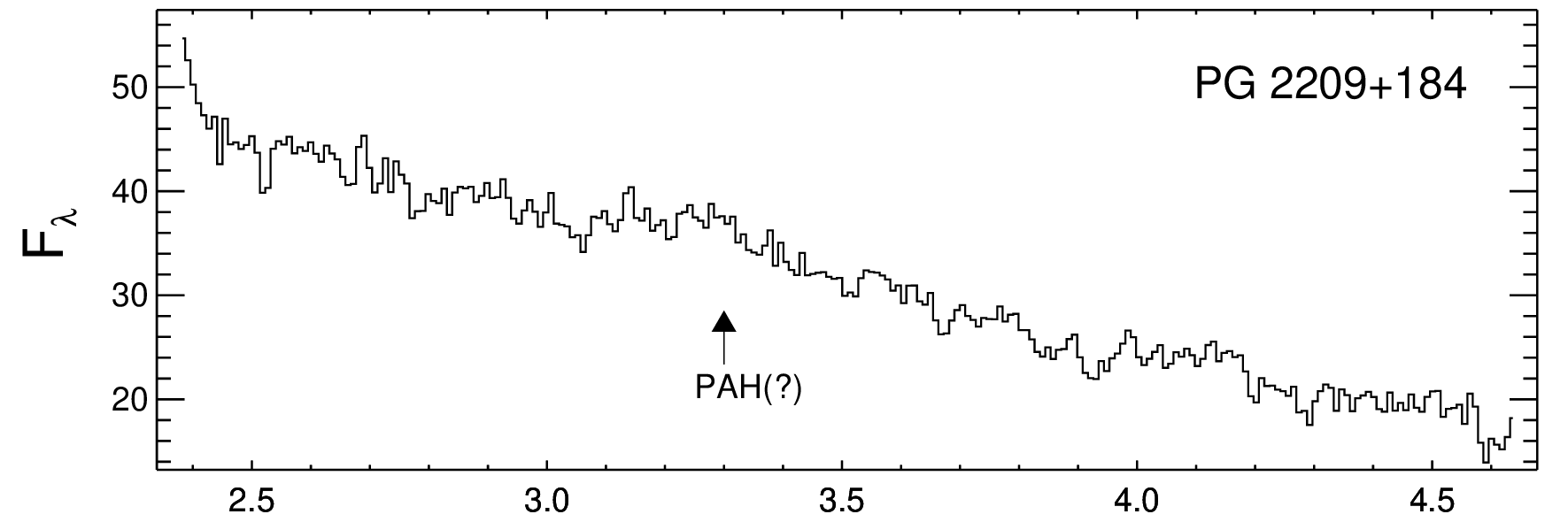}
\includegraphics[scale=0.20]{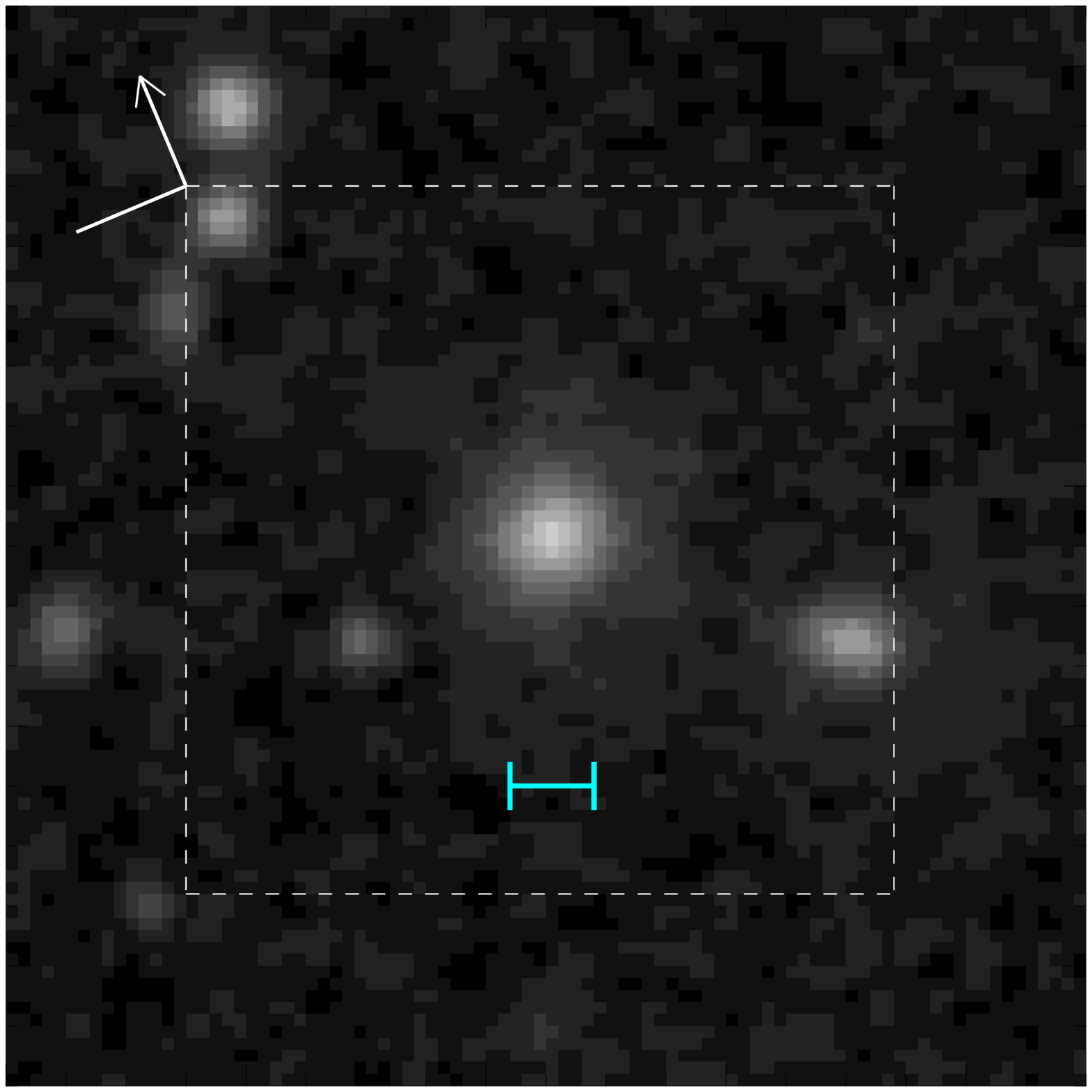}\\
\includegraphics[width=12cm,height=4cm]{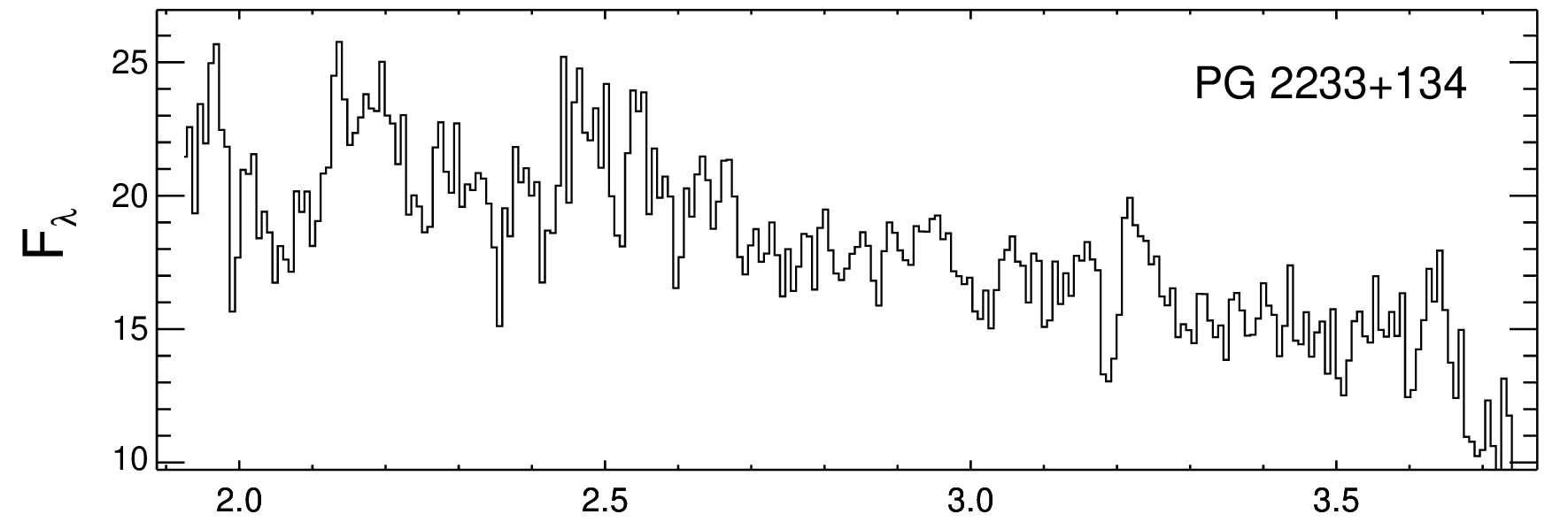}
\includegraphics[scale=0.20]{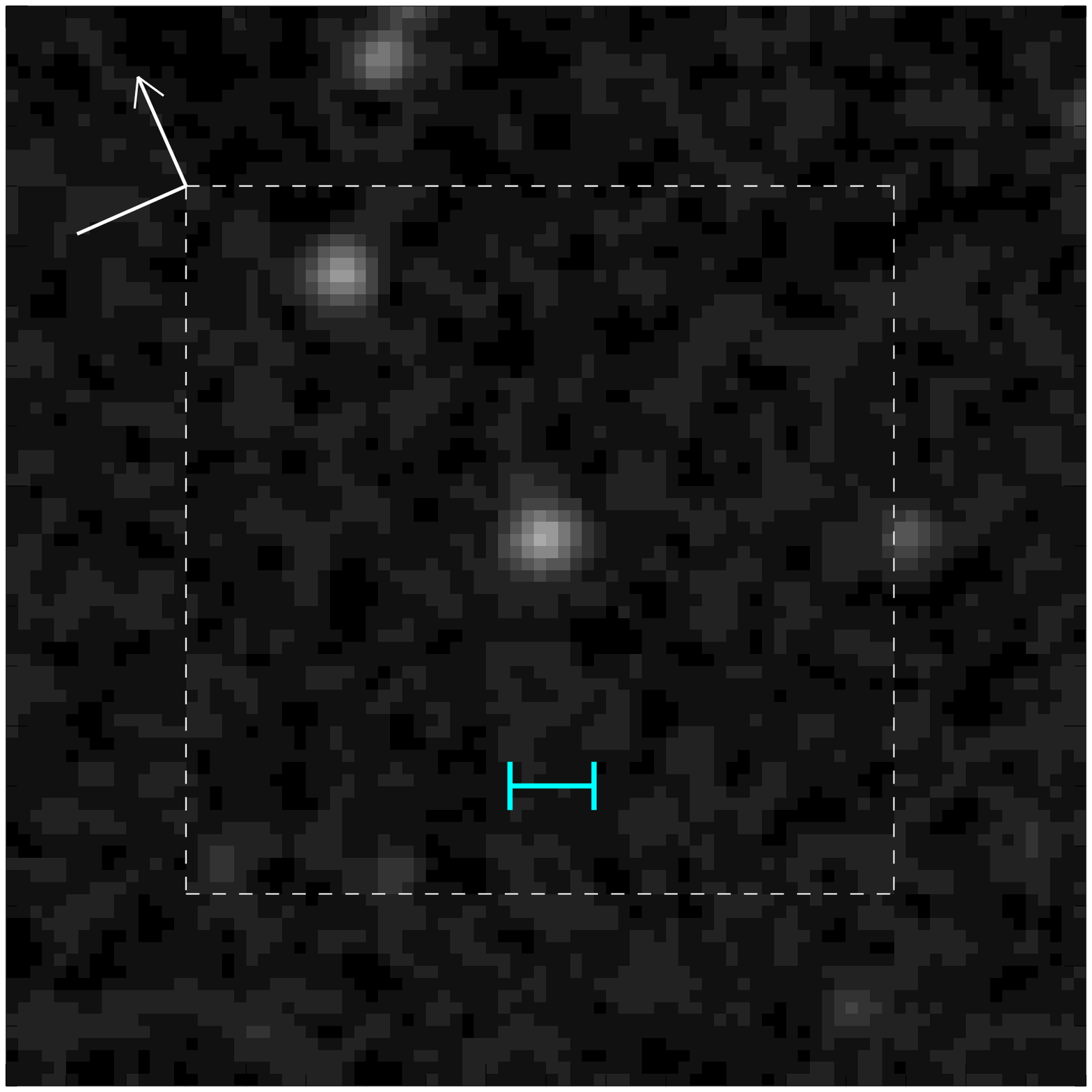}\\
\includegraphics[width=12cm,height=4cm]{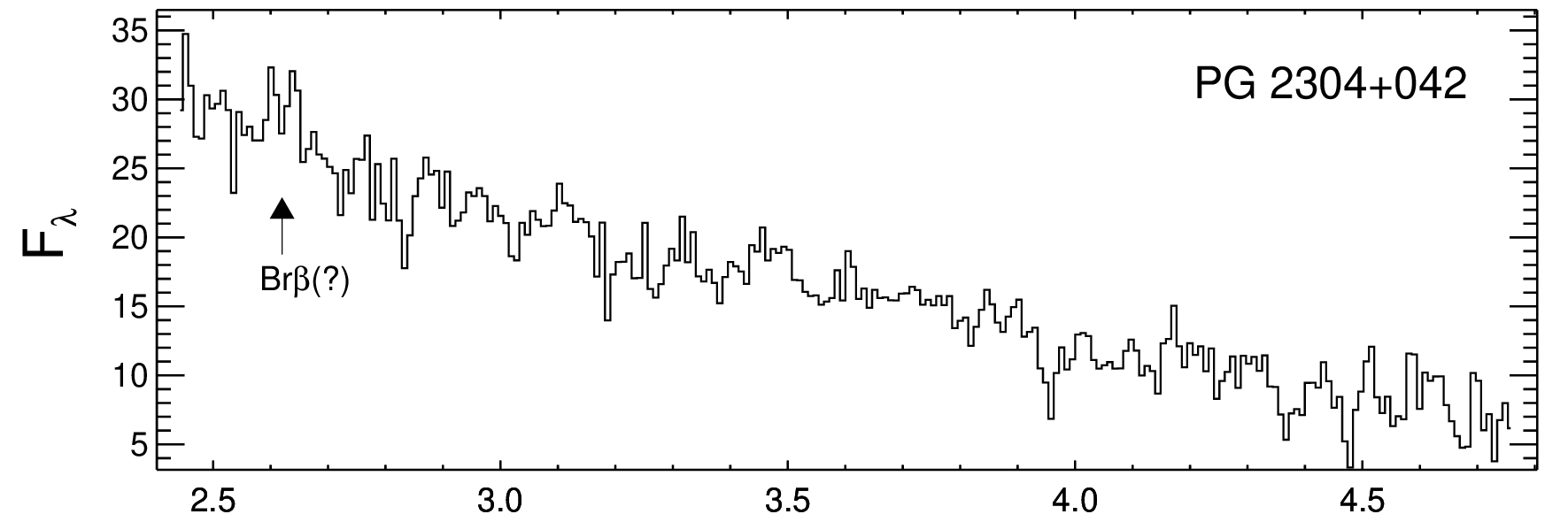}
\includegraphics[scale=0.20]{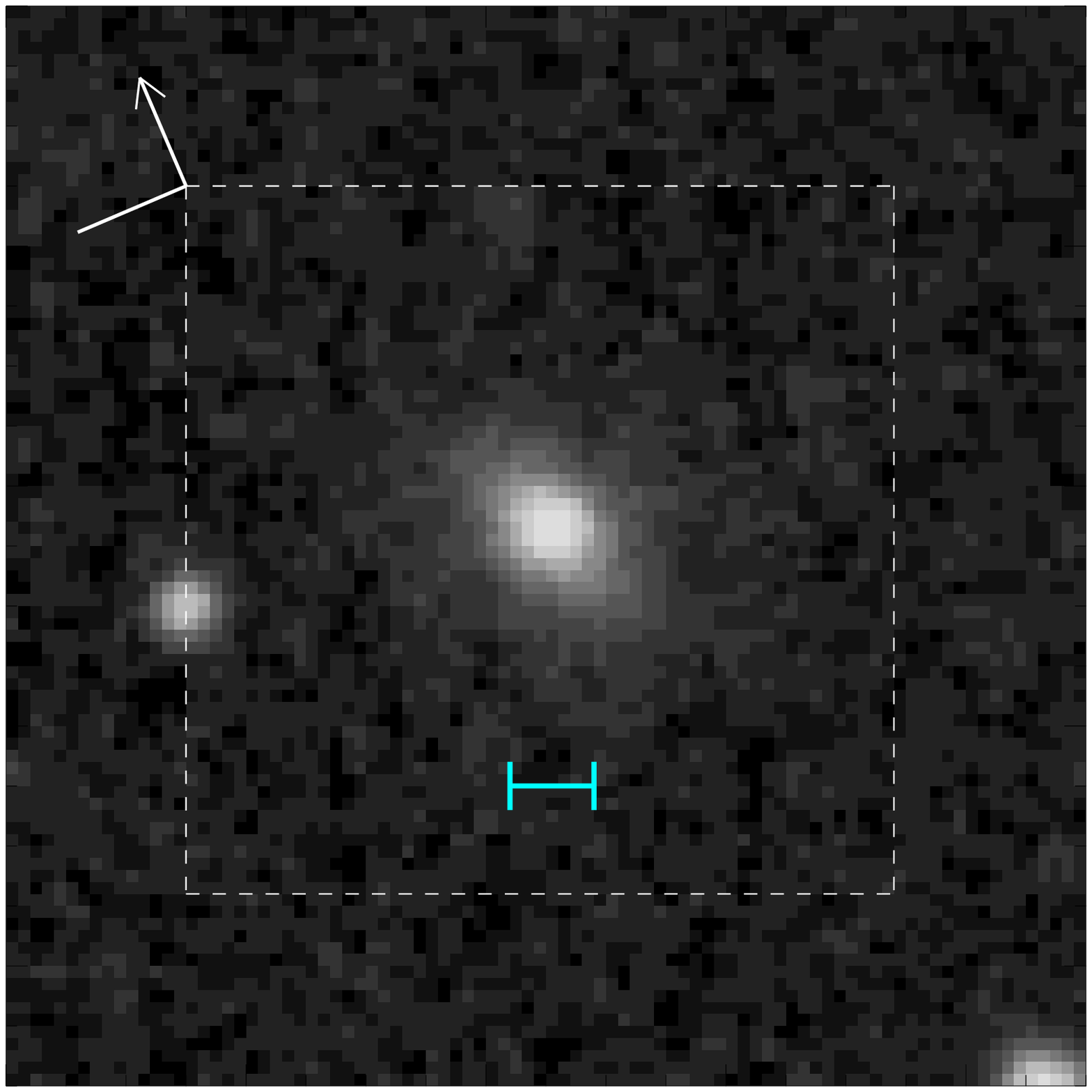}\\
\includegraphics[width=12cm,height=4cm]{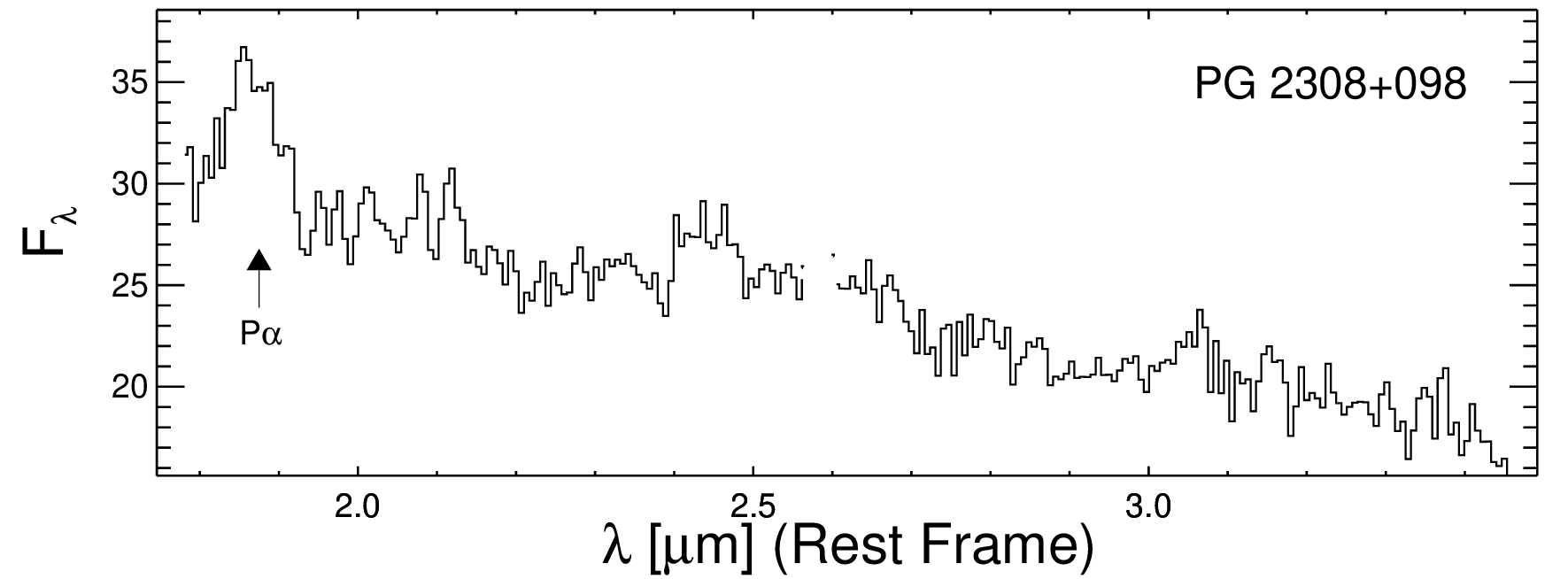}
\includegraphics[scale=0.20]{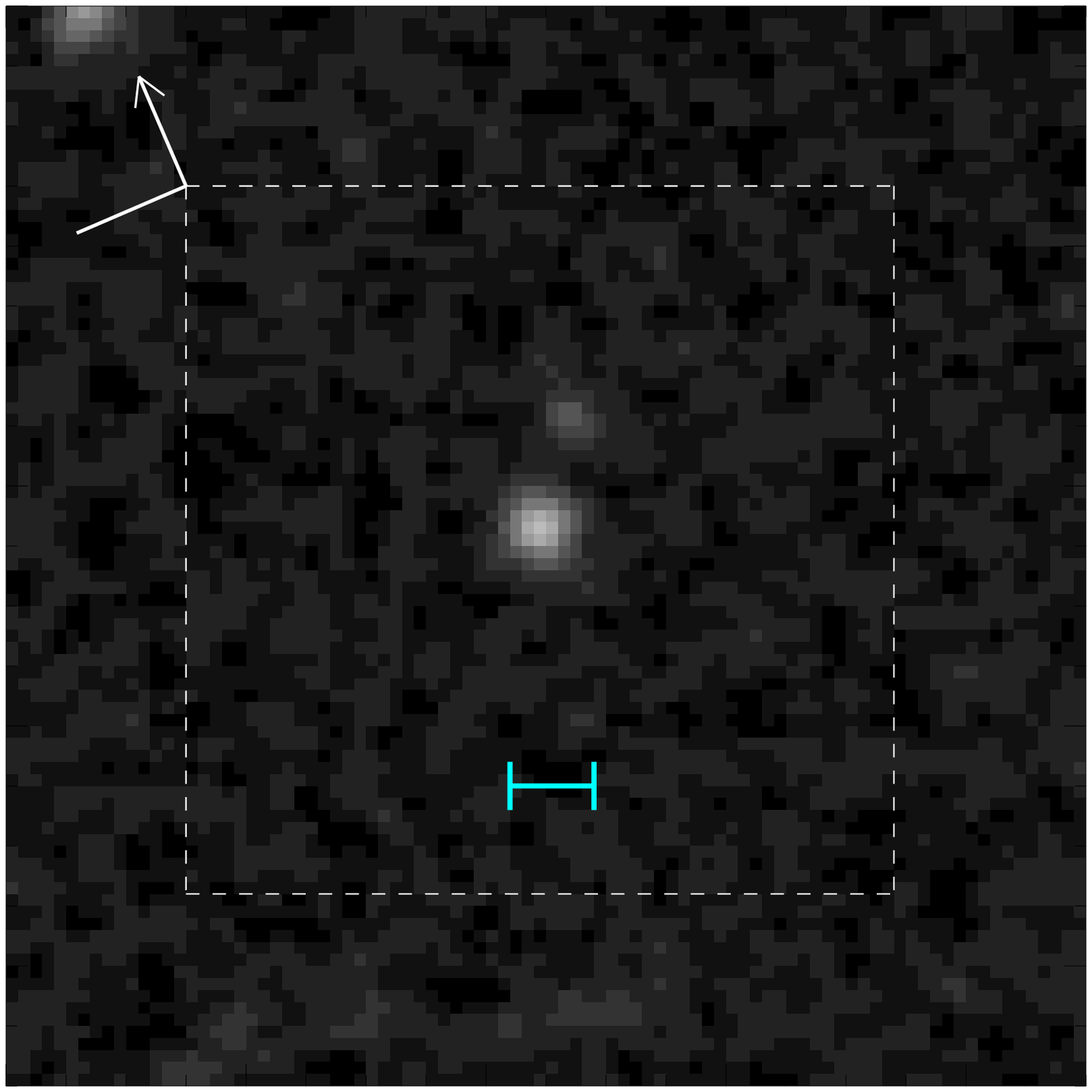}\\
\caption{Continued}
\end{figure}
\clearpage

\begin{figure}
\figurenum{5}
\includegraphics[width=12cm,height=4cm]{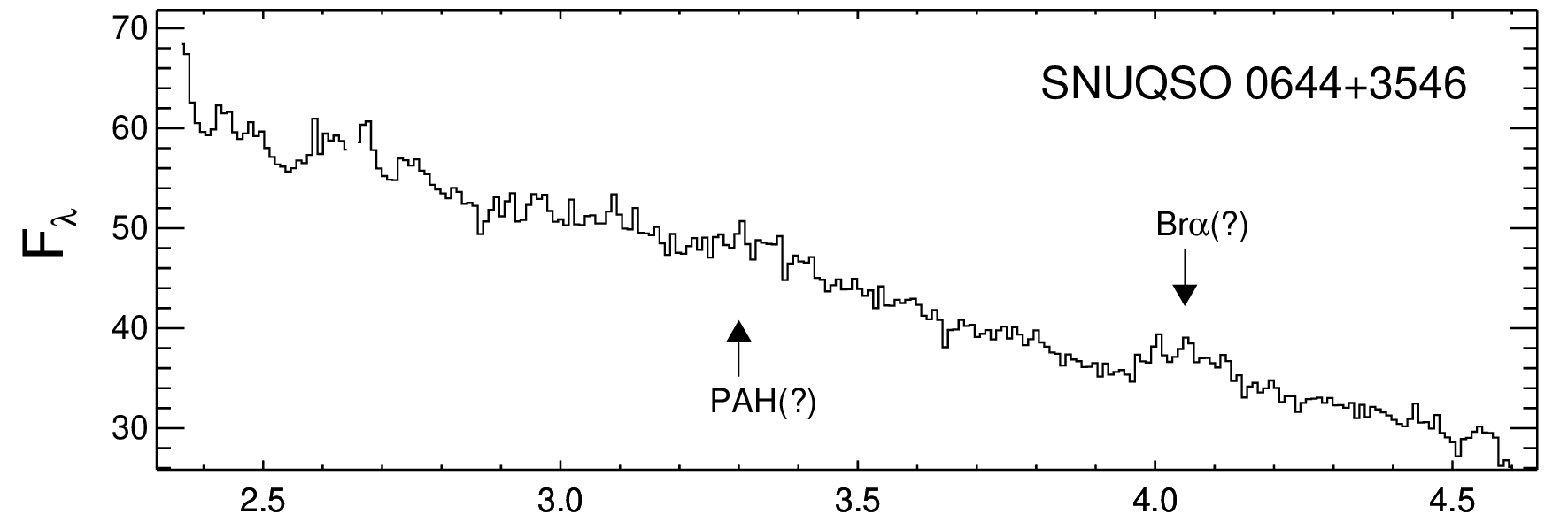}
\includegraphics[scale=0.20]{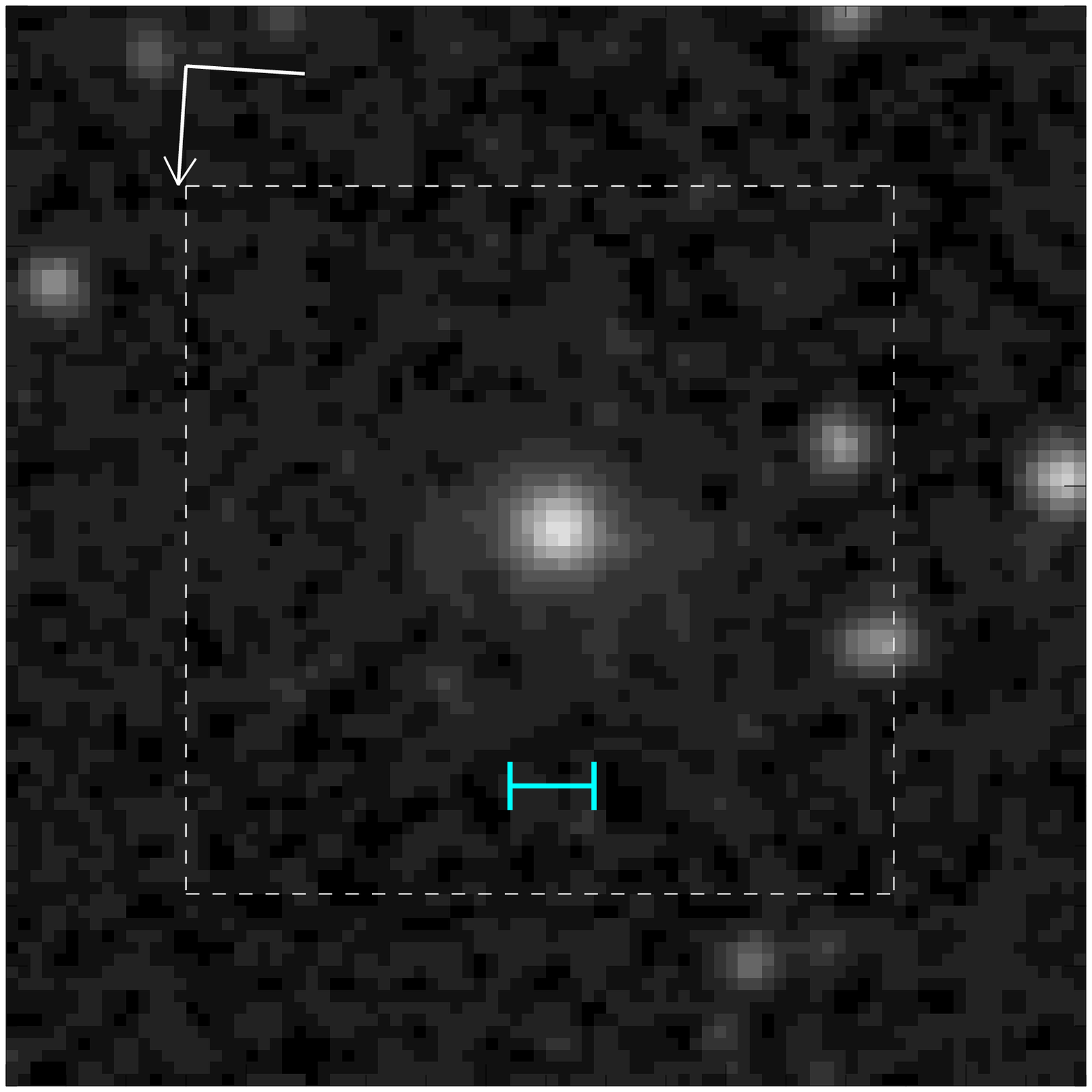}\\
\includegraphics[width=12cm,height=4cm]{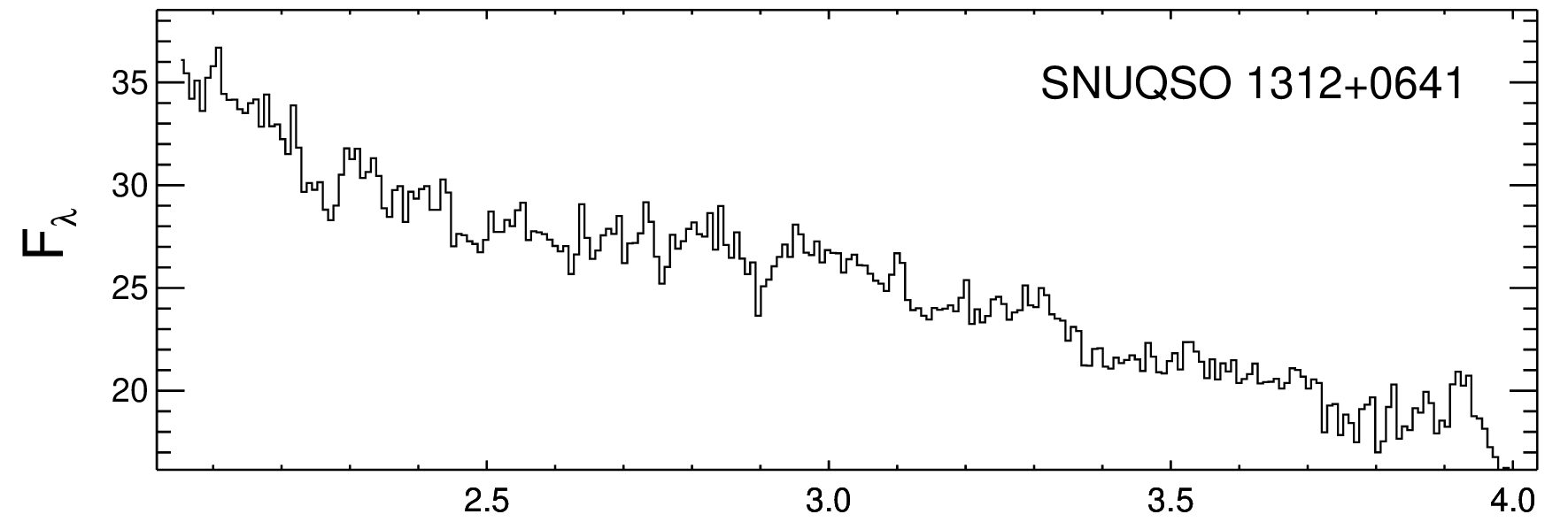}
\includegraphics[scale=0.20]{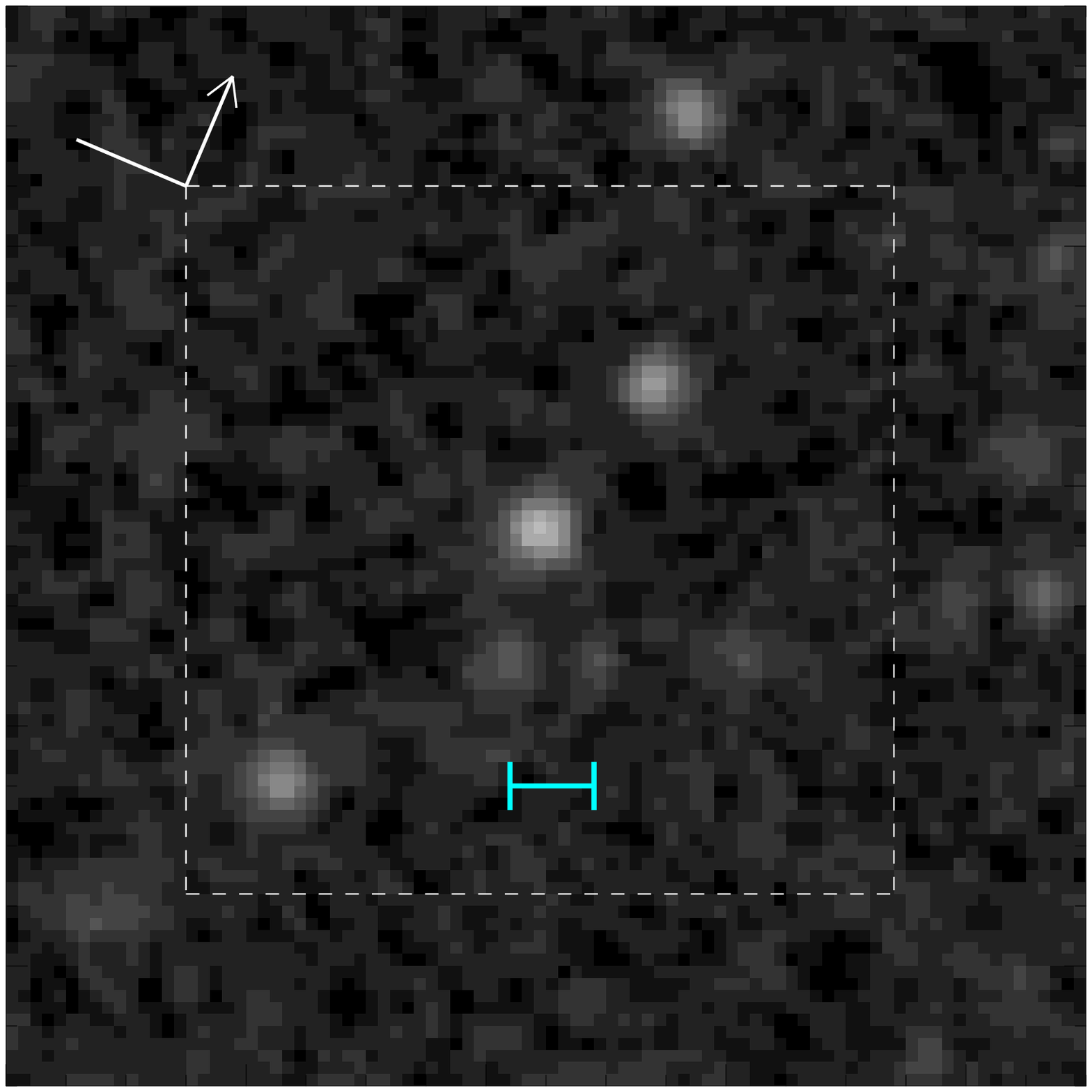}\\
\includegraphics[width=12cm,height=4cm]{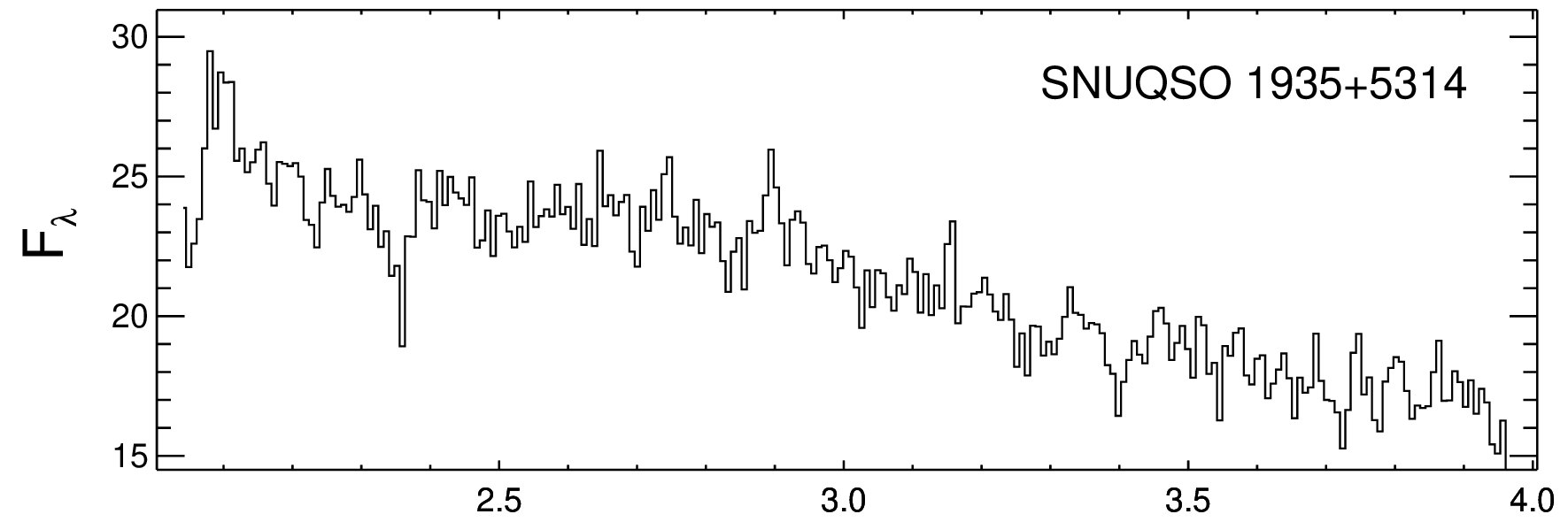}
\includegraphics[scale=0.20]{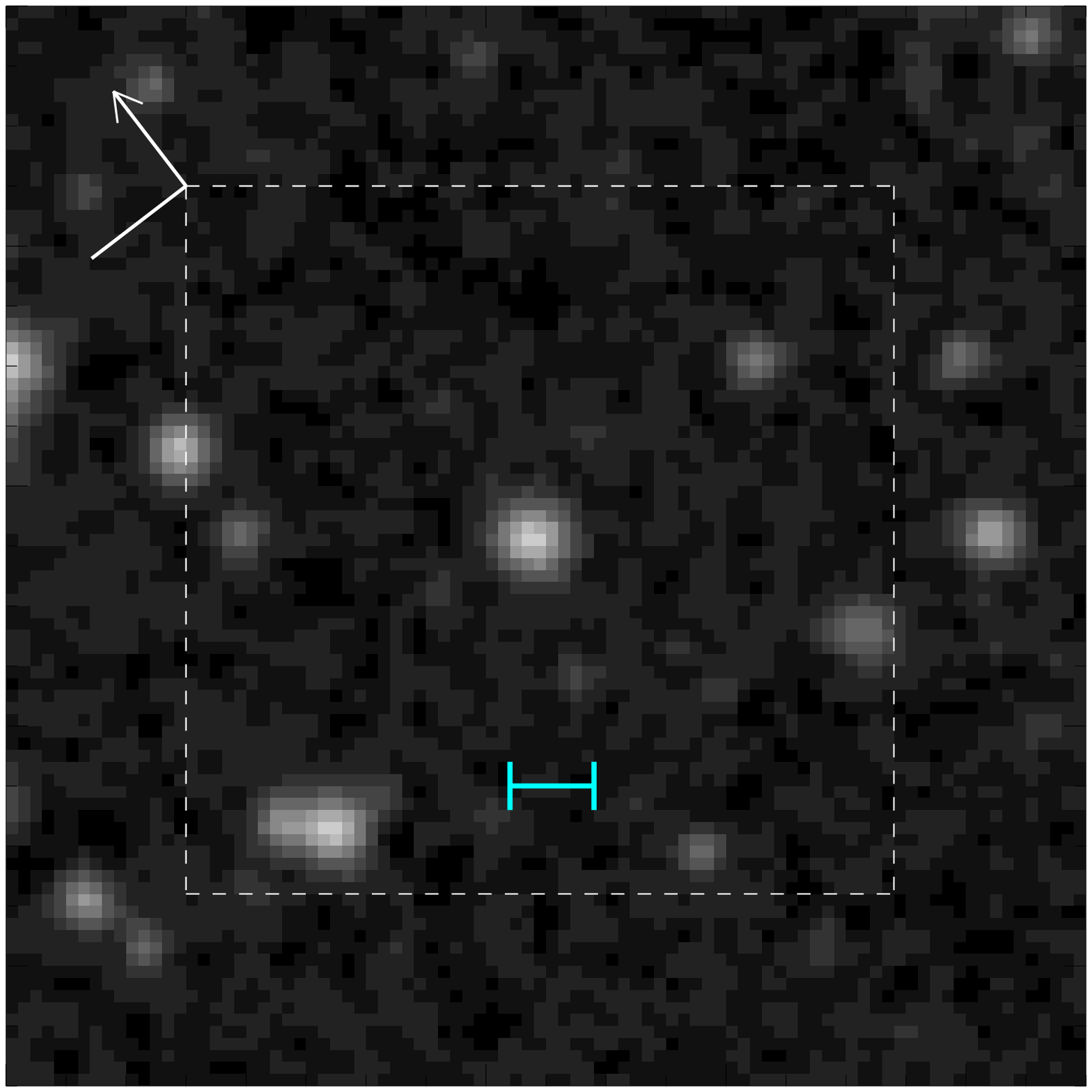}\\
\includegraphics[width=12cm,height=4cm]{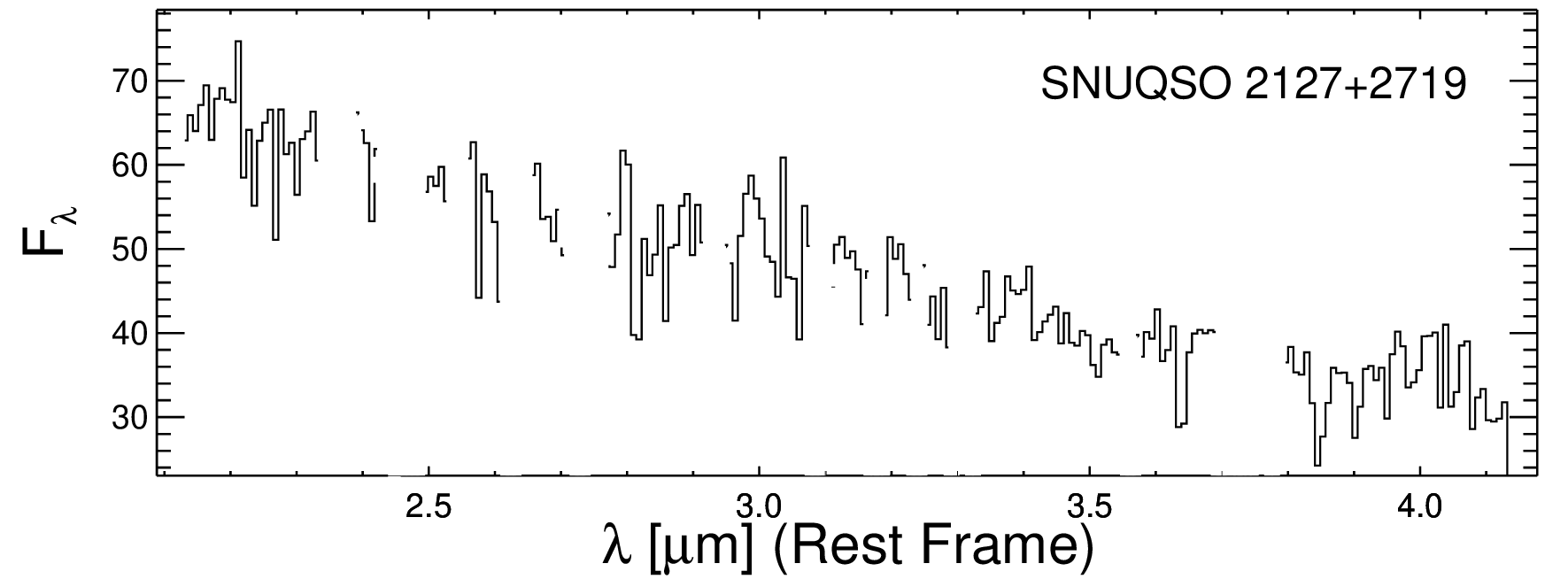}
\includegraphics[scale=0.20]{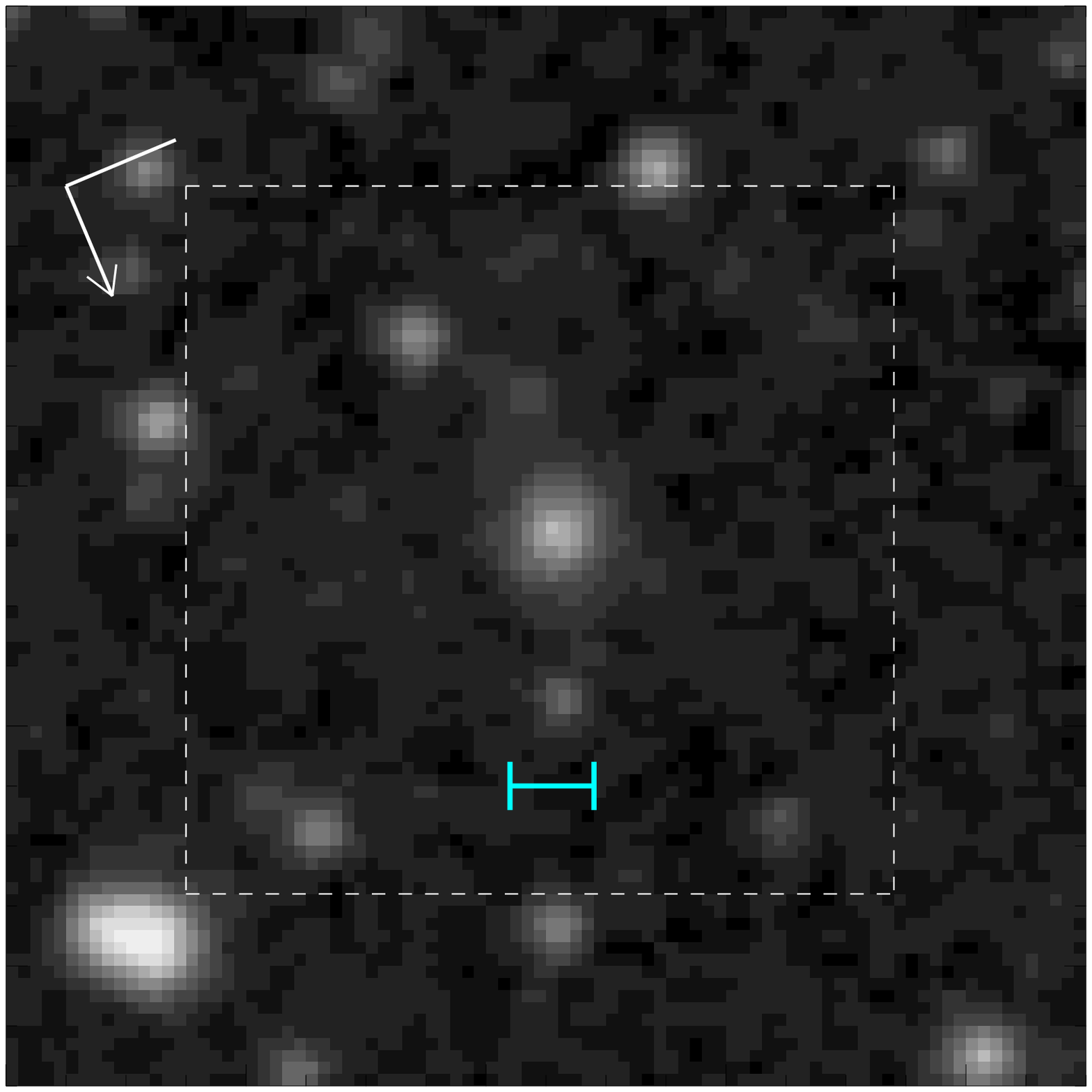}\\
\caption{Continued}
\end{figure}
\clearpage

\clearpage
%%%%%%%%%%%%%%%%%%%%%%%%%%%Table%%%%%%%%%%%%%%%%%%%%%%%%%%%%%%%%%%%%
\LongTables % optionally
%\begin{turnpage}
%\begin{landscape}
 \begin{deluxetable}{cccccccc}
 
% \rotate
 \tabletypesize{\scriptsize}
 \tablewidth{0pt}
 \tablenum{1}
 \tablecaption{Object List\label{tbl1}}
 \tablehead{
 \colhead{}& \colhead{$\alpha$}& \colhead{$\delta$}&
 \colhead{Redshift}& \colhead{$K$}& \colhead{$\log(M_{\rm BH})$}&
 \colhead{Exposure Time}& \colhead{Observed Dates} \\
 \colhead{Objects}& \colhead{(J2000.0)}& \colhead{(J2000.0)}&
 \colhead{$z$}& \colhead{(mag)}& \colhead{$(M_{\rm \odot})$}&
 \colhead{(s)}& \colhead{}}
 \startdata
 Mrk\,335&		 00 06 19.5&   $+$20 12 10& 0.025&		   10.59&   7.11\tablenotemark{a}&
 1188&        2008$-$06$-$30\\
 PG\,0026$+$129&   00 29 13.6&   $+$13 16 03& 0.142&        12.11&   8.56\tablenotemark{a}&
 1100&        2008$-$07$-$02, 03\\
 PG\,0052$+$251&   00 54 52.1&   $+$25 25 38& 0.155&        12.23&   8.53\tablenotemark{a}&
 1144&        2008$-$07$-$13, 14\\
 Fairall\,9&       01 23 45.7&   $-$58 48 20& 0.047&        11.11&   8.37\tablenotemark{a}&
 1188&        2008$-$12$-$01, 02\\
 Mrk\,590& 		02 14 33.5&   $-$00 46 00& 0.026&        10.57&   7.64\tablenotemark{a}&
 660&          2009$-$07$-$24\\
 
 3C\,120&          04 33 11.0&   $+$05 21 15& 0.033&        10.57&   7.71\tablenotemark{a}&
 1188&        2009$-$02$-$25\\
 Ark\,120&         05 16 11.4&   $-$00 08 59& 0.032&        10.22&   8.14\tablenotemark{a}&
 1056&        2008$-$09$-$09\\
 Mrk\,79&          07 42 32.7&   $+$49 48 34& 0.022&        10.68&   7.68\tablenotemark{a}&
 1188&        2008$-$10$-$10\\
 PG\,0804$+$761&   08 10 58.6&   $+$76 02 42& 0.100&        10.91&   8.80\tablenotemark{a}&
 1188&        2008$-$10$-$04\\
 PG\,0844$+$349&   08 47 42.4&   $+$34 45 04& 0.064&        12.01&   7.93\tablenotemark{a}&
 1188&        2009$-$04$-$24\\
 
 Mrk\,110&         09 25 12.8&   $+$52 17 10& 0.035&        12.25&   7.36\tablenotemark{a}&
 1056&        2008$-$10$-$29, 2009$-$04$-$25\\
 NGC\,3516&        11 06 47.4&   $+$72 34 06& 0.008&        9.61&   7.46\tablenotemark{b}&
 1188&        2008$-$10$-$25, 26\\
 NGC\,3783&        11 39 01.7&    $-$37 44 18&  0.009&        10.01&   7.44\tablenotemark{a}&
 1188&        2008$-$07$-$03\\
 NGC\,4051&        12 03 09.6&   $+$44 31 52& 0.002&        10.01&   6.20\tablenotemark{b}&
 748&             2009$-$05$-$29, 30\\
 NGC\,4151&        12 10 32.5&   $+$39 24 20& 0.003&        8.51&   7.09\tablenotemark{a}&
 1056&        2008$-$06$-$03, 04\\
 
 PG\,1211$+$143&   12 14 17.7&   $+$14 03 12& 0.080&        11.29&   8.13\tablenotemark{a}&
 1144&        2009$-$12$-$18, 19\\
 3C\,273&          12 29 06.7&   $+$02 03 08& 0.158&        9.97&   8.91\tablenotemark{a}&
 1848&		2008$-$06$-$26, 27, 12$-$27\\
 PG\,1229$+$204&   12 32 03.6&   $+$20 09 29& 0.063&        12.14&   7.83\tablenotemark{a}&
 1188&        2009$-$06$-$20\\
 NGC\,4593&        12 39 39.4&   $-$05 20 39& 0.009&        10.03&   6.69\tablenotemark{a}&
 1144&        2008$-$07$-$02, 2009$-$01$-$01\\
 PG\,1307$+$085&   13 09 47.0&   $+$08 19 48& 0.155&        12.42&   8.61\tablenotemark{a}&
 1144&        2008$-$07$-$03, 04\\
 
 Mrk\,279&         13 53 03.4&   $+$69 18 29& 0.030&        10.39&   7.51\tablenotemark{a}&
 1144&        2008$-$11$-$14, 15, 16\\
 PG\,1411$+$442&   14 13 48.3&   $+$44 00 14& 0.089&        11.50&   8.61\tablenotemark{a}&
 1144&        2008$-$06$-$28, 29, 30\\
 NGC\,5548&        14 17 59.5&   $+$25 08 12& 0.017&        10.13&   7.61\tablenotemark{b}&
 1144&        2008$-$07$-$13, 14\\
 Mrk\,817&         14 36 22.0&   $+$58 47 39& 0.031&        10.90&   7.60\tablenotemark{a}&
 660&             2008$-$06$-$12\\
 PG\,1613$+$658&   16 13 57.1&   $+$65 43 09& 0.129&        11.41&   8.41\tablenotemark{a}&
 1144&        2008$-$06$-$07, 08\\
 
 PG\,1617$+$175&   16 20 11.2&   $+$17 24 27& 0.112&        12.41&   8.74\tablenotemark{a}&
 1188&        2008$-$08$-$21\\
 PG\,1700$+$518&   17 01 24.8&   $+$51 49 20& 0.292&        11.98&   8.85\tablenotemark{a}&
 924&        2008$-$08$-$15, 17\\
 3C\,390.3&        18 42 08.9&  $+$79 46 17& 0.056&        11.71&   8.42\tablenotemark{a}&
 1188&        2008$-$09$-$10\\
 Mrk\,509&         20 44 09.7&   $-$10 43 24& 0.034&        10.19&   8.12\tablenotemark{a}&
 1012&        2009$-$04$-$30\\
 PG\,2130$+$099&   21 32 27.8&   $+$10 08 19& 0.062&        10.66&   7.54\tablenotemark{c}&
 1100&        2009$-$05$-$20\\
 
 NGC\,7469&        23 03 15.6&   $+$08 52 26& 0.016&        9.762&   7.05\tablenotemark{a}&
 1100&        2008$-$06$-$10, 11\\
 
 PG\,0003$+$158&   00 05 59.2&     $+$16 09 49&    0.450&       13.53&     9.24\tablenotemark{d}&
 1144&        2008$-$06$-$30, 12$-$29, 30\\
 PG\,0007$+$106&   00 10 31.0&     $+$10 58 30&    0.089&       11.78&     8.69\tablenotemark{d}&
 1188&        2009$-$12$-$28\\
 PG\,0043$+$039&   00 45 47.3&     $+$04 10 24&    0.385&       13.59&     9.09\tablenotemark{d}&
 1012&        2010$-$01$-$02\\
 PG\,0049$+$171&   00 51 54.8&     $+$17 25 58&    0.064&       13.34&     8.31\tablenotemark{d}&
 1188&        2010$-$01$-$09\\
 PG\,0050$+$124&   00 53 34.9&     $+$12 41 36&    0.061&       10.35&     7.41\tablenotemark{d}&
 1144&        2008$-$07$-$08, 09\\
 
 PG\,0838$+$770&   08 44 45.3&     $+$76 53 10&    0.131&       13.15&     8.12\tablenotemark{d}&
 1188&        2008$-$10$-$06\\
 PG\,0923$+$129&   09 26 03.3&     $+$12 44 04&    0.029&       11.30&     8.56\tablenotemark{d}&
 396&                  2009$-$11$-$12\\
 PG\,0934$+$013&   09 37 01.0&     $+$01 05 43&    0.050&       13.50&     7.01\tablenotemark{d}&
 792&             2009$-$05$-$17\\
 PG\,0947$+$396&   09 50 48.4&     $+$39 26 51&    0.205&       12.76&     8.64\tablenotemark{d}&
 1188&        2009$-$11$-$07\\
 PG\,1011$-$040&   10 14 20.7&     $-$04 18 40&    0.058&       12.65&     7.28\tablenotemark{d}&
 1144&        2009$-$11$-$28\\
 
 PG\,1012$+$008&   10 14 54.9&     $+$00 33 37&    0.186&       13.08&     8.21\tablenotemark{d}&
 1100&        2009$-$11$-$27\\
 PG\,1022$+$519&   10 25 31.3&     $+$51 40 35&    0.044&       12.71&     7.11\tablenotemark{d}&
 396&                  2009$-$11$-$07\\
 PG\,1048$-$090&   10 51 29.9&     $-$09 18 10&    0.344&       13.95&     9.17\tablenotemark{d}&
 1188&        2009$-$06$-$08\\
 PG\,1048$+$342&   10 51 43.9&     $+$33 59 27&    0.167&       13.60&     8.33\tablenotemark{d}&
 704&             2009$-$11$-$22\\
 PG\,1049$-$005&   10 51 51.4&     $-$00 51 18&    0.359&       12.84&     9.15\tablenotemark{d}&
 1144&        2009$-$06$-$04\\
 
 PG\,1100$+$772&   11 04 13.7&     $+$76 58 58&    0.311&       13.05&     9.24\tablenotemark{d}&
 1188&        2008$-$10$-$18\\
 PG\,1103$-$006&   11 06 31.8&     $-$00 52 52&    0.423&       13.85&     9.29\tablenotemark{d}&
 792&             2009$-$06$-$08\\
 PG\,1114$+$445&   11 17 06.4&     $+$44 13 33&    0.143&       12.33&     8.56\tablenotemark{d}&
 1144&        2009$-$11$-$21, 22\\
 PG\,1115$+$407&   11 18 30.3&     $+$40 25 54&    0.154&       12.76&     7.63\tablenotemark{d}&
 1144&        2009$-$11$-$24\\
 PG\,1116$+$215&   11 19 08.7&     $+$21 19 18&    0.176&       11.54&     8.48\tablenotemark{d}&
 1056&        2009$-$12$-$03\\
 
 PG\,1121$+$422&   11 24 39.2&     $+$42 01 45&    0.225&       13.26&     8.00\tablenotemark{d}&
 396&                  2009$-$11$-$26\\
 PG\,1202$+$281&   12 04 42.1&     $+$27 54 12&    0.165&       12.87&     8.58\tablenotemark{d}&
 396&                  2009$-$12$-$11\\
 PG\,1216$+$069&   12 19 20.9&     $+$06 38 39&    0.331&       13.30&     9.16\tablenotemark{d}&
 616&             2009$-$12$-$24\\
 PG\,1244$+$026&   12 46 35.2&     $+$02 22 09&    0.048&       13.19&     6.49\tablenotemark{d}&
 1100&        2008$-$06$-$30, 12$-$31\\
 PG\,1259$+$593&   13 01 12.9&     $+$59 02 07&    0.477&       13.07&     8.88\tablenotemark{d}&
 1144&        2009$-$11$-$27, 28\\
 
 PG\,1302$-$102&   13 05 33.0&     $-$10 33 19&    0.278&       12.85&     8.84\tablenotemark{d}&
 1100&        2009$-$07$-$10, 11\\
 PG\,1322$+$659&   13 23 49.5&     $+$65 41 48&    0.168&       12.85&     8.25\tablenotemark{d}&
 1056&        2009$-$11$-$21, 22\\
 PG\,1351$+$236&   13 54 06.4&     $+$23 25 49&    0.055&       13.01&     8.53\tablenotemark{d}&
 792&             2008$-$07$-$09\\
 PG\,1402$+$261&   14 05 16.2&     $+$25 55 34&    0.164&       12.16&     7.91\tablenotemark{d}&
 1144&        2008$-$07$-$09\\
 PG\,1415$+$451&   14 17 00.7&     $+$44 56 06&    0.113&       12.23&     7.98\tablenotemark{d}&
 1144&        2008$-$12$-$29, 2009$-$06$-$28\\
 
 PG\,1416$-$129&   14 19 03.8&     $-$13 10 44&    0.129&       13.34&     9.01\tablenotemark{d}&
 1100&        2010$-$01$-$27\\
 PG\,1425$+$267&   14 27 35.6&     $+$26 32 15&    0.364&       13.57&     9.70\tablenotemark{d}&
 1188&        2008$-$07$-$16\\
 PG\,1427$+$480&   14 29 43.1&     $+$47 47 26&    0.220&       13.48&     8.05\tablenotemark{d}&
 1188&        2009$-$06$-$29, 07$-$01\\
 PG\,1448$+$273&   14 51 08.7&     $+$27 09 27&    0.065&       12.05&     6.94\tablenotemark{d}&
 1188&        2009$-$07$-$22, 23\\
 PG\,1501$+$106&   15 04 01.2&     $+$10 26 16&    0.036&       11.47&     8.49\tablenotemark{d}&
 1188&        2009$-$08$-$02\\
 
 PG\,1512$+$370&   15 14 43.0&     $+$36 50 50&    0.370&       13.69&     9.34\tablenotemark{d}&
 1100&        2008$-$07$-$22\\
 PG\,1519$+$226&   15 21 14.2&     $+$22 27 44&    0.137&       12.47&     7.91\tablenotemark{d}&
 1188&        2009$-$08$-$02, 03\\
 PG\,1534$+$580&   15 35 52.3&     $+$57 54 09&    0.029&       12.18&     8.17\tablenotemark{d}&
 1100&        2008$-$06$-$29, 12$-$31, 2009$-$01$-$01\\
 PG\,1535$+$547&   15 36 38.3&     $+$54 33 33&    0.038&       11.63&     7.16\tablenotemark{d}&
 704&             2008$-$07$-$06, 08\\
 PG\,1543$+$489&   15 45 30.2&     $+$48 46 09&    0.399&       13.14&     7.96\tablenotemark{d}&
 1188&        2009$-$01$-$17, 20, 21\\
 
 PG\,1545$+$210&   15 47 43.5&     $+$20 52 17&    0.264&       13.18&     9.28\tablenotemark{d}&
 616&             2008$-$08$-$11\\
 PG\,1626$+$554&   16 27 56.1&     $+$55 22 32&    0.133&       12.67&     8.46\tablenotemark{d}&
 1188&        2010$-$01$-$22\\
 PG\,1704$+$608&   17 04 41.4&     $+$60 44 31&    0.371&       12.43&     9.36\tablenotemark{d}&
 1100&        2008$-$07$-$23\\
 PG\,2112$+$059&   21 14 52.6&     $+$06 07 42&    0.466&       12.85&     8.97\tablenotemark{d}&
 1056&        2009$-$05$-$13, 14\\
 PG\,2209$+$184&   22 11 53.9&     $+$18 41 50&    0.070&       12.55&     8.73\tablenotemark{d}&
 1144&        2009$-$12$-$03, 05\\
 
 PG\,2233$+$134&   22 36 07.7&     $+$13 43 55&    0.325&       13.81&     8.00\tablenotemark{d}&
 1188&        2009$-$12$-$07, 08\\
 PG\,2304$+$042&   23 07 02.9&     $+$04 32 57&    0.042&       12.54&     8.53\tablenotemark{d}&
 748&             2009$-$12$-$12\\
 PG\,2308$+$098&   23 11 17.7&     $+$10 08 15&    0.433&       13.23&     9.56\tablenotemark{d}&
 1012&        2008$-$12$-$14\\
 SNUQSO\,0644$+$3546&	06 44 10.8&	$+$35 46 44&	0.077&	11.82&	  8.86\tablenotemark{e}&
 1188&        2008$-$10$-$01\\
 SNUQSO\,1312$+$0641&	13 12 04.7& $+$06 41 07& 	0.239&	13.20&	  9.09\tablenotemark{e}&
 1100&        2008$-$07$-$06, 2009$-$01$-$03\\
 SNUQSO\,1935$+$5314&	19 35 21.1& $+$53 14 11&	0.248&	13.14&   8.61\tablenotemark{e}&
 1188&	     2008$-$11$-$17, 2009$-$05$-$11\\
 SNUQSO\,2127$+$2719&	21 27 56.4& $+$27 19 05&	0.197&   12.63&   8.74\tablenotemark{e}&
 396&			     2009$-$05$-$26\\
 
 \enddata
 \tablenotetext{a}{The $M_{\rm BH}$ values are estimated from reverberation mapping method \citep{peterson04}.}
 \tablenotetext{b}{The $M_{\rm BH}$ values are estimated from reverberation mapping method \citep{denney10}.}
 \tablenotetext{c}{The $M_{\rm BH}$ values are estimated from reverberation mapping method \citep{grier08}.}
 \tablenotetext{d}{The $M_{\rm BH}$ values are based on $\mathrm{FWHM_{H\beta}}$ and $L_{\rm 5100}$ \citep{vestergaard06}.}
 \tablenotetext{e}{The $M_{\rm BH}$ values are based on $\mathrm{FWHM_{H\alpha}}$ and $L_{\mathrm{H\alpha}}$ \citep{lee08,im07}.}
 \end{deluxetable}
\clearpage
%\end{landscape}
%\end{turnpage}

\begin{figure}
\begin{center}
\centering
%\figurenum{99}
%\scalebox{1.1}{\plotone{table2.eps}}
\includegraphics[width=1.3\textwidth, angle =90 ]{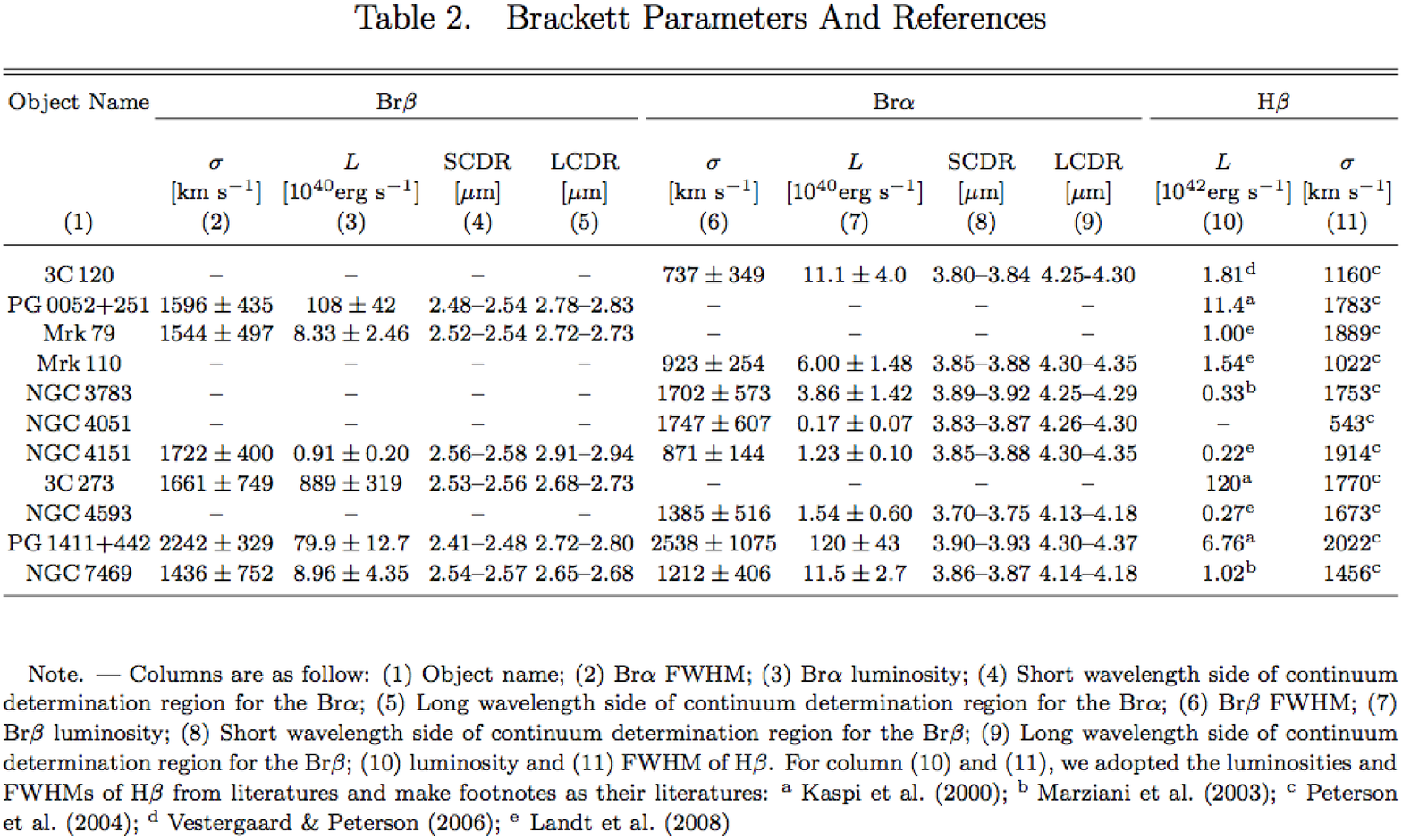}
%\caption{}
\end{center}
\end{figure}
\clearpage

\end{document}